\documentclass[lettersize,journal]{IEEEtran}
\usepackage{aliascnt}
\usepackage{cite}
\usepackage{wrapfig}
\usepackage{gensymb}
\usepackage{soul}
\usepackage{setspace} 
\newcounter{counterstepone}
\newcounter{counteryy}
\newcounter{counterycycy}
\newcounter{counteryty}
\newcounter{counterycy}
\newcounter{counterytty}
\newcounter{counteryttt}
\newcounter{countertyctt}
\newcounter{counterty}
\newcounter{countertt}
\IEEEoverridecommandlockouts
\newenvironment{talign}
 {\let\displaystyle\textstyle\align}
 {\endalign}
\usepackage{multirow}
\usepackage{enumitem}
\usepackage{pifont}
\usepackage{amsmath,amssymb,amsfonts,amsthm}
\usepackage{algorithmic}
\usepackage[ruled,linesnumbered]{algorithm2e}
\usepackage{graphicx}
\usepackage{textcomp}
\usepackage{caption}
\usepackage {epstopdf}
\usepackage[hidelinks]{hyperref}
\usepackage{float}
\usepackage[utf8]{inputenc}
\usepackage{tabularx}
\usepackage{booktabs}
\usepackage{cases}
\usepackage{xcolor}
\usepackage{authblk}
\usepackage{cleveref}
\usepackage[labelformat=brace]{subcaption}

\usepackage{bm}
\def\BibTeX{{\rm B\kern-.05em{\sc i\kern-.025em b}\kern-.08em
    T\kern-.1667em\lower.7ex\hbox{E}\kern-.125emX}}

\title{Human-Centric Resource Allocation in the\\[-5pt]
Metaverse over Wireless Communications\\[-5pt]} 
\usepackage{graphicx}
\usepackage{mathrsfs}

\newtheoremstyle{slanted}
{0em plus 0em minus 0em}%   Space above
  {0em plus 0em minus 0em}%   Space below
  {\em}%  Body font
  {}%          Indent amount (empty = no indent, \parindent = para indent)
  {\bfseries}% Thm head font
  {.}%         Punctuation after thm head
  { }%     Space after thm head: " " = normal interword space;
  {}%          Thm head spec (can be left empty, meaning `normal')

\theoremstyle{slanted}
\newtheorem{assumption}{Assumption}

\theoremstyle{slanted}

\theoremstyle{slanted}

\theoremstyle{slanted}

\theoremstyle{slanted}

\theoremstyle{slanted}
\newtheorem{proposition}{Proposition}

\theoremstyle{slanted}
\newtheorem{lemma}{Lemma}

\usepackage{enumitem}
\renewenvironment{thebibliography}[1]{
  \begin{oldthebibliography}{#1}
    \setlength{\itemsep}{0em}
    \setlength{\parskip}{0em}
}
{
  \end{oldthebibliography}
}
\begin{document}

% \setcitestyle{number}
\author{Jun~Zhao,
        Liangxin~Qian,
        Wenhan~Yu\vspace{-30pt}

        \thanks{Jun Zhao is a faculty member in the School of Computer Science and Engineering, Nanyang Technological University (NTU), Singapore. Liangxin~Qian and
        Wenhan~Yu are both PhD students supervised by Jun Zhao. \newline Emails: JunZhao@ntu.edu.sg, \{qian0080, wenhan002\}@e.ntu.edu.sg \\ Corresponding author: Jun Zhao
}   
        }
% \author{Jun~Zhao,~\IEEEmembership{Member, IEEE},
%         Liangxin~Qian,
%         Wenhan~Yu
%         }
% \IEEEpubid{\begin{minipage}{\textwidth} 
%  aaa \\ b
% \end{minipage}}
 \maketitle

% \cfoot{\thepage}
% \renewcommand{\headrulewidth}{0.4pt}
% \renewcommand{\footrulewidth}{0pt}

%  \pagestyle{plain} \thispagestyle{plain}
\begin{abstract}

The Metaverse will provide numerous immersive applications for human users, by consolidating technologies like extended reality (XR), video streaming, and cellular networks. Optimizing wireless communications to enable the human-centric  Metaverse is important to satisfy the demands of mobile users. In this paper, we formulate the optimization of the system utility-cost ratio (UCR) for the Metaverse over wireless networks. Our human-centric utility measure for virtual reality (VR) applications of the Metaverse represents users' perceptual assessment of the VR video quality as a function of the data rate and the video resolution and is learned from real datasets. The variables jointly optimized in our problem include the allocation of both communication and computation resources as well as VR video resolutions. The system cost in our problem comprises the energy consumption and delay and is \mbox{non-convex} with respect to the optimization variables. To solve the  \mbox{non-convex} optimization, we develop a novel fractional programming technique, which contributes to optimization theory and has broad applicability beyond our paper. Our proposed algorithm for the system UCR optimization is computationally efficient and finds a stationary point to the constrained optimization. Through extensive simulations, our algorithm is demonstrated to outperform other approaches.

% effectively achieve improvements in the utility-cost ratio or QoE. 

% regression

% a computational complexity of polylogarithmic (i.e., polynomials of logarithms).

% 1. Human-Centric

% 2. joint communication and computation optimization

% 3. propose a new method to for minimization problems involving the sum-of-ratios
% %computation offloading  problem. Previous studies ? 

% Our paper contributes to \textit{not just} the problem formulation and solving for human-centric Metaverse, \textit{but also} a novel optimization technique for sum-of-ratios.
\end{abstract}

%computation offloading for mobile-edge server computing in emerging  wireless networks with Reconfigurable intelligent surfaces
\begin{IEEEkeywords}

Metaverse, human-centric, 
%Quality of Experience, 
 resource allocation, virtual reality, wireless communications.\vspace{-15pt}
\end{IEEEkeywords}

\section{Introduction}
% \subsection{Background}
 The Metaverse is expected to offer a myriad of opportunities for mobile users to interact with the immersive virtual world~\cite{wang2022survey}. In various Augmented/Virtual Reality (AR/VR) applications for the Metaverse, humans are at the core since users judge whether the AR/VR videos or games provide a satisfying Quality of Experience (QoE)~\cite{mourtzis2022human}. Compared with the traditional Quality of Service (QoS) that measures the objective service performance (e.g., bit rate, data accuracy), QoE as a utility measure concerns the enjoyment of users~\cite{xu2022full}. Providing satisfying utilities to multiple users in a resource-constrained system requires allocating resources wisely. In this paper, we formulate and solve human-centric resource allocation for VR in the
Metaverse over wireless communications. Our goal is
 to reduce the Metaverse system's cost in terms of delay and energy, as well as to enhance the human-centric utilities of 
 mobile users accessing the Metaverse via wireless networks. Tackling this problem also motivates us to propose a new optimization technique.

 % The Metaverse, as a mirror of the physical world, should be human-centric~\cite{wang2022survey,xu2022full}. In particular, 

 % Among those, various applications are based on Virtual Reality (VR); e.g., VR video streaming, gaming, and live broadcasting.

%  Advanced technologies have offered a myriad of opportunities for users for interacting with and enjoying
% the immersive virtual world, known as the Metaverse. Among those, various applications are based on
% Virtual Reality (VR); e.g., VR video streaming, gaming, and live broadcasting. The Metaverse, as a mirror
% of the physical world and a futuristic concept allowing everyone to access it anywhere and anytime, is
% human-centric by design [1], [2]. 

\textbf{Studied problem.} \label{Studiedproblem} Our researched system consists of one Metaverse Server (MS) and multiple VR Users (VUs). We consider downlink wireless communications, where the MS sends to each VU the corresponding VR video via frequency division multiple access (FDMA). The MS solves the system utility-cost ratio (UCR) optimization by allocating 1) communication resources (i.e., bandwidth and transmission power) for the MS's communication with each VU, and 2) the MS's computation resources for processing the videos to be sent to VUs, as well as deciding 3) the video resolution for all VUs, and 4) the CPU frequencies for the VUs. Then, the MS uses the allocated computation resource to process each VR video with the selected resolution, and transmits the videos to the VUs with the decided bandwidth and transmission power. Each VU receives VR frames of a video and  processes the frames with the arranged CPU frequency. To perform the UCR optimization, the MS knows the human-centric utilities of all VUs and how the energy or delay depends on the optimization variables.  The system cost is a weighted sum of the energy consumption and delay. For energy, we take into account both the MS and VUs. The energy usage on the MS comprises those  for video processing and transmission, whereas the energy of VUs is  for video processing. The delay computation  includes the processing on the MS, the wireless transmission, and the processing on each VU. Next, we discuss VUs' human-centric utilities.

\textbf{Human-centric utility.} Prior resource allocation studies~\cite{luo2020hfel,zhan2021l4l,yu2020joint,zhou2022icdcs,lu2022cost,qian2019optimal} for wireless communications typically do not consider human-centric utilities. Incorporating subjective user perception into the design is critical for the development of VR and the Metaverse, as it provides valuable insights into how the technologies can be improved to deliver the best possible experiences for VUs. A recent work~\cite{mourtzis2022human} also argues the importance of developing the Metaverse to be human-centric. In our paper, the human-centric utility for each VU is learned from the VU's perceptual assessment of the VR video quality as a function of the data rate and the video resolution, as illustrated by a recent dataset reporting users' evaluation of watching $360^\circ$ VR videos~\cite{ssv360}. Then UCR is the ratio of all VUs' sum human-centric utilities to the system cost.

Our \textbf{contributions} include the problem formulation, a novel fractional programming technique, and an efficient optimization algorithm, as listed below.
% Our contributions can be categorized but not limited to the following aspects:
\begin{itemize}
    \item To the best of our knowledge, our work is the first in the literature to consider the optimization of the system utility-cost ratio (UCR) for the Metaverse over wireless communications. Our work is also among pioneering studies that incorporate human-centric utility for Metaverse optimization.
\item We propose a novel technique for fractional programming (FP), where the objective to be minimized is the sum of a convex function and a series of non-convex ratios with convex numerators and concave denominators. FP of the above kind cannot be addressed by prior 
%state-of-the-art
work~\cite{jong2012efficient,shen2018fractional} (viz., Section~\ref{opt-sum-of-ratios}). Our  technique contributes to optimization theory and is applicable to many other problems.    
      % \item We formulate the novel UCR in the human-centric Metaverse framework, embedding the subject user perception in the model to more precisely specify the user experience, and taking the system cost into consideration. The subjective test results are incorporated into the utility function design.
    \item Our UCR optimization is difficult to solve due to the following two aspects: 1) the objective function being the sum of a complicated function and a sequence of non-convex fractions, and 2)  jointly deciding five vector variables (for bandwidth, transmission power, video resolution, allocation of the Metaverse server's computing resource, CPU frequencies of VR users, respectively). Despite the  challenges, we propose an efficient algorithm by leveraging our novel FP technique above and carefully identifying a roadmap (on Pages~\pageref{roadmapvariables} and~\pageref{step-by-step}) to solve the variables step-by-step.  
    % We create a comprehensive joint optimization system model including video resolutions, computation, and network resources to elevate the user experience and ease the computation burden when facing the unified and user-centric Metaverse.
    % \item We design a novel optimization method when dealing with heterogeneous optimization objects. Specifically, we formulate a new sum of ratios problem and analyze it by KKT conditions to find a globally optimized solution for UCR.
    \item Simulations demonstrate the superiority of our algorithm over
 other baselines. The human-centric utilities used in the simulations are learnt from real-world data including a recent VR dataset~\cite{ssv360}.
    % We conduct systematic experiments for evaluating our proposed method. Our proposed approach offers outstanding gains in UCR performance compared to other baselines (e.g., partial optimization methods) and is able to meet the needs of a wide range of user scenarios.
\end{itemize}

% VR video applications in the Metaverse

% fractional programming .

\textbf{Roadmap.} This work is organized as follows. We review related work in Section~\ref{sec: related work}. The system model is presented in Section~\ref{sec: problem formulation}. We propose a novel technique for fractional programming (FP) in Section~\ref{opt-sum-of-ratios}. Using this FP technique, we analyze how to solve the UCR optimization for the Metaverse over wireless communications in
Sections~\ref{sec: optimization} and~\ref{kktproblem5}. Based on the analysis, Section~\ref{sec: algorithm} presents our algorithm for the UCR optimization, as well as its performance including solution quality,  convergence, and time complexity.
% (iv) We provide a detailed derivation of how to solve the formulated problem and give its time complexity from Section~\ref{opt-sum-of-ratios} to Section~\ref{sec: algorithm}. (v) 
We model human-centric utilities from real datasets in Section~\ref{sec:modelutility}, and use the obtained utility functions to provide simulation results in Section~\ref{sec: experiment}. We conclude the paper in Section~\ref{sec: conclusion}.

% compare our proposed approach with baselines and analyze the results in depth in Section~\ref{sec: experiment}. (vi) Finally, we summarize our work and draw conclusions in Section~\ref{sec: conclusion}.

\section{Related Work} \label{sec: related work}

We discuss related work from the following aspects: optimization in wireless networks, the Metaverse over wireless communications, human-centric utility, and the fractional programming technique.

\textbf{Optimization for wireless networks.} Many studies have addressed optimization related to delay, energy, or utility for wireless networks, as discussed below.
Optimizing the system cost, defined as the weighted sum of system delay and system energy consumption, is investigated in~\cite{luo2020hfel,zhan2021l4l} for wireless federated learning, in~\cite{yu2020joint} for {UAV}-enabled mobile edge computing, in~\cite{lu2022cost} for 5G networks. 
% Minimizing the cost (i.e., the weighted sum of  delay and energy) among all users is considered in~\cite{yang2021joint}.
% In~\cite{ke2020deep}, bandwidth is also incorporated into the cost in addition to delay and energy.
 In~\cite{qian2019optimal}, the difference between the utility and the energy consumption in a heterogeneous network is maximized.  There are also papers on ratio optimization to improve the system's performance. The ratio is often energy efficiency (EE)~\cite{jiang2013relation,mobihocfullversion} or computation efficiency (CE)~\cite{hu2021computation,hu2022energy}, which denotes the ratio of the number of transmitted or computed bits to energy consumption. EE or CE above can be understood as $\frac{\textnormal{utility}}{\textnormal{energy}}$, but surprisingly there seems no existing work in communication/network publications on optimizing $\frac{\textnormal{utility}}{\textnormal{energy}+\textnormal{delay}}$ like our paper, although we have conducted an extensive literature survey. Moreover, we consider human-centric utility for the Metaverse, which further increases the novelty of our studied problem. 
 
\textbf{Metaverse over wireless communications.} Recently, researching the Metaverse over wireless communications and networks has become an emerging topic. Recent papers~\cite{wang2022survey,xu2022full} have surveyed Metaverse research from different aspects:~\cite{wang2022survey} focusing on fundamental underlying technologies as well as security/privacy issues,~\cite{xu2022full} on how edge computing empowers the Metaverse. In addition to surveys~\cite{wang2022survey,xu2022full} above, we discuss  representative technical work~\cite{meng2022sampling,yu2022asynchronous,wang2023semantic,jiang2022reliable,ren2022quantum} below. In~\cite{meng2022sampling}, sampling, communication and prediction are co-designed to minimize the communication load for synchronizing a real-world device and its digital model in the Metaverse.
Yu~\textit{et~al.}~\cite{yu2022asynchronous} optimize the delay and reliability of wireless Metaverse using deep reinforcement learning. Contest theory is utilized in~\cite{wang2023semantic} for the Metaverse with semantic communications, while game theory is applied in~\cite{jiang2022reliable,ren2022quantum} for the vehicular Metaverse. In~\cite{mobihocfullversion} led by the current paper's first author, fractional programming (FP) is leveraged for energy efficiency optimization of the Metaverse subject to physical-layer security of wireless communications. In addition to the difference in terms of   problem formulation compared with  ours,~\cite{mobihocfullversion} allocates  communication resources only without optimizing computing resources and video resolutions. Also~\cite{mobihocfullversion} does not use our novel FP technique  
% to be presented in  
of Section~\ref{opt-sum-of-ratios}.

\textbf{Network Utility Maximization (NUM).} Our work is related to the research on network utility maximization (NUM)~\cite{palomar2006tutorial}. For a network of users, NUM considers that all users act altruistically to maximize the total network utility~\cite{chen2014social},  defined as the sum of all users' individual
utilities. In the classical NUM problem by Kelly~\textit{et~al.}~\cite{kelly1998rate}, the goal is to allocate traffic rates to users in order to maximize the total network utility subject to resource constraints (e.g., link capacity limitations). Since then, various NUM problems have been investigated in the literature~\cite{heydaribeni2019distributed,gu2019fairness,wang2020constrained}.  
% consider a classical NUM problem for rate control in communication networks.  
The total network utility is also referred to as the social welfare in~\cite{gong2017social}, where game theory is adopted to solve the problem. Despite the relevance of NUM research to our work, we emphasize that our objective is optimizing the ratio $\frac{\mathcal{U}}{\mathcal{C}}$ of the total system utility $\mathcal{U}$ to the total system cost $\mathcal{C}$, rather than just maximizing $\mathcal{U}$. The optimization of the fraction $\frac{\mathcal{U}}{\mathcal{C}}$ is more challenging than that of $\mathcal{U}$ due to the \mbox{non-convexity} of the fraction.

\textbf{Human-centric utility.} When utility is referred to as video quality, it can be measured using objective or subjective assessment methods. The subjective quality assessment (SQA) results in human-centric perceptual utility since human subjects are asked for their opinions directly. Higher human-centric utility means better Quality of Experience (QoE), which is in contrast with the traditional notion of Quality of Service (QoS) that quantifies the objective performance of the system. The survey~\cite{chen2014qos} covers human-centric utility for traditional 2D video applications. Human-centric design for Augmented/Virtual Reality (AR/VR) and the Metaverse has received much interest recently.
A 2023 survey~\cite{cao2023mobile} systematically reviews human-centric mobile AR. Elwardy~\textit{et~al.}~\cite{ssv360} report SQA of users watching 360\degree~videos when wearing HTC Vive Pro VR headsets. In~\cite{mourtzis2022human}, the human-centric nature of the Metaverse and using it for personalized value creation are discussed.
In the current paper on
VR for the Metaverse, we model human-centric utility functions of VR users from SQA video datasets~\cite{ssv360,Netflix}, as elaborated on in Section~\ref{sec:modelutility} later. The logarithmic function form will be adopted, which has been used in~\cite{yang2012crowdsourcing} for crowdsourcing, in~\cite{deng2020wireless} for mobile edge computing, and in~\cite{lyu2021service} for space-air-ground integrated networks.

\iffalse

The QoE-based scheduling strategies and optimizations for wireless networks have been presented by many previous works. Hu \textit{et al.}~\cite{HuQoE} jointly optimize the computing scheduling, UAV trajectory, and bandwidth allocation to guarantee the QoE of users in the downlink transmission. However, their QoE merely consists of two simple objective metrics, throughput and delay. Li \textit{et al.}~\cite{LiQoE} optimizes the QoE regarding point cloud video streaming. In their paper, the QoE is formulated as the number of points in users' views. Song \textit{et al.}~\cite{SongQoE} optimize the QoE model in the Internet of Vehicles, where they study the communication between the vehicles and roadside units through reinforcement learning. However, their QoE formulation is established on the ratio of the number of interesting files to the number of total files. The QoE models in the above-mentioned works are all simple and based on objective quality assessment. In this paper, we involve the subjective test results as a guide to our QoE modeling.

\fi

\textbf{Fractional programming (FP)}. In this paper, we present a novel FP technique and use it
% ~\st{FP is used in our paper} 
to transform a non-convex optimization problem into parametric convex optimization. The detailed comparison between our work and other FP papers~\cite{jong2012efficient,shen2018fractional} is deferred to Section~\ref{opt-sum-of-ratios}.

% FP is used in our paper to transform a non-convex optimization problem into parametric convex optimization. We present a novel FP technique which FP papers~\cite{jong2012efficient,shen2018fractional} do not cover. 

\section{System Model} \label{sec: problem formulation}

% \subsection{Communication Scenario} \label{sec-scenario}

Our studied system consists of one Metaverse Server (MS) and $N$ VR Users (VUs), indexed by $\mathcal{N} = \left\{1, 2, \cdots, \mathit{N} \right\}$. In downlink wireless communications, the MS sends to each VU the corresponding VR video via frequency division multiple access (FDMA) so that communications do not interfere.

\begin{figure*}[t] 
\centering
\setlength{\abovecaptionskip}{-0.1cm}
\includegraphics[width=0.9\linewidth]{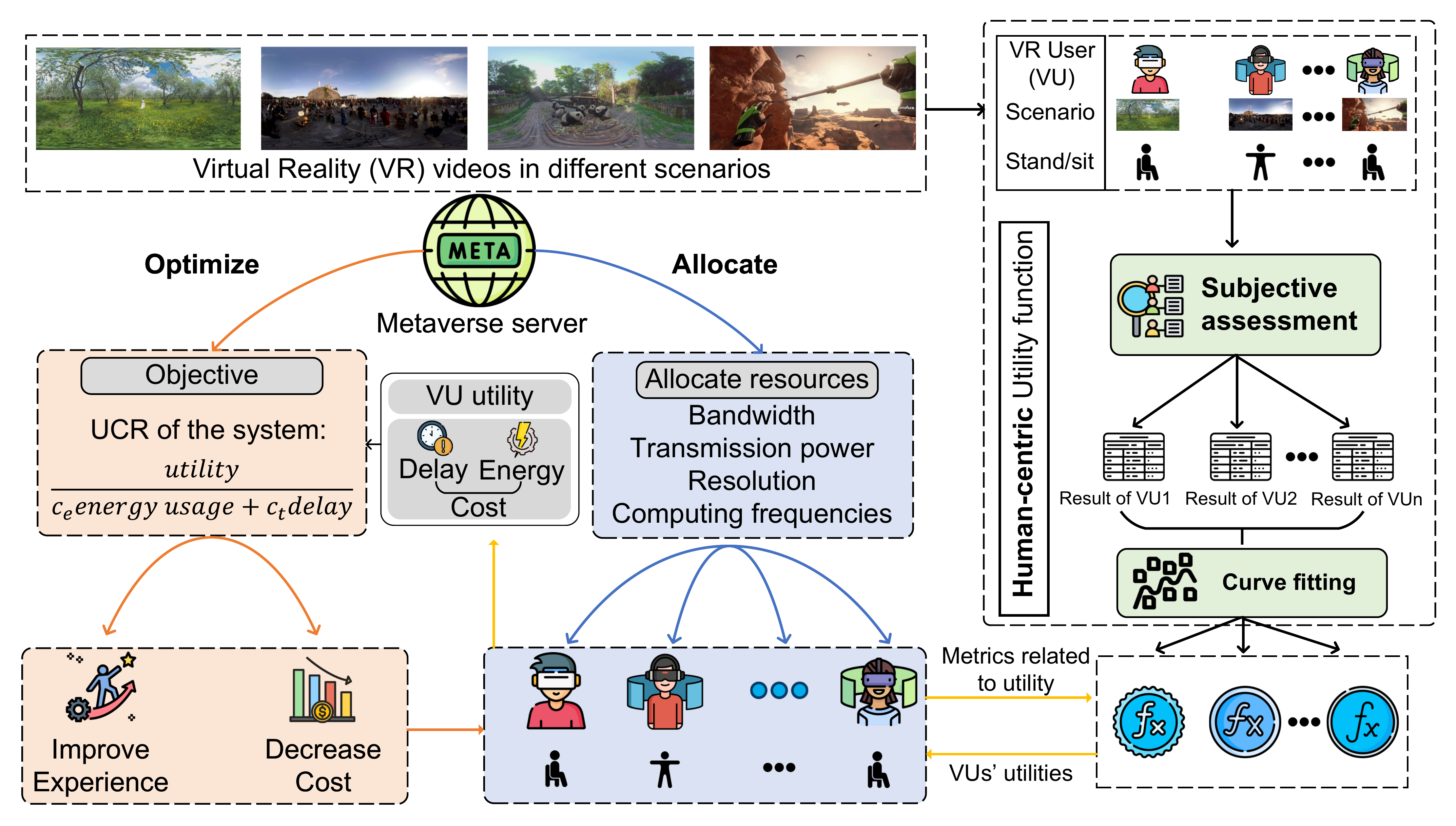}
\caption{Optimizing the system utility-cost ratio (UCR), in a system of one Metaverse Server (MS) and $N$ VR Users (VUs), by deciding communication and computation resources as well as VR video resolutions.}\vspace{-0.4cm}
\label{fig:model}
\end{figure*}

We have overviewed the system operation in the ``Studied problem'' paragraph on Page~\pageref{Studiedproblem}. As already stated, our goal is to optimize the system utility-cost ratio (UCR), by deciding communication resources (i.e., bandwidth and transmission power) and computation resources (i.e., the MS's computation allocation and the VUs' CPU frequencies), as well as VR video resolutions. Figure~\ref{fig:model} illustrates the system model.

% We do not repeat overviewing the system operation, which has been given in the ``Studied problem'' paragraph on Page~\pageref{Studiedproblem}.

Note that before the video transmission, there are message exchanges between the MS and VUs for control purpose; e.g., each VU informs the MS of its maximum CPU frequency and utility function, and the MS notifies the obtained CPU frequency for each VU from the system utility-cost ratio (UCR) optimization. We ignore the overhead of the control information since it is much smaller than the video data sizes. Below we first introduce notations, which are used to define the system utility and cost in Sections~\ref{subsectionutility} and~\ref{Delayenergy}. Then we formalize the UCR optimization in Section~\ref{OptimizationProblem}.

\iffalse

We first provide an overview of the system operation, before describing the optimization problem in detail. The MS solves the system utility-cost ratio (UCR) optimization by allocating 1) communication resources (i.e., bandwidth and transmission power) for the MS's communication with each VU, and 2) the MS's computation resources for processing the videos to be sent to VUs, as well as deciding 3) the video resolution for all VUs, and 4) the CPU frequencies for the VUs.
% and deciding the computation frequencies (i.e., CPU frequencies) for the server and VUs. 
Then, the MS uses the allocated computation resource to process each VR video with the selected resolution, and transmits the videos to the VUs with the decided bandwidth and transmission power. Each VU receives VR frames of a video and  processes the frames with the arranged CPU frequency. 
% , we assume that it can be reliably transmitted through the in-band channel, and neglect the overhead for its transmission. 

\fi

% This paper studies optimizing the UCR of the system, which is the trade-off between utility and cost, and the formulated human-perceived utility is related to the transmission rate, while the cost contains the energy cost and delay cost. Therefore, the system model is introduced by the sequence of transmission rate, human-perceived utility, delay, energy consumption, and problem formulation.

%For the convenience of the reader, we will give below a notation table of the system parameters.

% \subsection{Achievable downlink transmission rate}

For communications via  FDMA,
we define $\boldsymbol{b}=[b_1, b_2, \ldots, b_N]$, $\boldsymbol{p}=[p_1, p_2, \ldots, p_N]$ as the bandwidths and transmission powers used for the MS to communicate with VUs. With $g_n$ being the channel attenuation from MS to VU $n$, the achievable rate from MS to VU $n$ is given by the function notation below:
\begin{talign}
    r_n(b_n, p_n) = b_n \log_2(1+\frac{g_np_n}{\sigma^2b_n}). \label{shannon} 
\end{talign} 
% and $r_n(b_n,p_n)$ represents $r_n$ as a function of $(b_n,p_n)$, and such function notation is used throughout the paper.

~\vspace{-20pt}

\subsection{Modeling the human-centric utilities of VR users}\label{subsectionutility}

% The human-perceived utility

% More details are given in Section~\ref{sec:modelutility}.

Based on the subjective test in~\cite{ssv360}, for each VU $n$, we formulate the human-centric utility as $U_n(r_n,s_n)$, a function of the transmission rate $r_n$ and resolution $s_n$ satisfying Assumption~\ref{Uassumption:non-decreasing} below. 
\begin{assumption} \label{Uassumption:non-decreasing}
$U_n(r_n,s_n)$ is \mbox{non-decreasing} in $r_n$ and $s_n$, concave in $r_n$, and concave in $s_n$. 
\end{assumption}
The vector $\boldsymbol{s}=[s_1, r_2, \ldots, s_N]$ gives the resolutions of VR frames for the VUs. The system utility, defined as the sum of all $N$ VUs' human-centric utilities, is given by
\begin{talign}
\mathcal{U}(\boldsymbol{b},\boldsymbol{p},\boldsymbol{s}) = & \sum_{n \in \mathcal{N}} U_n(r_n(b_n,p_n),s_n). \label{eq:utility}
\end{talign}
% This utility function is obtained based on real datasets in Section~\ref{secmodelutility}. Here, we give two basic assumptions of the utility function.
%$\mathcal{U}(\boldsymbol{b},\boldsymbol{p},\boldsymbol{s},\boldsymbol{f}^{\textnormal{MS}},\boldsymbol{f}^{\textnormal{VU}})$. However, based on our data, we can only have $\mathcal{U}(\boldsymbol{b},\boldsymbol{p},\boldsymbol{s})$. We will provide further analysis and discussion for $\mathcal{U}(\boldsymbol{b},\boldsymbol{p},\boldsymbol{s})$ in the subsequent Section \ref{secmodelutility}.

% \begin{assumption}\label{Uassumption:concave}
% $U_n(r_n,s_n)$ is concave in $r_n$ and is also concave in $s_n$. 
% \end{assumption}

Our analysis and algorithm use Assumption~\ref{Uassumption:non-decreasing}, and do not need $U_n(r_n,s_n)$'s joint concavity in $r_n$ and $s_n$, though the expression of $U_n(r_n,s_n)$ in Section~\ref{sec:modelutility} from real datasets is  jointly concave in $r_n$\vspace{-10pt} and $s_n$.

% \begin{wrapfigure}{r}{0.3\textwidth}
% \vspace{255pt}\captionsetup{justification=centering,font=footnotesize}
% \includegraphics[width=0.3\textwidth]{figure9.pdf}
%  \vspace{-41pt} \caption{The timeline.} \label{timeline} \vspace{-430pt}
% \end{wrapfigure}

\subsection{System cost comprising delay and energy consumption\vspace{-5pt}}\label{Delayenergy}

\begin{figure}[H]
\centering
\includegraphics[width=0.5\textwidth]{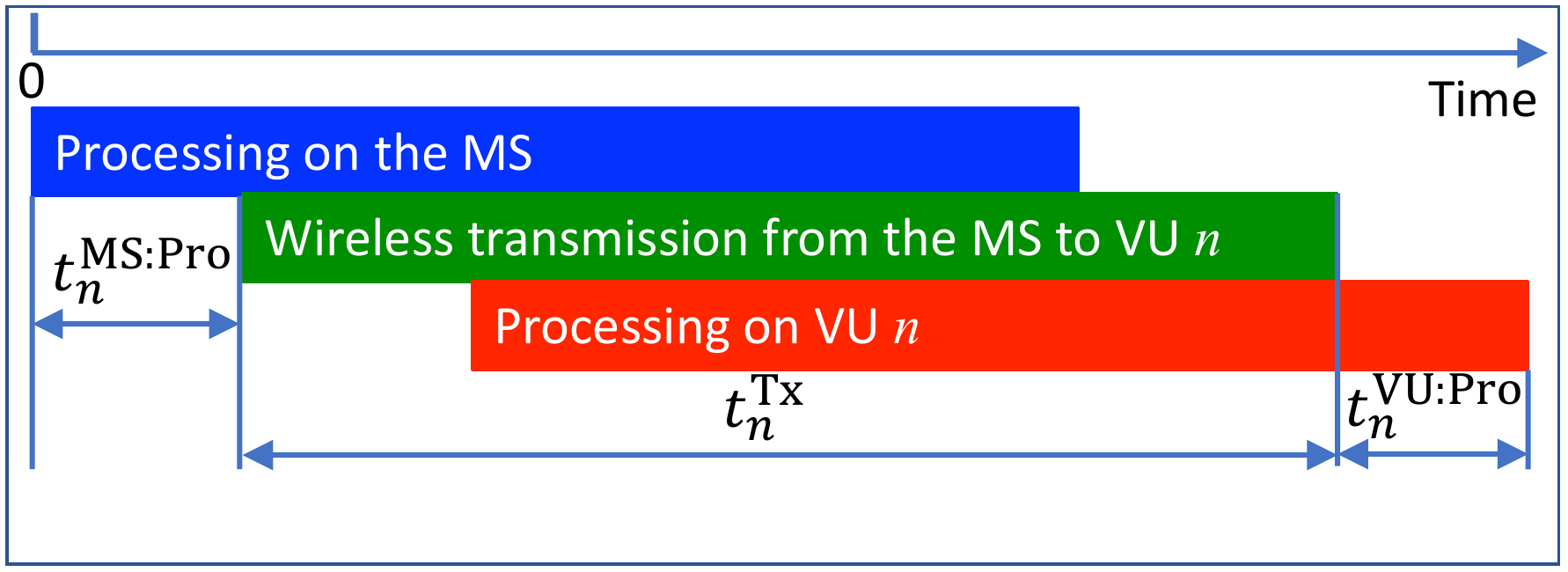}\vspace{-5pt}  \caption{The timeline.\vspace{-10pt}} \label{timeline} 
\end{figure}

We start with defining some notations. For each $n\in\mathcal{N}$,
let $f_n^{\textnormal{MS}}$ be the MS's computational resource allocated to process the frames for VU $n$.  Such allocation of computing resources is also considered in~\cite{feng2021min} for edge computing. The CPU frequency of VU $n$ is denoted by $f_n^{\textnormal{VU}}$. Then
$\boldsymbol{f}^{\textnormal{MS}}:=[f_1^{\textnormal{MS}}, f_2^{\textnormal{MS}}, \ldots, f_N^{\textnormal{MS}}]$ and $\boldsymbol{f}^{\textnormal{VU}}:=[f_1^{\textnormal{VU}}, f_2^{\textnormal{VU}}, \ldots, f_N^{\textnormal{VU}}]$.  
About the frames for VU $n$, let
$\mu_n$ be the number of bits per pixel, and $\nu_n>1$ be the compression ratio. 
The MS will generate a VR video of $\Lambda_n$ frames for XU $n$. Let $\mathcal{A}_n(s_n,\Lambda_n) $ (resp., $  \mathcal{B}_n(s_n,\Lambda_n)$) be the number of CPU cycles on MS (resp., VU $n$) to process a part of those $\Lambda_n$ frames before (resp., after) wireless transmission. While later frames of the VR video are yet to be generated, earlier frames can be transmitted from the MS to each VU $n$. Similarly, while later frames of the VR video are yet to be received, VU $n$ can process earlier frames which have already been accepted. Hence, the following three stages partially overlap: processing at the MS, wireless transmission from the MS to VU $n$, and processing at VU $n$, as shown in Fig.~\ref{timeline}. 

Then we define the following:\label{defineAB}
% In our system, the delay encompasses three parts:
\begin{itemize}
    \item the time used on the MS to generate and process frames \\for VU $n$ before wireless transmission: $t_n^{\textnormal{MS}:\textnormal{Pro}}(s_n,f_n^{\textnormal{MS}}) = \frac{\mathcal{A}_n(s_n,\Lambda_n)}{f_n^{\textnormal{MS}}}$, 
    \item the time expended to transmit all $\Lambda_n $ VR frames from the MS to VU $n$: $t_n^{\textnormal{Tx}}(b_n,p_n, s_n) = \frac{s_n\textcolor{black}{\mu_n}  {\color{black}\Lambda_n}}{r_n(b_n,p_n) \textcolor{black}{\nu_n}}$,
    \item the time cost on VU $n$ for processing frames after wireless transmission: $t_n^{\textnormal{VU}:\textnormal{Pro}}(s_n,f_n^{\textnormal{VU}}) = \frac{\mathcal{B}_n(s_n,\Lambda_n)}{f_n^{\textnormal{VU}}}$,
\end{itemize}
 We will set the expressions of $\mathcal{A}_n(\cdot), \mathcal{B}_n(\cdot)$ in Section~\ref{sec: experiment} on simulations. 
% Suppose the VR video for XU $n$ consists of $\Lambda_n$ frames. 
Thus, the delay for VU $n$ is
\begin{talign}
t_n(b_n,p_n,s_n,f_n^{\textnormal{MS}},f_n^{\textnormal{VU}})  & = t_n^{\textnormal{MS}:\textnormal{Pro}}(s_n,f_n^{\textnormal{MS}})+t_n^{\textnormal{Tx}}(b_n,p_n, s_n) \nonumber \\&+t_n^{\textnormal{VU}:\textnormal{Pro}}(s_n,f_n^{\textnormal{VU}}). \label{eq:userdelay}
\end{talign}
Then, we let the maximum of all VUs' delays be the system delay:
\begin{talign}
\mathcal{T}(\boldsymbol{b},\boldsymbol{p},\boldsymbol{s},\boldsymbol{f}^{\textnormal{MS}},\boldsymbol{f}^{\textnormal{VU}})  & = \max_{n \in \mathcal{N}}t_n(b_n,p_n,s_n,f_n^{\textnormal{MS}},f_n^{\textnormal{VU}}) .\label{eq:systemdelay}
\end{talign}

% $t_n^{\textnormal{MS}:\textnormal{Pro}}(s_n,f_n^{\textnormal{MS}}) + \Lambda_n \max\{t_n^{\textnormal{MS}:\textnormal{Pro}}(s_n,f_n^{\textnormal{MS}}),t_n^{\textnormal{Tx}}(b_n,p_n, s_n)\}+t_n^{\textnormal{VU}:\textnormal{Pro}}(s_n,f_n^{\textnormal{VU}})$

% \subsection{Energy consumption} \label{secEnergyconsumption}
Let $\kappa^{\textnormal{MS}}, \kappa_n^{\textnormal{VU}}$ be MS's and VU $n$'s effective switched capacitance. From the process of generating $\Lambda_n$ frames at MS to rendering them at VU $n$, the following energy will be consumed:  \label{defineFG}
% As mentioned in Section~\ref{sec-scenario}, the energy spent on the MS is for (i) each VUs' data processing: $E_n^{\textnormal{MS}:\textnormal{Pro}}$, and the energy used by each VU includes (i) energy used for data processing $E_n^{\textnormal{VU}:\textnormal{Pro}}$; (ii) energy used for data transmission $E_n^{\textnormal{VU}:\textnormal{Tx}}$. Specifically, they are formulated as:
\begin{itemize}
    \item energy spent on MS to process $\Lambda_n$ VR frames for VU $n$: $E_n^{\textnormal{MS}:\textnormal{Pro}}(s_n,f_n^{\textnormal{MS}}) =  \kappa^{\textnormal{MS}}{\color{black}\mathcal{F}_n(s_n,\Lambda_n)}({f_n^{\textnormal{MS}}})^2$,
    \item energy spent for transmitting $\Lambda_n$ VR frames from MS to VU $n$: $E_n^{\textnormal{Tx}}(b_n,p_n, s_n) = \frac{(p_n +p_n^{\textnormal{cir}})  s_n\textcolor{black}{\mu_n}{\color{black}\Lambda_n}}{r_n(b_n,p_n)\textcolor{black}{\nu_n}}$,
    \item energy spent on VU $n$ to process $\Lambda_n$ VR frames: $E_n^{\textnormal{VU}:\textnormal{Pro}}(s_n,f_n^{\textnormal{VU}}) =  \kappa_n^{\textnormal{VU}}{\color{black}\mathcal{G}_n(s_n,\Lambda_n)} (f_n^{\textnormal{VU}})^2$,
\end{itemize}
where we note that  $\mathcal{F}_n(s_n,\Lambda_n) $ (resp., $  \mathcal{G}_n(s_n,\Lambda_n)$) is different from $\mathcal{A}_n(s_n,\Lambda_n) $ (resp., $  \mathcal{B}_n(s_n,\Lambda_n)$) above, since the latter is only before (resp., after) wireless transmission as shown in Fig.~\ref{timeline}, while the former considers CPU cycles to process $\Lambda_n$ frames. The notations above highlight the dependence on $\Lambda_n$, but we do not write $\Lambda_n$ in delay and energy functions as $\Lambda_n$ is not optimized.
% The VR video for XU $n$ consists of $\Lambda_n$ frames. 
The total consumed energy is:
\begin{talign}
&\mathcal{E}(\boldsymbol{b},\boldsymbol{p},\boldsymbol{s},\boldsymbol{f}^{\textnormal{MS}},\boldsymbol{f}^{\textnormal{VU}}) =    \sum_{n\in\mathcal{N}}E_n^{\textnormal{MS}:\textnormal{Pro}}(s_n,f_n^{\textnormal{MS}}) \nonumber \\
&+\sum_{n\in\mathcal{N}}E_n^{\textnormal{Tx}}(b_n,p_n,s_n)+\sum_{n \in \mathcal{N}}E_n^{\textnormal{VU}:\textnormal{Pro}}(s_n,f_n^{\textnormal{VU}}) \label{eq:systemenergy}.
% \nonumber\\
% = & E^{\textnormal{MS}:\textnormal{Pro}}(\boldsymbol{s},\boldsymbol{f}^{\textnormal{MS}})+E^{\textnormal{Tx}}(\boldsymbol{b},\boldsymbol{p},\boldsymbol{s})+ E_n^{\textnormal{VU}:\textnormal{Pro}}(\boldsymbol{s},\boldsymbol{f}^{\textnormal{VU}})  
\end{talign}

The system cost is a weighted sum of the system delay in Eq.~(\ref{eq:systemdelay}) and energy consumption in Eq.~(\ref{eq:systemenergy}):
\begin{talign}
\mathcal{C}(\boldsymbol{b},\boldsymbol{p},\boldsymbol{s},\boldsymbol{f}^{\textnormal{MS}},\boldsymbol{f}^{\textnormal{VU}})  & = c_{\hspace{0.5pt}\textnormal{e}}\mathcal{E}(\boldsymbol{b},\boldsymbol{p},\boldsymbol{s},\boldsymbol{f}^{\textnormal{MS}},\boldsymbol{f}^{\textnormal{VU}}) \nonumber \\
&+c_{\hspace{0.8pt}\textnormal{t}}\mathcal{T}(\boldsymbol{b},\boldsymbol{p},\boldsymbol{s},\boldsymbol{f}^{\textnormal{MS}},\boldsymbol{f}^{\textnormal{VU}}), \label{eq:systemcost}
\end{talign}
where $c_\textnormal{e} $ and $ c_\textnormal{t}$ are the weight parameters for energy and delay, respectively. With the utility in Eq.~(\ref{eq:utility}) and the cost of the whole system in Eq.~(\ref{eq:systemcost}), we present the optimization problem in the next section.\vspace{-8pt}

\subsection{Optimization problem\vspace{-5pt}} \label{OptimizationProblem}
The aim is to maximize the utility-cost ratio (UCR) of the system as follows:
 \begin{subequations} \label{problem1} 
\begin{talign}
\textnormal{Problem $\mathbb{P}_{1}$}:~&\max_{\boldsymbol{b},\boldsymbol{p},\boldsymbol{s},\boldsymbol{f}^{\textnormal{MS}},\boldsymbol{f}^{\textnormal{VU}}} \quad \frac{\mathcal{U}(\boldsymbol{b},\boldsymbol{p},\boldsymbol{s})}{\mathcal{C}(\boldsymbol{b},\boldsymbol{p},\boldsymbol{s},\boldsymbol{f}^{\textnormal{MS}},\boldsymbol{f}^{\textnormal{VU}})} \tag{\ref{problem1}}\\[-2pt]
\textrm{s.t.} \quad  &\sum_{n \in \mathcal{N}} b_n \leq b_{\textnormal{max}}, \label{constraintbn} \\[-2pt] 
   & \sum_{n \in \mathcal{N}} p_n \leq p_{\textnormal{max}} , \label{constraintpn} 
  % \\ & s_n \in \mathcal{S}_n,~\forall n \in \mathcal{N}, \label{constraintsn} 
    \\& s_n \in \mathbb{S}_n,~\forall n \in \mathcal{N},\text{ where the set $\mathbb{S}_n$} \nonumber \\
    &\text{ can be continuous or discrete},  \label{constraintsn} 
  \\[-2pt] & \sum_{n \in \mathcal{N}} f_n^{\textnormal{MS}} \leq f_{\textnormal{max}}^{\textnormal{MS}}, \label{constraintfMS} \\[-2pt] & f_n^{\textnormal{VU}} \leq f_{n,\textnormal{max}}^{\textnormal{VU}},~\forall n \in \mathcal{N}. \label{constraintfn} 
\end{talign}
\end{subequations}
% The problem in Eq.~(\ref{problem}) means we maximize the utility-cost ratio $\frac{U}{C}$ as the UCR of the system. 
Constraints (\ref{constraintbn}), (\ref{constraintpn}), and (\ref{constraintfMS}) mean the sum-limit of the bandwidth, power, and computing resources of the MS. Constraint (\ref{constraintsn}) gives  the range of the resolution, and (\ref{constraintfn}) sets the CPU frequency limit of each VU. Our approach to solving $\mathbb{P}_{1}$ will use a fractional programming technique presented next.\vspace{-8pt}

\section{Our Proposed Technique for Fractional Programming (FP)\vspace{-5pt}} \label{opt-sum-of-ratios}

In this section, we will first formulate the FP problem and then explain how our proposed FP technique differs from those in the state-of-the-art work~\cite{jong2012efficient,shen2018fractional}.

% We discuss state of the art~\cite{jong2012efficient} and~\cite{shen2018fractional} 

% We will propose a novel technique for fractional programming (FP). The studied FP problem is as follows. 

\textbf{Fractional programming (FP) problem.}
% In this section, a novel fractional programming method is introduced.
% We explain the studied fractional programming (FP) problem as follows.
 Let $A_n(\boldsymbol{x}), B_n(\boldsymbol{x}), G(\boldsymbol{x})$ be functions of variable(s) $\boldsymbol{x}$, and these functions have definitions on a convex set $\mathcal{S}$, which is a subset of a real vector space. Also, for $\boldsymbol{x} \in \mathcal{S}$, we have $A_n(\boldsymbol{x}) \geq 0$ and $B_n(\boldsymbol{x}) > 0$. Then we consider:
 \begin{talign}
& \text{\mbox{\texttt{FP-problem}}:optimizing $H(\boldsymbol{x}) : = G(\boldsymbol{x}) + \sum_{n=1}^N\frac{A_n(\boldsymbol{x})}{B_n(\boldsymbol{x})}$} \nonumber \\
&\text{subject to $\boldsymbol{x} \in \mathcal{S}$.} \label{FP-whole}   
\end{talign}
  Two specific instances of \texttt{FP-problem} above are as follows:
\begin{talign}
& \text{\mbox{\texttt{FP-maximization}}: maximizing $H(\boldsymbol{x})$ subject to $\boldsymbol{x} \in \mathcal{S}$,} \nonumber \\
& \text{for concave $A_n(\boldsymbol{x})$ and convex $B_n(\boldsymbol{x})$,} \label{FP-maximization} \\
& \text{\mbox{\texttt{FP-minimization}}: minimizing $H(\boldsymbol{x})$ subject to $\boldsymbol{x} \in \mathcal{S}$,} \nonumber \\
& \text{for convex  $A_n(\boldsymbol{x})$ and concave $B_n(\boldsymbol{x})$.} \label{FP-minimization}   
\end{talign}
% \begin{itemize}[leftmargin=15pt]
% \item[\ding{172}] 
% \item[\ding{173}]    
% \end{itemize}
Note that the above problem formulations~(\ref{FP-whole})~(\ref{FP-maximization})~(\ref{FP-minimization}) cover constrained optimization where the constraints are convex so that we can incorporate the constraints into defining $\mathcal{S}$.

\begin{table*}[]
\caption{A comparison of our paper and other work \cite{jong2012efficient,shen2018fractional} on fractional programming (FP).}
\label{tab:my-table}
\centering\begin{tabular}{|l|l|}
\hline
Paper                                 & Contribution to fractional programming                                                                                                                                                                                                                                                  \\ \hline
Jong \cite{jong2012efficient}         & \begin{tabular}[c]{@{}l@{}}Find the global solution to \mbox{\texttt{FP-maximization}} in~(\ref{FP-maximization})\\ and~\mbox{\texttt{FP-minimization}} in~(\ref{FP-minimization}), both only in the special case of $G(\boldsymbol{x})\equiv 0$, \\ but fail to tackle any problem in the case of $G(\boldsymbol{x})\not\equiv 0$.\end{tabular} \\ \hline
Shen and Yu~\cite{shen2018fractional} & \begin{tabular}[c]{@{}l@{}}Find a stationary point for~\mbox{\texttt{FP-maximization}} in~(\ref{FP-maximization}) \\ for both cases of $G(\boldsymbol{x})\equiv 0$ and $G(\boldsymbol{x})\not\equiv 0$, \\ but fail to tackle~\mbox{\texttt{FP-minimization}}.\end{tabular}                                                \\ \hline
Our current work                      & \begin{tabular}[c]{@{}l@{}}Find a stationary point for~\mbox{\texttt{FP-minimization}} in~(\ref{FP-minimization})\\ for both cases of $G(\boldsymbol{x})\equiv 0$ and $G(\boldsymbol{x})\not\equiv 0$.\end{tabular}                                                                                                    \\ \hline
\end{tabular}
\end{table*}

\textbf{An overview of our contribution in FP technique.} With problems defined in~(\ref{FP-whole})~(\ref{FP-maximization})~(\ref{FP-minimization}) above, Table~\ref{tab:my-table} compares our novel FP technique and those in \cite{jong2012efficient,shen2018fractional}. We present the details in the paragraphs below, describing~\cite{jong2012efficient},~\cite{shen2018fractional}  and our work, respectively.

% \textbf{Prior work on FP and our new result.} 

\textbf{Prior work~\cite{jong2012efficient} on FP.} 
% We discuss~\cite{jong2012efficient} and~\cite{shen2018fractional} below, which present state of the art in FP. 
In case of $G(\boldsymbol{x})\hspace{-2pt}\equiv\hspace{-2pt} 0$,~\cite{jong2012efficient} optimizes $ \sum_{n=1}^N\frac{A_n(\boldsymbol{x})}{B_n(\boldsymbol{x})}$, which is referred to as the sum of ratios (SoR). Then \mbox{\texttt{FP-maximization}} (resp.,~\mbox{\texttt{FP-minimization}}) under $G(\boldsymbol{x})\equiv 0$ can be referred to \mbox{\texttt{SoR-maximization}} (resp.,~\mbox{\texttt{SoR-minimization}}). Via a transform into parametric convex optimization problems,~\cite{jong2012efficient} obtains a global optimum for \mbox{\texttt{SoR-maximization}} (i.e., maximizing $\sum_{n=1}^N\frac{A_n(\boldsymbol{x})}{B_n(\boldsymbol{x})}$ for concave $A_n(\boldsymbol{x})$ and convex $B_n(\boldsymbol{x})$) and \mbox{\texttt{SoR-minimization}} (i.e., maximizing $\sum_{n=1}^N\frac{A_n(\boldsymbol{x})}{B_n(\boldsymbol{x})}$ for convex $A_n(\boldsymbol{x})$ and concave $B_n(\boldsymbol{x})$). However, the approach of~\cite{jong2012efficient} is only applicable to the case of $G(\boldsymbol{x})\equiv 0$. The reason is that although the original SoR optimization and the transformed problem find
 the same optimal solution for the variable(s), the optimal objective-function values of the two problems are different.

\textbf{Prior work~\cite{shen2018fractional} on FP.} 
 To address cases of $G(\boldsymbol{x})\equiv 0$ and $G(\boldsymbol{x})\not\equiv 0$, in the breakthrough work~\cite{shen2018fractional}, Shen and Yu transform each $\frac{A_n(\boldsymbol{x})}{B_n(\boldsymbol{x})}$ into $J_n(\boldsymbol{x},y_n)\hspace{-2pt}:=$\,\mbox{$2y_n \sqrt{A_n(\boldsymbol{x})} - {y_n}^2 B_n(\boldsymbol{x}) $,}  and prove that for concave $A_n(\boldsymbol{x})$ and convex $B_n(\boldsymbol{x})$, \mbox{\texttt{FP-maximization}} in~(\ref{FP-maximization}) is the same as maximizing $V(\boldsymbol{x},\boldsymbol{y}):=G(\boldsymbol{x}) + \sum_{n=1}^N J_n(\boldsymbol{x},y_n)$ subject to $\boldsymbol{x} \in \mathcal{S}$ and $y_n \in \mathbb{R}$ (the set of real numbers). Then alternating optimization (AO) is adopted to optimize $\boldsymbol{x}$ and $\boldsymbol{y}:=[y_1,\ldots,y_N]$ in an alternating manner, since $V(\boldsymbol{x},\boldsymbol{y})$ is concave in $\boldsymbol{x}$ and concave in $\boldsymbol{y}$, despite not being jointly concave in them. This AO algorithm leads to a stationary point for \mbox{\texttt{FP-maximization}}.  Note that~\cite{shen2018fractional} tackles only \mbox{\texttt{FP-maximization}} and does not address \mbox{\texttt{FP-minimization}}. $J_n(\boldsymbol{x},y_n)$ above can not be used for \mbox{\texttt{FP-minimization}}, since  the minimum of $J_n(\boldsymbol{x},y_n)$ is $-\infty$.  
 % However, the result of~\cite{shen2018fractional} does not apply to \mbox{FP-minimization} in~(\ref{FP-minimization}). 

\textbf{Our new technique for FP.} Based on the above discussions, \cite{jong2012efficient,shen2018fractional} do not cover \mbox{\texttt{FP-minimization}} in~(\ref{FP-minimization}) with $G(\boldsymbol{x})\not\equiv 0$.
  To fill this gap, our paper proposes the following technique for \mbox{\texttt{FP-minimization}}  in both cases of $G(\boldsymbol{x})\equiv 0$ and $G(\boldsymbol{x})\not\equiv 0$. Specifically, we transform each $\frac{A_n(\boldsymbol{x})}{B_n(\boldsymbol{x})}$ into $K_n(\boldsymbol{x},y_n):=[A_n(\boldsymbol{x})]^2y_n + \frac{1}{4[B_n(\boldsymbol{x})]^2y_n} $, and prove that for convex  $A_n(\boldsymbol{x})$ and concave $B_n(\boldsymbol{x})$, \mbox{\texttt{FP-minimization}} in~(\ref{FP-minimization}) is the same as minimizing $W(\boldsymbol{x},\boldsymbol{y}):=G(\boldsymbol{x}) + \sum_{n=1}^N K_n(\boldsymbol{x},y_n)$ subject to $\boldsymbol{x} \in \mathcal{S}$ and $y_n \in \mathbb{R}^+$. The above holds because with $\boldsymbol{y}^{\#}(\boldsymbol{x})$ denoting $\boldsymbol{y}\in(\mathbb{R}^+)^N$ which minimizes $W(\boldsymbol{x},\boldsymbol{y})$ given $\boldsymbol{x}$ (i.e., $y_n^{\#}(\boldsymbol{x}):=\frac{1}{2A_n(\boldsymbol{x})B_n(\boldsymbol{x})} $), the partial derivative of $W(\boldsymbol{x},\boldsymbol{y})$ with respect to $\boldsymbol{x}$ at $\boldsymbol{y}$ being $\boldsymbol{y}^{\#}(\boldsymbol{x})$ is the same as the  derivative of $H(\boldsymbol{x})$ with respect to $\boldsymbol{x}$, where the computations are straightforward and shown in the Appendix of our full version~\cite{full}. Then \mbox{\texttt{FP-minimization}} in~(\ref{FP-minimization}) can be tackled by optimizing $\boldsymbol{x}$ and $\boldsymbol{y}$ in an alternating manner to minimize $W(\boldsymbol{x},\boldsymbol{y})$, since $W(\boldsymbol{x},\boldsymbol{y})$ is convex in $\boldsymbol{x}$ (for convex $G(\boldsymbol{x}) $) and convex in $\boldsymbol{y}$ (for any $G(\boldsymbol{x}) $), despite not being jointly convex in them. For non-convex $G(\boldsymbol{x}) $, optimizing $W(\boldsymbol{x},\boldsymbol{y})$ with respect to $\boldsymbol{x}$ can employ techniques such as difference-of-convex programming or successive convex approximation~\cite{zhou2021solving}. 
  % Using the terminology of stationary points in~\cite{shen2018fractional} for constrained optimization, '
The above alternating optimization finds a stationary point for \mbox{\texttt{FP-minimization}} in~(\ref{FP-minimization}).\label{opt-sum-of-ratios2} Finally, we note that while our proposed FP technique will be used to solve the current paper's Problem $\mathbb{P}_{1}$ of Section~\ref{OptimizationProblem}, the technique  can also be applied to many other FP problems~\cite{shen2018fractional} beyond our paper. \vspace{-8pt}

\section[Solve the Optimization Problem P1]{Solve the Optimization Problem $\mathbb{P}_{1}$\vspace{-5pt}} \label{sec: optimization}

We present our method of solving $\mathbb{P}_{1}$ of (\ref{problem1}) below. The denominator in the objective function of (\ref{problem1}), i.e., the system cost, is given by (\ref{eq:systemcost}) and involves a ``maximize'' term from the system delay in (\ref{eq:systemdelay}). We add an auxiliary variable $T$ to circumvent that ``maximize'' so that $\mathbb{P}_{1}$ is transformed into Problem $\mathbb{P}_{2}$:
 \begin{subequations}  \label{problemP2}
\begin{talign}
\textnormal{Problem $\mathbb{P}_{2}$}:~&\max_{\boldsymbol{b},\boldsymbol{p},\boldsymbol{s},\boldsymbol{f}^{\textnormal{MS}},\boldsymbol{f}^{\textnormal{VU}},T} \quad \frac{\mathcal{U}(\boldsymbol{b},\boldsymbol{p},\boldsymbol{s})}{c_{\hspace{0.5pt}\textnormal{e}}\mathcal{E}(\boldsymbol{b},\boldsymbol{p},\boldsymbol{s},\boldsymbol{f}^{\textnormal{MS}},\boldsymbol{f}^{\textnormal{VU}})+c_{\hspace{0.8pt}\textnormal{t}}T} \tag{\ref{problemP2}} \\
\textrm{s.t.} \quad & t_n(b_n,p_n,s_n,f_n^{\textnormal{MS}},f_n^{\textnormal{VU}}) \leq T,~\forall n \in \mathcal{N}, \label{constraintTtau} \\  \quad & \textnormal{(\ref{constraintbn})}, \textnormal{(\ref{constraintpn})}, \textnormal{(\ref{constraintsn})}, \textnormal{(\ref{constraintfMS})}, \textnormal{(\ref{constraintfn})}.
\end{talign}
\end{subequations}

% In Problem $\mathbb{P}_{2}$, $\mathcal{E}(\boldsymbol{b},\boldsymbol{p},\boldsymbol{s},\boldsymbol{f}^{\textnormal{MS}},\boldsymbol{f}^{\textnormal{VU}})$ is the total energy consumed by the MS and VUs and $T$ means the maximum delay (i.e., system delay) throughout the whole data transmission and processing phases.

\renewcommand{\thealgocf}{A\arabic{algocf}}

\begin{algorithm*}
\caption{Solve Problem $\mathbb{P}_{2}$ and hence Problem $\mathbb{P}_{1}$ for the system UCR optimization.}
\label{algo:MN}

Initialize $i \leftarrow -1$ and for all $n \in \mathcal{N}$: $b_n^{(0)} = \frac{b_{\max}}{N}$, $p_n^{(0)} = \frac{p_{\max}}{N}$, $s_n^{(0)} = \frac{s_n^{\textnormal{min}} + s_n^{\textnormal{max}}}{2}$, $(f_n^{\textnormal{MS}})^{(0)} = \frac{f_{\textnormal{max}}^{\textnormal{MS}}}{N}$, $(f_n^{\textnormal{VU}})^{(0)} = f_{n,\textnormal{max}}^{\textnormal{VU}}$;

$T^{(0)}\leftarrow \max_{n \in \mathcal{N}}t_n(b_n^{(0)},p_n^{(0)},s_n^{(0)},(f_n^{\textnormal{MS}})^{(0)},(f_n^{\textnormal{VU}})^{(0)}) $;

\Repeat{the relative difference between $y^{(i+1)}$ (i.e., $\frac{\mathcal{U}(\boldsymbol{b}^{(i+1)},\boldsymbol{p}^{(i+1)},\boldsymbol{s}^{(i+1)})}{c_{\hspace{0.5pt}\textnormal{e}}\mathcal{E}(\boldsymbol{b}^{(i+1)},\boldsymbol{p}^{(i+1)},\boldsymbol{s}^{(i+1)},(\boldsymbol{f}^{\textnormal{MS}})^{(i+1)},(\boldsymbol{f}^{\textnormal{VU}})^{(i+1)})+c_{\hspace{0.8pt}\textnormal{t}}T^{(i+1)}}$) and $y^{(i)}$ is no greater than $\epsilon_1$ for a small positive number $\epsilon_1$ (i.e., $\frac{y^{(i+1)}}{y^{(i)}}- 1 \leq \epsilon_1$)}{
%\textit{ } \label{outermost}

Let $i \leftarrow i+1$;

$y^{(i)} \leftarrow \frac{\mathcal{U}(\boldsymbol{b}^{(i)},\boldsymbol{p}^{(i)},\boldsymbol{s}^{(i)})}{c_{\hspace{0.5pt}\textnormal{e}}\mathcal{E}(\boldsymbol{b}^{(i)},\boldsymbol{p}^{(i)},\boldsymbol{s}^{(i)},(\boldsymbol{f}^{\textnormal{MS}})^{(i)},(\boldsymbol{f}^{\textnormal{VU}})^{(i)})+c_{\hspace{0.8pt}\textnormal{t}}T^{(i)}}$; \label{setyi}

//Lines 8--25 solve $\mathbb{P}_{3}(y^{(i)})$

 Initialize $j \leftarrow -1$, 
 
 %$s_n^{(i,j)}$

$[\boldsymbol{b}^{(i,0)},\boldsymbol{p}^{(i,0)},\boldsymbol{s}^{(i,0)},(\boldsymbol{f}^{\textnormal{MS}})^{(i,0)},(\boldsymbol{f}^{\textnormal{VU}})^{(i,0)},T^{(i,0)}] \leftarrow [\boldsymbol{b}^{(i)},\boldsymbol{p}^{(i)},\boldsymbol{s}^{(i)},(\boldsymbol{f}^{\textnormal{MS}})^{(i)},(\boldsymbol{f}^{\textnormal{VU}})^{(i)},T^{(i)}]$;

\Repeat{the relative difference between $H_{\mathbb{P}_{3}}(\boldsymbol{b}^{(i,j+1)},\boldsymbol{p}^{(i,j+1)},\boldsymbol{s}^{(i,j+1)},(\boldsymbol{f}^{\textnormal{MS}})^{(i,j+1)},(\boldsymbol{f}^{\textnormal{VU}})^{(i,j+1)},T^{(i,j+1)}\mid y^{(i)})$ and $H_{\mathbb{P}_{3}}(\boldsymbol{b}^{(i,j)},\boldsymbol{p}^{(i,j)},\boldsymbol{s}^{(i,j)},(\boldsymbol{f}^{\textnormal{MS}})^{(i,j)},(\boldsymbol{f}^{\textnormal{VU}})^{(i,j)},T^{(i,j)}\mid y^{(i)})$ is no greater than $\epsilon_2$ for a small positive number $\epsilon_2$ (i.e., $\frac{H_{\mathbb{P}_{3}}(\boldsymbol{b}^{(i,j+1)},\boldsymbol{p}^{(i,j+1)},\boldsymbol{s}^{(i,j+1)},(\boldsymbol{f}^{\textnormal{MS}})^{(i,j+1)},(\boldsymbol{f}^{\textnormal{VU}})^{(i,j+1)},T^{(i,j+1)}\mid y^{(i)})}{H_{\mathbb{P}_{3}}(\boldsymbol{b}^{(i,j)},\boldsymbol{p}^{(i,j)},\boldsymbol{s}^{(i,j)},(\boldsymbol{f}^{\textnormal{MS}})^{(i,j)},(\boldsymbol{f}^{\textnormal{VU}})^{(i,j)},T^{(i,j)}\mid y^{(i)})}- 1 \leq \epsilon_2$) \label{P3convergence}}{

% \textit{//} \label{mid-level}

//Lines 13--22 solve $\mathbb{P}_{4}(y^{(i)},\boldsymbol{s}^{(i,j)})$

Let $j \leftarrow j+1$.

Initialize $\ell \leftarrow -1$;

$[\boldsymbol{b}^{(i,j,0)},\boldsymbol{p}^{(i,j,0)}] \leftarrow [\boldsymbol{b}^{(i,j)},\boldsymbol{p}^{(i,j)}]$;

\Repeat{$\ell \geq 1$ and the relative difference between the optimal objective-function values for Problem $\mathbb{P}_{5}( \boldsymbol{z}^{(i,j,\ell)},y^{(i)},\boldsymbol{s}^{(i,j)})$ and Problem $\mathbb{P}_{5}( \boldsymbol{z}^{(i,j,\ell-1)},y^{(i)},\boldsymbol{s}^{(i,j)})$  is no greater than $\epsilon_3$ for a small positive number $\epsilon_3$ (i.e., $\frac{V^*_{\mathbb{P}_{5}}( \boldsymbol{z}^{(i,j,\ell)},y^{(i)},\boldsymbol{s}^{(i,j)})}{V^*_{\mathbb{P}_{5}}( \boldsymbol{z}^{(i,j,\ell-1)},y^{(i)},\boldsymbol{s}^{(i,j)})}- 1 \leq \epsilon_3$)}{

% \textit{//}\label{innermost}

Let $\ell \leftarrow \ell+1$.

Set $z_n^{(i,j,\ell)} \leftarrow  \frac{1}{2 \cdot (p_n^{(i,j,\ell)} +p_n^{\textnormal{cir}})  s_n^{(i)} \textcolor{black}{\mu_n}{\color{black}\Lambda_n} \cdot r_n(b_n^{(i,j,\ell)},p_n^{(i,j,\ell)}) \textcolor{black}{\nu_n}}$

Obtain $[\boldsymbol{b}^{(i,j,\ell+1)},\boldsymbol{p}^{(i,j,\ell+1)},(\boldsymbol{f}^{\textnormal{MS}})^{(i,j,\ell+1)},(\boldsymbol{f}^{\textnormal{VU}})^{(i,j,\ell+1)},T^{(i,j,\ell+1)}]$ through solving Problem $\mathbb{P}_{5}(\boldsymbol{z}^{(i,j,\ell)},y^{(i)},\boldsymbol{s}^{(i,j)})$ according to Algorithm~\ref{algo:P5}, and denote the resulting optimal objective-function value of  $\mathbb{P}_{5}( \boldsymbol{z}^{(i,j,\ell)},y^{(i)},\boldsymbol{s}^{(i,j)})$ by $V^*_{\mathbb{P}_{5}}( \boldsymbol{z}^{(i,j,\ell)},y^{(i)},\boldsymbol{s}^{(i,j)})$; \label{solveP5}

}

Set $[\boldsymbol{b}^{(i,j+1)},\boldsymbol{p}^{(i,j+1)},(\boldsymbol{f}^{\textnormal{MS}})^{(i,j+1)},(\boldsymbol{f}^{\textnormal{VU}})^{(i,j+1)},T^{(i,j+1)}]  \leftarrow [\boldsymbol{b}^{(i,j,\ell+1)},\boldsymbol{p}^{(i,j,\ell+1)},(\boldsymbol{f}^{\textnormal{MS}})^{(i,j,\ell+1)},(\boldsymbol{f}^{\textnormal{VU}})^{(i,j,\ell+1)},T^{(i,j,\ell+1)}]$, which we consider as a solution to $\mathbb{P}_{4}(y^{(i)},\boldsymbol{s}^{(i,j)})$;

Set $\boldsymbol{s}$ as $\boldsymbol{s}^{(i,j+1)}$ denoting the optimal solution of Problem $\mathbb{P}_{6}(\boldsymbol{b}^{(i,j+1)},\boldsymbol{p}^{(i,j+1)},(\boldsymbol{f}^{\textnormal{MS}})^{(i,j+1)},(\boldsymbol{f}^{\textnormal{VU}})^{(i,j+1)},T^{(i,j+1)},y^{(i)})$, which is obtained from~(\ref{setoptimalsn}).

} 

Set $[\boldsymbol{b}^{(i+1)},\boldsymbol{p}^{(i+1)},\boldsymbol{s}^{(i+1)},(\boldsymbol{f}^{\textnormal{MS}})^{(i+1)},(\boldsymbol{f}^{\textnormal{VU}})^{(i+1)},T^{(i+1)}]\leftarrow[\boldsymbol{b}^{(i,j+1)},\boldsymbol{p}^{(i,j+1)},\boldsymbol{s}^{(i,j+1)},(\boldsymbol{f}^{\textnormal{MS}})^{(i,j+1)},(\boldsymbol{f}^{\textnormal{VU}})^{(i,j+1)},T^{(i,j+1)}]$, which we consider as a solution to Problem $\mathbb{P}_{3}(y^{(i)})$;

}

Return $[\boldsymbol{b}^{(i+1)},\boldsymbol{p}^{(i+1)},\boldsymbol{s}^{(i+1)},(\boldsymbol{f}^{\textnormal{MS}})^{(i+1)},(\boldsymbol{f}^{\textnormal{VU}})^{(i+1)},T^{(i+1)}]$ as a solution to Problem $\mathbb{P}_{2}$, which means $[\boldsymbol{b}^{(i+1)},\boldsymbol{p}^{(i+1)},\boldsymbol{s}^{(i+1)},(\boldsymbol{f}^{\textnormal{MS}})^{(i+1)},(\boldsymbol{f}^{\textnormal{VU}})^{(i+1)}]$ is a solution to Problem $\mathbb{P}_{1}$;
\end{algorithm*}

The subsections below present our steps for solving $\mathbb{P}_{1}$. These steps together induce our Algorithm~\ref{algo:MN} on Page~\pageref{algo:MN}, which can be better understood after readers have finished all subsections below.\vspace{-8pt}

% To help explain our steps of solving $\mathbb{P}_{1}$ in the subsections below, 
% we present Algorithm~\ref{algo:MN} on Page~\pageref{algo:MN}  to solve  $\mathbb{P}_{1}$, where some parts can be better understood only after 

\subsection{Dinkelbach's transform for the ratio optimization} \label{secDinkelbach}

% However, Problem $\mathbb{P}_{2}$ is a sum-of-concave-convex-ratios optimization problem, which is non-convex and hard to be solved by conventional optimization methods. Therefore, another auxiliary variable $y$ is needed to transform Problem $\mathbb{P}_{2}$ into a parametric convex programming problem.  Assuming $y$ is given, we can reformulate Problem $\mathbb{P}_{2}$ and its objective function as Problem $\mathbb{P}_{3}$ and $H_{\mathbb{P}_{3}}$ as follows:

%\begin{spacing}{1.5}

For $\mathbb{P}_{2}$ maximizing a ratio (i.e., $\frac{\textnormal{numerator of (\ref{problemP2})}}{\textnormal{denominator of (\ref{problemP2})}}$), we use Dinkelbach's transform~\cite{shen2018fractional} to transform $\mathbb{P}_{2}$ into a series of parametric optimization $\mathbb{P}_{3}(y)$ which maximizes ``\mbox{$\textnormal{numerator of (\ref{problemP2})}\hspace{-2pt} -\hspace{-2pt}y\hspace{-1pt}\cdot \hspace{-1pt} \textnormal{denominator of (\ref{problemP2})}$}'' subject to $\mathbb{P}_{2}$'s constraints, where solving the current $\mathbb{P}_{3}(y)$ decides ``$y$'' used in the next $\mathbb{P}_{3}(y)$. For an optimization problem $\mathbb{P}_{i}$, let $H_{\mathbb{P}_{i}}$ denote its objective function. Then $H_{\mathbb{P}_{3}}$ and $\mathbb{P}_{3}$ are as follows:
\begin{talign}
H_{\mathbb{P}_{3}}&(\boldsymbol{b},\boldsymbol{p},\boldsymbol{s},\boldsymbol{f}^{\textnormal{MS}},\boldsymbol{f}^{\textnormal{VU}},T\mid y): \nonumber \\
&=\textnormal{numerator of (\ref{problemP2})} -y\cdot \textnormal{denominator of (\ref{problemP2})} \nonumber \\
& =\mathcal{U}(\boldsymbol{b},\boldsymbol{p},\boldsymbol{s}) -  y \cdot (c_{\hspace{0.5pt}\textnormal{e}}\mathcal{E}(\boldsymbol{b},\boldsymbol{p},\boldsymbol{s},\boldsymbol{f}^{\textnormal{MS}},\boldsymbol{f}^{\textnormal{VU}})+c_{\hspace{0.8pt}\textnormal{t}} T),
\end{talign}
\begin{talign}
\textnormal{Pro}&\textnormal{blem}~\textnormal{$\mathbb{P}_{3}(y)$}: \nonumber \\
&\max_{\boldsymbol{b},\boldsymbol{p},\boldsymbol{s},\boldsymbol{f}^{\textnormal{MS}},\boldsymbol{f}^{\textnormal{VU}},T} \quad H_{\mathbb{P}_{3}}(\boldsymbol{b},\boldsymbol{p},\boldsymbol{s},\boldsymbol{f}^{\textnormal{MS}},\boldsymbol{f}^{\textnormal{VU}},T\mid y) \\
&\textrm{s.t.} \quad \textnormal{(\ref{constraintbn})}, \textnormal{(\ref{constraintpn})}, \textnormal{(\ref{constraintsn})}, \textnormal{(\ref{constraintfMS})}, \textnormal{(\ref{constraintfn})},  \textnormal{(\ref{constraintTtau})} . \nonumber
\end{talign}
%\end{subequations}

The process of using $\mathbb{P}_{3}$ to solve $\mathbb{P}_{2}$ is as follows. For ease of explanation, we denote $[\boldsymbol{b},\boldsymbol{p},\boldsymbol{s},\boldsymbol{f}^{\textnormal{MS}},\boldsymbol{f}^{\textnormal{VU}},T]$ by $\boldsymbol{x}$, and write the objective function of $\mathbb{P}_{3}(y)$ as $U(\boldsymbol{x})-y\cdot C(\boldsymbol{x})$. Then starting from a feasible $\boldsymbol{x}^{(0)}$ at initialization, we set $\boldsymbol{y}^{(0)}$ as $\frac{U(\boldsymbol{x}^{(0)})}{C(\boldsymbol{x}^{(0)})}$. Then we solve $\mathbb{P}_{3}(\boldsymbol{y}^{(0)})$, denote the obtained solution as $\boldsymbol{x}^{(1)}$, set $\boldsymbol{y}^{(1)}$ as $\frac{U(\boldsymbol{x}^{(1)})}{C(\boldsymbol{x}^{(1)})}$. This process continues iteratively: in the $(i+1)$th iteration, $\boldsymbol{y}^{(i)}$ is set as $\frac{U(\boldsymbol{x}^{(i)})}{C(\boldsymbol{x}^{(i)})}$ (given by Line~\ref{setyi} of Algorithm~\ref{algo:MN} on Page~\pageref{algo:MN}), and $\boldsymbol{x}^{(i+1)}$ is obtained from solving $\mathbb{P}_{3}(\boldsymbol{y}^{(i)})$. As stated in~\cite{shen2018fractional}, the above process converges and does not lose optimality; i.e., under global optimization of each $P_3(y)$ (not achieved in our current paper, as discussed later), global optimization of $P_2$ is also achieved. 

We will solve each
 $\mathbb{P}_{3}(y)$ by alternating optimizing (AO) $\boldsymbol{s}$ and $\boldsymbol{b},\boldsymbol{p},\boldsymbol{f}^{\textnormal{MS}},\boldsymbol{f}^{\textnormal{VU}}, T$. Section~\ref{Optimizingbp} optimizes $\boldsymbol{b},\boldsymbol{p},\boldsymbol{f}^{\textnormal{MS}},\boldsymbol{f}^{\textnormal{VU}},T $ given $\boldsymbol{s}$, while Section~\ref{Optimizings} optimizes $\boldsymbol{s}$ given $\boldsymbol{b},\boldsymbol{p},\boldsymbol{f}^{\textnormal{MS}},\boldsymbol{f}^{\textnormal{VU}},T $. AO means looping through these two steps until convergence; i.e., the relative difference between the objective-function values of consecutive iterations is no more than the error tolerance, \vspace{-5pt} as shown in  Line~\ref{P3convergence} of Algorithm~\ref{algo:MN}.
 
 % That means we first fix $\boldsymbol{s}$ and optimize $\boldsymbol{b},\boldsymbol{p},\boldsymbol{f}^{\textnormal{MS}},\boldsymbol{f}^{\textnormal{VU}}, T$. The obtained optimal $\boldsymbol{b_{\ast}},\boldsymbol{p_{\ast}},\boldsymbol{f_{\ast}}^{\textnormal{MS}},\boldsymbol{f_{\ast}}^{\textnormal{VU}}, T_{\ast}$ are then seen as the given parameters and used to optimize $\boldsymbol{s}$. By looping through these two steps until the $y$ value converges, the optimal solutions to Problem $\mathbb{P}_{3}$ can be obtained. Next, we will elaborate on how to perform alternate optimization in detail.

\subsection[Optimizing b, p, fMS, fVU, T given s for Problem P3(y)]{Optimizing $\boldsymbol{b},\boldsymbol{p},\boldsymbol{f}^{\textnormal{MS}},\boldsymbol{f}^{\textnormal{VU}},T $ given $\boldsymbol{s}$ for Problem $\mathbb{P}_{3}(y)$} \label{Optimizingbp} \label{AO1}

% \subsubsection{\textbf{Using our proposed transform}}

% \begin{talign}
% \textnormal{Problem $\mathbb{P}_{4}(y,\boldsymbol{s})$}:~\max_{\boldsymbol{b},\boldsymbol{p},\boldsymbol{f}^{\textnormal{MS}},\boldsymbol{f}^{\textnormal{VU}},T} \quad & \mathcal{U}(\boldsymbol{b},\boldsymbol{p},\boldsymbol{s}) - y \cdot (c_{\hspace{0.5pt}\textnormal{e}}\mathcal{E}(\boldsymbol{b},\boldsymbol{p},\boldsymbol{s},\boldsymbol{f}^{\textnormal{MS}},\boldsymbol{f}^{\textnormal{VU}})+c_{\hspace{0.8pt}\textnormal{t}} T) \\
% \textrm{s.t.} \quad &  \textnormal{(\ref{constraintbn})}, \textnormal{(\ref{constraintpn})}, \textnormal{(\ref{constraintfMS})}, \textnormal{(\ref{constraintfn})},  \textnormal{(\ref{constraintTtau})} . \nonumber
% \end{talign}

For Problem $\mathbb{P}_{3}(y)$, given $\boldsymbol{s}$, optimizing $\boldsymbol{b},\boldsymbol{p},\boldsymbol{f}^{\textnormal{MS}},\boldsymbol{f}^{\textnormal{VU}},T $ means the following optimization: 

\begin{talign}
&\textnormal{Problem $\mathbb{P}_{4}(y,\boldsymbol{s})$}: \nonumber \\
&\max_{\boldsymbol{b},\boldsymbol{p},\boldsymbol{f}^{\textnormal{MS}},\boldsymbol{f}^{\textnormal{VU}},T}~ F(\boldsymbol{b},\boldsymbol{p},\boldsymbol{f}^{\textnormal{MS}},\boldsymbol{f}^{\textnormal{VU}} \mid \boldsymbol{s}, y) - yc_{\hspace{0.5pt}\textnormal{e}} \cdot E^{\textnormal{Tx}}(\boldsymbol{b},\boldsymbol{p},\boldsymbol{s}) \label{P4obj} \\[-2pt]
&\textrm{s.t.}~   \textnormal{(\ref{constraintbn})}, \textnormal{(\ref{constraintpn})}, \textnormal{(\ref{constraintfMS})}, \textnormal{(\ref{constraintfn})},  \textnormal{(\ref{constraintTtau})} . \nonumber
\\[-2pt]
&\textnormal{where } F(\boldsymbol{b},\hspace{-2pt}\boldsymbol{p},\hspace{-2pt}\boldsymbol{f}^{\textnormal{MS}},\hspace{-2pt}\boldsymbol{f}^{\textnormal{VU}},\hspace{-2pt} T\mid \boldsymbol{s},\hspace{-2pt} y) \hspace{-2pt} = \hspace{-2pt} \mathcal{U}(\boldsymbol{b},\hspace{-2pt}\boldsymbol{p},\hspace{-2pt}\boldsymbol{s}) \hspace{-2pt} \nonumber \\
&~~~~~~~~~~~~~~~~~~~~~~~~~~~~~~~~~~~~-\hspace{-2pt} y \hspace{-2pt}\cdot\hspace{-2pt} [c_{\hspace{0.5pt}\textnormal{e}}\hspace{-2pt} \cdot \hspace{-2pt}(\sum\limits_{n\in\mathcal{N}}E_n^{\textnormal{MS}:\textnormal{Pro}}(s_n,\hspace{-2pt}f_n^{\textnormal{MS}})\hspace{-2pt}\nonumber \\
&~~~~~~~~~~~~~~~~~~~~~~~~~~~~~~~~~~~~+\hspace{-2pt}\sum\limits_{n \in \mathcal{N}}E_n^{\textnormal{VU}:\textnormal{Pro}}(s_n,\hspace{-2pt}f_n^{\textnormal{VU}}))\hspace{-2pt}+\hspace{-2pt}c_{\hspace{0.8pt}\textnormal{t}} T]  , \nonumber
\\[-2pt]
& \text{and } E^{\textnormal{Tx}}(\boldsymbol{b},\boldsymbol{p},\boldsymbol{s})  :=\sum_{n\in\mathcal{N}}E_n^{\textnormal{Tx}}(b_n,p_n,s_n) \nonumber \\
&~~~~~~~~~~~~~~~~~~~~=\sum_{n \in \mathcal{N}} \frac{(p_n +p_n^{\textnormal{cir}})  s_n\textcolor{black}{\mu_n}{\color{black}\Lambda_n}}{r_n(b_n,p_n)\textcolor{black}{\nu_n}} .\label{fractions}
\end{talign}

Eq.~(\ref{fractions}) has a  summation of \mbox{non-convex} ratios, which we address using our fractional programming technique of Section~\ref{opt-sum-of-ratios}, as detailed soon. Note that we cannot use the sum-of-ratios approach in~\cite{shen2018fractional}, since the objective function in~(\ref{P4obj}) includes not just the  sum of ratios, but also $F(\boldsymbol{b},\hspace{-2pt}\boldsymbol{p},\hspace{-2pt}\boldsymbol{f}^{\textnormal{MS}},\hspace{-2pt}\boldsymbol{f}^{\textnormal{VU}},\hspace{-2pt} T\mid \boldsymbol{s},\hspace{-2pt} y)$.

\subsection[Leveraging our fractional programming technique to solve P4(y, s)]{Leveraging our fractional programming technique to solve $\mathbb{P}_{4}(y,\boldsymbol{s})$\vspace{-5pt}} \label{FPsec}

% $F(\boldsymbol{b},\boldsymbol{p},\boldsymbol{f}^{\textnormal{MS}},\boldsymbol{f}^{\textnormal{VU}},T \mid \boldsymbol{s}, y)$ is jointly concave in $\boldsymbol{b},\boldsymbol{p},\boldsymbol{f}^{\textnormal{MS}},\boldsymbol{f}^{\textnormal{VU}},T$ given $\boldsymbol{s}, y$. 
We utilize our fractional programming technique of Section~\ref{opt-sum-of-ratios} to transform $\mathbb{P}_{4}$ into a series of $\mathbb{P}_{5}$:
%can find a stationary point.
% The optimal $p_n$ given $b_n, f_n^{\textnormal{VU}}, s_n$ is
% \begin{talign}
% p_n & = \frac{N_0b_n}{g_n}\Big(2^{\frac{d_n}{b_n \cdot (T - \frac{\mathcal{A}_n(s_n,\Lambda_n)}{f_n^{\textnormal{MS}}} - \frac{\mathcal{B}_n(s_n,\Lambda_n)}{f_n^{\textnormal{VU}}})}} - 1 \Big)  \nonumber  \\ & 
% \end{talign}
\begin{talign}
&\textnormal{$\mathbb{P}_{5}(\boldsymbol{z},y,\boldsymbol{s})$}:\hspace{-10pt}\max\limits_{\boldsymbol{b},\boldsymbol{p},\boldsymbol{f}^{\textnormal{MS}},\boldsymbol{f}^{\textnormal{VU}},T} \hspace{-5pt}  F(\boldsymbol{b},\hspace{-2pt}\boldsymbol{p},\hspace{-2pt}\boldsymbol{f}^{\textnormal{MS}},\hspace{-2pt}\boldsymbol{f}^{\textnormal{VU}},\hspace{-2pt}T  \mid \boldsymbol{s}, \hspace{-2pt}y) \nonumber \\
&- yc_{\hspace{0.5pt}\textnormal{e}} \hspace{-2pt}\cdot \hspace{-2pt}\sum_{n \in \mathcal{N}} \left\{[(p_n +p_n^{\textnormal{cir}})  s_n\textcolor{black}{\mu_n}{\color{black}\Lambda_n}]^2z_n \hspace{-2pt}+\hspace{-2pt} \frac{1}{4(r_n(b_n,p_n)\textcolor{black}{\nu_n})^2z_n}\right\} \label{eqP5} \\[-5pt]
&\textrm{s.t.} \quad  \textnormal{(\ref{constraintbn})}, \textnormal{(\ref{constraintpn})}, \textnormal{(\ref{constraintfMS})}, \textnormal{(\ref{constraintfn})},  \textnormal{(\ref{constraintTtau})} , 
\end{talign} 
where we introduce the auxiliary $\boldsymbol{z}:=[z_1,z_2,\ldots,z_N]$ with $z_n > 0$. We solve $\mathbb{P}_{5}(\boldsymbol{z},y,\boldsymbol{s})$ in Section~\ref{kktproblem5}.
% The objective function in ?? is jointly concave in $\boldsymbol{b},\boldsymbol{p},\boldsymbol{f}^{\textnormal{MS}},\boldsymbol{f}^{\textnormal{VU}},T$ given $\boldsymbol{s}, y$.

The process of using $\mathbb{P}_{5}(\boldsymbol{z},y,\boldsymbol{s})$ to solve $\mathbb{P}_{4}(y,\boldsymbol{s})$ is as follows. For ease of explanation, we denote $[\boldsymbol{b},\boldsymbol{p},\boldsymbol{f}^{\textnormal{MS}},\boldsymbol{f}^{\textnormal{VU}},T]$ by $\boldsymbol{\chi}$, and write the objective function of $\mathbb{P}_{5}(\boldsymbol{z},y,\boldsymbol{s})$ as $A(\boldsymbol{\chi})-\sum_{n \in \mathcal{N}}(B_n(\boldsymbol{\chi})z_n+\frac{C_n(\boldsymbol{\chi})}{z_n})$. Then starting from a feasible $\boldsymbol{\chi}^{(0)}$ at initialization, we set $z_n^{(0)}$ as $\sqrt{\frac{C_n(\boldsymbol{\chi}^{(0)})}{B_n(\boldsymbol{\chi}^{(0)})}}$ (i.e., optimizing the above objective function with respect to $\boldsymbol{z}$ given $\boldsymbol{\chi}=\boldsymbol{\chi}^{(0)}$). Then we solve $\mathbb{P}_{5}(\boldsymbol{z}^{(0)},y,\boldsymbol{s})$, denote the obtained solution as $\boldsymbol{\chi}^{(1)}$, set $z_n^{(1)}$ as $\sqrt{\frac{C_n(\boldsymbol{\chi}^{(1)})}{B_n(\boldsymbol{\chi}^{(1)})}}$. This process continues iteratively: in the $(\ell+1)$th iteration, $z_n^{(\ell)}$ is set as $\sqrt{\frac{C_n(\boldsymbol{\chi}^{(\ell)})}{B_n(\boldsymbol{\chi}^{(\ell)})}}$, and $\boldsymbol{\chi}^{(\ell+1)}$  is obtained from solving $\mathbb{P}_{5}(\boldsymbol{z}^{(\ell)},y,\boldsymbol{s})$. As explained in Section~\ref{FPsec}, the above process is alternating optimization and thus converges. We will discuss its performance in  Section~\ref{kktproblem5}.

% From~\cite{shen2018fractional}, the above process converges and does not lose optimality; i.e., under global optimization of each $P_3(y)$ (unfortunately not achieved in our  paper, as discussed in Section~\ref{sec: algorithm}), global optimization of $P_2$ is also achieved. 

\subsection[Optimizing s given b, p, fMS, fVU, T for Problem P3(y)]{Optimizing $\boldsymbol{s}$ given $\boldsymbol{b},\boldsymbol{p},\boldsymbol{f}^{\textnormal{MS}},\boldsymbol{f}^{\textnormal{VU}},T $ for Problem $\mathbb{P}_{3}(y)$}\label{Optimizings}

% We consider $\mathbb{S}_n$ in~(\ref{constraintsn}) as $[s_n^{\textnormal{min}}, s_n^{\textnormal{max}}]$ below. If $\mathbb{S}_n$ is discrete, one way is to consider a continuous interval for optimizing $s_n$ first and then map the optimized solution of $s_n$ to one of the discrete values.
% For continuous $\mathbb{S}_n$ above,

When $\mathbb{S}_n$ in~(\ref{constraintsn}) is $[s_n^{\textnormal{min}}, s_n^{\textnormal{max}}]$,
given $\boldsymbol{b},\boldsymbol{p},\boldsymbol{f}^{\textnormal{MS}},\boldsymbol{f}^{\textnormal{VU}},T $, optimizing $\boldsymbol{s}$ for $\mathbb{P}_{3}(y)$ means the following:
\begin{talign}
&\textnormal{Problem $\mathbb{P}_{6}(\boldsymbol{b},\boldsymbol{p},\boldsymbol{f}^{\textnormal{MS}},\boldsymbol{f}^{\textnormal{VU}},T,y)$}: \nonumber \\
&\max_{\boldsymbol{s}} \quad  \mathcal{U}(\boldsymbol{b},\boldsymbol{p},\boldsymbol{s}) -  y \cdot (c_{\hspace{0.5pt}\textnormal{e}}\mathcal{E}(\boldsymbol{b},\boldsymbol{p},\boldsymbol{s},\boldsymbol{f}^{\textnormal{MS}},\boldsymbol{f}^{\textnormal{VU}})+c_{\hspace{0.8pt}\textnormal{t}} T) \\[-5pt]
&\textrm{s.t.} \quad  s_n^{\textnormal{min}} \leq s_n \leq \min\{  s_n^{\textnormal{max}},\,v_n(b_n,p_n,f_n^{\textnormal{MS}},f_n^{\textnormal{VU}},T) \},\nonumber \\
&~~~~~~\forall n \in \mathcal{N},  \nonumber
\end{talign}
where $v_n(b_n,p_n,f_n^{\textnormal{MS}},f_n^{\textnormal{VU}},T)$ is defined as $s_n$ which makes $t_n(b_n,p_n,s_n,f_n^{\textnormal{MS}},f_n^{\textnormal{VU}})$ equal $T$.

Assuming $\mathcal{A}_n(s_n,\Lambda_n) $, $  \mathcal{B}_n(s_n,\Lambda_n)$, $\mathcal{F}_n(s_n,\Lambda_n) $, and $  \mathcal{G}_n(s_n,\Lambda_n)$ of Section~\ref{Delayenergy} to be convex in $s_n$ (which hold in our simulations in Section~\ref{sec: experiment}), $\mathbb{P}_{6}(\boldsymbol{b},\boldsymbol{p},\boldsymbol{f}^{\textnormal{MS}},\boldsymbol{f}^{\textnormal{VU}},T,y)$ is convex optimization for $\boldsymbol{s}$, for which the Karush--Kuhn--Tucker (KKT) conditions give a global optimum. Using the KKT conditions, we obtain that with $s_n^{\#}$ denoting the maximum point of the function  $V_n(s_n):=U_n(r_n(b_n,p_n),s_n)-  y c_{\hspace{0.5pt}\textnormal{e}}\times \big(\kappa^{\textnormal{MS}}{\color{black}\mathcal{F}_n(s_n,\Lambda_n)}({f_n^{\textnormal{MS}}})^2+\frac{(p_n +p_n^{\textnormal{cir}})  s_n\textcolor{black}{\mu_n}{\color{black}\Lambda_n}}{r_n(b_n,p_n)\textcolor{black}{\nu_n}}+\kappa_n^{\textnormal{VU}}{\color{black}\mathcal{G}_n(s_n,\Lambda_n)} (f_n^{\textnormal{VU}})^2\big)$ that is concave with respect to $s_n \in (0,\infty)$, the optimal solution of $s_n$ to  $\mathbb{P}_{6}(\boldsymbol{b},\boldsymbol{p},\boldsymbol{f}^{\textnormal{MS}},\boldsymbol{f}^{\textnormal{VU}},T,y)$ is  
\begin{talign}
\begin{cases}
\text{for continuous $\mathbb{S}_n=[s_n^{\textnormal{min}}, s_n^{\textnormal{max}}]$: }\\\widetilde{s}_n:= \max\{s_n^{\textnormal{min}} ,\min\{s_n^{\#}, ~s_n^{\textnormal{max}},~v_n(b_n,p_n,f_n^{\textnormal{MS}},f_n^{\textnormal{VU}},T)\}\}, \\  \text{for discrete $\mathbb{S}_n$: one of $s_n^{\textnormal{up}}$ and $s_n^{\textnormal{low}}$ which gives a higher }\nonumber \\[-5pt]
\text{~~~~~~~~~~~~~~~~~~~$V_n(\cdot)$, for $s_n^{\textnormal{up}}$ (resp., $s_n^{\textnormal{low}}$) denoting the} \\[-5pt] \text{~~~~~~~~~~~~~~~~~~~smallest (resp., largest) $s_n$ in $\mathbb{S}_n$ greater}\nonumber \\[-5pt] \text{~~~~~~~~~~~~~~~~~~~(resp., less) than $\widetilde{s}_n$.}
\end{cases} \label{setoptimalsn}
\end{talign}

\subsection{Putting the above together: Our Algorithm~\ref{algo:MN} on Page~\pageref{algo:MN}\vspace{-5pt}} \label{Putting}

Based on the above, we present Algorithm~\ref{algo:MN} on Page~\pageref{algo:MN} to solve $\mathbb{P}_{1}$. Algorithm~\ref{algo:MN} consists of three levels of iterations: the outermost iteration based on Dinkelbach's transform in Section~\ref{secDinkelbach}, the mid-level iteration for alternating optimization based on Sections~\ref{AO1} and~\ref{Optimizings}, and the innermost iteration using our fractional programming technique as discussed in Section~\ref{FPsec}. In Algorithm~\ref{algo:MN}'s pseudocode, Line 3 represents the outermost iteration based on Dinkelbach's transform, which solves a series of Problem $\mathbb{P}_{3}(y)$ for iteratively-updated $y$, in order to solve Problem $\mathbb{P}_{2}$ at convergence. Line 9 corresponding to
 the mid-level iteration is to alternating solve Problem $\mathbb{P}_{4}(y,\boldsymbol{s})$ and  Problem $\mathbb{P}_{6}(\boldsymbol{b},\boldsymbol{p},\boldsymbol{f}^{\textnormal{MS}},\boldsymbol{f}^{\textnormal{VU}},T,y)$, in order to resolve Problem $\mathbb{P}_{3}(y)$ at convergence. In Line 14, the innermost iteration is executed to solve Problem $\mathbb{P}_{5}(\boldsymbol{z},y,\boldsymbol{s})$ for iteratively-updated $\boldsymbol{z}$, in order to settle Problem $\mathbb{P}_{4}(y,\boldsymbol{s})$ at convergence.

% , as also noted by the comments in  Lines~\ref{outermost},~\ref{mid-level}, and~\ref{innermost} of Algorithm~\ref{algo:MN}.

We defer the solution quality and time complexity of  Algorithm~\ref{algo:MN}  to Section~\ref{sec: algorithm} after  explaining in Section~\ref{kktproblem5} below how each $\textnormal{$\mathbb{P}_{5}(\boldsymbol{z},y,\boldsymbol{s})$}$ in Line~\ref{solveP5} of Algorithm~\ref{algo:MN} is solved.\vspace{-10pt}

\section[Global Optimization of Problem P5]{Global Optimization of Problem $\mathbb{P}_{5}(\boldsymbol{z},y,\boldsymbol{s})$ in~(\ref{eqP5})\vspace{-5pt}} \label{kktproblem5}

%Solving the KKT conditions (\ref{Stationaritybn})--(\ref{Dualfeasibility}) for convex optimization of $\mathbb{P}_{5}(\boldsymbol{z},y,\boldsymbol{s})$

% \subsubsection[]{\textbf{KKT conditions for convex optimization of $\mathbb{P}_{5}(\boldsymbol{z},y,\boldsymbol{s})$}} 

The transmission rate $r_n(b_n,p_n)$ is  jointly concave in $b_n$ and $p_n$~\cite{zhou2022icdcs}. Then from the composition rule in Eq.~(3.11) of~\cite{boyd2004convex} and our Assumption~\ref{Uassumption:non-decreasing} on Page~\pageref{Uassumption:non-decreasing}, $\mathcal{U}(\boldsymbol{b},\boldsymbol{p},\boldsymbol{s})$ in Eq.~(\ref{eq:utility}) is jointly concave in $\boldsymbol{b}$ and $\boldsymbol{p}$. Thus, $\mathbb{P}_{5}(\boldsymbol{z},y,\boldsymbol{s})$ belongs to convex optimization.  The CVX tool~\cite{boyd2004convex} can be used to solve it. However, the worst-case complexity of global convex optimization grows exponentially with the problem size $N$ from Section~1.4.2 of~\cite{boyd2004convex}. Below we analyze the KKT conditions~\cite{boyd2004convex} to globally optimize $\mathbb{P}_{5}(\boldsymbol{z},y,\boldsymbol{s})$.

% The constraints \textnormal{(\ref{constraintbn})}, \textnormal{(\ref{constraintpn})}, \textnormal{(\ref{constraintfMS})}, \textnormal{(\ref{constraintfn})}, and  \textnormal{(\ref{constraintTtau})} are all convex. 

% We define $[\boldsymbol{b}^*,\boldsymbol{p}^*,(\boldsymbol{f}^{\textnormal{MS}})^*,(\boldsymbol{f}^{\textnormal{VU}})^*,T^*]$ as a globally optimal solution to Problem $\mathbb{P}_{5}$. 

\newcommand{\blackasymbol}{\rule{0.5em}{0.7em}}

The Lagrange function of $\mathbb{P}_{5}(\boldsymbol{z},y,\boldsymbol{s})$ is given below, where $\alpha, \beta, \gamma, \boldsymbol{\delta}, \boldsymbol{\zeta}$ denote  the multipliers:
\begin{talign}
&L_{\mathbb{P}_{5}}(\boldsymbol{b},\boldsymbol{p},\boldsymbol{f}^{\textnormal{MS}},\boldsymbol{f}^{\textnormal{VU}},T, \alpha,\beta,\gamma,\boldsymbol{\delta}  ,\boldsymbol{\zeta}  \mid \boldsymbol{z},y, \boldsymbol{s}) \nonumber \\ & = - F(\boldsymbol{b},\boldsymbol{p},\boldsymbol{f}^{\textnormal{MS}},\boldsymbol{f}^{\textnormal{VU}},T  \mid y, \boldsymbol{s}) \nonumber \\ & + yc_{\hspace{0.5pt}\textnormal{e}} \cdot \sum_{n \in \mathcal{N}} \left\{[(p_n +p_n^{\textnormal{cir}})  s_n\textcolor{black}{\mu_n}{\color{black}\Lambda_n}]^2z_n + \frac{1}{4(r_n(b_n,p_n)\textcolor{black}{\nu_n})^2z_n}\right\} \nonumber\\ &   + \alpha \cdot \big( \sum_{n \in \mathcal{N}} b_n - b_{\textnormal{max}}\big) + \beta \cdot \big( \sum_{n \in \mathcal{N}} p_n - p_{\textnormal{max}}\big) \nonumber \\ & + \gamma \cdot \big(\sum_{n \in \mathcal{N}} f_n^{\textnormal{MS}} - f_{\textnormal{max}}^{\textnormal{MS}}\big) +  \sum_{n \in \mathcal{N}} \big[ \delta_n \cdot (f_n^{\textnormal{VU}} -f_{n,\textnormal{max}}^{\textnormal{VU}} )\big]  \nonumber \\ & +  \sum_{n \in \mathcal{N}} \big[\zeta_n \cdot (t_n(b_n,p_n,s_n,f_n^{\textnormal{MS}},f_n^{\textnormal{VU}}) - T)\big]. 
\end{talign} 

Abbreviating $L_{\mathbb{P}_{5}}(\boldsymbol{b},\boldsymbol{p},\boldsymbol{f}^{\textnormal{MS}},\boldsymbol{f}^{\textnormal{VU}},T, \alpha,\beta,\gamma,\boldsymbol{\delta}  ,\boldsymbol{\zeta}  \mid \boldsymbol{z},y, \boldsymbol{s}),\boldsymbol{f}^{\textnormal{VU}},T  \mid \boldsymbol{s}, y)$, $U_n(r_n(b_n,p_n),s_n)$ and $r_n(b_n,p_n)$ as $L_{\mathbb{P}_{5}}$, $U_n$ and $r_n$ for simplicity, we present the KKT conditions of  $\mathbb{P}_{5}(\boldsymbol{z},y,\boldsymbol{s})$ as (\ref{Stationaritybn})-(\ref{Dualfeasibility}) below.   
% \begin{itemize}
% \item \textbf{Stationarity:} 

\hspace{-3pt}\textbf{\textbullet~Stationarity: }
\begin{talign} &\forall n \in \mathcal{N}:
\frac{\partial L_{\mathbb{P}_{5}}}{\partial b_n} = 0, \text{ meaning } \nonumber \\ & ~~~~~~~~~ -\frac{\partial U_n}{\partial r_n} \cdot \frac{\partial r_n}{\partial b_n}  - \frac{y c_{\hspace{0.5pt}\textnormal{e}}}{2{r_n}^3\textcolor{black}{{\nu_n}^2}z_n} \frac{\partial r_n}{\partial b_n} \nonumber \\& ~~~~~~~~~+ \alpha - \zeta_n 
 \frac{s_n\textcolor{black}{\mu_n}  {\color{black}\Lambda_n}}{{r_n}^2\textcolor{black}{\nu_n}}\frac{\partial r_n}{\partial b_n} = 0; \label{Stationaritybn} \\[-5pt]&  \forall n \hspace{-2pt} \in \hspace{-2pt} \mathcal{N}:\hspace{-2pt} \frac{\partial L_{\mathbb{P}_{5}}}{\partial p_n} \hspace{-2pt} =\hspace{-2pt}  0, \text{ meaning } \hspace{-2pt} \nonumber \\
 & ~~~~~~~~~-\frac{\partial U_n}{\partial r_n} \hspace{-2pt} \cdot\hspace{-2pt}  \frac{\partial r_n}{\partial p_n} \hspace{-2pt}  + \hspace{-2pt}  2y c_{\hspace{0.5pt}\textnormal{e}}z_n(s_n\textcolor{black}{\mu_n}{\color{black}\Lambda_n})^2(p_n \hspace{-2pt} +\hspace{-2pt} p_n^{\textnormal{cir}}) \hspace{-2pt}  \nonumber \\
 & ~~~~~~~~~-\hspace{-2pt}  \frac{y c_{\hspace{0.5pt}\textnormal{e}}}{2{r_n}^3\textcolor{black}{{\nu_n}^2}z_n}\frac{\partial r_n}{\partial p_n} \hspace{-2pt} +\hspace{-2pt}  \beta \hspace{-2pt} - \hspace{-2pt} \zeta_n 
 \frac{s_n\textcolor{black}{\mu_n}  {\color{black}\Lambda_n}}{{r_n}^2\textcolor{black}{\nu_n}}\frac{\partial r_n}{\partial p_n} \hspace{-2pt} =\hspace{-2pt}  0;\label{Stationaritypn} \\& \forall n \in \mathcal{N}:\frac{\partial L_{\mathbb{P}_{5}}}{\partial f_n^{\textnormal{MS}}} = 0, \text{ meaning } \nonumber \\ &~~~~~~~~~y c_{\hspace{0.5pt}\textnormal{e}} \cdot2 \kappa^{\textnormal{MS}} {\color{black}\mathcal{F}_n(s_n,\Lambda_n)}{f_n^{\textnormal{MS}}}  + \gamma - \zeta_n \frac{\mathcal{A}_n(s_n,\Lambda_n)}{(f_n^{\textnormal{MS}})^2}= 0;\label{partialLpartialfMS}\\&  \forall n \in \mathcal{N}:\frac{\partial L_{\mathbb{P}_{5}}}{\partial f_n^{\textnormal{VU}}} = 0, \text{ meaning } \nonumber \\ &~~~~~~~~~y c_{\hspace{0.5pt}\textnormal{e}} \cdot2 \kappa_n^{\textnormal{VU}} {\color{black}\mathcal{G}_n(s_n,\Lambda_n)}{f_n^{\textnormal{VU}}}   + \delta_n - \zeta_n  \frac{{\color{black}\mathcal{B}_n(s_n,\Lambda_n)}}{(f_n^{\textnormal{VU}})^2} = 0 ; \label{partialLpartialfVU}\\[-5pt]& \frac{\partial L_{\mathbb{P}_{5}}}{\partial T} = 0, \text{ meaning } \sum_{n \in \mathcal{N}} \zeta_n = y c_{\hspace{0.8pt}\textnormal{t}}. \label{partialLpartialT} \\
% \end{talign}
% % \item \textbf{Complementary slackness:}
% \begin{talign} 
&\textbf{\textbullet~Complementary slackness: } \nonumber \\ &\alpha \cdot \big( \sum_{n \in \mathcal{N}} b_n - b_{\textnormal{max}}\big) = 0; \label{Complementaryalpha} \\& \beta \cdot \big( \sum_{n \in \mathcal{N}} p_n - p_{\textnormal{max}}\big) = 0; \label{Complementarybeta}  \\&  \gamma \cdot \big(\sum_{n \in \mathcal{N}} f_n^{\textnormal{MS}} - f_{\textnormal{max}}^{\textnormal{MS}}\big)  = 0;\label{Complementarygamma} \\&   \delta_n \cdot (f_n^{\textnormal{VU}} -f_{n,\textnormal{max}}^{\textnormal{VU}} )= 0\textnormal{ for all }n \in \mathcal{N};  \label{Complementarydelta}\\& \zeta_n \cdot (t_n(b_n,p_n,s_n,f_n^{\textnormal{MS}},f_n^{\textnormal{VU}}) - T) = 0\textnormal{ for all }n \in \mathcal{N}. \label{Complementarytime} \\
% \end{talign}
% % \item \textbf{Primal feasibility:}
% \begin{talign} 
&\textbf{\textbullet~Primal feasibility: }\textnormal{(\ref{constraintbn})}, \textnormal{(\ref{constraintpn})}, \textnormal{(\ref{constraintfMS})}, \textnormal{(\ref{constraintfn})},  \textnormal{(\ref{constraintTtau})}. \label{Primalfeasibility} \\
% \end{talign}
% % \item \textbf{Dual feasibility:}
% \begin{talign} 
&\textbf{\textbullet~Dual feasibility: }\nonumber\\ 
&\begin{array}{l}
\text{(\ref{Dualfeasibility}a):~} \alpha \geq 0; ~~~\text{(\ref{Dualfeasibility}b):~}\beta \geq 0;~~~\text{(\ref{Dualfeasibility}c):~}\gamma\geq 0;   \\
\text{(\ref{Dualfeasibility}d):~}\delta_n\geq 0 \textnormal{ for all }n \in \mathcal{N};\text{(\ref{Dualfeasibility}e):~}\zeta_n\geq 0 \textnormal{ for all }n \in \mathcal{N}.
\end{array} \label{Dualfeasibility}
\end{talign}

~\vspace{-12pt}

We now analyze the KKT conditions of (\ref{Stationaritybn})--(\ref{Dualfeasibility}), to solve $\mathbb{P}_{5}(\textnormal{\footnotesize$\bigstar$})$, where ``$\textnormal{\footnotesize$\bigstar$}$'' denotes ``$\boldsymbol{z},y,\boldsymbol{s}$'' from now on for notation simplicity. Among the Lagrange multipliers, it is clear from (\ref{Stationaritybn}) that 
\begin{talign}
\alpha  > 0 .\label{Stationaritypn3alpha}
\end{talign}
%  while we are not sure about $\beta,\gamma,\boldsymbol{\delta} $. The reason for $\alpha  > 0  $ is that in~(\ref{Stationaritybn}), we have: 
%  \begin{itemize}
%      \item $\frac{\partial U_n}{\partial r_n} \geq 0$ since $U_n(r_n,s_n)$ is a \mbox{non-decreasing} function of $r_n$ from Assumption~\ref{Uassumption:non-decreasing} on Page~\pageref{Uassumption:non-decreasing},
%       \item $\frac{\partial r_n}{\partial b_n} > 0$ from $r_n(b_n, p_n) = b_n \log_2(1+\frac{g_np_n}{\sigma^2b_n})$ with $b_n>0$ and $p_n>0$ (specifically, $\frac{\partial r_n}{\partial b_n}$ is given by $
%     \log_2(1+\vartheta_n)
%      -\frac{\vartheta_n}{(1+\vartheta_n)\ln2}$ for ${\vartheta_n : = \frac{g_np_n}{\sigma_n^2b_n}}$, and it is straightforward to prove that $\log_2(1+x) - \frac{x}{(1+x) \ln 2}$ is positive   for $x>0$),
%      % as shown in Result ``(ii)'' in Appendix~\ref{seclemsolvepngivenalphaandbeta} of~\cite{full}),
%      \item  $\zeta_n\geq 0$ from (\ref{Dualfeasibility}e), and
%        \item  $r_n > 0, y>0, z_n >0, s_n >0$.
%  \end{itemize}
Then using (\ref{Stationaritypn3alpha}) in (\ref{Complementaryalpha}), we obtain 
\begin{talign}
 \sum_{n \in \mathcal{N}} b_n = b_{\textnormal{max}}.    \label{Complementaryalpha2}
\end{talign}
Thus, (\ref{Complementaryalpha}) \textnormal{(\ref{constraintbn})} (\ref{Dualfeasibility}a) can be replaced by (\ref{Stationaritypn3alpha}) (\ref{Complementaryalpha2}). Hence,
\begin{talign}
\begin{array}{l}\textnormal{the  KKT conditions of  Problem $\mathbb{P}_{5}(\textnormal{\footnotesize$\bigstar$})$, which include }   \\  \textnormal{(\ref{Stationaritybn})--(\ref{Complementarytime}), (\ref{Primalfeasibility}) (i.e., \textnormal{(\ref{constraintbn})}, \textnormal{(\ref{constraintpn})}, \textnormal{(\ref{constraintfMS})}, \textnormal{(\ref{constraintfn})},  \textnormal{(\ref{constraintTtau})}), and (\ref{Dualfeasibility}) } \\ 
\textnormal{(i.e., (\ref{Dualfeasibility}a), (\ref{Dualfeasibility}b), (\ref{Dualfeasibility}c), (\ref{Dualfeasibility}d), (\ref{Dualfeasibility}e))},    \\  \textnormal{can be expressed as $\mathcal{S}_{KKT}$, which denotes the collection of}    \\\textnormal{(\ref{Stationaritybn})--(\ref{partialLpartialT}),  (\ref{Complementarybeta})--(\ref{Complementarytime}), \textnormal{(\ref{constraintpn})}, \textnormal{(\ref{constraintfMS})}, \textnormal{(\ref{constraintfn})},  \textnormal{(\ref{constraintTtau})}, (\ref{Dualfeasibility}b), (\ref{Dualfeasibility}c)}, \\ \textnormal{(\ref{Dualfeasibility}d), (\ref{Dualfeasibility}e), (\ref{Stationaritypn3alpha}), and (\ref{Complementaryalpha2}).} \end{array} \nonumber \\ &  \label{KKTP5}
\end{talign}

 \textbf{Identifying a roadmap to compute the variables step-by-step.} \label{roadmapvariables} Given ``$\boldsymbol{z},y, \boldsymbol{s}$'' (denoted by ``$\textnormal{\small$\bigstar$}$'' below), we will find $\boldsymbol{b},\boldsymbol{p},\boldsymbol{f}^{\textnormal{MS}},\boldsymbol{f}^{\textnormal{VU}},T, \alpha,\beta,\gamma,\boldsymbol{\delta}  ,\boldsymbol{\zeta}$ to satisfy the KKT conditions $\mathcal{S}_{KKT}$ defined above.
 
 % (defined in (\ref{KKTP5})) of  Problem $\mathbb{P}_{5}(\textnormal{\footnotesize$\bigstar$})$.

 % \subsection{A roadmap to compute the variables step-by-step} \label{roadmapx}
 
 % which can be categorized as follows:
 % \begin{itemize}
 %     \item Equalities: (\ref{Stationaritybn})--(\ref{partialLpartialT}),  (\ref{Complementarybeta})--(\ref{Complementarytime}) and (\ref{Complementaryalpha}) (i.e.,  (\ref{Complementaryalpha2}));
 %       \item Inequalities: (\ref{Primalfeasibility}) (i.e., \textnormal{(\ref{constraintbn})}, \textnormal{(\ref{constraintpn})}, \textnormal{(\ref{constraintfMS})}, \textnormal{(\ref{constraintfn})},  \textnormal{(\ref{constraintTtau})}) and (\ref{Dualfeasibility}).
 % \end{itemize}

\addtocounter{equation}{1}
 \setcounter{counterstepone}{\value{equation}}
 
\addtocounter{equation}{1}
 \setcounter{counteryy}{\value{equation}}
 
\addtocounter{equation}{1} \setcounter{counterycycy}{\value{equation}}

\addtocounter{equation}{1} \setcounter{counteryty}{\value{equation}}

\addtocounter{equation}{1} \setcounter{counterycy}{\value{equation}}

\addtocounter{equation}{1} \setcounter{counterytty}{\value{equation}}

\addtocounter{equation}{1} \setcounter{counteryttt}{\value{equation}}

\addtocounter{equation}{1} \setcounter{countertyctt}{\value{equation}}

\addtocounter{equation}{1} \setcounter{counterty}{\value{equation}}

 \addtocounter{equation}{1} \setcounter{countertt}{\value{equation}}

We will partition the  KKT conditions given in $\mathcal{S}_{KKT}$ of~(\ref{KKTP5}) for  $\mathbb{P}_{5}(\textnormal{\footnotesize$\bigstar$})$ into the sets $\mathcal{S}_{1.1},  \mathcal{S}_{1.2.1}, \mathcal{S}_{1.2.2.1},$ $ \mathcal{S}_{1.2.2.2}, \mathcal{S}_{2.1}, \mathcal{S}_{2.2}$ defined below to enable a step-by-step approach, in order to solve for the variables: \label{step-by-step}
\begin{spacing}{1}
\begin{talign}
\begin{cases}\textnormal{\textbf{Step 1:} Considering  $\boldsymbol{\zeta}$ as an parameter, find $[\boldsymbol{b},\boldsymbol{p},\boldsymbol{f}^{\textnormal{MS}}$,}\\ \textnormal{$\boldsymbol{f}^{\textnormal{VU}},\alpha,\beta,\gamma,\boldsymbol{\delta}]$ satisfying } \textnormal{$\mathcal{S}_{1.1} \cup  \mathcal{S}_{1.2.1} \cup \mathcal{S}_{1.2.2.1} \cup \mathcal{S}_{1.2.2.2}$} \\ \textnormal{as functions of $[\boldsymbol{\zeta},\textnormal{\small$\bigstar$}]$ through  Steps 1.1 and 1.2 below, }\\  \textnormal{where} \mathcal{S}_{1.1} \cup  \mathcal{S}_{1.2.1} \cup \mathcal{S}_{1.2.2.1} \cup \mathcal{S}_{1.2.2.2} = \\ \bigg\{\begin{array}{l}
\textnormal{(\ref{Stationaritybn}), (\ref{Stationaritypn}),  (\ref{partialLpartialfMS}), (\ref{partialLpartialfVU}), (\ref{Complementarybeta}), (\ref{Complementarygamma}), (\ref{Complementarydelta}),}  \\[-5pt] \textnormal{(\ref{constraintpn})}, 
 \textnormal{(\ref{constraintfMS}), \textnormal{(\ref{constraintfn})},   (\ref{Dualfeasibility}b), (\ref{Dualfeasibility}c), (\ref{Dualfeasibility}d), (\ref{Stationaritypn3alpha}), (\ref{Complementaryalpha2})}\end{array}\bigg\}, 
\hfill \textnormal{(\number\value{counterstepone})\hspace{42pt}}\\ 
\textnormal{for $\mathcal{S}_{1.1}$, $\mathcal{S}_{1.2.1}$, $\mathcal{S}_{1.2.2.1}$ and $\mathcal{S}_{1.2.2.2}$ defined in~(\number\value{counteryy}),~(\number\value{counteryty}),}\\\textnormal{(\number\value{counterytty}) and~(\number\value{counteryttt}) below,} \\
\begin{cases}
\textnormal{\textbf{Step 1.1}: Find $\boldsymbol{f}^{\textnormal{MS}},\boldsymbol{f}^{\textnormal{VU}},\gamma,\boldsymbol{\delta}$ satisfying $\mathcal{S}_{1.1}$ as}\\ \textnormal{ functions of $[\boldsymbol{\zeta},\textnormal{\small$\bigstar$}]$:} \\ \mathcal{S}_{1.1}:=\big\{\textnormal{(\ref{partialLpartialfMS}), (\ref{partialLpartialfVU}), (\ref{Complementarygamma}), (\ref{Complementarydelta}), \textnormal{(\ref{constraintfMS})}, \textnormal{(\ref{constraintfn})}, (\ref{Dualfeasibility}c), (\ref{Dualfeasibility}d)}\big\}, 
\hfill \\ ~~~~~~~~~~~~~~~~~~~~~~~~~~~~~~~~~~~~~~~~~~~~~~~~~~~~~~~~~~~\textnormal{(\number\value{counteryy})\hspace{29pt}} 
\\
\textnormal{\textbf{Step 1.2}: Find $\boldsymbol{b},\boldsymbol{p},\alpha,\beta$ satisfying} \\ \textnormal{ $\mathcal{S}_{1.2.1} \cup \mathcal{S}_{1.2.2.1} \cup \mathcal{S}_{1.2.2.2}$ as functions  of $[\boldsymbol{\zeta},\textnormal{\small$\bigstar$}]$ through }\\ \textnormal{Steps 1.2.1, 1.2.2.1, and 1.2.2.2 below, where }\\ \mathcal{S}_{1.2.1} \cup \mathcal{S}_{1.2.2.1} \cup \mathcal{S}_{1.2.2.2}=\\\big\{\textnormal{(\ref{Stationaritybn}), (\ref{Stationaritypn}),  (\ref{Complementarybeta}), \textnormal{(\ref{constraintpn})}, (\ref{Dualfeasibility}b), (\ref{Stationaritypn3alpha}), (\ref{Complementaryalpha2})}\big\}, 
\hfill \textnormal{(\number\value{counterycycy})\hspace{55pt}} \\ \textnormal{for $\mathcal{S}_{1.2.1}$, $\mathcal{S}_{1.2.2.1}$ and $\mathcal{S}_{1.2.2.2}$ defined in~(\number\value{counteryty}),~(\number\value{counterytty})} \\ \textnormal{and~(\number\value{counteryttt}) below,} 
\\
\begin{cases}
\textnormal{\textbf{Step 1.2.1}: Considering $[\alpha,\beta]$ as parameters, find} \\ \textnormal{ $\boldsymbol{b},\boldsymbol{p}$  satisfying $\mathcal{S}_{1.2.1}$ as functions of $[\alpha,\beta,\boldsymbol{\zeta},\textnormal{\small$\bigstar$}]$,}\\ \textnormal{ for $\mathcal{S}_{1.2.1}:=\big\{\textnormal{(\ref{Stationaritybn}), 
(\ref{Stationaritypn})}\big\}$,}
\hfill 
\textnormal{(\number\value{counteryty})\hspace{15pt}}  
\\
\textnormal{\textbf{Step 1.2.2}: Using  results of Step 1.2.1, find $[\alpha,\beta]$} \\ \textnormal{ satisfying $\mathcal{S}_{1.2.2.1} \cup \mathcal{S}_{1.2.2.2}$ as functions  of}\\ \textnormal{ $[\boldsymbol{\zeta},\textnormal{\small$\bigstar$}]$ through Steps 1.2.2.1 and 1.2.2.2 below, }  
\\
\textnormal{where } \mathcal{S}_{1.2.2.1} \cup \mathcal{S}_{1.2.2.2}=\\ \big\{\textnormal{(\ref{Complementarybeta}), \textnormal{(\ref{constraintpn})}, (\ref{Dualfeasibility}b), (\ref{Stationaritypn3alpha}), (\ref{Complementaryalpha2})}\big\}, 
\hfill \textnormal{(\number\value{counterycy})\hspace{15pt}} \\ \textnormal{for $\mathcal{S}_{1.2.2.1}$ and $\mathcal{S}_{1.2.2.2}$ defined in~(\number\value{counterytty}) and~(\number\value{counteryttt}) below,}  
\\ \begin{cases}
\textnormal{\textbf{Step 1.2.2.1:} considering $\beta$ as a parameter, find $\alpha$} \\ \textnormal{satisfying $\mathcal{S}_{1.2.2.1}$ as a function of $[\beta,\boldsymbol{\zeta},\textnormal{\small$\bigstar$}]$, for}\\ \textnormal{ $\mathcal{S}_{1.2.2.1}:=\big\{\textnormal{(\ref{Stationaritypn3alpha}), (\ref{Complementaryalpha2})}\big\}$,}
\hfill 
\textnormal{(\number\value{counterytty})\hspace{1pt}} 
\\
\textnormal{\textbf{Step 1.2.2.2}: using  results of Step 1.2.2.1, find $\beta$ } \\ \textnormal{satisfying $\mathcal{S}_{1.2.2.2}$ as a function of $[\boldsymbol{\zeta},\textnormal{\small$\bigstar$}]$,}\\ \textnormal{for $\mathcal{S}_{1.2.2.2}:=\big\{\textnormal{(\ref{Complementarybeta})}, \textnormal{(\ref{constraintpn})}, \textnormal{(\ref{Dualfeasibility}b)}\big\}$,}\hfill 
\textnormal{(\number\value{counteryttt})\hspace{1pt}} \\
\end{cases}
\end{cases} 
\end{cases}\\
\textnormal{\textbf{Step 2:} Using  results of Steps 1.1 and 1.2, find $[T,\boldsymbol{\zeta}]$} \\ \textnormal{ satisfying $\mathcal{S}_{2.1} \cup \mathcal{S}_{2.2}$ as a function  of ``$\textnormal{\small$\bigstar$}$'' through Steps }\\ \textnormal{ 2.1 and 2.2 below, where $\mathcal{S}_{2.1} \cup \mathcal{S}_{2.2}:=$} \\ \textnormal{$\big\{\textnormal{(\ref{partialLpartialT}), (\ref{Complementarytime}),  (\ref{constraintTtau}), (\ref{Dualfeasibility}e)}\big\}$,}\hspace{115pt}\textnormal{(\number\value{countertyctt})~} \\ \textnormal{for $\mathcal{S}_{2.1}$ and $\mathcal{S}_{2.2}$ defined in~(\number\value{counterty}) and~(\number\value{countertt}) below,} 
\\ \begin{cases}
\textnormal{\textbf{Step 2.1}: considering $T$ as a parameter, find $\boldsymbol{\zeta}$ } \\ \textnormal{satisfying $\mathcal{S}_{2.1}$ as a function of $[T,\textnormal{\small$\bigstar$}]$,}\\ \textnormal{for $\mathcal{S}_{2.1}:=\big\{\textnormal{(\ref{Complementarytime}),  (\ref{constraintTtau}), (\ref{Dualfeasibility}e)}\big\}$,}\hfill 
\textnormal{(\number\value{counterty})\hspace{-14pt}} 
\\
\textnormal{\textbf{Step 2.2}: using  results of Steps 1.1 and 1.2, find $T$ } \\ \textnormal{satisfying $\mathcal{S}_{2.2}$ as a function of ``$\textnormal{\small$\bigstar$}$'',}   \\ \textnormal{for $\mathcal{S}_{2.2}:=\{\textnormal{(\ref{partialLpartialT})}\}$,} \hfill 
\textnormal{(\number\value{countertt})\hspace{-14pt}}  
\end{cases} 
\end{cases}  \nonumber  \\ & \label{roadmap} 
\end{talign}  
\end{spacing}

Steps 1.1, 1.2.1, 1.2.2.1, 1.2.2.2, 2.1, and 2.2 use $\mathcal{S}_{1.1} ,  \mathcal{S}_{1.2.1} , \mathcal{S}_{1.2.2.1} , \mathcal{S}_{1.2.2.2}, \mathcal{S}_{2.1} , $ and $ \mathcal{S}_{2.2}$, respectively. Each step uses a subset of the KKT conditions $\mathcal{S}_{KKT}$ (defined in (\ref{KKTP5})) of Problem  $\mathbb{P}_{5}(\textnormal{\footnotesize$\bigstar$})$, and all steps combined together utilize all the conditions, since it holds from (\number\value{counteryy}), (\number\value{counteryty}), (\number\value{counterytty}), (\number\value{counteryttt}),  (\number\value{counterty})  and (\number\value{countertt}) that
\begin{talign}\label{partition}
& \mathcal{S}_{1.1} \cup  \mathcal{S}_{1.2.1} \cup \mathcal{S}_{1.2.2.1} \cup \mathcal{S}_{1.2.2.2}\cup \mathcal{S}_{2.1} \cup \mathcal{S}_{2.2}  \\ \nonumber &=   \textnormal{KKT conditions $\mathcal{S}_{KKT}$ in (\ref{KKTP5}).} 
% \nonumber \\ &  \quad \textnormal{~i.e., }\begin{cases}
%      \textnormal{Equalities: (\ref{Stationaritybn})--(\ref{partialLpartialT}),  (\ref{Complementarybeta})--(\ref{Complementarytime}) and 
%      %(\ref{Complementaryalpha}) (i.e., 
%      (\ref{Complementaryalpha2})),}  \\
%  \textnormal{Inequalities: (\ref{Primalfeasibility}) (i.e., \textnormal{(\ref{constraintbn})}, \textnormal{(\ref{constraintpn})}, \textnormal{(\ref{constraintfMS})}, \textnormal{(\ref{constraintfn})},  \textnormal{(\ref{constraintTtau})}) and (\ref{Dualfeasibility}).}
%\end{cases}  
\end{talign}
% The result~(\ref{partition}) above means that
% For $\mathcal{S}_{1.1} ,  \mathcal{S}_{1.2.1} , \mathcal{S}_{1.2.2.1} , \mathcal{S}_{1.2.2.2}, \mathcal{S}_{2.1} , \mathcal{S}_{2.2}$ together form a partition of the KKT conditions given in~(\ref{KKTP5}) for  Problem $\mathbb{P}_{5}(\textnormal{\footnotesize$\bigstar$})$. 

% The roadmap of the steps on Page~\pageref{roadmap}  illustrates our approach in an intuitive way. To present our approach in a formal way, we need to 

We introduce notations to denote the computed results in the steps. The goal is to obtain  
\begin{talign}
\begin{array}{l}\textnormal{$[\boldsymbol{b}^{\#}(\textnormal{\small$\bigstar$}),\boldsymbol{p}^{\#}(\textnormal{\small$\bigstar$}),(\boldsymbol{f}^{\textnormal{MS}})^{\#}(\textnormal{\small$\bigstar$}),(\boldsymbol{f}^{\textnormal{VU}})^{\#}(\textnormal{\small$\bigstar$}),T^{\#}(\textnormal{\small$\bigstar$}), \alpha^{\#}(\textnormal{\small$\bigstar$})$,} \\ \textnormal{$\beta^{\#}(\textnormal{\small$\bigstar$}),\gamma^{\#}(\textnormal{\small$\bigstar$}),\boldsymbol{\delta}^{\#}(\textnormal{\small$\bigstar$})  ,\boldsymbol{\zeta}^{\#}(\textnormal{\small$\bigstar$})]$,} \textnormal{denoting a solution of} \\ \textnormal{$[\boldsymbol{b},\boldsymbol{p},\boldsymbol{f}^{\textnormal{MS}},\boldsymbol{f}^{\textnormal{VU}},T, \alpha,\beta,\gamma,\boldsymbol{\delta},\boldsymbol{\zeta}]$ to the KKT conditions }\\ \textnormal{$\mathcal{S}_{KKT}$ in (\ref{KKTP5});}\\ \textnormal{i.e., $[\boldsymbol{b}^{\#}(\textnormal{\small$\bigstar$}),\boldsymbol{p}^{\#}(\textnormal{\small$\bigstar$}),(\boldsymbol{f}^{\textnormal{MS}})^{\#}(\textnormal{\small$\bigstar$}),(\boldsymbol{f}^{\textnormal{VU}})^{\#}(\textnormal{\small$\bigstar$}),T^{\#}(\textnormal{\small$\bigstar$})]$} \\ \textnormal{denotes a global optimum to  $\mathbb{P}_{5}(\textnormal{\small$\bigstar$})$.}   \end{array} \label{sharpcondition}   
\end{talign}

% To better understand additional notations, we present them in several propositions below. For better 
%  understanding, 
 % we let the numbering of the propositions be consistent with the steps. Specifically, 
For X being 1.1, 1.2, 1.2.1, 1.2.2.1, 1.2.2.2, 2.1, or 2.2, Proposition X
 %1.1 (resp., 1.2.1, 1.2.2.1, etc.) 
  below is a formal presentation of Step X
  %1.1 (resp., 1.2.1, 1.2.2.1, etc.) 
  above. For better clarity, we also present the following table to help understand notations.

\begin{table*}[!h]
\caption{Notes for notations, where {``$\,\blackasymbol\,$'' denotes a wildcard symbol} hereinafter for convenience.\vspace{-4pt}}
\label{tab:my-table}
\centering\begin{tabular}{|l|l|}
\hline
Notations                                                                                                                                                                   & Notes                                                                              \\ \hline
$\textnormal{\small$\bigstar$}$                                                                                                                                             & Represent ``$\boldsymbol{z},y, \boldsymbol{s}$''                                                \\ \hline
$\acute{\blackasymbol}(\cdot)$ notations                                                                                                                                    & Defined in Propositions~1.1 and~1.2                                               \\ \hline
\begin{tabular}[c]{@{}l@{}}~\\[-12pt] $\blackasymbol^{\#}(\cdot)$, $\widetilde{\blackasymbol}(\cdot)$, $\breve{\blackasymbol}(\cdot)$, $\grave{\blackasymbol}(\cdot)$\end{tabular} & Defined in (\ref{sharpcondition}), and Propositions~1.2.1, 1.2.2.1, and 2.1, respectively \\ \hline
\end{tabular}
\vspace{-15pt}\end{table*}

\newtheorem*{proposition1.1}{Proposition 1.1}

\begin{proposition1.1} \label{proposition1.1x}  
We have the following results which formally explain Step 1.1 of Page~\pageref{roadmap}.\\(i) Given ``$\textnormal{\small$\bigstar$}$'', if in $\mathcal{S}_{1.1}$ defined in (\number\value{counteryy}), we substitute  $\boldsymbol{\zeta}$ with $\boldsymbol{\zeta}^{\#}(\textnormal{\small$\bigstar$})$ defined in~(\ref{sharpcondition}), then\newline $[\boldsymbol{f}^{\textnormal{MS}},\boldsymbol{f}^{\textnormal{VU}},\gamma,\boldsymbol{\delta}]$ satisfying $\mathcal{S}_{1.1}$ is $[(\boldsymbol{f}^{\textnormal{MS}})^{\#}(\textnormal{\small$\bigstar$}),(\boldsymbol{f}^{\textnormal{VU}})^{\#}(\textnormal{\small$\bigstar$}),\gamma^{\#}(\textnormal{\small$\bigstar$}),\boldsymbol{\delta}^{\#}(\textnormal{\small$\bigstar$})]$  defined in~(\ref{sharpcondition}). \newline
(ii) Given ``$\textnormal{\small$\bigstar$}$'' and $\boldsymbol{\zeta}$, let the solution of $[\boldsymbol{f}^{\textnormal{MS}},\boldsymbol{f}^{\textnormal{VU}},\gamma,\boldsymbol{\delta}]$ to $\mathcal{S}_{1.1}$ be$[\acute{\boldsymbol{f}}^{\textnormal{MS}}(\boldsymbol{\zeta}\mid \textnormal{\small$\bigstar$}),\acute{\boldsymbol{f}}^{\textnormal{VU}}(\boldsymbol{\zeta}\mid \textnormal{\small$\bigstar$}),\acute{\gamma}(\boldsymbol{\zeta}\mid \textnormal{\small$\bigstar$}),\acute{\boldsymbol{\delta}}(\boldsymbol{\zeta}\mid \textnormal{\small$\bigstar$})]$. Then
\begin{talign}
% \resizebox{0.93\hsize}{!}\hspace{-2pt}
&\textnormal{$[\acute{\boldsymbol{f}}^{\textnormal{MS}}(\boldsymbol{\zeta}^{\#}(\textnormal{\small$\bigstar$})\mid \textnormal{\small$\bigstar$}),\acute{\boldsymbol{f}}^{\textnormal{VU}}(\boldsymbol{\zeta}^{\#}(\textnormal{\small$\bigstar$})\mid \textnormal{\small$\bigstar$}),\acute{\gamma}(\boldsymbol{\zeta}^{\#}(\textnormal{\small$\bigstar$})\mid \textnormal{\small$\bigstar$})$,} \nonumber \\ &\textnormal{$\acute{\boldsymbol{\delta}}(\boldsymbol{\zeta}^{\#}(\textnormal{\small$\bigstar$})\mid \textnormal{\small$\bigstar$})]$ = $[(\boldsymbol{f}^{\textnormal{MS}})^{\#}(\textnormal{\small$\bigstar$}),(\boldsymbol{f}^{\textnormal{VU}})^{\#}(\textnormal{\small$\bigstar$}),\gamma^{\#}(\textnormal{\small$\bigstar$}),\boldsymbol{\delta}^{\#}(\textnormal{\small$\bigstar$})]$.}\label{eqproposition1.1x}  
\end{talign}
\end{proposition1.1}

\noindent
\textbf{Proof of Proposition~1.1:} Given ``$\textnormal{\small$\bigstar$}$'' and $\boldsymbol{\zeta}$, the conditions in $\mathcal{S}_{1.1}$ of (\number\value{counteryy}) are necessary and sufficient to decide $[\boldsymbol{f}^{\textnormal{MS}},\boldsymbol{f}^{\textnormal{VU}},\gamma,\boldsymbol{\delta}]$. Since setting $[\boldsymbol{\zeta},\boldsymbol{f}^{\textnormal{MS}},\boldsymbol{f}^{\textnormal{VU}},\gamma,\boldsymbol{\delta}]$ as  $[\boldsymbol{\zeta}^{\#}(\textnormal{\small$\bigstar$}),(\boldsymbol{f}^{\textnormal{MS}})^{\#}(\textnormal{\small$\bigstar$}),(\boldsymbol{f}^{\textnormal{VU}})^{\#}(\textnormal{\small$\bigstar$}),\gamma^{\#}(\textnormal{\small$\bigstar$}),\boldsymbol{\delta}^{\#}(\textnormal{\small$\bigstar$})]$ satisfies  $\mathcal{S}_{1.1}$ due to~(\ref{sharpcondition}), Results (i) and (ii) of Proposition~1.1 clearly hold. \qed

 \newtheorem*{proposition1.2}{Proposition 1.2}

\begin{proposition1.2} \label{proposition1.2}
We have the following results, which formally explain Step 1.2 of Page~\pageref{roadmap}.\\(i) Given ``$\textnormal{\small$\bigstar$}$'', if in $\mathcal{S}_{1.2.1} \cup \mathcal{S}_{1.2.2.1} \cup \mathcal{S}_{1.2.2.2}$ defined in~(\number\value{counterycycy}), we substitute  $\boldsymbol{\zeta}$ with $\boldsymbol{\zeta}^{\#}(\textnormal{\small$\bigstar$})$ defined in~(\ref{sharpcondition}),  then $[\boldsymbol{b},\boldsymbol{p},\alpha,\beta]$ satisfying $\mathcal{S}_{1.2.1} \cup \mathcal{S}_{1.2.2.1} \cup \mathcal{S}_{1.2.2.2}$ is $[\boldsymbol{b}^{\#}(\textnormal{\small$\bigstar$}),\boldsymbol{p}^{\#}(\textnormal{\small$\bigstar$}),\alpha^{\#}(\textnormal{\small$\bigstar$}),\beta^{\#}(\textnormal{\small$\bigstar$})]$ defined in~(\ref{sharpcondition}). \\ (ii)
 Given ``$\textnormal{\small$\bigstar$}$'' and $\boldsymbol{\zeta}$, let the solution of $[\boldsymbol{b},\boldsymbol{p},\alpha,\beta]$ to $\mathcal{S}_{1.2.1} \cup \mathcal{S}_{1.2.2.1} \cup \mathcal{S}_{1.2.2.2}$\\ be $[\acute{\boldsymbol{b}}(\boldsymbol{\zeta}\mid \textnormal{\small$\bigstar$}),\acute{\boldsymbol{p}}(\boldsymbol{\zeta}\mid \textnormal{\small$\bigstar$}),\acute{\alpha}(\boldsymbol{\zeta}\mid \textnormal{\small$\bigstar$}),\acute{\beta}(\boldsymbol{\zeta}\mid \textnormal{\small$\bigstar$})]$. Then
\begin{talign}
&\textnormal{$[\acute{\boldsymbol{b}}(\boldsymbol{\zeta}^{\#}(\textnormal{\small$\bigstar$})\mid \textnormal{\small$\bigstar$}),\acute{\boldsymbol{p}}(\boldsymbol{\zeta}^{\#}(\textnormal{\small$\bigstar$})\mid \textnormal{\small$\bigstar$}),\acute{\alpha}(\boldsymbol{\zeta}^{\#}(\textnormal{\small$\bigstar$})\mid \textnormal{\small$\bigstar$})$,}  \textnormal{$\acute{\beta}(\boldsymbol{\zeta}^{\#}(\textnormal{\small$\bigstar$})\mid \textnormal{\small$\bigstar$})]$} \nonumber \\ &= \textnormal{$[\boldsymbol{b}^{\#}(\textnormal{\small$\bigstar$}),\boldsymbol{p}^{\#}(\textnormal{\small$\bigstar$}),\alpha^{\#}(\textnormal{\small$\bigstar$}),\beta^{\#}(\textnormal{\small$\bigstar$})]$.}\label{eqproposition1.2v2}
\end{talign}
\end{proposition1.2}

\noindent
\textbf{Proof of Proposition~1.2:} Given ``$\textnormal{\small$\bigstar$}$'' and $\boldsymbol{\zeta}$, the conditions in $\mathcal{S}_{1.2.1} \cup \mathcal{S}_{1.2.2.1} \cup \mathcal{S}_{1.2.2.2}$ of~(\number\value{counterycycy}) are necessary and sufficient to decide $[\boldsymbol{b},\boldsymbol{p},\alpha,\beta]$. Since setting $[\boldsymbol{\zeta},\boldsymbol{b},\boldsymbol{p},\alpha,\beta]$ as  $[\boldsymbol{\zeta}^{\#}(\textnormal{\small$\bigstar$}),\boldsymbol{b}^{\#}(\textnormal{\small$\bigstar$}),\boldsymbol{p}^{\#}(\textnormal{\small$\bigstar$}),\alpha^{\#}(\textnormal{\small$\bigstar$}),\beta^{\#}(\textnormal{\small$\bigstar$})]$ satisfies $\mathcal{S}_{1.2.1} \cup \mathcal{S}_{1.2.2.1} \cup \mathcal{S}_{1.2.2.2}$ due to~(\ref{sharpcondition}), Results (i) and (ii) of Proposition~1.2 clearly hold. \qed
 
% Recall from Proposition~1.2 that
% given ``$\textnormal{\small$\bigstar$}$'' and $\boldsymbol{\zeta}$, the solution of $[\boldsymbol{b},\boldsymbol{p},\alpha,\beta]$ to $\mathcal{S}_{1.2}$ is \\$[\acute{\boldsymbol{b}}(\boldsymbol{\zeta}\mid \textnormal{\small$\bigstar$}),\acute{\boldsymbol{p}}(\boldsymbol{\zeta}\mid \textnormal{\small$\bigstar$}),\acute{\alpha}(\boldsymbol{\zeta}\mid \textnormal{\small$\bigstar$}),\acute{\beta}(\boldsymbol{\zeta}\mid \textnormal{\small$\bigstar$})]$, which we analyze below.  

\newtheorem*{proposition1.2.1}{Proposition 1.2.1}

\begin{proposition1.2.1} \label{proposition1.2.1}
We have the following result which formally explains Step 1.2.1 of Page~\pageref{roadmap}.\\ 
 (i) Given ``$\textnormal{\small$\bigstar$}$'' and $\boldsymbol{\zeta}$, if in $\mathcal{S}_{1.2.1}$ defined in (\number\value{counteryty}), we substitute  $[\alpha,\beta]$ with $[\acute{\alpha}(\boldsymbol{\zeta}\mid \textnormal{\small$\bigstar$}),\acute{\beta}(\boldsymbol{\zeta}\mid \textnormal{\small$\bigstar$})]$ defined in Proposition~1.2, then $[\boldsymbol{b},\boldsymbol{p}]$ satisfying $\mathcal{S}_{1.2.1}$ is $[\acute{\boldsymbol{b}}(\boldsymbol{\zeta}\mid \textnormal{\small$\bigstar$}),\acute{\boldsymbol{p}}(\boldsymbol{\zeta}\mid \textnormal{\small$\bigstar$})]$ defined in Proposition~1.2. 
% \begin{talign}
% \mathcal{S}_{1.2.1}:=\big\{\textnormal{(\ref{Stationaritybn}), (\ref{Stationaritypn})}\big\}. \label{setS1.1}   
% \end{talign} 
\\ (ii) Given ``$\textnormal{\small$\bigstar$}$'' and $[\alpha,\beta,\boldsymbol{\zeta}]$, let the solution of $[\boldsymbol{b},\boldsymbol{p}]$ to $\mathcal{S}_{1.2.1}$ be $[\widetilde{\boldsymbol{b}}(\alpha,\beta,\boldsymbol{\zeta}\mid \textnormal{\small$\bigstar$}),\widetilde{\boldsymbol{p}}(\alpha,\beta,\boldsymbol{\zeta}\mid \textnormal{\small$\bigstar$})]$. Then
\begin{talign}
&\textnormal{$[\widetilde{\boldsymbol{b}}(\acute{\alpha}(\boldsymbol{\zeta}\mid \textnormal{\small$\bigstar$}),\acute{\beta}(\boldsymbol{\zeta}\mid \textnormal{\small$\bigstar$}),\boldsymbol{\zeta}\mid \textnormal{\small$\bigstar$}),\widetilde{\boldsymbol{p}}(\acute{\alpha}(\boldsymbol{\zeta}\mid \textnormal{\small$\bigstar$}),\acute{\beta}(\boldsymbol{\zeta}\mid \textnormal{\small$\bigstar$}),\boldsymbol{\zeta}\mid \textnormal{\small$\bigstar$})]$} \nonumber \\  &= \textnormal{ $[\acute{\boldsymbol{b}}(\boldsymbol{\zeta}\mid \textnormal{\small$\bigstar$}),\acute{\boldsymbol{p}}(\boldsymbol{\zeta}\mid \textnormal{\small$\bigstar$})]$.}\label{eqproposition1.2.1}
\end{talign}
\end{proposition1.2.1}

\noindent
\textbf{Proof of Proposition~1.2.1:} Given ``$\textnormal{\small$\bigstar$}$'' and $[\alpha,\beta,\boldsymbol{\zeta}]$, the conditions in $\mathcal{S}_{1.2.1}$ of~(\number\value{counteryty}) are necessary and sufficient to decide $[\boldsymbol{b},\boldsymbol{p}]$. Since setting $[\boldsymbol{b},\boldsymbol{p},\alpha,\beta]$ as  $[\acute{\boldsymbol{b}}(\boldsymbol{\zeta}\mid \textnormal{\small$\bigstar$}),\acute{\boldsymbol{p}}(\boldsymbol{\zeta}\mid \textnormal{\small$\bigstar$}),\acute{\alpha}(\boldsymbol{\zeta}\mid \textnormal{\small$\bigstar$}),\acute{\beta}(\boldsymbol{\zeta}\mid \textnormal{\small$\bigstar$})]$ satisfies $\mathcal{S}_{1.2.1}$ by  the  $\acute{\blackasymbol}(\cdot)$ notations in Proposition~1.2, Results (i) and (ii) of Proposition~1.2.1 clearly hold.  \qed

\newtheorem*{proposition1.2.2}{Proposition 1.2.2}

\begin{proposition1.2.2} \label{proposition1.2.2}
We have the following result which formally explains Step 1.2.2 of Page~\pageref{roadmap}.\\ Given ``$\textnormal{\small$\bigstar$}$'' and $[\alpha,\beta,\boldsymbol{\zeta}]$, if in $\mathcal{S}_{1.2.2.1} \cup \mathcal{S}_{1.2.2.2}$ given in (\number\value{counterycy}), we substitute  $[\boldsymbol{b},\boldsymbol{p}]$ \\ with $[\widetilde{\boldsymbol{b}}(\alpha,\beta,\boldsymbol{\zeta}\mid \textnormal{\small$\bigstar$}),\widetilde{\boldsymbol{p}}(\alpha,\beta,\boldsymbol{\zeta}\mid \textnormal{\small$\bigstar$})]$ defined in Proposition~1.2.1, then $[\alpha,\beta]$ satisfying $\mathcal{S}_{1.2.2.1} \cup \mathcal{S}_{1.2.2.2}$  is $[\acute{\alpha}(\boldsymbol{\zeta}\mid \textnormal{\small$\bigstar$}),\acute{\beta}(\boldsymbol{\zeta}\mid \textnormal{\small$\bigstar$})]$ defined in Proposition~1.2.
% \begin{talign}
% \mathcal{S}_5:=\big\{\textnormal{(\ref{Complementarybeta}), (\ref{Complementaryalpha2}), \textnormal{(\ref{constraintbn})}, \textnormal{(\ref{constraintpn})}, (\ref{Dualfeasibility}a), (\ref{Dualfeasibility}b)}\big\}. \label{setS5}   
% \end{talign}
\end{proposition1.2.2}

\noindent
\textbf{Proof of Proposition~1.2.2:} Given ``$\textnormal{\small$\bigstar$}$'' and $[\alpha,\beta,\boldsymbol{\zeta}]$, if in $\mathcal{S}_{1.2.2.1} \cup \mathcal{S}_{1.2.2.2}$ of (\number\value{counterycy}), we substitute  $[\boldsymbol{b},\boldsymbol{p}]$  with $[\widetilde{\boldsymbol{b}}(\alpha,\beta,\boldsymbol{\zeta}\mid \textnormal{\small$\bigstar$}),\widetilde{\boldsymbol{p}}(\alpha,\beta,\boldsymbol{\zeta}\mid \textnormal{\small$\bigstar$})]$ defined in Proposition~1.2.1, the conditions in $\mathcal{S}_{1.2.2.1} \cup \mathcal{S}_{1.2.2.2}$ of (\number\value{counterycy}) are necessary and sufficient to decide $[\alpha,\beta]$. Since setting $[\boldsymbol{b},\boldsymbol{p},\alpha,\beta]$ as \\ $[\widetilde{\boldsymbol{b}}(\acute{\alpha}(\boldsymbol{\zeta}\mid \textnormal{\small$\bigstar$}),\acute{\beta}(\boldsymbol{\zeta}\mid \textnormal{\small$\bigstar$}),\boldsymbol{\zeta}\mid \textnormal{\small$\bigstar$}),\widetilde{\boldsymbol{p}}(\acute{\alpha}(\boldsymbol{\zeta}\mid \textnormal{\small$\bigstar$}),\acute{\beta}(\boldsymbol{\zeta}\mid \textnormal{\small$\bigstar$}),\boldsymbol{\zeta}\mid \textnormal{\small$\bigstar$}),\acute{\alpha}(\boldsymbol{\zeta}\mid \textnormal{\small$\bigstar$}),\acute{\beta}(\boldsymbol{\zeta}\mid \textnormal{\small$\bigstar$})]$ \\ (i.e., $[\acute{\boldsymbol{b}}(\boldsymbol{\zeta}\mid \textnormal{\small$\bigstar$}),\acute{\boldsymbol{p}}(\boldsymbol{\zeta}\mid \textnormal{\small$\bigstar$}),\acute{\alpha}(\boldsymbol{\zeta}\mid \textnormal{\small$\bigstar$}),\acute{\beta}(\boldsymbol{\zeta}\mid \textnormal{\small$\bigstar$})]$ according to~(\ref{eqproposition1.2.1})) satisfies $\mathcal{S}_{1.2.2.1} \cup \mathcal{S}_{1.2.2.2}$ by the 
 definition of the  $\acute{\blackasymbol}(\cdot)$ notations in Proposition~1.2, Proposition~1.2.2 clearly follows. \qed

Despite Proposition~1.2.2, simultaneously solving for $[\alpha,\beta]$ to $\mathcal{S}_{1.2.2.1} \cup \mathcal{S}_{1.2.2.2}$ is challenging. Instead, we solve for $\alpha$ as a function of $\beta$ first and then decide $\beta$ in Propositions~1.2.2.1 and~1.2.2.2 below.

\newtheorem*{proposition1.2.2.1}{Proposition 1.2.2.1}

\begin{proposition1.2.2.1} \label{proposition1.2.2.1}
We have the following results which formally explain Step 1.2.2.1 of Page~\pageref{roadmap}.\\(i) Given ``$\textnormal{\small$\bigstar$}$'' and $\boldsymbol{\zeta}$, if in $\mathcal{S}_{1.2.2.1}$ defined in (\number\value{counterytty}), we substitute  $\boldsymbol{b}$ with  $\widetilde{\boldsymbol{b}}(\alpha,\acute{\beta}(\boldsymbol{\zeta}\mid \textnormal{\small$\bigstar$}),\boldsymbol{\zeta}\mid \textnormal{\small$\bigstar$})$, then $\alpha$ satisfying $\mathcal{S}_{1.2.2.1}$  is $\acute{\alpha}(\boldsymbol{\zeta}\mid \textnormal{\small$\bigstar$})$, where the  $\acute{\blackasymbol}(\cdot)$ and $\widetilde{\blackasymbol}(\cdot)$ notations are defined in Propositions~1.2 and 1.2.1.
% \begin{talign}
% \mathcal{S}_{1.2.2.1}:=\big\{\textnormal{(\ref{Complementaryalpha2}), \textnormal{(\ref{constraintbn})},  (\ref{Dualfeasibility}a)}\big\}. \label{setS6}   
% \end{talign}
\\(ii) Given ``$\textnormal{\small$\bigstar$}$'' and $\boldsymbol{\zeta}$, if in $\mathcal{S}_{1.2.2.1}$, we substitute  $\boldsymbol{b}$  with $\widetilde{\boldsymbol{b}}(\alpha,\beta,\boldsymbol{\zeta}\mid \textnormal{\small$\bigstar$})$ defined in Proposition~1.2.1, let the solution of $\alpha$ to $\mathcal{S}_{1.2.2.1}$ be $\breve{\alpha}(\beta,\iffalse T,\fi\boldsymbol{\zeta}\mid \textnormal{\small$\bigstar$})$. Then
\begin{talign}
&\textnormal{$\breve{\alpha}(\acute{\beta}(\boldsymbol{\zeta}\mid \textnormal{\small$\bigstar$}),\iffalse T,\fi\boldsymbol{\zeta}\mid \textnormal{\small$\bigstar$})$ equals $\acute{\alpha}(\boldsymbol{\zeta}\mid \textnormal{\small$\bigstar$})$, and}\label{eqproposition1.2.2.1v1} \\
% \end{talign}
% and
% \begin{talign}
&\textnormal{$[\widetilde{\boldsymbol{b}}(\breve{\alpha}(\acute{\beta}(\boldsymbol{\zeta}\mid \textnormal{\small$\bigstar$}),\iffalse T,\fi\boldsymbol{\zeta}\mid \textnormal{\small$\bigstar$}),\acute{\beta}(\boldsymbol{\zeta}\mid \textnormal{\small$\bigstar$}),\boldsymbol{\zeta}\mid \textnormal{\small$\bigstar$}),\widetilde{\boldsymbol{p}}(\breve{\alpha}(\acute{\beta}(\boldsymbol{\zeta}\mid \textnormal{\small$\bigstar$}),\iffalse T,\fi\boldsymbol{\zeta}\mid \textnormal{\small$\bigstar$}),$} \nonumber \\ &\textnormal{$\acute{\beta}(\boldsymbol{\zeta}\mid \textnormal{\small$\bigstar$}),\boldsymbol{\zeta}\mid \textnormal{\small$\bigstar$})]$ equals $[\acute{\boldsymbol{b}}(\boldsymbol{\zeta}\mid \textnormal{\small$\bigstar$}),\acute{\boldsymbol{p}}(\boldsymbol{\zeta}\mid \textnormal{\small$\bigstar$})]$.}\label{eqproposition1.2.2.1v2}
\end{talign}
\end{proposition1.2.2.1}

\noindent
\textbf{Proof of Proposition~1.2.2.1:} Given ``$\textnormal{\small$\bigstar$}$'' and $\boldsymbol{\zeta}$, if in $\mathcal{S}_{1.2.2.1}$ defined in (\number\value{counterytty}), we substitute  $\boldsymbol{b}$ with  $\widetilde{\boldsymbol{b}}(\alpha,\acute{\beta}(\boldsymbol{\zeta}\mid \textnormal{\small$\bigstar$}),\boldsymbol{\zeta}\mid \textnormal{\small$\bigstar$})$, then  the conditions in $\mathcal{S}_{1.2.2.1}$ of (\number\value{counterytty}) are necessary and sufficient to decide $\alpha$. Since setting $\alpha$ as $\acute{\alpha}(\boldsymbol{\zeta}\mid \textnormal{\small$\bigstar$})$ and setting  $\boldsymbol{b}$ as   $\widetilde{\boldsymbol{b}}(\acute{\alpha}(\boldsymbol{\zeta}\mid \textnormal{\small$\bigstar$}),\acute{\beta}(\boldsymbol{\zeta}\mid \textnormal{\small$\bigstar$}),\boldsymbol{\zeta}\mid \textnormal{\small$\bigstar$})$ (i.e., $\acute{\boldsymbol{b}}(\boldsymbol{\zeta}\mid \textnormal{\small$\bigstar$})$ according to~(\ref{eqproposition1.2.1})) satisfies $\mathcal{S}_{1.2.2.1}$ by the 
 definition of the  $\acute{\blackasymbol}(\cdot)$ notations in Proposition~1.2, Proposition~1.2.2.1 clearly follows. In particular, after we have~(\ref{eqproposition1.2.2.1v1}), we further obtain~(\ref{eqproposition1.2.2.1v2}) from (\ref{eqproposition1.2.1}) and~(\ref{eqproposition1.2.2.1v1}).\qed

\newtheorem*{proposition1.2.2.2}{Proposition 1.2.2.2}

\begin{proposition1.2.2.2} \label{proposition1.2.2.2}
We have the following result which formally explains Step 1.2.2.2 of Page~\pageref{roadmap}.\\Given ``$\textnormal{\small$\bigstar$}$'', if in $\mathcal{S}_{1.2.2.2}$ defined in (\number\value{counteryttt}), we substitute  $\boldsymbol{p}$ with $\widetilde{\boldsymbol{p}}(\breve{\alpha}(\beta,\iffalse T,\fi\boldsymbol{\zeta}\mid \textnormal{\small$\bigstar$}),\beta,\boldsymbol{\zeta}\mid \textnormal{\small$\bigstar$})$, then $\beta$ satisfying $\mathcal{S}_{1.2.2.2}$  is $\acute{\beta}(\boldsymbol{\zeta}\mid \textnormal{\small$\bigstar$})$,  where   $\breve{\blackasymbol}(\cdot)$, $\widetilde{\blackasymbol}(\cdot)$,  and $\acute{\blackasymbol}(\cdot)$ notations are defined in Propositions~1.2.2.1, 1.2.1, and 1.2.
% \begin{talign}
% \mathcal{S}_{1.2.2.2}:=\big\{\textnormal{(\ref{Complementarybeta})}, \textnormal{(\ref{constraintpn})},\textnormal{(\ref{Dualfeasibility}b)}\big\}. \label{setS5}   
% \end{talign}
\end{proposition1.2.2.2}

\noindent
\textbf{Proof of Proposition~1.2.2.2:} Given ``$\textnormal{\small$\bigstar$}$'', if in $\mathcal{S}_{1.2.2.2}$ defined in (\number\value{counteryttt}), we substitute  $\boldsymbol{p}$ with $\widetilde{\boldsymbol{p}}(\breve{\alpha}(\beta,\iffalse T,\fi\boldsymbol{\zeta}\mid \textnormal{\small$\bigstar$}),\beta,\boldsymbol{\zeta}\mid \textnormal{\small$\bigstar$})$, then the    conditions in $\mathcal{S}_{1.2.2.2}$ of (\number\value{counteryttt}) are necessary and sufficient to decide $\beta$. Since setting $\beta$ as $\acute{\beta}(\boldsymbol{\zeta}\mid \textnormal{\small$\bigstar$})$ and setting $\boldsymbol{p}$ as $\widetilde{\boldsymbol{p}}(\breve{\alpha}(\acute{\beta}(\boldsymbol{\zeta}\mid \textnormal{\small$\bigstar$}),\iffalse T,\fi\boldsymbol{\zeta}\mid \textnormal{\small$\bigstar$}),\acute{\beta}(\boldsymbol{\zeta}\mid \textnormal{\small$\bigstar$}),\boldsymbol{\zeta}\mid \textnormal{\small$\bigstar$})$ (i.e., $\acute{\boldsymbol{p}}(\boldsymbol{\zeta}\mid \textnormal{\small$\bigstar$})$ based on~(\ref{eqproposition1.2.2.1v1}) and~(\ref{eqproposition1.2.2.1v2})) satisfies $\mathcal{S}_{1.2.2.2}$ by the  $\acute{\blackasymbol}(\cdot)$ notations in Proposition~1.2, Proposition~1.2.2.2 clearly follows. \qed

 \newtheorem*{proposition2}{Proposition 2}

\begin{proposition2} \label{proposition2}
We have the following result which formally explains Step 2 of Page~\pageref{roadmap}.\\
Given ``$\textnormal{\small$\bigstar$}$'', if in $\mathcal{S}_{2.1} \cup \mathcal{S}_{2.2}$ defined in (\number\value{countertyctt}), we substitute  $[\boldsymbol{b},\boldsymbol{p},\boldsymbol{f}^{\textnormal{MS}},\boldsymbol{f}^{\textnormal{VU}},\alpha,\beta,\gamma,\boldsymbol{\delta}]$  with \\$[\acute{\boldsymbol{b}}(\boldsymbol{\zeta}\mid \textnormal{\small$\bigstar$}),\acute{\boldsymbol{p}}(\boldsymbol{\zeta}\mid \textnormal{\small$\bigstar$}),\acute{\boldsymbol{f}}^{\textnormal{MS}}(\boldsymbol{\zeta}\mid \textnormal{\small$\bigstar$}),\acute{\boldsymbol{f}}^{\textnormal{VU}}(\boldsymbol{\zeta}\mid \textnormal{\small$\bigstar$}),\acute{\alpha}(\boldsymbol{\zeta}\mid \textnormal{\small$\bigstar$}),\acute{\beta}(\boldsymbol{\zeta}\mid \textnormal{\small$\bigstar$}),\acute{\gamma}(\boldsymbol{\zeta}\mid \textnormal{\small$\bigstar$}),\acute{\boldsymbol{\delta}}(\boldsymbol{\zeta}\mid \textnormal{\small$\bigstar$})]$ defined in Propositions~1.1 and~1.2, then $[\boldsymbol{\zeta},T]$ satisfying $\mathcal{S}_{2.1} \cup \mathcal{S}_{2.2}$  is $[\boldsymbol{\zeta}^{\#}(\textnormal{\small$\bigstar$}),T^{\#}(\textnormal{\small$\bigstar$})]$ defined in~(\ref{sharpcondition}).
% \begin{talign}
% \mathcal{S}_{2.1} \cup \mathcal{S}_{2.2}:=\big\{\textnormal{(\ref{partialLpartialT}), (\ref{Complementarytime}),  (\ref{constraintTtau}), (\ref{Dualfeasibility}e)}\big\}. \label{setS3}   
% \end{talign} 
\end{proposition2}

Despite Proposition~2, simultaneously solving for $[\boldsymbol{\zeta},T]$ to $\mathcal{S}_{1.2.2.1} \cup \mathcal{S}_{1.2.2.2}$ is challenging. Instead, we solve for $\zeta$ as a function of $T$ first and then decide $T$ in Propositions~2.1 and~2.2 below.

 \newtheorem*{proposition21}{Proposition 2.1}

\begin{proposition21}
We have the following result which formally explains Step 2.1 of Page~\pageref{roadmap}.\\
(i) Given ``$\textnormal{\small$\bigstar$}$'', if in $\mathcal{S}_{2.1}$ defined in (\number\value{counterty}), we substitute  $[T,\boldsymbol{b},\boldsymbol{p},\boldsymbol{f}^{\textnormal{MS}},\boldsymbol{f}^{\textnormal{VU}}]$ with \\ $[T^{\#}(\textnormal{\small$\bigstar$}),\acute{\boldsymbol{b}}(\boldsymbol{\zeta}\mid \textnormal{\small$\bigstar$}),\acute{\boldsymbol{p}}(\boldsymbol{\zeta}\mid \textnormal{\small$\bigstar$}),\acute{\boldsymbol{f}}^{\textnormal{MS}}(\boldsymbol{\zeta}\mid \textnormal{\small$\bigstar$}),\acute{\boldsymbol{f}}^{\textnormal{VU}}(\boldsymbol{\zeta}\mid \textnormal{\small$\bigstar$})]$, then $\boldsymbol{\zeta}$ satisfying $\mathcal{S}_{2.1}$  is $\boldsymbol{\zeta}^{\#}(\textnormal{\small$\bigstar$})$, where the  $\blackasymbol^{\#}(\cdot)$ and $\acute{\blackasymbol}(\cdot)$ notations are defined in (\ref{sharpcondition}) and Proposition~1.2 respectively.
% \begin{talign}
% \mathcal{S}_{1.2.2.1}:=\big\{\textnormal{(\ref{Complementaryalpha2}), \textnormal{(\ref{constraintbn})},  (\ref{Dualfeasibility}a)}\big\}. \label{setS6}   
% \end{talign}
\\(ii) Given ``$\textnormal{\small$\bigstar$}$'', if in $\mathcal{S}_{2.1}$ defined in (\number\value{counterty}), we substitute  $[\boldsymbol{b},\boldsymbol{p},\boldsymbol{f}^{\textnormal{MS}},\boldsymbol{f}^{\textnormal{VU}}]$ with \\ $[\acute{\boldsymbol{b}}(\boldsymbol{\zeta}\mid \textnormal{\small$\bigstar$}),\acute{\boldsymbol{p}}(\boldsymbol{\zeta}\mid \textnormal{\small$\bigstar$}),\acute{\boldsymbol{f}}^{\textnormal{MS}}(\boldsymbol{\zeta}\mid \textnormal{\small$\bigstar$}),\acute{\boldsymbol{f}}^{\textnormal{VU}}(\boldsymbol{\zeta}\mid \textnormal{\small$\bigstar$})]$,  let the solution of $\boldsymbol{\zeta}$ to $\mathcal{S}_{2.1}$ be $\grave{\boldsymbol{\zeta}}(T\mid \textnormal{\small$\bigstar$})$. Then
\begin{talign}
&{\textnormal{$\grave{\boldsymbol{\zeta}}(T^{\#}(\textnormal{\small$\bigstar$})\mid \textnormal{\small$\bigstar$})$ equals $\boldsymbol{\zeta}^{\#}(\textnormal{\small$\bigstar$})$, and}}\label{eqproposition2.1v1} \\
% \end{talign}
% and\vspace{-10pt}
% \begin{talign}
&[\acute{\boldsymbol{b}}(\grave{\boldsymbol{\zeta}}(T^{\#}(\textnormal{{\tiny$\bigstar$}})|\textnormal{{\tiny$\bigstar$}})|\textnormal{{\tiny$\bigstar$}}),\acute{\boldsymbol{p}}(\grave{\boldsymbol{\zeta}}(T^{\#}(\textnormal{{\tiny$\bigstar$}})|\textnormal{{\tiny$\bigstar$}})|\textnormal{{\tiny$\bigstar$}}),\acute{\boldsymbol{f}}^{\textnormal{MS}}(\grave{\boldsymbol{\zeta}}(T^{\#}(\textnormal{{\tiny$\bigstar$}})|\textnormal{{\tiny$\bigstar$}})|\textnormal{{\tiny$\bigstar$}}),\nonumber \\
&\acute{\boldsymbol{f}}^{\textnormal{VU}}(\grave{\boldsymbol{\zeta}}(T^{\#}(\textnormal{{\tiny$\bigstar$}})|\textnormal{{\tiny$\bigstar$}})|\textnormal{{\tiny$\bigstar$}}), \acute{\alpha}(\grave{\boldsymbol{\zeta}}(T^{\#}(\textnormal{{\tiny$\bigstar$}})|\textnormal{{\tiny$\bigstar$}})|\textnormal{{\tiny$\bigstar$}}),\acute{\beta}(\grave{\boldsymbol{\zeta}}(T^{\#}(\textnormal{{\tiny$\bigstar$}})|\textnormal{{\tiny$\bigstar$}})|\textnormal{{\tiny$\bigstar$}}), \nonumber \\
&\acute{\gamma}(\grave{\boldsymbol{\zeta}}(T^{\#}(\textnormal{{\tiny$\bigstar$}})|\textnormal{{\tiny$\bigstar$}})|\textnormal{{\tiny$\bigstar$}}),\acute{\boldsymbol{\delta}}(\grave{\boldsymbol{\zeta}}(T^{\#}(\textnormal{{\tiny$\bigstar$}})|\textnormal{{\tiny$\bigstar$}})|\textnormal{{\tiny$\bigstar$}})] \nonumber\\
&\textnormal{equals $[\boldsymbol{b}^{\#}(\textnormal{\small$\bigstar$}),\boldsymbol{p}^{\#}(\textnormal{\small$\bigstar$}),(\boldsymbol{f}^{\textnormal{MS}})^{\#}(\textnormal{\small$\bigstar$}),(\boldsymbol{f}^{\textnormal{VU}})^{\#}(\textnormal{\small$\bigstar$}),$}\nonumber \\
&\textnormal{$\alpha^{\#}(\textnormal{\small$\bigstar$}),\beta^{\#}(\textnormal{\small$\bigstar$}),\gamma^{\#}(\textnormal{\small$\bigstar$}),\boldsymbol{\delta}^{\#}(\textnormal{\small$\bigstar$})]$.}\label{eqproposition2.1v2}
\end{talign}\end{proposition21}

\noindent
\textbf{Proof of Proposition~2.1:} Given ``$\textnormal{\small$\bigstar$}$'', if in $\mathcal{S}_{2.1}$ defined in (\number\value{counterty}), we substitute  $[T,\boldsymbol{b},\boldsymbol{p},\boldsymbol{f}^{\textnormal{MS}},\boldsymbol{f}^{\textnormal{VU}}]$ with $[T^{\#}(\textnormal{\small$\bigstar$}),\acute{\boldsymbol{b}}(\boldsymbol{\zeta}\mid \textnormal{\small$\bigstar$}),\acute{\boldsymbol{p}}(\boldsymbol{\zeta}\mid \textnormal{\small$\bigstar$}),\acute{\boldsymbol{f}}^{\textnormal{MS}}(\boldsymbol{\zeta}\mid \textnormal{\small$\bigstar$}),\acute{\boldsymbol{f}}^{\textnormal{VU}}(\boldsymbol{\zeta}\mid \textnormal{\small$\bigstar$})]$, then  the conditions in $\mathcal{S}_{2.1}$ of (\number\value{counterty}) are necessary and sufficient to decide $\boldsymbol{\zeta}$. Since setting $[\boldsymbol{\zeta},T,\boldsymbol{b},\boldsymbol{p},\boldsymbol{f}^{\textnormal{MS}},\boldsymbol{f}^{\textnormal{VU}}]$ as $[\boldsymbol{\zeta}^{\#}(\textnormal{\small$\bigstar$}),T^{\#}(\textnormal{\small$\bigstar$}),\acute{\boldsymbol{b}}(\boldsymbol{\zeta}^{\#}(\textnormal{\small$\bigstar$})\mid \textnormal{\small$\bigstar$}),\acute{\boldsymbol{p}}(\boldsymbol{\zeta}^{\#}(\textnormal{\small$\bigstar$})\mid \textnormal{\small$\bigstar$}),\acute{\boldsymbol{f}}^{\textnormal{MS}}(\boldsymbol{\zeta}^{\#}(\textnormal{\small$\bigstar$})\mid \textnormal{\small$\bigstar$}),\acute{\boldsymbol{f}}^{\textnormal{VU}}(\boldsymbol{\zeta}^{\#}(\textnormal{\small$\bigstar$})\mid \textnormal{\small$\bigstar$})]$ (i.e., $[\boldsymbol{\zeta}^{\#}(\textnormal{\small$\bigstar$}),T^{\#}(\textnormal{\small$\bigstar$}),\boldsymbol{b}^{\#}(\textnormal{\small$\bigstar$}),\boldsymbol{p}^{\#}(\textnormal{\small$\bigstar$}),(\boldsymbol{f}^{\textnormal{MS}})^{\#}(\textnormal{\small$\bigstar$}),(\boldsymbol{f}^{\textnormal{VU}})^{\#}(\textnormal{\small$\bigstar$})]$ according to~(\ref{eqproposition1.1x}) and~(\ref{eqproposition1.2v2})) satisfies $\mathcal{S}_{2.1}$ by the $\blackasymbol^{\#}(\cdot)$ notations in (\ref{sharpcondition}), Proposition~2.1 clearly follows. \qed

\newtheorem*{proposition22}{Proposition 2.2}

\begin{proposition22}
We have the following result which formally explains Step 2.2 of Page~\pageref{roadmap}.\\
Given ``$\textnormal{\small$\bigstar$}$'', if in $\mathcal{S}_{2.2}$ defined in (\number\value{countertt}), we substitute  $[\boldsymbol{\zeta},\boldsymbol{b},\boldsymbol{p},\boldsymbol{f}^{\textnormal{MS}},\boldsymbol{f}^{\textnormal{VU}}]$ with \\$[\grave{\boldsymbol{\zeta}}(T\mid \textnormal{\small$\bigstar$}),\acute{\boldsymbol{b}}(\grave{\boldsymbol{\zeta}}(T\mid \textnormal{\small$\bigstar$})\mid \textnormal{\small$\bigstar$}),\acute{\boldsymbol{p}}(\grave{\boldsymbol{\zeta}}(T\mid \textnormal{\small$\bigstar$})\mid \textnormal{\small$\bigstar$}),\acute{\boldsymbol{f}}^{\textnormal{MS}}(\grave{\boldsymbol{\zeta}}(T\mid \textnormal{\small$\bigstar$})\mid \textnormal{\small$\bigstar$}),\acute{\boldsymbol{f}}^{\textnormal{VU}}(\grave{\boldsymbol{\zeta}}(T\mid \textnormal{\small$\bigstar$})\mid \textnormal{\small$\bigstar$})]
$, then $T$ satisfying $\mathcal{S}_{2.2}$  is $T^{\#}(\textnormal{\small$\bigstar$})$, where $\acute{\blackasymbol}(\cdot)$, $\grave{\blackasymbol}(\cdot)$, and $\blackasymbol^{\#}(\cdot)$ notations are  in Proposition~1.2, Proposition~2.1, and (\ref{sharpcondition}).
\end{proposition22}

\noindent
\textbf{Proof of Proposition~2.2:} Given ``$\textnormal{\small$\bigstar$}$'', if in $\mathcal{S}_{2.2}$ defined in (\number\value{countertt}), we substitute  $[\boldsymbol{\zeta},\boldsymbol{b},\boldsymbol{p},\boldsymbol{f}^{\textnormal{MS}},\boldsymbol{f}^{\textnormal{VU}}]$ with \\$[\grave{\boldsymbol{\zeta}}(T\mid \textnormal{\small$\bigstar$}),\acute{\boldsymbol{b}}(\grave{\boldsymbol{\zeta}}(T\mid \textnormal{\small$\bigstar$})\mid \textnormal{\small$\bigstar$}),\acute{\boldsymbol{p}}(\grave{\boldsymbol{\zeta}}(T\mid \textnormal{\small$\bigstar$})\mid \textnormal{\small$\bigstar$}),\acute{\boldsymbol{f}}^{\textnormal{MS}}(\grave{\boldsymbol{\zeta}}(T\mid \textnormal{\small$\bigstar$})\mid \textnormal{\small$\bigstar$}),\acute{\boldsymbol{f}}^{\textnormal{VU}}(\grave{\boldsymbol{\zeta}}(T\mid \textnormal{\small$\bigstar$})\mid \textnormal{\small$\bigstar$})]
$, then  the conditions in $\mathcal{S}_{2.2}$ of (\number\value{countertt}) are necessary and sufficient to decide $T$. Since setting $[\boldsymbol{\zeta},T,\boldsymbol{b},\boldsymbol{p},\boldsymbol{f}^{\textnormal{MS}},\boldsymbol{f}^{\textnormal{VU}}]$ as \\$[\grave{\boldsymbol{\zeta}}(T^{\#}(\textnormal{\small$\bigstar$})\mid \textnormal{\small$\bigstar$}),\acute{\boldsymbol{b}}(\grave{\boldsymbol{\zeta}}(T^{\#}(\textnormal{\small$\bigstar$})\mid \textnormal{\small$\bigstar$})\mid \textnormal{\small$\bigstar$}),\acute{\boldsymbol{p}}(\grave{\boldsymbol{\zeta}}(T^{\#}(\textnormal{\small$\bigstar$})\mid \textnormal{\small$\bigstar$})\mid \textnormal{\small$\bigstar$}),\acute{\boldsymbol{f}}^{\textnormal{MS}}(\grave{\boldsymbol{\zeta}}(T^{\#}(\textnormal{\small$\bigstar$})\mid \textnormal{\small$\bigstar$})\mid \textnormal{\small$\bigstar$}),\acute{\boldsymbol{f}}^{\textnormal{VU}}(\grave{\boldsymbol{\zeta}}(T^{\#}(\textnormal{\small$\bigstar$})\mid \textnormal{\small$\bigstar$})\mid \textnormal{\small$\bigstar$})]
$ (i.e., $[\boldsymbol{\zeta}^{\#}(\textnormal{\small$\bigstar$}),T^{\#}(\textnormal{\small$\bigstar$}),\boldsymbol{b}^{\#}(\textnormal{\small$\bigstar$}),\boldsymbol{p}^{\#}(\textnormal{\small$\bigstar$}),(\boldsymbol{f}^{\textnormal{MS}})^{\#}(\textnormal{\small$\bigstar$}),(\boldsymbol{f}^{\textnormal{VU}})^{\#}(\textnormal{\small$\bigstar$})]$ according to~(\ref{eqproposition1.1x})~(\ref{eqproposition1.2v2}) and~(\ref{eqproposition2.1v1})) satisfies $\mathcal{S}_{2.2}$ by the 
 definition of the $\blackasymbol^{\#}(\cdot)$ notations in (\ref{sharpcondition}), Proposition~2.2 clearly follows. \qed

\begin{figure*}
\begin{talign}
& \small\textnormal{\textbullet~Computing $[\boldsymbol{b}^{\#}(\textnormal{\small$\bigstar$}),\boldsymbol{p}^{\#}(\textnormal{\small$\bigstar$}),(\boldsymbol{f}^{\textnormal{MS}})^{\#}(\textnormal{\small$\bigstar$}),(\boldsymbol{f}^{\textnormal{VU}})^{\#}(\textnormal{\small$\bigstar$}),T^{\#}(\textnormal{\small$\bigstar$})]$}\nonumber \\ &\small\textnormal{which denotes a globally optimal solution to Problem $\mathbb{P}_{5}(\textnormal{\small$\bigstar$})$ as defined in~(\ref{sharpcondition})}  \nonumber \\ & \stackrel{\textnormal{Via (\ref{eqproposition2.1v2})}}{\Longleftrightarrow}   \begin{cases} \small\underline{\textnormal{\textbullet~\textbf{Algorithm} $\text{1.1}$:~Computing $[\acute{\boldsymbol{f}}^{\textnormal{MS}}(\boldsymbol{\zeta}\mid \textnormal{\small$\bigstar$}),\acute{\boldsymbol{f}}^{\textnormal{VU}}(\boldsymbol{\zeta}\mid \textnormal{\small$\bigstar$}),\acute{\gamma}(\boldsymbol{\zeta}\mid \textnormal{\small$\bigstar$}),\acute{\boldsymbol{\delta}}(\boldsymbol{\zeta}\mid \textnormal{\small$\bigstar$})]$ defined in Proposition~1.1,}}  \nonumber \\  \small\underline{\textnormal{\textbullet~\textbf{Algorithm} $\text{1.2}$:~Computing $[\acute{\boldsymbol{b}}(\boldsymbol{\zeta}\mid \textnormal{\small$\bigstar$}),\acute{\boldsymbol{p}}(\boldsymbol{\zeta}\mid \textnormal{\small$\bigstar$})]$ defined in Proposition~1.2,}} \nonumber \\  \small\stackrel{\textnormal{Via (\ref{eqproposition1.2.1})}}{\Longleftrightarrow} \textnormal{\textbullet~Computing $[\widetilde{\boldsymbol{b}}(\acute{\alpha}(\boldsymbol{\zeta}\mid \textnormal{\small$\bigstar$}),\acute{\beta}(\boldsymbol{\zeta}\mid \textnormal{\small$\bigstar$}),\boldsymbol{\zeta}\mid \textnormal{\small$\bigstar$}),\widetilde{\boldsymbol{p}}(\acute{\alpha}(\boldsymbol{\zeta}\mid \textnormal{\small$\bigstar$}),\acute{\beta}(\boldsymbol{\zeta}\mid \textnormal{\small$\bigstar$}),\boldsymbol{\zeta}\mid \textnormal{\small$\bigstar$})]$,} \nonumber \\   \Longleftrightarrow \begin{cases}
\small\underline{\textnormal{\textbullet~\textbf{Algorithm} $\text{1.2.1}$:~Computing $[\widetilde{\boldsymbol{b}}(\alpha,\beta,\boldsymbol{\zeta}\mid \textnormal{\small$\bigstar$}),\widetilde{\boldsymbol{p}}(\alpha,\beta,\boldsymbol{\zeta}\mid \textnormal{\small$\bigstar$})]$ defined in Proposition~1.2.1,}}  \\  \small\textnormal{\textbullet~Computing $[\acute{\alpha}(\boldsymbol{\zeta}\mid \textnormal{\small$\bigstar$}),\acute{\beta}(\boldsymbol{\zeta}\mid \textnormal{\small$\bigstar$})]$ defined in Proposition~1.2} \\  \stackrel{\substack{\textnormal{Via (\ref{eqproposition1.2.2.1v1}) and} \\
\textnormal{Proposition 1.2.2.2}}}{\Longleftrightarrow} \begin{cases}
\small\underline{\textnormal{\textbullet~\textbf{Algorithm} $\text{1.2.2.1}$: }} \begin{array}{l} 
\small\underline{\textnormal{Computing $\breve{\alpha}(\beta,\iffalse T,\fi\boldsymbol{\zeta}\mid \textnormal{\small$\bigstar$})$ defined in Proposition~1.2.2.1}} \\
\small\textnormal{(which needs Algorithm $\text{1.2.1}$  above),}\end{array} \\
\small\underline{\textnormal{\textbullet~\textbf{Algorithm} $\text{1.2.2.2}$:}}~\begin{array}{l}\small\underline{\textnormal{use Algorithm $\text{1.2.2.1}$  to compute $\acute{\beta}(\boldsymbol{\zeta}\mid \textnormal{\small$\bigstar$})$}}\\
\small{\textnormal{according to Proposition~1.2.2.2.}}\end{array}
\end{cases}
\end{cases} \nonumber \\    \small\underline{\textnormal{\textbullet~\textbf{Algorithm} $\text{2.1}$:~Computing $\grave{\boldsymbol{\zeta}}(T\mid \textnormal{\small$\bigstar$})$ defined in Proposition 2.1}}\textnormal{, which needs}\nonumber \\ \small\textnormal{$[\acute{\boldsymbol{b}}(\boldsymbol{\zeta}\mid \textnormal{\small$\bigstar$}),\acute{\boldsymbol{p}}(\boldsymbol{\zeta}\mid \textnormal{\small$\bigstar$}),\acute{\boldsymbol{f}}^{\textnormal{MS}}(\boldsymbol{\zeta}\mid \textnormal{\small$\bigstar$}),\acute{\boldsymbol{f}}^{\textnormal{VU}}(\boldsymbol{\zeta}\mid \textnormal{\small$\bigstar$})]$; i.e., Algorithms $\text{1.1}, \text{1.2.1}, \text{1.2.2.1}, \text{1.2.2.2}$  above,} \nonumber \\ \small\underline{\textnormal{\textbullet~\textbf{Algorithm} $\text{2.2}$:~Computing $T^{\#}(\textnormal{\small$\bigstar$})$ according to Proposition 2.2,}} \\
\small\textnormal{which needs Algorithm $\text{2.1}$ as well as Algorithms $\text{1.1}, \text{1.2.1}, \text{1.2.2.1}, \text{1.2.2.2}$ above.}   \vspace{-10pt}
\end{cases}
\end{talign}
\vspace{0pt}\caption{Our procedure to solve Problem $\mathbb{P}_{5}(\textnormal{\small$\bigstar$})$.\vspace{-16pt}}\label{fig1}\end{figure*}

Based on the above propositions, Fig.~\ref{fig1} and Algorithm~\ref{algo:P5} present our procedure to solve  $\mathbb{P}_{5}(\textnormal{\small$\bigstar$})$. For clarity, the function ``Alg-Solve-$x(\cdot)$''  is to compute $x(\cdot)$; e.g., $\textnormal{Alg-Solve-}T^{\#}(\textnormal{\small$\bigstar$})$ obtains $T^{\#}(\textnormal{\small$\bigstar$})$. \vspace{10pt}

\renewcommand{\thealgocf}{A\arabic{algocf}}

 %\begin{minipage}{1\linewidth}
\begin{algorithm}
       
\caption{Given $\boldsymbol{z},y,\boldsymbol{s}$ (written as ``$\textnormal{\small$\bigstar$}$'' below), find a globally optimal solution to Problem $\mathbb{P}_{5}(\textnormal{\small$\bigstar$})$, denoted as $[\boldsymbol{b}^{\#}(\textnormal{\small$\bigstar$}),\boldsymbol{p}^{\#}(\textnormal{\small$\bigstar$}),(\boldsymbol{f}^{\textnormal{MS}})^{\#}(\textnormal{\small$\bigstar$}),(\boldsymbol{f}^{\textnormal{VU}})^{\#}(\textnormal{\small$\bigstar$}),T^{\#}(\textnormal{\small$\bigstar$})]$.}
\label{algo:P5}

Use $ \textnormal{Alg-Solve-}T^{\#}(\textnormal{\small$\bigstar$})$ of Algorithm 2.2 on Page~\pageref{alg2.2} to obtain $T^{\#}(\textnormal{\small$\bigstar$})$;

Use Algorithm 2.1 on Page~\pageref{alg2.1} to obtain $\grave{\boldsymbol{\zeta}}(T^{\#}(\textnormal{\small$\bigstar$})\mid \textnormal{\small$\bigstar$})$, which is $\boldsymbol{\zeta}^{\#}(\textnormal{\small$\bigstar$})$ from~(\ref{eqproposition2.1v1});

\mbox{Use Algorithm 1.2 on Page~\pageref{alg1.2} to obtain} $[\acute{\boldsymbol{b}}(\boldsymbol{\zeta}^{\#}(\textnormal{\small$\bigstar$})\mid \textnormal{\small$\bigstar$}), \acute{\boldsymbol{p}}(\boldsymbol{\zeta}^{\#}(\textnormal{\small$\bigstar$})\mid \textnormal{\small$\bigstar$}),\acute{\alpha}(\boldsymbol{\zeta}^{\#}(\textnormal{\small$\bigstar$})\mid \textnormal{\small$\bigstar$})]$, which equals $[\boldsymbol{b}^{\#}(\textnormal{\small$\bigstar$}),\boldsymbol{p}^{\#}(\textnormal{\small$\bigstar$})]$ from~(\ref{eqproposition1.2v2}); \textit{//Comment: Algorithm 1.2 uses Algorithms 1.2.1, 1.2.2.1, and 1.2.2.2.}

Use Algorithm 1.1 on Page~\pageref{alg1.1} to obtain $[\acute{\boldsymbol{f}}^{\textnormal{MS}}(\boldsymbol{\zeta}^{\#}(\textnormal{\small$\bigstar$})\mid \textnormal{\small$\bigstar$}),\acute{\boldsymbol{f}}^{\textnormal{VU}}(\boldsymbol{\zeta}^{\#}(\textnormal{\small$\bigstar$})\mid \textnormal{\small$\bigstar$})]$, which equals $[(\boldsymbol{f}^{\textnormal{MS}})^{\#}(\textnormal{\small$\bigstar$}),(\boldsymbol{f}^{\textnormal{VU}})^{\#}(\textnormal{\small$\bigstar$})]$ from~(\ref{eqproposition1.1x});

\end{algorithm}

~\vspace{-5pt}

For X being 1.1, 1.2, 1.2.1, 1.2.2.1, 1.2.2.2, 2.1, or 2.2,
Algorithm X will be presented for the computation in Proposition~X, as shown in Fig. 3 and explained in detail below.
% From the above, we will establish the following to solve Problem $\mathbb{P}_{5}(\textnormal{\small$\bigstar$})$: Algorithms \text{1.1}, \text{1.2.1}, \text{1.2.2.1}, \text{1.2.2.2}, \text{2.1}, \text{2.2}, which we detail below. 

% For better organization, we use a bullet to indicate the beginning of each computation part.

%  \textbf{Explaining how we decide the order of solving for different variables.} blabla

%  \textbf{Defining notations for Steps 2.1--2.5.} Table~\ref{talbenotation}.

% \subsubsection[]{\textbf{Solving for Table~\ref{talbenotation}'s notations used for Steps 2.1--2.5}} \label{solvetalbenotation}~

% We now ?. For clarity, we explicitly write the 
%  section number before each part. 

 \textbf{Algorithm $\text{1.1}$:~Computing $[\acute{\boldsymbol{f}}^{\textnormal{MS}}(\boldsymbol{\zeta}\mid \textnormal{\small$\bigstar$}),\acute{\boldsymbol{f}}^{\textnormal{VU}}(\boldsymbol{\zeta}\mid \textnormal{\small$\bigstar$}),\acute{\gamma}(\boldsymbol{\zeta}\mid \textnormal{\small$\bigstar$}),\acute{\boldsymbol{\delta}}(\boldsymbol{\zeta}\mid \textnormal{\small$\bigstar$})]$ defined in Proposition~1.1 using $\mathcal{S}_{1.1}$ of (\number\value{counteryy}).} \label{algexplain1.1} Among $\mathcal{S}_{1.1}$, $[\acute{\boldsymbol{f}}^{\textnormal{MS}}(\boldsymbol{\zeta}\mid \textnormal{\small$\bigstar$}),\acute{\gamma}(\boldsymbol{\zeta}\mid \textnormal{\small$\bigstar$})]$ is $[\boldsymbol{f}^{\textnormal{MS}},\gamma]$ satisfying (\ref{partialLpartialfMS})  (\ref{Complementarygamma})  \textnormal{(\ref{constraintfMS})}  (\ref{Dualfeasibility}c), while $[\acute{\boldsymbol{f}}^{\textnormal{VU}}(\boldsymbol{\zeta}\mid \textnormal{\small$\bigstar$}),\acute{\boldsymbol{\delta}}(\boldsymbol{\zeta}\mid \textnormal{\small$\bigstar$})]$ is $[\boldsymbol{f}^{\textnormal{VU}},\boldsymbol{\delta}]$ satisfying  (\ref{partialLpartialfVU})  (\ref{Complementarydelta})   \textnormal{(\ref{constraintfn})} (\ref{Dualfeasibility}d). From (\ref{partialLpartialfMS}) and (\ref{partialLpartialfVU}), with $\text{PositiveRoot}(\mathbb{E})$ for an equation $\mathbb{E}$ denoting the positive root of $\mathbb{E}$, we have
\begin{talign}
&f_n^{\textnormal{MS}} \nonumber \\ &= \text{PositiveRoot}(\zeta_n \frac{\mathcal{A}_n(s_n,\Lambda_n)}{x^2} - y c_{\hspace{0.5pt}\textnormal{e}} \cdot2 \kappa^{\textnormal{MS}} \mathcal{F}_n(s_n,\Lambda_n)x = \gamma), \label{fnmseq} \\[-8pt] &f_n^{\textnormal{VU}} \nonumber \\ &= \text{PositiveRoot}( \zeta_n  \frac{\mathcal{B}_n(s_n,\Lambda_n)}{x^2}  - y c_{\hspace{0.5pt}\textnormal{e}} \cdot2 \kappa_n^{\textnormal{VU}} \mathcal{G}_n(s_n,\Lambda_n)x  = \delta_n). \label{fnvueq}   
\end{talign}

 % $   \sqrt[3]{\frac{\zeta_n}{2y c_{\hspace{0.5pt}\textnormal{e}}  \kappa^{\textnormal{MS}}} }   $

To obtain $[\boldsymbol{f}^{\textnormal{VU}},\boldsymbol{\delta}]$ satisfying  (\ref{partialLpartialfVU})  (\ref{Complementarydelta})   \textnormal{(\ref{constraintfn})} (\ref{Dualfeasibility}d), we have the following two cases:
 \begin{itemize}
\item If $\text{PositiveRoot}( \zeta_n  \frac{\mathcal{B}_n(s_n,\Lambda_n)}{x^2}  - y c_{\hspace{0.5pt}\textnormal{e}} \cdot2 \kappa_n^{\textnormal{VU}} \mathcal{G}_n(s_n,\Lambda_n)x  = 0) > f_{n,\textnormal{max}}^{\textnormal{VU}}$; i.e., if $ \sqrt[3]{\frac{\mathcal{B}_n(s_n,\Lambda_n)\zeta_n}{2\mathcal{G}_n(s_n,\Lambda_n) y c_{\hspace{0.5pt}\textnormal{e}} \kappa_n^{\textnormal{VU}}}}  > f_{n,\textnormal{max}}^{\textnormal{VU}} $,   setting $\delta_n $ as $0$ violates \textnormal{(\ref{constraintfn})}. This  with   (\ref{Dualfeasibility}d) means $\delta_n>0$, which with   (\ref{Complementarydelta})  induces $f_n^{\textnormal{VU}} =f_{n,\textnormal{max}}^{\textnormal{VU}} $;
 \item If $\text{PositiveRoot}( \zeta_n  \frac{\mathcal{B}_n(s_n,\Lambda_n)}{x^2}  - y c_{\hspace{0.5pt}\textnormal{e}} \cdot2 \kappa_n^{\textnormal{VU}} \mathcal{G}_n(s_n,\Lambda_n)x  = 0) \leq f_{n,\textnormal{max}}^{\textnormal{VU}}$; i.e., if $\sqrt[3]{\frac{\mathcal{B}_n(s_n,\Lambda_n)\zeta_n}{2\mathcal{G}_n(s_n,\Lambda_n) y c_{\hspace{0.5pt}\textnormal{e}} \kappa_n^{\textnormal{VU}}}}  \leq f_{n,\textnormal{max}}^{\textnormal{VU}} $, then setting $\delta_n $ as $0$ and $f_n^{\textnormal{VU}} = \sqrt[3]{\frac{\mathcal{B}_n(s_n,\Lambda_n)\zeta_n}{2\mathcal{G}_n(s_n,\Lambda_n) y c_{\hspace{0.5pt}\textnormal{e}} \kappa_n^{\textnormal{VU}}}}  $ satisfies (\ref{partialLpartialfVU})  (\ref{Complementarydelta})   \textnormal{(\ref{constraintfn})} (\ref{Dualfeasibility}d). 
 \end{itemize}

To obtain $[\boldsymbol{f}^{\textnormal{MS}},\gamma]$ satisfying (\ref{partialLpartialfMS})  (\ref{Complementarygamma})  \textnormal{(\ref{constraintfMS})}  (\ref{Dualfeasibility}c), we discuss the following two cases:
 \begin{itemize}
\item[{\color{blue}\ding{182}}]  
 If $\text{PositiveRoot}(\zeta_n \frac{\mathcal{A}_n(s_n,\Lambda_n)}{x^2} - y c_{\hspace{0.5pt}\textnormal{e}} \cdot2 \kappa^{\textnormal{MS}} \mathcal{F}_n(s_n,\Lambda_n)x = 0) > f_{\textnormal{max}}^{\textnormal{MS}}$; i.e., if $\sum_{n \in \mathcal{N}} \sqrt[3]{\frac{\mathcal{A}_n(s_n,\Lambda_n)\zeta_n}{2\mathcal{F}_n(s_n,\Lambda_n)y c_{\hspace{0.5pt}\textnormal{e}}  \kappa^{\textnormal{MS}}} } > f_{\textnormal{max}}^{\textnormal{MS}}$, then setting $\gamma$ as $0$ violates \textnormal{(\ref{constraintfMS})}. This with   (\ref{Dualfeasibility}c) means $\gamma>0$, which is used in   (\ref{Complementarygamma}) to induce
 \begin{talign}
 &\sum_{n \in \mathcal{N}} \text{PositiveRoot}(\zeta_n \frac{\mathcal{A}_n(s_n,\Lambda_n)}{x^2} \nonumber \\ &- y c_{\hspace{0.5pt}\textnormal{e}} \cdot2 \kappa^{\textnormal{MS}} \mathcal{F}_n(s_n,\Lambda_n)x = \gamma) = f_{\textnormal{max}}^{\textnormal{MS}}  .  \label{fmseq}
 \end{talign}
 After obtaining the desired $\gamma$, we use it in~(\ref{fnmseq}) to get $f_n^{\textnormal{MS}}$.
\item[{\color{blue}\ding{183}}] 
  If $\text{PositiveRoot}(\zeta_n \frac{\mathcal{A}_n(s_n,\Lambda_n)}{x^2} - y c_{\hspace{0.5pt}\textnormal{e}} \cdot2 \kappa^{\textnormal{MS}} \mathcal{F}_n(s_n,\Lambda_n)x = 0) \leq f_{\textnormal{max}}^{\textnormal{MS}}$; i.e., if $\sum_{n \in \mathcal{N}} \sqrt[3]{\frac{\mathcal{A}_n(s_n,\Lambda_n)\zeta_n}{2\mathcal{F}_n(s_n,\Lambda_n)y c_{\hspace{0.5pt}\textnormal{e}}  \kappa^{\textnormal{MS}}} }  \leq  f_{\textnormal{max}}^{\textnormal{MS}}$, then setting $\gamma$ as $0$ and $f_n^{\textnormal{MS}} = \sqrt[3]{\frac{\mathcal{A}_n(s_n,\Lambda_n)\zeta_n}{2\mathcal{F}_n(s_n,\Lambda_n)y c_{\hspace{0.5pt}\textnormal{e}}  \kappa^{\textnormal{MS}}} } $ satisfies (\ref{partialLpartialfMS})  (\ref{Complementarygamma})  \textnormal{(\ref{constraintfMS})}  (\ref{Dualfeasibility}c). 
 \end{itemize}
 
\renewcommand{\thealgocf}{\mbox{1.1 explained  above}}

\begin{algorithm}
\caption{\newline$\textnormal{Alg-Solve-}\acute{\boldsymbol{f}}^{\textnormal{MS}}(\boldsymbol{\zeta}\hspace{-2pt}\mid\hspace{-2pt} \textnormal{\small$\bigstar$}) \text{-and-} \acute{\boldsymbol{f}}^{\textnormal{VU}}(\boldsymbol{\zeta}\hspace{-2pt}\mid\hspace{-2pt} \textnormal{\small$\bigstar$})$.
% , which finds $\widetilde{b}_n(\alpha, \beta,\iffalse T,\fi\boldsymbol{\zeta}\mid \boldsymbol{z},y, \boldsymbol{s})$ via bisection search according to~?.
} \label{alg1.1}
\label{algo:bandwidth}

Set $\acute{f}_n^{\textnormal{VU}}(\boldsymbol{\zeta}\mid \textnormal{\small$\bigstar$})  $ as $ \min \big\{ f_{n,\textnormal{max}}^{\textnormal{VU}},~ \sqrt[3]{\frac{\mathcal{B}_n(s_n,\Lambda_n)\zeta_n}{2\mathcal{G}_n(s_n,\Lambda_n) y c_{\hspace{0.5pt}\textnormal{e}} \kappa_n^{\textnormal{VU}}}}\big\}   $;

\textbf{if} $\sum_{n \in \mathcal{N}} \sqrt[3]{\frac{\mathcal{A}_n(s_n,\Lambda_n)\zeta_n}{2\mathcal{F}_n(s_n,\Lambda_n)y c_{\hspace{0.5pt}\textnormal{e}}  \kappa^{\textnormal{MS}}} }  \leq  f_{\textnormal{max}}^{\textnormal{MS}}$, then set $\acute{f}_n^{\textnormal{MS}}(\boldsymbol{\zeta}\mid \textnormal{\small$\bigstar$})$ as $\sqrt[3]{\frac{\mathcal{A}_n(s_n,\Lambda_n)\zeta_n}{2\mathcal{F}_n(s_n,\Lambda_n)y c_{\hspace{0.5pt}\textnormal{e}}  \kappa^{\textnormal{MS}}} } $;

\textbf{else} with function $C(\cdot)$ being set as $\sum_{n \in \mathcal{N}} \text{PositiveRoot}(\zeta_n \frac{\mathcal{A}_n(s_n,\Lambda_n)}{x^2} - y c_{\hspace{0.5pt}\textnormal{e}} \cdot2 \kappa^{\textnormal{MS}} \mathcal{F}_n(s_n,\Lambda_n)x = \gamma)$, and $C_{\textnormal{target}}$ being set as $f_{\textnormal{max}}^{\textnormal{MS}} $, run~$\textnormal{Standard-Bisection-Search}([\boldsymbol{\zeta}, \textnormal{\small$\bigstar$}],\textnormal{function~}C(\cdot),\newline C_{\textnormal{target}},0,f_{\textnormal{max}}^{\textnormal{MS}})$ based on Algorithm~\ref{algo:bisection} on Page~\pageref{algo:bisection} to obtain the desired $\gamma$ satisfying Eq.~(\ref{fmseq}),  
%set this $\gamma$ as $\widetilde{p}_n(\alpha, \beta,\iffalse T,\fi\boldsymbol{\zeta}\mid \boldsymbol{z},y, \boldsymbol{s})$;
% find the solution of $\gamma$ to $\sum_{n \in \mathcal{N}} \text{PositiveRoot}(\zeta_n \frac{\mathcal{A}_n(s_n,\Lambda_n)}{x^2} - y c_{\hspace{0.5pt}\textnormal{e}} \cdot2 \kappa^{\textnormal{MS}} \mathcal{A}_n(s_n,\Lambda_n)x = \gamma) = f_{\textnormal{max}}^{\textnormal{MS}} $ using the bisection search?, 

 let the obtained $\gamma$ be $\acute{\gamma}$, and set set $\acute{f}_n^{\textnormal{MS}}(\boldsymbol{\zeta}\mid \textnormal{\small$\bigstar$})$ as $\text{PositiveRoot}(\zeta_n \frac{\mathcal{A}_n(s_n,\Lambda_n)}{x^2} - y c_{\hspace{0.5pt}\textnormal{e}} \cdot2 \kappa^{\textnormal{MS}} \mathcal{F}_n(s_n,\Lambda_n)x = \acute{\gamma})$ for each $n \in \mathcal{N}$;

Return $\acute{\boldsymbol{f}}^{\textnormal{VU}}(\boldsymbol{\zeta}\mid \textnormal{\small$\bigstar$})$  (resp., $\acute{\boldsymbol{f}}^{\textnormal{MS}}(\boldsymbol{\zeta}\mid \textnormal{\small$\bigstar$})$) as $[\acute{f}_n^{\textnormal{VU}}(\boldsymbol{\zeta}\mid \textnormal{\small$\bigstar$})  |_{n \in \mathcal{N}}]$ (resp., $[\acute{f}_n^{\textnormal{MS}}(\boldsymbol{\zeta}\mid \textnormal{\small$\bigstar$})|_{n \in \mathcal{N}}]$);  
\end{algorithm}

~

\textbf{Algorithm $\text{1.2}$:~Computing $[\acute{\boldsymbol{b}}(\boldsymbol{\zeta}\mid \textnormal{\small$\bigstar$}),\acute{\boldsymbol{p}}(\boldsymbol{\zeta}\mid \textnormal{\small$\bigstar$})]$ defined in Proposition~1.2.} \label{algexplain1.2} 
% The details are provided in the pseudocode of Algorithm $\text{1.2}$ on Page~\pageref{alg1.2}, where it is shown that 
As shown in the pseudocode, Algorithm $\text{1.2}$  calls Algorithms $\text{1.2.1}$, 1.2.2.1, and 1.2.2.2 detailed below.\vspace{5pt}

\renewcommand{\thealgocf}{\mbox{1.2 explained  above}} \label{alg1.2}

\begin{algorithm}
\caption{\newline $\textnormal{Alg-Solve-}\acute{\boldsymbol{b}}(\boldsymbol{\zeta}\mid \textnormal{\small$\bigstar$})\text{-and-}\acute{\boldsymbol{p}}(\boldsymbol{\zeta}\mid \textnormal{\small$\bigstar$})$.}
\label{alg1.1}
\label{algo:bandwidth}

Run $\textnormal{Alg-Solve-}\acute{\beta}(\boldsymbol{\zeta}\mid \textnormal{\small$\bigstar$})$ of Algorithm 1.2.2.2 to obtain $\acute{\beta}(\boldsymbol{\zeta}\mid \textnormal{\small$\bigstar$})$;

Use the just obtained $\acute{\beta}(\boldsymbol{\zeta}\mid \textnormal{\small$\bigstar$})$ as an input to $\textnormal{Alg-Solve-}\breve{\alpha}(\beta,\iffalse T,\fi\boldsymbol{\zeta}\mid \textnormal{\small$\bigstar$})$ of Algorithm 1.2.2.1 to get $\breve{\alpha}(\acute{\beta}(\boldsymbol{\zeta}\mid \textnormal{\small$\bigstar$}),\iffalse T,\fi\boldsymbol{\zeta}\mid \textnormal{\small$\bigstar$})$, which equals $\acute{\alpha}(\boldsymbol{\zeta}\mid \textnormal{\small$\bigstar$})$ according to~(\ref{eqproposition1.2.2.1v1});

For each $n \in \mathcal{N}$, use Lines 2 and 1's obtained $\acute{\alpha}(\boldsymbol{\zeta}\mid \textnormal{\small$\bigstar$})$ and $\acute{\beta}(\boldsymbol{\zeta}\mid \textnormal{\small$\bigstar$})$ as inputs to $\textnormal{Alg-Solve-}\widetilde{b}_n(\alpha, \beta,\iffalse T,\fi\boldsymbol{\zeta}\mid \textnormal{\small$\bigstar$})$ (resp., $\textnormal{Alg-Solve-}\widetilde{p}_n(\alpha, \beta,\iffalse T,\fi\boldsymbol{\zeta}\mid \textnormal{\small$\bigstar$})$) of Algorithm 1.2.1 to get $\widetilde{b}_n(\acute{\alpha}(\boldsymbol{\zeta}\mid \textnormal{\small$\bigstar$}), \acute{\beta}(\boldsymbol{\zeta}\mid \textnormal{\small$\bigstar$}),\iffalse T,\fi\boldsymbol{\zeta}\mid \textnormal{\small$\bigstar$})$ (resp., $\widetilde{p}_n(\acute{\alpha}(\boldsymbol{\zeta}\mid \textnormal{\small$\bigstar$}), \acute{\beta}(\boldsymbol{\zeta}\mid \textnormal{\small$\bigstar$}),\iffalse T,\fi\boldsymbol{\zeta}\mid \textnormal{\small$\bigstar$})$), which equals $\acute{b}_n(\boldsymbol{\zeta}\mid \textnormal{\small$\bigstar$})$ (resp., $\acute{p}_n(\boldsymbol{\zeta}\mid \textnormal{\small$\bigstar$})$) according to~(\ref{eqproposition1.2.1}); \vspace{3pt}
 
Return $\acute{\boldsymbol{b}}(\boldsymbol{\zeta}\mid \textnormal{\small$\bigstar$})$  (resp., $\acute{\boldsymbol{p}}(\boldsymbol{\zeta}\mid \textnormal{\small$\bigstar$})$) as $[\acute{b}_n(\boldsymbol{\zeta}\mid \textnormal{\small$\bigstar$})|_{n \in \mathcal{N}}]$ (resp., $[\acute{p}_n(\boldsymbol{\zeta}\mid \textnormal{\small$\bigstar$})|_{n \in \mathcal{N}}]$); 

\end{algorithm}

~

\textbf{Algorithm $\text{1.2.1}$:~Computing $[\widetilde{\boldsymbol{b}}(\alpha,\beta,\boldsymbol{\zeta}\mid \textnormal{\small$\bigstar$}),\widetilde{\boldsymbol{p}}(\alpha,\beta,\boldsymbol{\zeta}\mid \textnormal{\small$\bigstar$})]$ defined in Proposition~1.2.1 using $\mathcal{S}_{1.2.1}$ of (\number\value{counteryty}).} \label{algexplain1.2.1part1} Recall that $\mathcal{S}_{1.2.1}$ includes (\ref{Stationaritybn}) (\ref{Stationaritypn}).
From (\ref{Stationaritybn}) (\ref{Stationaritypn}) and (\ref{shannon}),
% (i.e., $r_n(b_n, p_n) = b_n \log_2(1+\frac{g_np_n}{\sigma^2b_n})$),
with $\vartheta_n$ defined by
\begin{talign} 
\vartheta_n : = \frac{g_np_n}{\sigma_n^2b_n} , \label{solvepsinx}
\end{talign}
we obtain that $[\widetilde{b}_n(\alpha, \beta,\iffalse T,\fi\boldsymbol{\zeta}\mid \textnormal{\small$\bigstar$}),\widetilde{p}_n(\alpha, \beta,\iffalse T,\fi\boldsymbol{\zeta}\mid \textnormal{\small$\bigstar$})]$ is the solution of $[b_n,  p_n]$ to
\begin{talign} 
&\big(\frac{\partial U_n(r_n, s_n)}{\partial r_n}   + \frac{y c_{\hspace{0.5pt}\textnormal{e}}}{2z_n{r_n}^3\textcolor{black}{{\nu_n}^2}}  + 
 \frac{\zeta_n s_n\textcolor{black}{\mu_n}  {\color{black}\Lambda_n}}{{r_n}^2\textcolor{black}{\nu_n}}\big)\times\nonumber \\ &\big( 
    \log_2(1+\vartheta_n)
     -\frac{\vartheta_n}{(1+\vartheta_n)\ln2}\big) = \alpha, \textnormal{ and} \label{Stationaritybn3}\\ 
&\big(\frac{\partial U_n(r_n, s_n)}{\partial r_n}   + \frac{y c_{\hspace{0.5pt}\textnormal{e}}}{2z_n{r_n}^3\textcolor{black}{{\nu_n}^2}}  + 
 \frac{\zeta_n s_n\textcolor{black}{\mu_n}  {\color{black}\Lambda_n}}{{r_n}^2\textcolor{black}{\nu_n}}\big) \cdot \frac{g_n}{\sigma_n^2(1+\vartheta_n)\ln2} \nonumber \\ &  = \beta + 2(p_n +p_n^{\textnormal{cir}})y c_{\hspace{0.5pt}\textnormal{e}}z_n(s_n\textcolor{black}{\mu_n}{\color{black}\Lambda_n})^2.\label{Stationaritypn3}
\end{talign}
With (\ref{Stationaritybn3}) divided by (\ref{Stationaritypn3}), it holds that
% \begin{talign} 
% \frac{~~\frac{\partial r_n}{\partial b_n}~~}{~~\frac{\partial r_n}{\partial p_n}~~} = \frac{\alpha}{\beta + 2(p_n +p_n^{\textnormal{cir}})yz_n{s_n}^2}.  \label{Stationaritypn5}
% \end{talign}
% The left hand side of (\ref{Stationaritypn5}) can also be computed as follows. In view of $r_n(b_n, p_n) = b_n \log_2(1+\frac{g_np_n}{\sigma^2b_n})$, we obtain
\begin{talign} 
 \frac{
    \log_2(1+\vartheta_n)
     -\frac{\vartheta_n}{(1+\vartheta_n)\ln2}}{
    \frac{g_n}{\sigma_n^2(1+\vartheta_n)\ln2}} = \frac{\alpha}{\beta + 2(p_n +p_n^{\textnormal{cir}})y c_{\hspace{0.5pt}\textnormal{e}}z_n(s_n\textcolor{black}{\mu_n}{\color{black}\Lambda_n})^2}.  \label{solvepsin}
\end{talign}
Based on (\ref{solvepsin}), we get
\begin{talign}
    &\vartheta_n \hspace{-2pt} = \hspace{-2pt} \psi_n(p_n,\alpha,\beta \mid \textnormal{\small$\bigstar$}), \textnormal{ for } \textstyle{\psi_n(p_n,\alpha,\beta \mid \textnormal{\small$\bigstar$}) \hspace{-2pt}} \nonumber \\ &\textstyle{:=\hspace{-2pt}  \exp\big\{1\hspace{-2pt} +\hspace{-2pt} W\big(\frac{1}{e}(\frac{g_n\alpha}{[\beta + 2(p_n +p_n^{\textnormal{cir}})y c_{\hspace{0.5pt}\textnormal{e}}z_n(s_n\textcolor{black}{\mu_n}{\color{black}\Lambda_n})^2]\sigma_n^2}\hspace{-2pt} -\hspace{-2pt} 1)\big)\big\}\hspace{-2pt} -\hspace{-2pt} 1}. \label{definepsin}
\end{talign}
Then we can substitute $\vartheta_n $ from~(\ref{definepsin}) into (\ref{Stationaritybn3}) and (\ref{Stationaritypn3}), to decide $[\boldsymbol{b},\boldsymbol{p}]$, as shown in Lemma~\ref{lemmatilde} below.

\begin{lemma} \label{lemmatilde}
Given $[\alpha, \beta,\iffalse T,\fi\boldsymbol{\zeta}, \textnormal{\small$\bigstar$}]$, we know from Proposition~1.2.1 that $\widetilde{p}_n(\alpha, \beta,\iffalse T,\fi\boldsymbol{\zeta}\mid \textnormal{\small$\bigstar$})$ and $\widetilde{b}_n(\alpha, \beta,\iffalse T,\fi\boldsymbol{\zeta}\mid \textnormal{\small$\bigstar$})$ denote the values of $p_n$ and $b_n$ satisfying~(\ref{Stationaritybn}) and~(\ref{Stationaritypn}). We define $\psi_n(p_n,\alpha,\beta \mid \textnormal{\small$\bigstar$})$ in~(\ref{definepsin}), and define
\begin{talign}
\overline{r}_n(p_n,\alpha,\beta \mid \textnormal{\small$\bigstar$}) := \frac{g_np_n  \log_2(1+ \psi_n(p_n,\alpha,\beta \mid \textnormal{\small$\bigstar$}))}{\sigma_n^2\psi_n(p_n,\alpha,\beta \mid \textnormal{\small$\bigstar$})} . \label{eqdefinernoverline}
\end{talign} 
Then  $\widetilde{p}_n(\alpha, \beta,\iffalse T,\fi\boldsymbol{\zeta}\mid \textnormal{\small$\bigstar$})$ is a solution of $p_n$ to
\begin{talign}
&\Bigg[ \begin{array}{l}
     \big(\frac{\partial U_n(r_n, s_n)}{\partial r_n}\big)|_{r_n=\overline{r}_n(p_n,\alpha,\beta \mid \textnormal{\small$\bigstar$})}  \\ + \frac{y c_{\textnormal{e}}}{2z_n \cdot [\overline{r}_n(p_n,\alpha,\beta \mid \textnormal{\small$\bigstar$})]^3\textcolor{black}{{\nu_n}^2}}   + \frac{\zeta_n s_n\textcolor{black}{\mu_n}  {\color{black}\Lambda_n}}{ [\overline{r}_n(p_n,\alpha,\beta \mid \textnormal{\small$\bigstar$})]^2\textcolor{black}{\nu_n}}       
\end{array}  \Bigg]  \times \nonumber \\ &  \textstyle{\left[
    \log_2\big(1 + \psi_n(p_n,\alpha,\beta \mid \textnormal{\small$\bigstar$})\big)
      - \frac{\psi_n(p_n,\alpha,\beta \mid \textnormal{\small$\bigstar$})}{(1+\psi_n(p_n,\alpha,\beta \mid \textnormal{\small$\bigstar$}))\ln2}\right]}   =   \alpha. \label{solvepngivenalphaandbeta}
\end{talign} 
% \begin{talign}
%  &\bigg(\frac{\partial U_n(r_n, s_n)}{\partial r_n}\bigg)|_{r_n=\frac{g_np_n}{\sigma_n^2\psi_n(p_n,\alpha,\beta \mid \boldsymbol{z},y, \boldsymbol{s})}  \log_2(1+ \psi_n(p_n,\alpha,\beta \mid \boldsymbol{z},y, \boldsymbol{s}))}   + \frac{y}{2z_n}\cdot\bigg(\frac{\sigma_n^2\psi_n(p_n,\alpha,\beta \mid \boldsymbol{z},y, \boldsymbol{s})}{g_np_n\log_2(1+ \psi_n(p_n,\alpha,\beta \mid \boldsymbol{z},y, \boldsymbol{s}))}\bigg)^3  \nonumber \\ &+ \zeta_n s_n\cdot\bigg(\frac{\sigma_n^2\psi_n(p_n,\alpha,\beta \mid \boldsymbol{z},y, \boldsymbol{s})}{g_np_n\log_2(1+ \psi_n(p_n,\alpha,\beta \mid \boldsymbol{z},y, \boldsymbol{s}))}\bigg)^2 =  \frac{\alpha}{
%     \log_2\big(1+\psi_n(p_n,\alpha,\beta \mid \boldsymbol{z},y, \boldsymbol{s})\big)
%      -\frac{\psi_n(p_n,\alpha,\beta \mid \boldsymbol{z},y, \boldsymbol{s})}{(1+\psi_n(p_n,\alpha,\beta \mid \boldsymbol{z},y, \boldsymbol{s}))\ln2}}. \label{solvepngivenalphaandbeta}
% \end{talign}
~\\
With $\widetilde{p}_n(\alpha, \beta,\iffalse T,\fi\boldsymbol{\zeta}\mid \textnormal{\small$\bigstar$})$ decided according to~(\ref{solvepngivenalphaandbeta}) above, the corresponding $b_n$ satisfying~(\ref{Stationaritybn}) and~(\ref{Stationaritypn}) is  
\begin{talign}
\widetilde{b}_n(\alpha, \beta,\iffalse T,\fi\boldsymbol{\zeta}\mid \textnormal{\small$\bigstar$})   =  \frac{g_n\widetilde{p}_n(\alpha, \beta,\iffalse T,\fi\boldsymbol{\zeta}\mid \textnormal{\small$\bigstar$})}{\sigma_n^2\psi_n(\widetilde{p}_n(\alpha, \beta,\iffalse T,\fi\boldsymbol{\zeta}\mid \textnormal{\small$\bigstar$}),\alpha,\beta \mid \textnormal{\small$\bigstar$})}. \label{solvebngivenalphaandbeta}
\end{talign} 
\end{lemma}

\noindent\textbf{Proof of Lemma~\ref{lemmatilde}:} First, for $p_n$ and $b_n$ satisfying~(\ref{Stationaritybn}) and~(\ref{Stationaritypn}), we have already explained above that $\frac{g_np_n}{\sigma_n^2b_n} $ (i.e., $\vartheta_n$ defined in~(\ref{solvepsinx})) equals $\psi_n(p_n,\alpha,\beta \mid \textnormal{\small$\bigstar$})$ defined in~(\ref{definepsin}). Then $b_n$ equals $\frac{g_np_n}{\sigma_n^2 \cdot \psi_n(p_n,\alpha,\beta \mid \textnormal{\small$\bigstar$})} $, and $r_n(b_n, p_n) $ denoting $ b_n \log_2(1+\frac{g_np_n}{\sigma^2b_n})$ equals the right hand side of~(\ref{eqdefinernoverline}), which we denote as $\overline{r}_n(p_n,\alpha,\beta \mid \textnormal{\small$\bigstar$})$ in~(\ref{eqdefinernoverline}) for notation simplicity. Then letting $r_n$ be $\overline{r}_n(p_n,\alpha,\beta \mid \textnormal{\small$\bigstar$})$ in~(\ref{Stationaritybn3}), we obtain~(\ref{solvepngivenalphaandbeta}) which will be used to solve $p_n$. With $\widetilde{p}_n(\alpha, \beta,\iffalse T,\fi\boldsymbol{\zeta}\mid \textnormal{\small$\bigstar$})$ denoting the obtained $p_n$, the corresponding $b_n$ is given by (\ref{solvebngivenalphaandbeta}). Hence, we have proved that $\widetilde{p}_n(\alpha, \beta,\iffalse T,\fi\boldsymbol{\zeta}\mid \textnormal{\small$\bigstar$})$ and $\widetilde{b}_n(\alpha, \beta,\iffalse T,\fi\boldsymbol{\zeta}\mid \textnormal{\small$\bigstar$})$, denoting $p_n$ and $b_n$ satisfying~(\ref{Stationaritybn}) and~(\ref{Stationaritypn}), are given by (\ref{solvepngivenalphaandbeta}) and~(\ref{solvebngivenalphaandbeta}), respectively. \vspace{5pt}
\qed

Then we use  (\ref{solvepngivenalphaandbeta}) and~(\ref{solvebngivenalphaandbeta}) of Lemma~\ref{lemmatilde} to solve $\widetilde{p}_n(\alpha, \beta,\iffalse T,\fi\boldsymbol{\zeta}\mid \textnormal{\small$\bigstar$})$ and $\widetilde{b}_n(\alpha, \beta,\iffalse T,\fi\boldsymbol{\zeta}\mid \textnormal{\small$\bigstar$})$, respectively. They are considered as two subprocedures of Algorithm 1.2.1, as described below.

\renewcommand{\thealgocf}{1.2.1's Subprocedure 1 via Eq.~(\ref{solvepngivenalphaandbeta}) for transmission power}

\begin{algorithm}
\caption{\mbox{$\textnormal{Alg-Solve-}\widetilde{p}_n(\alpha, \beta,\iffalse T,\fi\boldsymbol{\zeta}\mid \textnormal{\small$\bigstar$})$.}
% , where ``$(\textnormal{$\blacktriangledown$})$'' means ``$(\alpha, \beta,\iffalse T,\fi\boldsymbol{\zeta}\mid \textnormal{\small$\bigstar$})$'' \begin{remark} \label{remarkalgo:transmitpower} Algorithm $\textnormal{Alg-Solve-}\widetilde{p}_n(\textnormal{$\blacktriangledown$})$ aims to find the solution of $p_n$ to (\ref{solvepngivenalphaandbeta}); i.e., $  \sum_{n \in \mathcal{N}} \widetilde{b}_n(\textnormal{$\blacktriangledown$})  = b_{\textnormal{max}}$, where $\widetilde{b}_n(\textnormal{$\blacktriangledown$})$ is defined in Table~\ref{talbenotation} and .\end{remark} to~(\ref{solvepngivenalphaandbeta}).
}
\label{algo:transmitpower}

With function $C(\cdot)$ being set as the left hand side of~(\ref{solvepngivenalphaandbeta}), and $C_{\textnormal{target}}$ being set as $\alpha$, run $\textnormal{Standard-Bisection-Search}([\alpha,\beta,\iffalse T,\fi\boldsymbol{\zeta},\textnormal{\small$\bigstar$}],\textnormal{function~}C(\cdot), \newline C_{\textnormal{target}},0,p_{\textnormal{max}})$ based on Algorithm~\ref{algo:bisection} on Page~\pageref{algo:bisection} to obtain the desired $p_n$ satisfying Eq.~(\ref{solvepngivenalphaandbeta}), and set this $p_n$ as $\widetilde{p}_n(\alpha, \beta,\iffalse T,\fi\boldsymbol{\zeta}\mid \textnormal{\small$\bigstar$})$;
% defined according to Eq.~(\ref{solvepngivenalphaandbeta}); 
% \newline //Comment: From Lemma~\ref{lemsolvepngivenalphaandbeta}, the left hand side of~(\ref{solvepngivenalphaandbeta}) given $[\alpha,\beta,\iffalse T,\fi\boldsymbol{\zeta},\textnormal{\small$\bigstar$}]$ is \mbox{decreasing} as $p_n$ increases. 

\end{algorithm}

~\vspace{5pt}

\renewcommand{\thealgocf}{ 1.2.1's Subprocedure  2 via Eq.~(\ref{solvebngivenalphaandbeta}) for bandwidth}

\begin{algorithm}
\caption{$\textnormal{Alg-Solve-}\widetilde{b}_n(\alpha,\beta,\iffalse T,\fi\boldsymbol{\zeta},\textnormal{\small$\bigstar$})$.
% , which finds $\widetilde{b}_n(\alpha, \beta,\iffalse T,\fi\boldsymbol{\zeta}\mid \textnormal{\small$\bigstar$})$ via bisection search according to~?.
}
\label{algo:bandwidth}

Run $\textnormal{Alg-Solve-}\widetilde{p}_n(\alpha,\beta,\iffalse T,\fi\boldsymbol{\zeta},\textnormal{\small$\bigstar$})$ (i.e., Algorithm~1.2.1's Subprocedure~1 above) to obtain $\widetilde{p}_n(\alpha, \beta,\iffalse T,\fi\boldsymbol{\zeta}\mid \textnormal{\small$\bigstar$})$;

Compute $\widetilde{b}_n(\alpha, \beta,\iffalse T,\fi\boldsymbol{\zeta}\mid \textnormal{\small$\bigstar$})$ according to Eq.~(\ref{solvebngivenalphaandbeta});

\end{algorithm}

~

 % (Pseudocode on Page~\pageref{algo:solvealphabreve} based on the following analysis)
 \textbf{Algorithm $\text{1.2.2.1}$:~Computing $\breve{\alpha}(\beta,\iffalse T,\fi\boldsymbol{\zeta}\mid \textnormal{\small$\bigstar$})$ defined in Proposition~1.2.2.1 using $\mathcal{S}_{1.2.2.1}$ of (\number\value{counterytty}).} \label{algexplain1.2.2.1}
% We have presented the relationship between the ``$(\textnormal{$\blacktriangledown$})$'' notations and the ``$(\textnormal{\small$\blacksquare$})$'' notations in~(\ref{relation-blacktriangledown-blacksquare}).
 Recall that $\mathcal{S}_{1.2.2.1}$ includes (\ref{Stationaritypn3alpha}) and (\ref{Complementaryalpha2}).   Proposition~1.2.2.1 defines  $\breve{\alpha}(\beta,\iffalse T,\fi\boldsymbol{\zeta}\mid \textnormal{\small$\bigstar$})$ as the solution of $\alpha$ to
 \begin{talign} 
 \sum_{n \in \mathcal{N}} \widetilde{b}_n(\alpha, \beta,\iffalse T,\fi\boldsymbol{\zeta}\mid \textnormal{\small$\bigstar$})  = b_{\textnormal{max}} \textnormal{ and } \alpha >0. \label{brevealpha}
\end{talign}

\renewcommand{\thealgocf}{1.2.2.1 explained  above}

\begin{algorithm}
\caption{\newline$\textnormal{Alg-Solve-}\breve{\alpha}(\beta,\iffalse T,\fi\boldsymbol{\zeta}\mid \textnormal{\small$\bigstar$})$.
% , where ``$(\textnormal{\small$\blacksquare$})$'' means ``$(\beta,\iffalse T,\fi\boldsymbol{\zeta}\mid \textnormal{\small$\bigstar$})$''. \begin{remark} \label{remarkalpha} Algorithm $\textnormal{Alg-Solve-}\breve{\alpha}(\textnormal{\small$\blacksquare$})$ aims to find the solution of $\alpha$ to (\ref{brevealpha}); i.e., $  \sum_{n \in \mathcal{N}} \widetilde{b}_n(\textnormal{$\blacktriangledown$})  = b_{\textnormal{max}}$, where $\widetilde{b}_n(\textnormal{$\blacktriangledown$})$ is defined in Table~\ref{talbenotation} and ``$(\textnormal{$\blacktriangledown$})$'' means ``$(\alpha, \beta,\iffalse T,\fi\boldsymbol{\zeta}\mid \textnormal{\small$\bigstar$})$''.\end{remark} 
}
\label{algo:solvealphabreve}

With function $C(\cdot)$ being set as $\sum_{n \in \mathcal{N}} \textnormal{Alg-Solve-}\widetilde{b}_n(\alpha,\beta,\iffalse T,\fi\boldsymbol{\zeta},\textnormal{\small$\bigstar$})$ (whose computation requires Algorithm~1.2.1's Subprocedure 2), and $C_{\textnormal{target}}$ being set as $b_{\textnormal{max}} $, run $\textnormal{Bisection-Search-with-No-Known-Upper-Bound}([\beta,\boldsymbol{z}, y, \boldsymbol{s}],\newline \textnormal{function~}C(\cdot),C_{\textnormal{target}})$ based on Algorithm~\ref{algo:BisectionSearchwithNoUpperBound} on Page~\pageref{algo:BisectionSearchwithNoUpperBound}
to obtain the desired $\alpha$ satisfying Eq.~(\ref{brevealpha}), and set this $\alpha$ as $\breve{\alpha}(\beta\mid \boldsymbol{z}, y, \boldsymbol{s})$;

\end{algorithm}

~

% After obtaining $\breve{\alpha}(\textnormal{\small$\blacksquare$})$ by solving~(\ref{brevealpha}), we let the $\alpha$ parameter in $[\widetilde{\boldsymbol{b}}(\blacktriangledown),\widetilde{\boldsymbol{p}}(\blacktriangledown),\widetilde{\boldsymbol{f}}^{\textnormal{MS}}(\blacktriangledown),\widetilde{\boldsymbol{f}}^{\textnormal{VU}}(\blacktriangledown), \widetilde{\gamma}(\blacktriangledown),\widetilde{\boldsymbol{\delta}}(\blacktriangledown) ]$ be $\breve{\alpha}(\textnormal{\small$\blacksquare$})$, and the corresponding results become $[\breve{\boldsymbol{b}}(\textnormal{\small$\blacksquare$}),\breve{\boldsymbol{p}}(\textnormal{\small$\blacksquare$}),\breve{\boldsymbol{f}}^{\textnormal{MS}}(\textnormal{\small$\blacksquare$}),\breve{\boldsymbol{f}}^{\textnormal{VU}}(\textnormal{\small$\blacksquare$}),\breve{\gamma}(\textnormal{\small$\blacksquare$}),\breve{\boldsymbol{\delta}}(\textnormal{\small$\blacksquare$})]$.  
 
\textbf{Algorithm $\text{1.2.2.2}$ (Pseudocode on Page~\pageref{algo:solvebeta} based on the following analysis):~Using Algorithm~$\text{1.2.2.1}$  to compute $\acute{\beta}(\boldsymbol{\zeta}\mid \textnormal{\small$\bigstar$})$ according to Proposition~1.2.2.2 using $\mathcal{S}_{1.2.2.2}$ of (\number\value{counteryttt}).} \label{algexplain1.2.2.2} Recall from (\number\value{counteryttt}) that $\mathcal{S}_{1.2.2.2}$ includes \textnormal{(\ref{Complementarybeta})}, \textnormal{(\ref{constraintpn})}, and \textnormal{(\ref{Dualfeasibility}b)}. Then $\acute{\beta}(\boldsymbol{\zeta}\mid \textnormal{\small$\bigstar$})$ is the solution of $\beta$ to  \vspace{5pt}
\begin{subnumcases}{}
\textnormal{based on (\ref{Complementarybeta}), we ensure }\nonumber \\ \textstyle{ \beta \cdot \big( \sum_{n \in \mathcal{N}} \widetilde{p}_n(\breve{\alpha}(\beta,\iffalse T,\fi\boldsymbol{\zeta}\mid \textnormal{\small$\bigstar$}),\beta,\boldsymbol{\zeta}\mid \textnormal{\small$\bigstar$}) - p_{\textnormal{max}}\big) = 0}; \label{solvebeta1} \\ \textnormal{based on (\ref{constraintpn}), we ensure } \nonumber \\ \textstyle{ \sum_{n \in \mathcal{N}} \widetilde{p}_n(\breve{\alpha}(\beta,\iffalse T,\fi\boldsymbol{\zeta}\mid \textnormal{\small$\bigstar$}),\beta,\boldsymbol{\zeta}\mid \textnormal{\small$\bigstar$})  \leq p_{\textnormal{max}}; }\label{solvebeta2}\\ \textnormal{based on (\ref{Dualfeasibility}b), we ensure } \beta \geq 0.  \label{solvebeta3}
\end{subnumcases}
% Based on~(\ref{relation-blacksquare-blacktriangle}), we obtain the ``$(\textnormal{$\blacktriangle$})$'' (i.e., ``$(\iffalse T,\fi\boldsymbol{\zeta}\mid \boldsymbol{z},y, \boldsymbol{s})$'') notations by enforcing (\ref{Complementarybeta}), (\ref{constraintpn}) and (\ref{Dualfeasibility}b) on the ``$(\textnormal{\small$\blacksquare$})$'' (i.e., ``$(\beta,\iffalse T,\fi\boldsymbol{\zeta}\mid \boldsymbol{z},y, \boldsymbol{s})$'') notations; i.e.,
% Given $[\boldsymbol{\zeta}, \textnormal{\small$\bigstar$}]$, the solution of $\beta$ to~(\ref{solvebeta1})--(\ref{solvebeta3}) is just $\acute{\beta}(\boldsymbol{\zeta}\mid \textnormal{\small$\bigstar$})$.  

~ \vspace{-5pt}

We have the following two cases:
\begin{itemize}
\item If $ \sum_{n \in \mathcal{N}} \widetilde{p}_n(\breve{\alpha}(0,\iffalse T,\fi\boldsymbol{\zeta}\mid \textnormal{\small$\bigstar$}),0,\boldsymbol{\zeta}\mid \textnormal{\small$\bigstar$})  > p_{\textnormal{max}}$, then the solution of $\beta$ to (\ref{solvebeta1})--(\ref{solvebeta3}) cannot be $0$  (which along with (\ref{solvebeta3}) means  $\beta>0$), since setting $\beta$ as $0$ violates (\ref{solvebeta2}). We use $\beta>0$ in (\ref{solvebeta1}) to get 
% We define $\hat{\beta}(\iffalse T,\fi\boldsymbol{\zeta}\mid \boldsymbol{z},y, \boldsymbol{s})$ as the solution of $\beta$ to
\begin{talign}
 \sum_{n \in \mathcal{N}} \widetilde{p}_n(\breve{\alpha}(\beta,\iffalse T,\fi\boldsymbol{\zeta}\mid \textnormal{\small$\bigstar$}),\beta,\boldsymbol{\zeta}\mid \textnormal{\small$\bigstar$})  = p_{\textnormal{max}}. \label{betahat}   
\end{talign}
% Line ? of Algorithm ? solves (\ref{betahat}) using the bisection method.
\item If $ \sum_{n \in \mathcal{N}} \widetilde{p}_n(\breve{\alpha}(0,\iffalse T,\fi\boldsymbol{\zeta}\mid \textnormal{\small$\bigstar$}),0,\boldsymbol{\zeta}\mid \textnormal{\small$\bigstar$})  \leq p_{\textnormal{max}}$, then $0$ is a solution of $\beta$ to (\ref{solvebeta1})--(\ref{solvebeta3}).\vspace{5pt}
\end{itemize}

\renewcommand{\thealgocf}{1.2.2.2 explained above}

\begin{algorithm}
\caption{$\textnormal{Alg-Solve-}\acute{\beta}(\boldsymbol{\zeta}\mid \textnormal{\small$\bigstar$})$.}
\label{algo:solvebeta}

\textbf{if} $ \sum_{n \in \mathcal{N}} \widetilde{p}_n(\breve{\alpha}(0,\iffalse T,\fi\boldsymbol{\zeta}\mid \textnormal{\small$\bigstar$}),0,\boldsymbol{\zeta}\mid \textnormal{\small$\bigstar$})  \leq p_{\textnormal{max}}$, then set $\acute{\beta}(\boldsymbol{\zeta}\mid \textnormal{\small$\bigstar$})$ as $0$;

\textbf{else} with function $C(\cdot)$ being set as $\sum_{n \in \mathcal{N}} \widetilde{p}_n(\breve{\alpha}(\beta,\iffalse T,\fi\boldsymbol{\zeta}\mid \textnormal{\small$\bigstar$}),\beta,\boldsymbol{\zeta}\mid \textnormal{\small$\bigstar$})$ (whose computation requires Algorithm~1.2.2.1 and Algorithm~1.2.1's Subprocedure 1), and $C_{\textnormal{target}}$ being set as $p_{\textnormal{max}} $, run $\textnormal{Bisection-Search-with-No-Known-Upper-Bound} \newline ([\boldsymbol{\zeta}, \textnormal{\small$\bigstar$}],\textnormal{function~}C(\cdot),C_{\textnormal{target}})$ based on Algorithm~\ref{algo:BisectionSearchwithNoUpperBound} on Page~\pageref{algo:BisectionSearchwithNoUpperBound} to obtain the desired $\beta$ satisfying Eq.~(\ref{betahat}), and set this $\beta$ as   $\acute{\beta}(\boldsymbol{\zeta}\mid \textnormal{\small$\bigstar$})$;

\end{algorithm}

~

\textbf{Algorithm $\text{2.1}$:~Computing $\grave{\boldsymbol{\zeta}}(T\mid \textnormal{\small$\bigstar$})$ defined in Proposition 2.1 using $\mathcal{S}_{2.1}$ defined in (\number\value{counterty}).} \label{algexplain2.1}  Recall from (\number\value{counterty}) that $\mathcal{S}_{2.1}$ includes (\ref{Complementarytime}),  (\ref{constraintTtau}), and (\ref{Dualfeasibility}e). Then for each $n \in \mathcal{N}$, Proposition~2.1 means  $\grave{{\zeta}}_n(T\mid \textnormal{\small$\bigstar$})$ is ${\zeta_n}$ which satisfies the following: 
 \begin{subnumcases}{}
% \textnormal{based on (\ref{partialLpartialT}), we ensure }\sum_{n \in \mathcal{N}} \zeta_n  = y c_{\hspace{0.8pt}\textnormal{t}}; \label{solveTzeta1} \\ 
\textnormal{based on (\ref{Complementarytime}), we ensure } \nonumber \\  \zeta_n \cdot (t_n(\acute{b}_n(\iffalse T,\fi\boldsymbol{\zeta}\mid \textnormal{\small$\bigstar$}),\acute{p}_n(\iffalse T,\fi\boldsymbol{\zeta}\mid \textnormal{\small$\bigstar$}),s_n,\acute{f}_n^{\textnormal{MS}}(\iffalse T,\fi\boldsymbol{\zeta}\mid \textnormal{\small$\bigstar$}),\nonumber \\\acute{f}_n^{\textnormal{VU}}(\iffalse T,\fi\boldsymbol{\zeta}\mid \textnormal{\small$\bigstar$})) - T) = 0; \label{solveTzeta2}\\ \textnormal{based on (\ref{constraintTtau}), we ensure }  \nonumber \\t_n(\acute{b}_n(\iffalse T,\fi\boldsymbol{\zeta}\mid \textnormal{\small$\bigstar$}),\acute{p}_n(\iffalse T,\fi\boldsymbol{\zeta}\mid \textnormal{\small$\bigstar$}),s_n,\acute{f}_n^{\textnormal{MS}}(\iffalse T,\fi\boldsymbol{\zeta}\mid \textnormal{\small$\bigstar$}),\nonumber \\ \acute{f}_n^{\textnormal{VU}}(\iffalse T,\fi\boldsymbol{\zeta}\mid \textnormal{\small$\bigstar$})) \leq T ;\label{solveTzeta3}\\ \textnormal{based on (\ref{Dualfeasibility}e), we ensure }\zeta_n  \geq 0.\label{solveTzeta4}
\end{subnumcases}

~ 

Given $n \in \mathcal{N}$, we can discuss two cases for ${\zeta_n}$ as follows:
\begin{itemize}[leftmargin=45pt]
\item[Case 1:] If setting $\zeta_n$ to $0$ violates~(\ref{solveTzeta3}), then $\zeta_n$ must be strictly positive, which is used in (\ref{solveTzeta2}) to show that the inequality in (\ref{solveTzeta3}) actually becomes equality in this case.
\item[Case 2:] If setting $\zeta_n$ to $0$ satisfies~(\ref{solveTzeta3}), then we can just set $\zeta_n$ as $0$.
\end{itemize}
Summarizing the above two cases, we know that after defining
\begin{talign}
&\mathring{\boldsymbol{\zeta}}^{(n)} := [\zeta_1, \zeta_2, \ldots, \zeta_{n-1}, 0,\zeta_{n+1}, \ldots, \zeta_N], \textnormal{ and} \label{mathringzetadefine}   
% \end{talign}
% and
% \begin{talign}
\\
&h_n(\boldsymbol{\zeta}\hspace{-2pt}\mid \hspace{-2pt}T) \hspace{-2pt} :=\hspace{-2pt} \begin{cases}
\hspace{-2pt}-\zeta_n,\textnormal{ if }t_n(\acute{b}_n(\mathring{\boldsymbol{\zeta}}^{(n)}\hspace{-2pt}\mid\hspace{-2pt} \textnormal{\small$\bigstar$}),\acute{p}_n(\mathring{\boldsymbol{\zeta}}^{(n)}\hspace{-2pt}\mid\hspace{-2pt} \textnormal{\small$\bigstar$}),s_n,\\ \hspace{25pt}\acute{f}_n^{\textnormal{MS}}(\mathring{\boldsymbol{\zeta}}^{(n)}\hspace{-2pt}\mid \hspace{-2pt}\textnormal{\small$\bigstar$}),\acute{f}_n^{\textnormal{VU}}(\mathring{\boldsymbol{\zeta}}^{(n)}\hspace{-2pt}\mid\hspace{-2pt} \textnormal{\small$\bigstar$})) \hspace{-2pt}\leq\hspace{-2pt} T,~~~\text{(\ref{definehn}a)} \\[-3pt]  \hspace{-2pt}t_n(\acute{b}_n(\iffalse T,\fi\boldsymbol{\zeta}\mid \textnormal{\small$\bigstar$}),\acute{p}_n(\iffalse T,\fi\boldsymbol{\zeta}\mid \textnormal{\small$\bigstar$}),s_n,\acute{f}_n^{\textnormal{MS}}(\iffalse T,\fi\boldsymbol{\zeta}\mid \textnormal{\small$\bigstar$}),\\ \hspace{30pt}\acute{f}_n^{\textnormal{VU}}(\iffalse T,\fi\boldsymbol{\zeta}\mid \textnormal{\small$\bigstar$})) - T, \textnormal{ otherwise,}~~~~~~~\text{(\ref{definehn}b)}
\end{cases}   \label{definehn}
\end{talign}
~\\setting $\boldsymbol{\zeta}$ as $\grave{\boldsymbol{\zeta}}(T\mid \textnormal{\small$\bigstar$})$ always ensures that $h_n(\boldsymbol{\zeta}\mid T)=0$. Letting $n$ iterate through $\mathcal{N}$ and defining
\begin{talign}
&\boldsymbol{h}(\boldsymbol{\zeta}\mid T) := [h_n(\boldsymbol{\zeta}\mid T)|_{n \in \mathcal{N}}], \label{definehn2r}
\end{talign}
 we know that \text{setting $\boldsymbol{\zeta}$ as $\grave{\boldsymbol{\zeta}}(T\mid \textnormal{\small$\bigstar$})$   ensures that }
\begin{talign}
\boldsymbol{h}(\boldsymbol{\zeta}\mid T) \textnormal{ equals the}  \textnormal{ $N$-dimensional zero vector } \boldsymbol{0}. \label{zetah}
\end{talign}
We define $\zeta_n^{\textnormal{upper}}$ such that \textnormal{when $\boldsymbol{\zeta}$ is $[0^{n-1},\zeta_n^{\textnormal{upper}},0^{N-n}]$,}
\begin{talign}
 h_n(\boldsymbol{\zeta}\mid T) = 0.  \label{zetahnupper}
\end{talign}

We will prove the following results and Lemma~\ref{lemsolvepngivenalphaandbetazeta}:
\begin{talign}
& \text{\hspace{2pt}~\textbullet~$\grave{{\zeta}}_n(T\mid \textnormal{\small$\bigstar$}) \leq \zeta_n^{\textnormal{upper}}$, which along with $\zeta_n \geq 0$ means} \nonumber \\ & \hspace{100pt}\text{ $\zeta_n \in [0,\zeta_n^{\textnormal{upper}}]$;}  \label{ding172} \\ & \begin{array}{l}\text{\textbullet~Given any $n \in \mathcal{N}$, for any $\zeta_1, \zeta_2, \ldots, \zeta_{n-1}, \zeta_{n+1}, \ldots, \zeta_N$, }
 \\ \text{we have~~$h_n([\zeta_1, \zeta_2, \ldots, \zeta_{n-1}, \zeta_n^{\textnormal{upper}},\zeta_{n+1}, \ldots, \zeta_N]\mid T) \leq 0$;}
\end{array}  \label{ding173}    \\ & \begin{array}{l}\text{\textbullet~Given any $n \in \mathcal{N}$,  for any $\zeta_1, \zeta_2, \ldots, \zeta_{n-1}, \zeta_{n+1}, \ldots, \zeta_N$, }
 \\ \text{we have~~$h_n([\zeta_1, \zeta_2, \ldots, \zeta_{n-1}, 0,\zeta_{n+1}, \ldots, \zeta_N]\mid T) \geq 0$;}  \\ \text{i.e., $h_n(\mathring{\boldsymbol{\zeta}}^{(n)}) \geq 0$ for $\mathring{\boldsymbol{\zeta}}^{(n)}$ defined in~(\ref{mathringzetadefine}).}
\end{array}  \label{ding174}    \end{talign} 
~\\

\begin{lemma}[Proved in the Appendix of our full version~\cite{full}] \label{lemsolvepngivenalphaandbetazeta}
Given $\boldsymbol{\zeta} $ and $ T$, given  $n \in \mathcal{N}$, given $\zeta_1, \zeta_2, \ldots, \zeta_{n-1}, \zeta_{n+1}, \ldots, \zeta_N$, then $h_n(\boldsymbol{\zeta}~\mid~T)$ is \mbox{non-increasing} as $\zeta_n$ increases. 
\end{lemma}

% \begin{itemize}[leftmargin=15pt]
% \item[(\ref{ding172})] 
% \item[(\ref{ding173})]  \\ .
% \item[(\ref{ding174})]  \\ 
% \end{itemize}

We prove the above Result (\ref{ding172}). From Lemma~\ref{lemsolvepngivenalphaandbetazeta} and~(\ref{zetahnupper}), it holds that
\begin{talign}
&h_n([0^{n-1},\zeta_n^{\textnormal{upper}},0^{N-n}]\mid T) = 0 = h_n(\boldsymbol{\zeta}\mid T) \nonumber \\
&\leq h_n([0^{n-1},\zeta_n,0^{N-n}]\mid T) \leq 0, \label{zetahnupper2}
\end{talign}
which with Lemma~\ref{lemsolvepngivenalphaandbetazeta} means $\grave{{\zeta}}_n(T\mid \textnormal{\small$\bigstar$}) \leq \zeta_n^{\textnormal{upper}}$.

The above Result (\ref{ding173}) clearly follows from Result (\ref{ding172}) and Lemma~\ref{lemsolvepngivenalphaandbetazeta}.

We prove the above Result (\ref{ding174}). From~(\ref{definehn}), it holds that
\begin{talign}
&h_n(\mathring{\boldsymbol{\zeta}}^{(n)})  := \left\{\begin{array}{l}
0,\textnormal{ if }t_n(\acute{b}_n(\mathring{\boldsymbol{\zeta}}^{(n)}\mid \textnormal{\small$\bigstar$}),\acute{p}_n(\mathring{\boldsymbol{\zeta}}^{(n)}\mid \textnormal{\small$\bigstar$}), \\
s_n,\acute{f}_n^{\textnormal{MS}}(\mathring{\boldsymbol{\zeta}}^{(n)}\mid \textnormal{\small$\bigstar$}),\acute{f}_n^{\textnormal{VU}}(\mathring{\boldsymbol{\zeta}}^{(n)}\mid \textnormal{\small$\bigstar$})) \leq T, \\  t_n(\acute{b}_n(\mathring{\boldsymbol{\zeta}}^{(n)}\mid \textnormal{\small$\bigstar$}),\acute{p}_n(\mathring{\boldsymbol{\zeta}}^{(n)}\mid \textnormal{\small$\bigstar$}),\\
s_n,\acute{f}_n^{\textnormal{MS}}(\mathring{\boldsymbol{\zeta}}^{(n)}\mid \textnormal{\small$\bigstar$}),\acute{f}_n^{\textnormal{VU}}(\mathring{\boldsymbol{\zeta}}^{(n)}\mid \textnormal{\small$\bigstar$})) - T,\\ \textnormal{ otherwise,}
\end{array}  \right\}  \geq 0.  \nonumber
\end{talign}
With the above Results (\ref{ding172}) (\ref{ding173}) (\ref{ding174}),
we apply the Poincar\'e--Miranda theorem~\cite{fonda2016generalizing}
and solve~(\ref{zetah}) to obtain $\grave{\boldsymbol{\zeta}}(T\mid \textnormal{\small$\bigstar$})$ using the multivariate bisection algorithm of~\cite{galvan2017multivariate}. The pseudocode is given as Algorithm 2.1 below.  Readers may wonder why the multivariate bisection is not used to jointly solve $[\boldsymbol{b},\boldsymbol{p},\boldsymbol{f}^{\textnormal{MS}},\boldsymbol{f}^{\textnormal{VU}},\alpha,\beta,\gamma,\boldsymbol{\delta}]$. The reason is that the conditions to use the multivariate bisection are quite strict; e.g., Results (\ref{ding172}) (\ref{ding173}) (\ref{ding174}) are \textbf{for any} $\zeta_1, \zeta_2, \ldots, \zeta_{n-1}, \zeta_{n+1}, \ldots, \zeta_N$ given any $n \in \mathcal{N}$. We do not have such strong conditions if we try to solve $[\boldsymbol{b},\boldsymbol{p},\boldsymbol{f}^{\textnormal{MS}},\boldsymbol{f}^{\textnormal{VU}},\alpha,\beta,\gamma,\boldsymbol{\delta}]$ together. 

% Given $\boldsymbol{z},y, \boldsymbol{s}$,
% suppose $T$ is also given. Then $\boldsymbol{\zeta}$ satisfying~(\ref{solveTzeta1})--(\ref{solveTzeta4}), denoted by $\grave{\boldsymbol{\zeta}}(T\mid \boldsymbol{z},y, \boldsymbol{s})$, is the solution to the following:

\renewcommand{\thealgocf}{2.1 explained above}
\begin{algorithm}
\caption{$\textnormal{Alg-Solve-}\grave{\boldsymbol{\zeta}}(T\mid \textnormal{\small$\bigstar$})$.
% , which finds $\grave{\boldsymbol{\zeta}}(T\mid \textnormal{\small$\bigstar$})$ via bisection search according to~?.
}
\label{alg2.1}

For $\boldsymbol{\zeta} :=[\zeta_1, \zeta_2, \ldots,  \zeta_N] \in \prod_{n \in \mathcal{N}} [0,\zeta_n^{\textnormal{upper}}]$ for $\zeta_n^{\textnormal{upper}}$ defined in~(\ref{zetahnupper}), given Results (\ref{ding172}) (\ref{ding173}) (\ref{ding174}), use the multivariate bisection algorithm proposed by~\cite{galvan2017multivariate}
to obtain $\boldsymbol{\zeta}$ which induces $\boldsymbol{h}(\boldsymbol{\zeta}\mid T)$ to be the $N$-dimensional zero vector $\boldsymbol{0}$, to achieve the tolerance level of  $\|\boldsymbol{h}(\boldsymbol{\zeta}\mid T)\|_2 \leq \epsilon_5$, where computing $\boldsymbol{h}(\boldsymbol{\zeta}\mid T)$ defined in (\ref{definehn}) and (\ref{definehn2r}) will call Algorithm 1.1 (resp., Algorithm 1.2) to compute $\acute{f}_n^{\textnormal{MS}}(\iffalse T,\fi\mathring{\boldsymbol{\zeta}}^{(n)}\mid \textnormal{\small$\bigstar$}) $ and $ \acute{f}_n^{\textnormal{VU}}(\iffalse T,\fi\mathring{\boldsymbol{\zeta}}^{(n)}\mid \textnormal{\small$\bigstar$})$ as well as $\acute{f}_n^{\textnormal{MS}}(\iffalse T,\fi\boldsymbol{\zeta}\mid \textnormal{\small$\bigstar$}) $ and $ \acute{f}_n^{\textnormal{VU}}(\iffalse T,\fi\boldsymbol{\zeta}\mid \textnormal{\small$\bigstar$})$ 
 (resp., $\acute{b}_n(\iffalse T,\fi\mathring{\boldsymbol{\zeta}}^{(n)}\mid \textnormal{\small$\bigstar$})$ and $\acute{p}_n(\iffalse T,\fi\mathring{\boldsymbol{\zeta}}^{(n)}\mid \textnormal{\small$\bigstar$})$ as well as $\acute{b}_n(\iffalse T,\fi\boldsymbol{\zeta}\mid \textnormal{\small$\bigstar$})$ and $\acute{p}_n(\iffalse T,\fi\boldsymbol{\zeta}\mid \textnormal{\small$\bigstar$})$) as inputs to the function $t_n()$ defined in~(\ref{eq:userdelay}), where $\mathring{\boldsymbol{\zeta}}^{(n)}$ is  defined in~(\ref{mathringzetadefine});

Return the obtained $\boldsymbol{\zeta}$ as $\grave{\boldsymbol{\zeta}}(T\mid \textnormal{\small$\bigstar$})$;
\end{algorithm}

~

\textbf{Algorithm $\text{2.2}$:~Computing $T^{\#}(\textnormal{\small$\bigstar$})$ according to Proposition 2.2 using $\mathcal{S}_{2.2}$ defined in (\number\value{countertt}).} \label{algexplain2.2} Recall from (\number\value{countertt}) that $\mathcal{S}_{2.2}$ includes (\ref{partialLpartialT}). Then Proposition~2.2 defines  $T^{\#}(\textnormal{\small$\bigstar$})$ as the solution of $T$ to 
% With $\grave{\boldsymbol{\zeta}}(T\mid \textnormal{\small$\bigstar$})$ defined above, $T^{\#}(\boldsymbol{z},y,\boldsymbol{s})$ (i.e., $T$ satisfying~(\ref{solveTzeta1})--(\ref{solveTzeta4})) is the solution to the following:
\begin{talign}
\sum_{n \in \mathcal{N}} \grave{\zeta}_n(T\mid \textnormal{\small$\bigstar$}) = y c_{\hspace{0.8pt}\textnormal{t}}.\label{zetah2}    
\end{talign}

\renewcommand{\thealgocf}{2.2 explained above}

\begin{algorithm}
\caption{$\textnormal{Alg-Solve-}T^{\#}(\textnormal{\small$\bigstar$})$.
% , which finds $T^{\#}(\textnormal{\small$\bigstar$})$ via bisection search according to~?.
}
\label{alg2.2}

% Initialize with a random $T_{ini} > 0$;

% Return $T_{ini} > 0$
% find $T_{up}:=2^i T_{ini}$ such that $i$ is the smallest integer 

Run $\textnormal{Bisection-Search-with-No-Known-Upper-Bound}$ $(\textnormal{\small$\bigstar$},\sum_{n \in \mathcal{N}} \grave{\zeta}_n(T\mid \textnormal{\small$\bigstar$}), y c_{\hspace{0.8pt}\textnormal{t}})$ based on Algorithm~\ref{algo:BisectionSearchwithNoUpperBound} on Page~\pageref{algo:BisectionSearchwithNoUpperBound}  to obtain the desired $T$ satisfying Eq.~(\ref{zetah2}), and set this $T$ as $T^{\#}(\textnormal{\small$\bigstar$})$, where computing $\sum_{n \in \mathcal{N}} \grave{\zeta}_n(T\mid \textnormal{\small$\bigstar$})$ will call Algorithm 2.1;
\end{algorithm}

% Based on the above analyses, we present the pseudocodes of the algorithms from Page ? to Page ?.

~

The bisection method is used repeatedly in the algorithms above. The pseudocodes are given below.

\renewcommand{\thealgocf}{A\arabic{algocf}}

\setcounter{algocf}{2}

\begin{algorithm}
\caption{\newline$\textnormal{Bisection-Search-with-No-Known-Upper-Bound}$ $(\boldsymbol{v},\textnormal{function~}C(u, \boldsymbol{v}),C_{\textnormal{target}})$, which returns $u \in [0,\infty)$ such that $C(u, \boldsymbol{v})$ equals (or is arbitrarily close to) $C_{\textnormal{target}}$ given $\boldsymbol{v}$, where $C(u, \boldsymbol{v})$ is \mbox{non-increasing} in $u$ given $\boldsymbol{v}$. 
% The function $\textnormal{Check}(C(u', \boldsymbol{v}),C_{\textnormal{target}})$ returns ``$>$'' (resp., ``$<$'', ``$=$'') if $C(u', \boldsymbol{v})$ is greater than (resp., less than, equal to) $C_{\textnormal{target}}$.
} 
\label{algo:BisectionSearchwithNoUpperBound}

Randomly pick $u^{(0)}$ from $(0,\infty)$;

\textbf{if} $C(u^{(0)}, \boldsymbol{v})=C_{\textnormal{target}}$:  return $u^{(0)}$ as the desired $u$;

\textbf{if} $C(u^{(0)}, \boldsymbol{v})<C_{\textnormal{target}}$  \newline//In this case, the solution $u$ is in $[0,u^{(0)})$

 \quad Use $\textnormal{Standard-Bisection-Search}(\boldsymbol{v},\textnormal{function~}C(u, \boldsymbol{v}),$ $C_{\textnormal{target}}, 0,u^{(0)})$ based on Algorithm~\ref{algo:bisection} to find the result, and return it as the desired $u$;

\textbf{if} $C(u^{(0)}, \boldsymbol{v})>C_{\textnormal{target}}$ 
\newline //In this case, the solution $u$ is in $(u^{(0)},\infty)$

 \quad Find $i\geq 0$ such that $C(u^{(0)}\cdot 2^i, \boldsymbol{v})>C_{\textnormal{target}}$ but $C(u^{(0)}\cdot 2^{i+1}, \boldsymbol{v}) \leq C_{\textnormal{target}}$  

\quad \textbf{if} $C(u^{(0)}\cdot 2^{i+1}, \boldsymbol{v}) = C_{\textnormal{target}}$:  return $u^{(0)}\cdot 2^{i+1}$  as the desired $u$;

\quad \textbf{elsif} $C(u^{(0)}\cdot 2^{i+1}, \boldsymbol{v}) < C_{\textnormal{target}}$

\quad \quad Use $\textnormal{Standard-Bisection-Search}(\boldsymbol{v},$\newline \hspace*{16pt}$\textnormal{function~}C(u, \boldsymbol{v}),C_{\textnormal{target}}, u^{(0)}\cdot 2^i,u^{(0)}\cdot 2^{i+1})$ \newline \hspace*{16pt} based on Algorithm~\ref{algo:bisection} to find the result, \newline \hspace*{16pt} and return it as the desired $u$;

\quad \textbf{endif} 

\textbf{endif} 

\end{algorithm}

\begin{algorithm}
\caption{\newline$\textnormal{Standard-Bisection-Search}$ $(\boldsymbol{v},\textnormal{function~}C(u, \boldsymbol{v}),C_{\textnormal{target}}, \textnormal{lower bound}~L_0, \textnormal{upper bound}~U_0)$, which returns $u \in [L_0,U_0]$ such that $C(u, \boldsymbol{v})$ equals (or is arbitrarily close to) $C_{\textnormal{target}}$ given $\boldsymbol{v}$, where $C(u, \boldsymbol{v})$ is \mbox{non-increasing} in $u$ given $\boldsymbol{v}$. 
% The function $\textnormal{Check}(C(u', \boldsymbol{v}),C_{\textnormal{target}})$ returns ``$>$'' (resp., ``$<$'', ``$=$'') if $C(u', \boldsymbol{v})$ is greater than (resp., less than, equal to) $C_{\textnormal{target}}$.
} 
\label{algo:bisection}

Initialize $B_{\textnormal{lower}} \leftarrow L_0$,  $B_{\textnormal{upper}} \leftarrow U_0$; 

\Repeat{$B_{\textnormal{upper}} -B_{\textnormal{lower}}$ is no greater than $\epsilon_4$ for a small positive number $\epsilon_4$}{

$u \leftarrow \frac{B_{\textnormal{lower}} + B_{\textnormal{upper}}}{2} $; 

\textbf{if} $C(u, \boldsymbol{v})=C_{\textnormal{target}}$:
 return $u$;

% \textbf{endif}

\textbf{if} $C(u, \boldsymbol{v})>C_{\textnormal{target}}$:
 $B_{\textnormal{lower}} \leftarrow  u$;

\textbf{else}: 
 $B_{\textnormal{upper}} \leftarrow  u$;

\textbf{endif}

} 

\end{algorithm}

~

To better understand bisection search in our algorithms above, 
we prove in the Appendix of our full paper~\cite{full} that the left-hand side of~(\ref{fmseq}) (resp.,~(\ref{solvepngivenalphaandbeta}),~(\ref{brevealpha}),~(\ref{betahat}),~(\ref{zetah2})) is \mbox{non-increasing} with respect to $\gamma$ (resp., $p_n$, $\alpha$, $\beta$, $T$). In simulations, the above often decreases so that there is a unique solution.

\section{Our Algorithm to Solve Problem $\mathbb{P}_{1}$} \label{sec: algorithm}

 Algorithm~\ref{algo:MN} has been presented on Page~\pageref{algo:MN}  to solve the system UCR optimization  $\mathbb{P}_{1}$. In Section~\ref{Putting}, we have also explained how the different building blocks in Sections~\ref{secDinkelbach},~\ref{AO1},~\ref{FPsec}, and~\ref{Optimizings} are combined together to produce Algorithm~\ref{algo:MN}'s pseudocode. We now discuss the performance of Algorithm~\ref{algo:MN}.

\textbf{Solution quality and convergence.} Algorithm~\ref{algo:MN} comprises three levels of iterations, with the innermost iteration from Line 17 containing Algorithm~\ref{algo:P5} in Line 20. Algorithm~\ref{algo:P5} obtains a global optimum for Problem $\mathbb{P}_{5}(\textnormal{\small$\bigstar$})$. The outermost iteration from Line 4 is based on Dinkelbach's transform and does not lose optimality. However, the mid-level iteration from Line 11 is based on alternating optimization and cannot guarantee local/global optimality. Hence, Algorithm~\ref{algo:MN} cannot guarantee local/global optimality for $\mathbb{P}_{1}$. Yet, using the terminology of stationary points in~\cite{shen2018fractional} for constrained optimization, Algorithm~\ref{algo:MN} finds a stationary point for $\mathbb{P}_{1}$. The convergence of Algorithm~\ref{algo:MN} is also clear from the above analysis.  
 
% \textbf{Convergence.} ?
 
\textbf{Time Complexity.} In Algorithm~\ref{algo:MN} and its subroutine Algorithm~\ref{algo:P5}, the bisection search is repeatedly used. Then the  complexity of Algorithm~\ref{algo:MN} is polylogarithmic in the error tolerance's reciprocal in various calls of the bisection search. Below we analyze the  complexities of Algorithms~\ref{algo:MN} and~\ref{algo:P5} with respect to the number $N$ of users.
% ? We first analyze the time complexity of Algorithm~\ref{algo:P5} for globally optimizing $\mathbb{P}_{5}(\textnormal{\small$\bigstar$})$.
 Line 1 of Algorithm~\ref{algo:P5} calls Algorithm 2.2, which calls Algorithm 2.1. Algorithm 2.1 called in the above and in Line 2 of Algorithm~\ref{algo:P5}  calls Algorithms 1.1 and 1.2. Algorithm 1.2 called in the above and in Line 3 of Algorithm~\ref{algo:P5} calls Algorithms 1.2.2.2, 1.2.2.1, and 1.2.1. For the multivariate bisection search in Algorithm 2.1, from Theorem 2.3 of~\cite{galvan2017multivariate}, to achieve the tolerance level of  $\|\boldsymbol{h}(\boldsymbol{\zeta}\mid T)\|_2 \leq \epsilon_5$, the number of iterations required is $\log_2 \frac{\sum_{n \in \mathcal{N}}\zeta_n^{\textnormal{upper}}}{\epsilon_5}$ for $\zeta_n^{\textnormal{upper}}$ define in~(\ref{zetahnupper}); i.e., logarithmic in $N$.
Computing $\widetilde{\boldsymbol{b}}(\alpha,\beta,\boldsymbol{\zeta}\mid \textnormal{\small$\bigstar$})$ and $\widetilde{\boldsymbol{p}}(\alpha,\beta,\boldsymbol{\zeta}\mid \textnormal{\small$\bigstar$})$ takes $\mathcal{O}(N)$. Then calculating $\breve{\alpha}(\beta,\iffalse T,\fi\boldsymbol{\zeta}\mid \textnormal{\small$\bigstar$})$ costs $\mathcal{O}(N)$. Thus, obtaining $\acute{\beta}(\boldsymbol{\zeta}\mid \textnormal{\small$\bigstar$})$ takes $\mathcal{O}(N)$. Finally, computing $\boldsymbol{h}(\boldsymbol{\zeta}\mid T)$ costs $\mathcal{O}(N)$. Line 1 of Alg.~\ref{algo:P5} takes $\mathcal{O}(N\log N)$. We analyze other lines of Alg.~\ref{algo:P5} similarly. Then Alg.~\ref{algo:P5} and each innermost iteration of Alg.~\ref{algo:MN} cost $\mathcal{O}(N\log N)$. In the outermost and mid-level iterations, each computation of the utility $\mathcal{U}()$ in~(\ref{eq:utility}) and the cost in~(\ref{eq:systemcost}) requires $\mathcal{O}(N)$. To summarize, Algorithm~\ref{algo:MN} takes $\mathcal{O}(N^2\log N)$.

% Obtaining $\boldsymbol{h}(\boldsymbol{\zeta}\mid T)$ requires $h_n(\boldsymbol{\zeta}\mid T)$ for $n \in \mathcal{N}$ .  The total complexity of $\mathbb{P}_{5}(\textnormal{\small$\bigstar$})$ is linear in the number $N$ of users and polylogarithmic in the error tolerance's reciprocal in the various calls of the bisection search.

\section{Modeling the Human-centric Utility from Real Data} \label{sec:modelutility}
%In terms of the expression of the human-perceived utility function $U_n(\cdot)$, we consider the datasets in \cite{elwardy2022ssv360} where five participants are involved to evaluate videos of four scenarios, i.e., ``FormationPace'', ``Alcatraz'', ``BloomingAppleOrchards'', and ``PandaBaseChengdu''. There are five resolution combinations in total: $7680\times3840$, $6144\times3072$, $4096\times2048$, $3600\times1800$, and $2048\times1024$. Additionally, the postures of the participants using the VR devices, i.e., seated and standing, are also taken into account, which makes the data richer and closer to the practical metaverse application scenario. 

We now model users' human-centric utilities in the Metaverse over wireless communications using two datasets~\cite{ssv360,Netflix} explained below, which are both based on real experiments of humans assessing videos.

\textbf{SSV360 dataset.} This dataset of~\cite{ssv360} captures users' evaluation of 360\degree~videos when wearing HTC Vive Pro Virtual Reality (VR) headsets. Each data point exhibits  a user's perceptual quality assessment of a 360\degree~scene of a given bitrate and a given video resolution, under standing or seated viewing (SSV). 
% The primary concern in the quality of experience (QoE) design is creating a human-perceived utility function. To accurately measure a user's subjective perception of video quality, we utilize the SSV360 dataset \cite{ssv360}, which is constructed by virtual-reality users in 360\degree~video scenarios. 
% In the dataset, having data points under different video bitrates yet the same resolution is due to different quantization parameters used in video compression. 

\textbf{Netflix dataset.} This dataset is a part of Netflix's Emmy Award-winning VMAF project~\cite{Netflix}. Each data point represents users' mean opinion score for a video at a given bitrate and a given resolution.

The wireless data rate needs to be large enough for users' smooth watching experience at the given video bitrate~\cite{liu2018edge}. We consider the  bitrate as a constant fraction (say $\theta$) of the wireless rate. Then substituting the bitrate $r_{\text{bitrate}}$ with the wireless rate $r_{\text{wireless}}$ just involves replacing $r_{\text{bitrate}}$ with $r_{\text{wireless}}/\theta$. Hence, for both datasets above, we perform curve-fitting with the bitrate and the resolution to obtain the utility functions. 

\textbf{Modeling human-centric utilities.} Based on the two datasets above, the human-centric utility of each user $n$, denoted by $U_n(r_n,s_n)$, is modeled as a function of the bitrate $r_n$ and the video resolution $s_n$. We adopt the logarithmic utility function, which is used in~\cite{yang2012crowdsourcing,deng2020wireless,lyu2021service} for various communication/network systems. The logarithmic function reflects users' diminishing marginal gain as the bitrate and the resolution  increase. Formally, we have $U_n(r_n,s_n)=\kappa_n \ln (1+ l^s_ns_n+l^r_nr_n)$ for coefficients $\kappa_n, l^s_n,l^r_n$, which are decided by fitting data. Since using a three-dimensional plot to show the two-variable function $U_n(r_n,s_n)$ is difficult for visual interpretation, we use the following transform to obtain a two-dimensional plot. Let $r_n ^{\max}$ (resp., $s_n^{\max}$) be the maximum  $r_n $ (resp.,  $s_n $) from the dataset. After defining $\alpha_n:=l^r_n r_n ^{\max}+l^s_n s_n^{\max}$, we  let $\frac{l^r_n}{\alpha_n}r_n+\frac{l^s_n}{\alpha_n}s_n$ be the $x$-axis coordinate, and plot $\kappa_n \ln (1+ \alpha_n x)$ as the $y$-axis coordinate, since it holds that $U_n(r_n,s_n)=\kappa_n \ln (1+ \alpha_n \cdot (\frac{l^r_n}{\alpha_n}r_n+\frac{l^s_n}{\alpha_n}s_n))=\kappa_n \ln (1+ \alpha_n x)$. With the above transformation, each data point's $x$-coordinate is between $0$ and $1$.

% Then $U_n(r_n,s_n)=\kappa_n \ln (1+ \alpha_n \cdot (\frac{l^r_n}{\alpha_n}r_n+\frac{l^s_n}{\alpha_n}s_n))$

% The $x$-axis is $(\frac{l^r_n}{\alpha_n}r_n+\frac{l^s_n}{\alpha_n}s_n)$, and the $y$-axis is $\kappa_n \ln (1+ \alpha_n x)$. 

% Under the specific user, scenario, and locomotion (standing or seated), the rating is most influenced by the resolution and bitrate~\cite{ssv360}. Therefore, the sum of these two is chosen to form the control variable: $l^s_ns_n+l^r_nr_n$, where $l^s_n,l^r_n$ are normalization parameters of resolution and bitrate for VU $n$ and the units of $s_n$ and $r_n$ are 100M pixels and 100M bps. Besides, we use $U_n(r_n,s_n)=\kappa_n \ln (1+ l^s_ns_n+l^r_nr_n)$ as the utility function??., 

% Use the data to learn $\kappa_n,l^s_n, l^r_n$ such that $U_n(r_n,s_n)=\kappa_n \ln (1+l^r_nr_n+ l^s_ns_n)$

\iffalse

This function reflects the diminishing return when increasing the resolution and bitrate for VU $n$, and $\kappa_n$ is the user-specific parameters for better fitting their real dataset. Since it is unlikely to quantify the influences of these two factors on the utility separately, we fine-tune the utility function parameters $\kappa_n$, and normalization parameters $l^s_n,l^r_n$ to approximate the data from SSV360. 

\fi

% and just  be a constant  the video bitrate times t and perform curve-fitting to see how it impacts a user's perceptual score, as shown in Fig.~\ref{fig:real_simul}(a).

\begin{figure*}[t]
\renewcommand{\thesubfigure}{\footnotesize(\alph{subfigure}}
\centering
\begin{subfigure}{0.45\textwidth}
\centering
\includegraphics[width=1\linewidth]{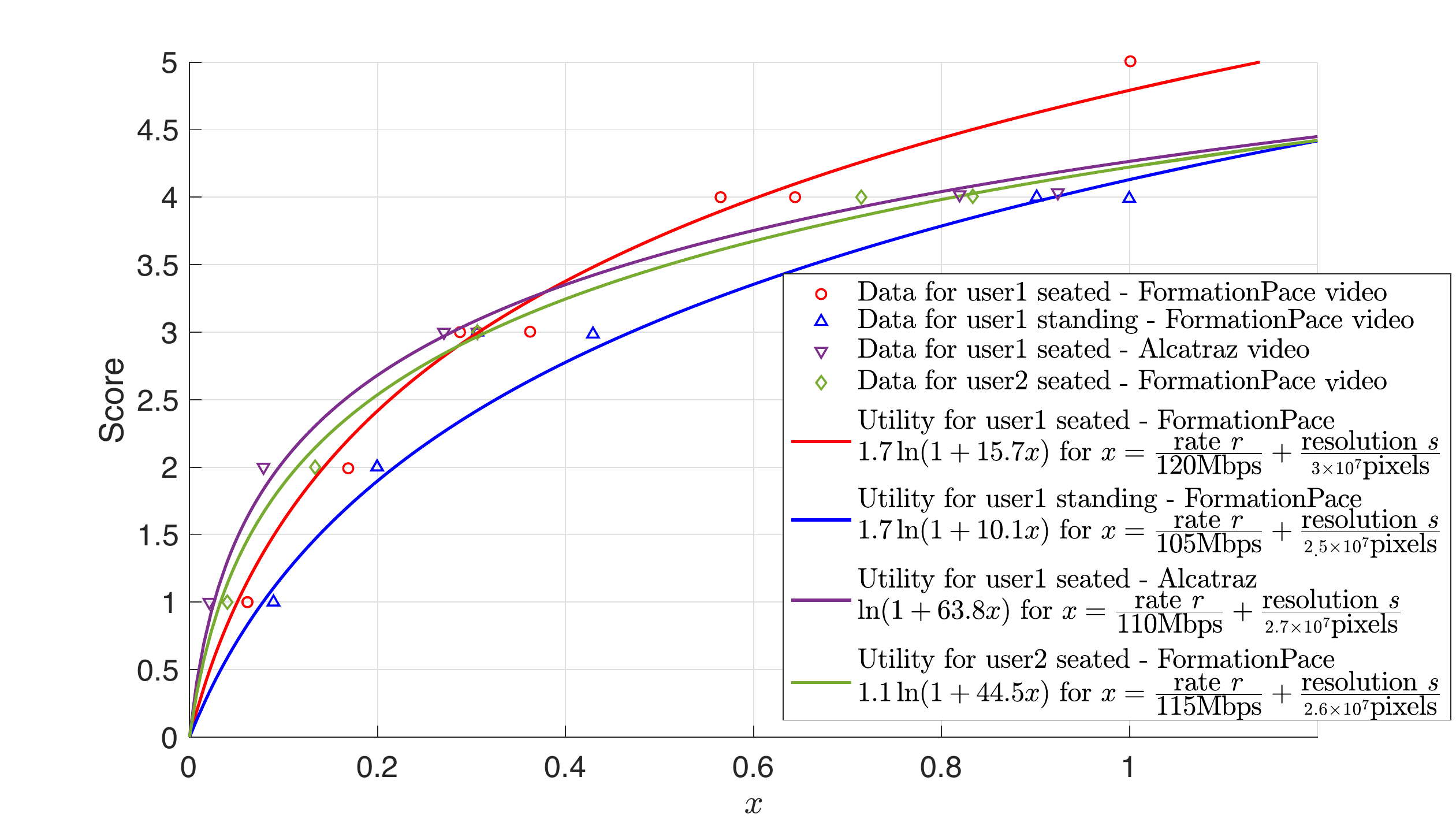}
\vspace{-7mm}
\subcaption{SSV360 dataset.}\label{fig:ssvjz}\vspace{-2mm}
\end{subfigure}%
\begin{subfigure}{0.45\textwidth}
\centering
\includegraphics[width=1\linewidth]{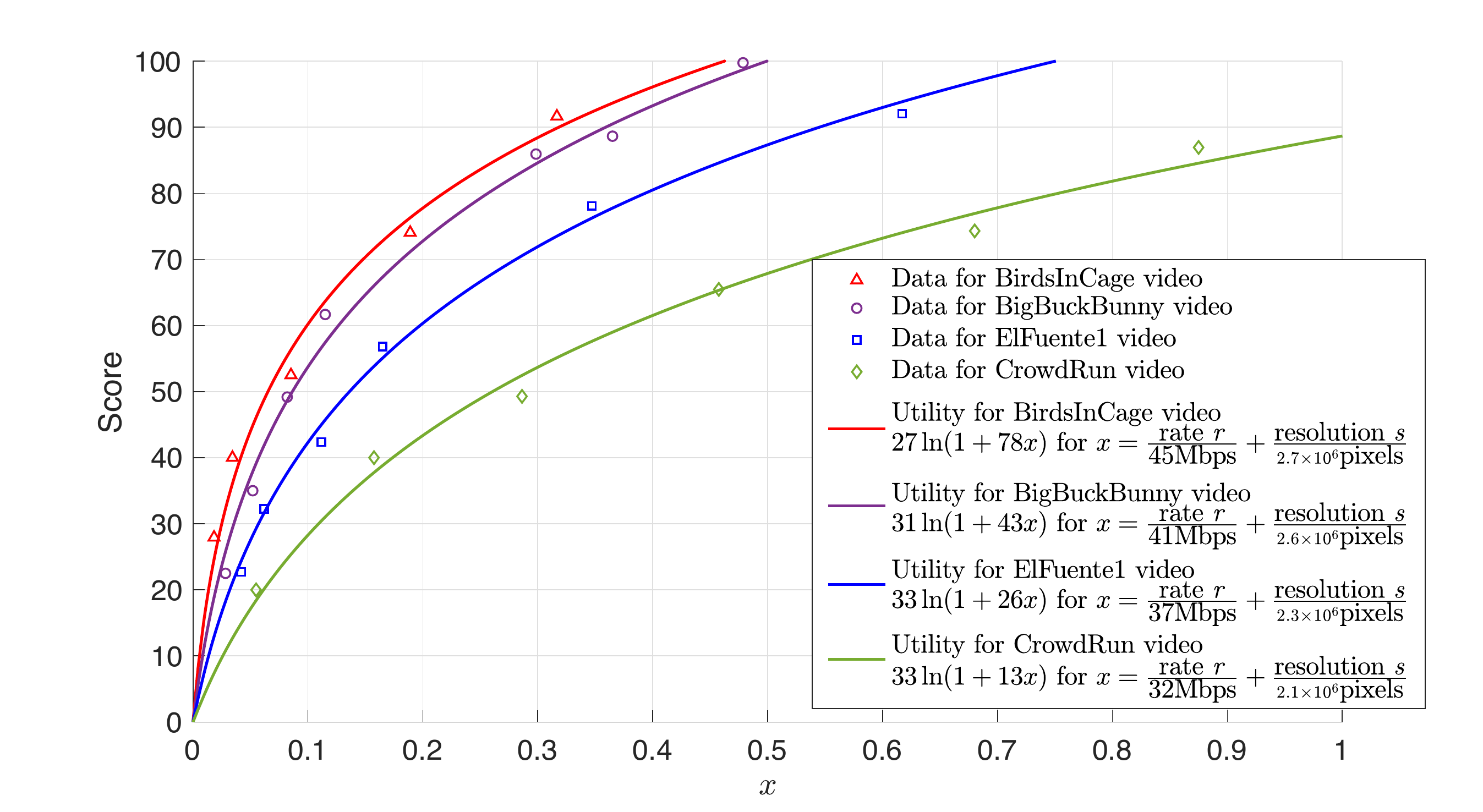}
\vspace{-7mm}
\subcaption{Netflix dataset.}\label{fig:resolutionjz}\vspace{-2.2mm}
\end{subfigure}%
\caption{Modeling the logarithmic human-centric utility functions from the SSV360 and \vspace{-10pt}Netflix datasets.}
\label{fig:Modelingutility}
\end{figure*}

In Fig.~\ref{fig:Modelingutility}(a) for the SSV360 dataset, the data and curves are about two users watching a 360\degree~video (``Alcatraz'' or ``FormationPace''~\cite{ssv360}) under seated or standing view. The score is an integer from~$1$ to~$5$ based on the well-known Absolute Category Rating. In Fig.~\ref{fig:Modelingutility}(b) for the Netflix dataset, the data and curves present users' average assessment (from $0$ to $100$) of  different videos (BirdsInCage, BigBuckBunny, ElFuente1, or CrowdRun~\cite{Netflix}). Both subfigures demonstrate that the curves of the logarithmic human-centric utility functions  fit the data. The specific expressions of the functions are provided in the \vspace{-8pt} legends.

\section{Simulations\vspace{-5pt}} \label{sec: experiment}

In this section on simulations,
we first describe the default settings  and then report various results.

% \subsection{Numerical settings} \label{sec-Numerical}

% The number of VUs $N$ is set from 4 to 20. We utilize 40 different utility functions in total and randomly allocate them for VUs.

\textbf{Default settings.}
We consider a macro-cell wireless channel model for urban areas. With $d_n$ denoting the distance between the Metaverse server (MS) and a virtual-reality user (VU) indexed by $n$, the path loss between them is $128.1+37.6\log d_n$ along with 8 decibels (dB) for the standard deviation of shadow fading~\cite{zhou2022icdcs}, where the unit of $d_n$ is kilometer. The power spectral density of Gaussian noise ${\sigma_n}^2$ is $-174$\,dBm/Hz (i.e., the thermal noise amount at 20\,\degree{C} room temperature). VUs are randomly located in a circle of radius 500m centered at the MU. 
The default total bandwidth $b_{\textnormal{max}}$ is 20GHz, and the total transmission power $p_{\textnormal{max}}$ is 30W. The effective switched capacitance $\kappa^{\textnormal{MS}}$ and $\kappa^{\textnormal{VU}}$ are set as $10^{-27}$. The number $\mu_n$ of bits per pixel is 16, and the compression rate $\nu_n$ is 100. The maximum CPU frequencies at the MS and VUs, $f_{\textnormal{max}}^{\textnormal{MS}}$ and $f_{n,\textnormal{max}}^{\textnormal{VU}}$, are 300GHz and 50GHz, respectively. The default  weights for  energy and delay are $c_e = 0.5$ and $c_t = 0.5$. {The default VU number is 5.} Based on measurements,
Section V of~\cite{liu2018edge} quantifies the computational complexity of processing a video frame of resolution $s_n$ as   $w(s_n) = (7 \times 10^{-10} \times {s_n}^{3/2} +  0.083) $ tera (i.e., trillion) floating-point operations (FLOPs). From Fig.~\ref{timeline} in Section~\ref{Delayenergy}, we know that $\mathcal{A}_n(s_n,\Lambda_n) $ (resp., $  \mathcal{B}_n(s_n,\Lambda_n)$) of Page~\pageref{defineAB} for delay is less than $\mathcal{F}_n(s_n,\Lambda_n) $ (resp., $  \mathcal{G}_n(s_n,\Lambda_n)$) of Page~\pageref{defineFG} for energy. In the simulations, we set both $\mathcal{A}_n(s_n,\Lambda_n) $ and $ \mathcal{B}_n(s_n,\Lambda_n)$  as
$\Lambda_n w(s_n) /30$, and set both $\mathcal{F}_n(s_n,\Lambda_n) $ and $ \mathcal{G}_n(s_n,\Lambda_n)$ as
$\Lambda_n w(s_n)$, which make all of them convex in $s_n$. For all VU $n$, we let $\Lambda_n$ be the same $\Lambda$. 
Then the optimization objective becomes a multiple of $\Lambda$, which thus has no impact. The above avoids considering the impact of heterogeneous $\Lambda_n$ for simplicity. Possible values for the resolution $s_n$ are $4096 \times 2160$, $3072 \times 1620$, $2048 \times 1080$, $1920 \times 1080$, and $1280 \times 720$ pixels, which are also referred to  as 4k, 3k, 2k, 1080p, and 720p. The SSV360 dataset~\cite{ssv360}  in Section~\ref{sec:modelutility}  includes the perceptual assessment of  users watching VR videos. We use those data for curve-fitting different logarithmic utility functions, and assign the functions to users in the simulations: one function for one user. 

% in \cite{zhou2022resource} and users are selected with equal probability from a circle of radius 500m. The parameters of utility functions for various VUs are derived from measured data in SSV360~\cite{ssv360}. We utilize 40 different utility functions in total.

% We consider both $\mathcal{A}_n(s_n,\Lambda_n), \mathcal{B}_n(s_n,\Lambda_n)$ as
% $(7 \times 10^{-10} \times {s_n}^{3/2} +  0.083) \times 10^{12} /30 $  floating point operations. The video for each XU $n$ comprises $\Lambda_n = 300$ VR frames. For simplicity, we just set $\mathcal{F}_n(s_n,\Lambda_n) $ (resp., $  \mathcal{G}_n(s_n,\Lambda_n)$)as $\Lambda_n \mathcal{A}_n(s_n,\Lambda_n) $ (resp., $  \Lambda_n \mathcal{B}_n(s_n,\Lambda_n)$). 

% We adopt the following function for ? proposed 

% With the frame rate being $29.97$ frames per second, the  

% the required number of CPU cycles per bit is 1000 \cite{you2018asynchronous}, thus, the CPU cycles of one frame at the MS and user side $\mathcal{A}_n(s_n,\Lambda_n), \mathcal{B}_n(s_n,\Lambda_n)$ can be $\frac{\text{resolution}\times24\times2\times1000}{\text{compression rate}}$.

% \subsection{Compared with other baselines}

\begin{figure*}[!t]
\centering
\begin{subfigure}{0.33\textwidth}
    \centering
    \includegraphics[width=1\linewidth]{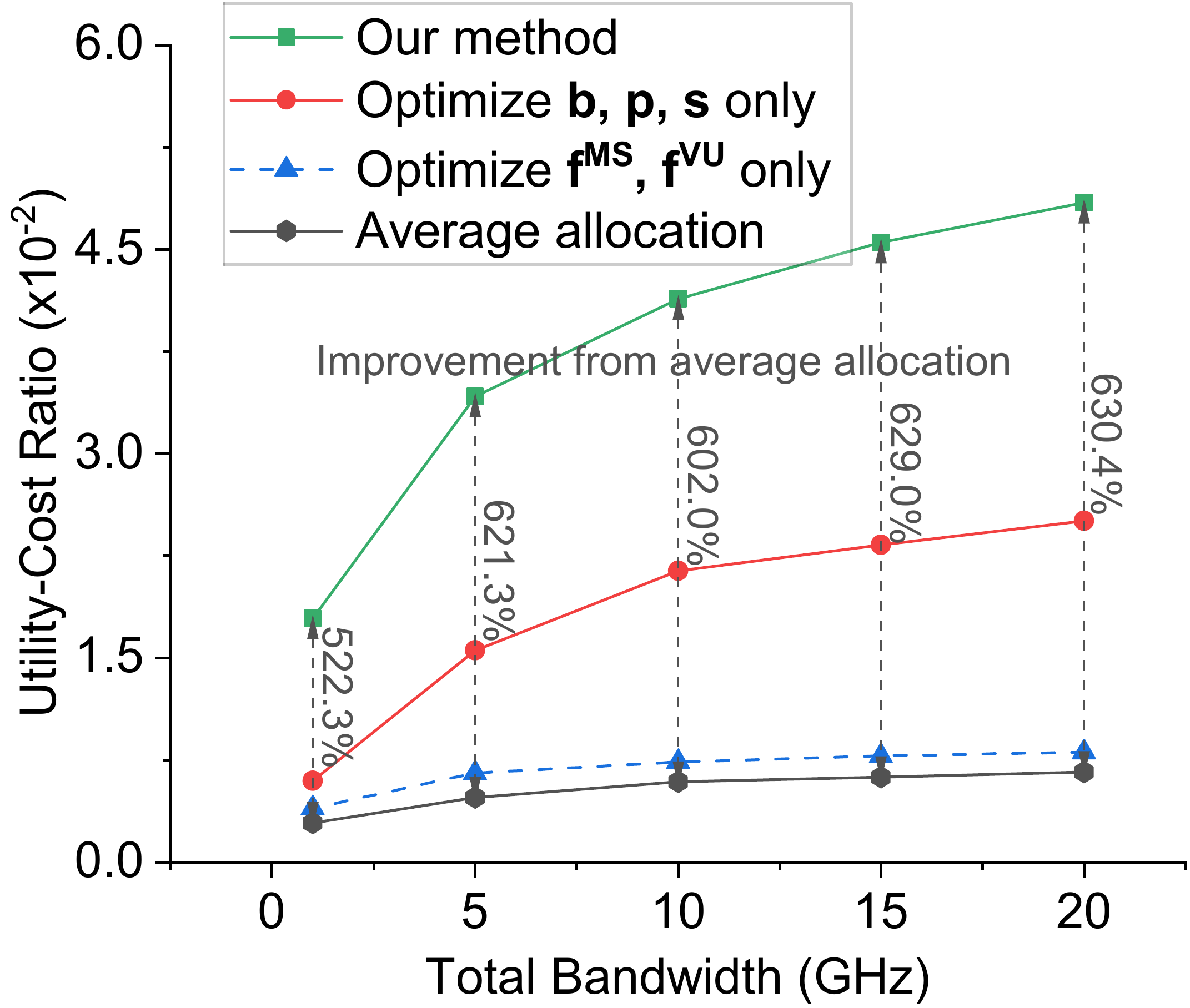}
    \label{fig:bandwidth}\vspace{-6mm}
    \subcaption{}\vspace{1mm}
\end{subfigure}%
\begin{subfigure}{0.33\textwidth}
    \centering
    \includegraphics[width=1\linewidth]{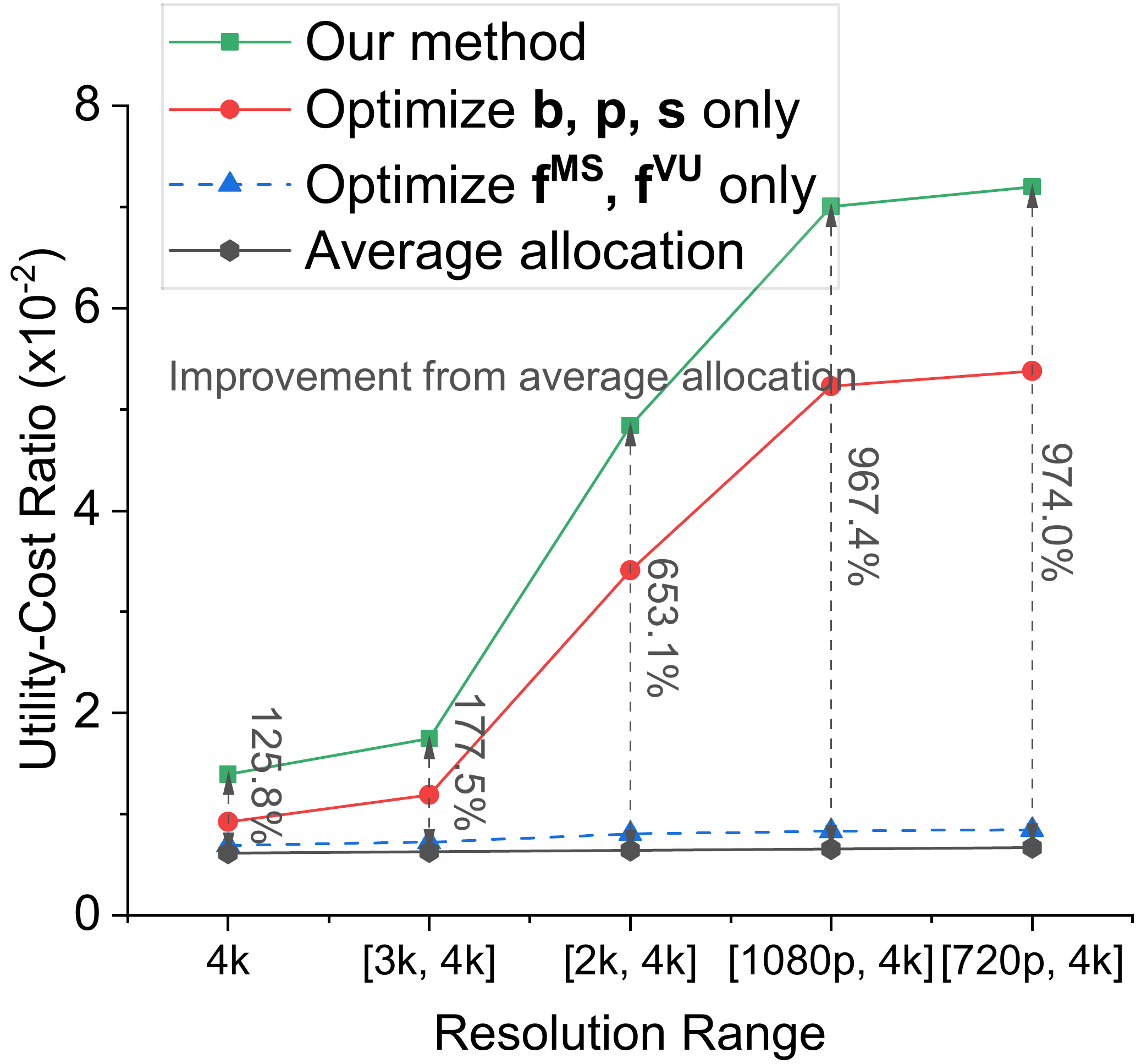}
    \label{fig:resolution}\vspace{-6mm}
    \subcaption{}\vspace{1mm}
\end{subfigure}%
\begin{subfigure}{0.33\textwidth}
\centering
\includegraphics[width=1\linewidth]{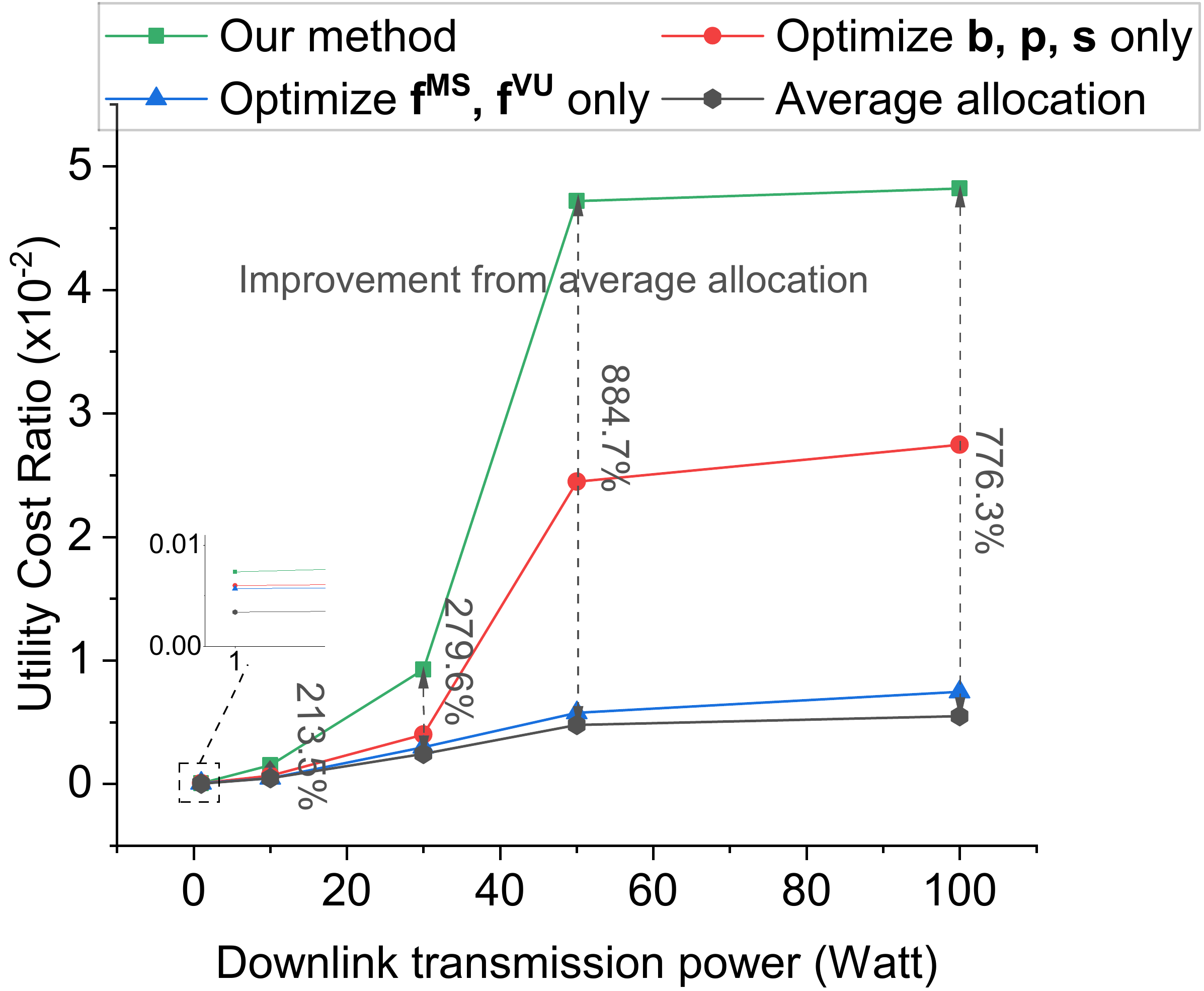}
\label{fig:power}\vspace{-6mm}
\subcaption{}\vspace{1mm}
\end{subfigure}%

\begin{subfigure}{0.33\textwidth}
\centering
\includegraphics[width=1\linewidth]{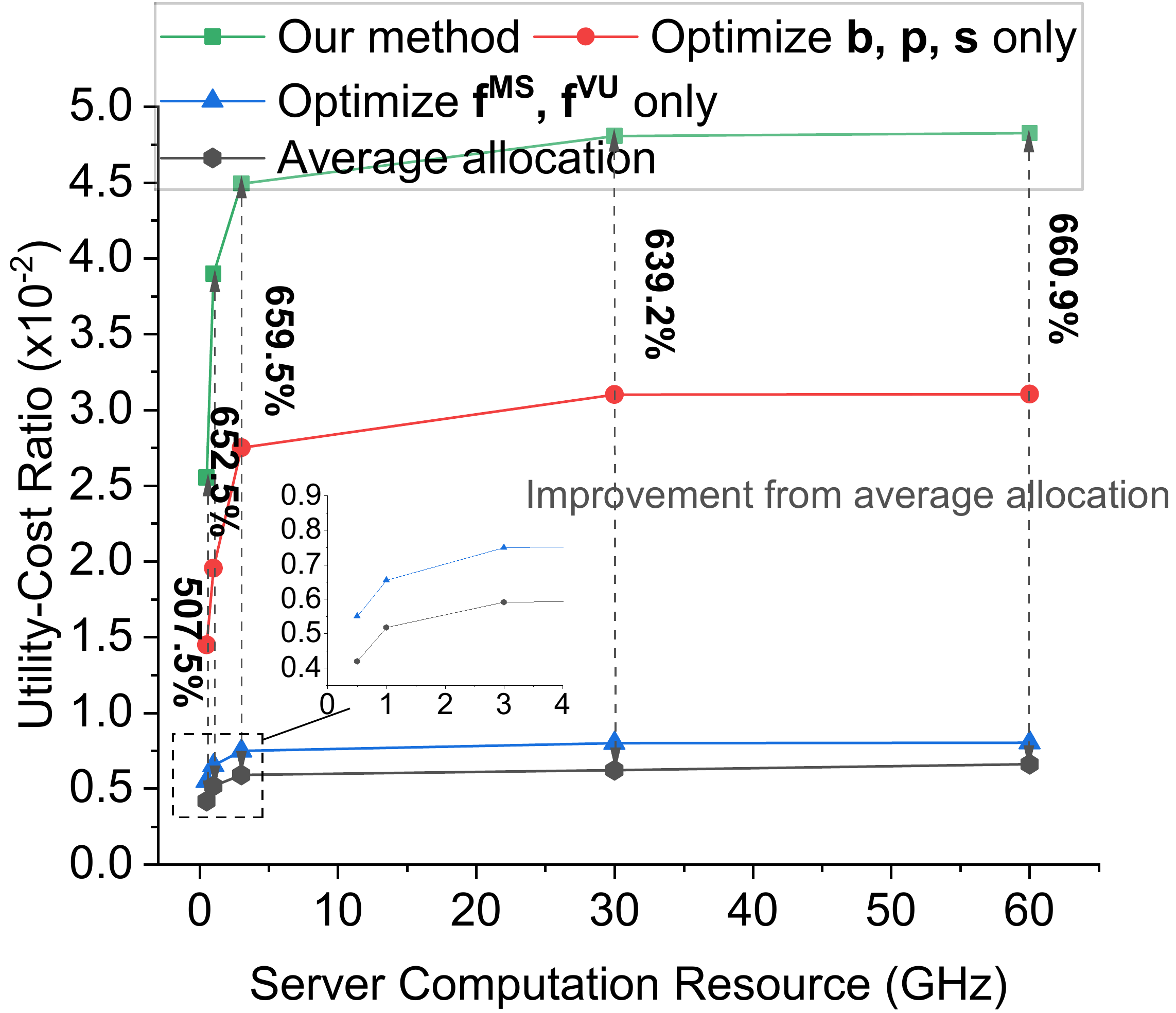}
\label{fig:fs}\vspace{-6mm}
\subcaption{}\vspace{-1mm}
\end{subfigure}%
\begin{subfigure}{0.33\textwidth}
\centering
\includegraphics[width=1\linewidth]{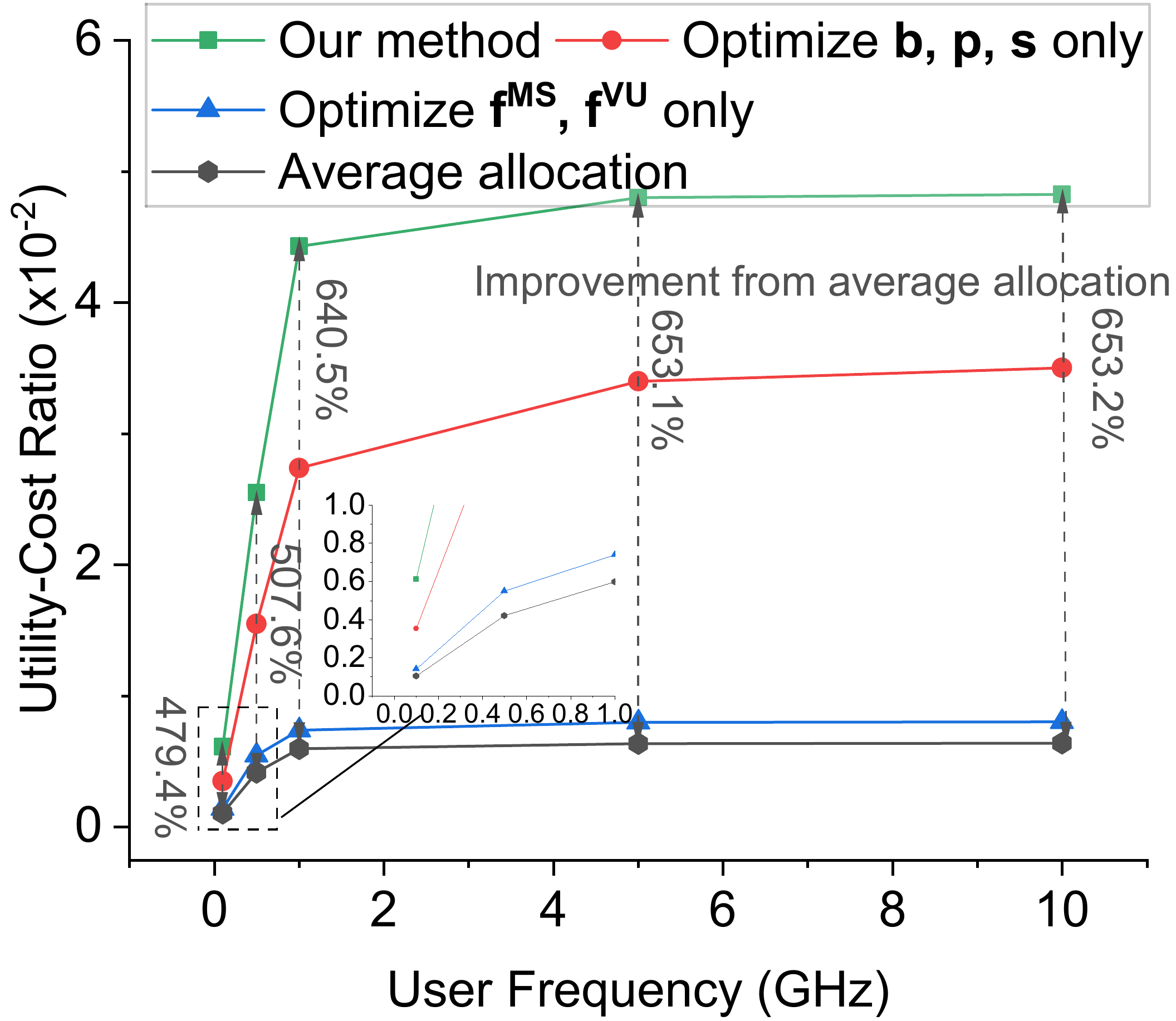}
\label{fig:fu}\vspace{-6mm}
\subcaption{}\vspace{-1mm}
\end{subfigure}%
\begin{subfigure}{0.33\textwidth}
\centering
\includegraphics[width=1\linewidth]{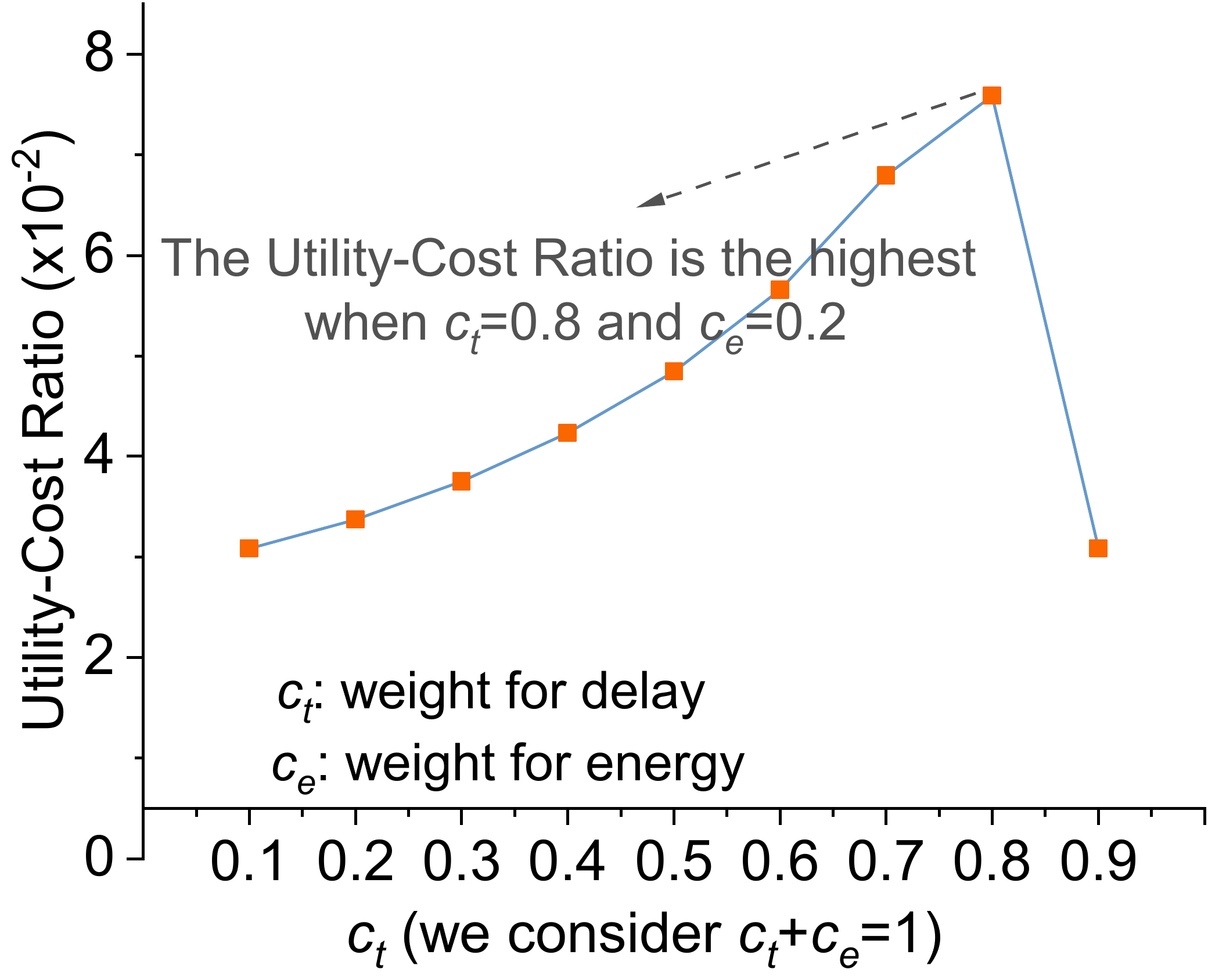}
\label{fig:cost}\vspace{-6mm}
\subcaption{}\vspace{-1mm}
\end{subfigure}%

\caption{The system utility-cost ratio (UCR) versus various parameters.}\vspace{-6mm}
\label{fig:ucrvsresource}
\end{figure*}

\textbf{Comparison with baselines.} For the simulation results, we first compare our algorithm with baselines: 
\begin{itemize}[leftmargin=0pt]
    \item \textit{\textbf{average allocation}}, which sets each $b_n$ as $\frac{b_{\textnormal{max}}}{N}$, each $p_n$ as $\frac{p_{\textnormal{max}}}{N}$, each $s_n$ as $2048 \times 1080$ (i.e., 2k resolution), each $f_n^{\textnormal{MS}}$ as $\frac{f_{\textnormal{max}}^{\textnormal{MS}}}{N}$, and each $f_n^{\textnormal{VU}}$ as $f_{n,\textnormal{max}}^{\textnormal{VU}}$; 
     \item   \textit{\textbf{optimize $\boldsymbol{b}$, $\boldsymbol{p}$, and $\boldsymbol{s}$ only}}, while setting each $f_n^{\textnormal{MS}}$ as $\frac{f_{\textnormal{max}}^{\textnormal{MS}}}{N}$ and each $f_n^{\textnormal{VU}}$ as $f_{n,\textnormal{max}}^{\textnormal{VU}}$, 
        \item \textit{\textbf{optimize $\boldsymbol{f^{\textnormal{MS}}}$ and $\boldsymbol{f^{\textnormal{VU}}}$ only}}, while setting each $b_n$ as $\frac{b_{\textnormal{max}}}{N}$, each $p_n$ as $\frac{p_{\textnormal{max}}}{N}$, and each $s_n$ as $2048 \times 1080$.
\end{itemize}
Various simulation results are plotted in the subfigures of Fig.~\ref{fig:ucrvsresource} for a detailed comparison and examining the impact of different parameters on the system utility-cost ratio (UCR). We discuss the results below.
\begin{itemize}[leftmargin=0pt]
    \item 
\textit{\textbf{UCR versus the total bandwidth.}}
% \subsubsection{UCR versus maximum bandwidth}
Here we vary the total bandwidth from 1GHz to 20GHz. In Fig.~\ref{fig:ucrvsresource}(a), larger bandwidth induces higher data rates, which reduce latency  and energy consumption, thus increasing the system UCR. In addition, the difference in UCR  between the proposed algorithm and the average allocation baseline rises from 522.3\% to 630.4\% as the total bandwidth increases.
 \item 
\textit{\textbf{UCR versus minimum resolution.}}
We fix the maximum resolution as $4096 \times 2160$ (4k) and change the minimum resolution from $1280\times720$ (720P) to $4096 \times 2160$. From Fig.~\ref{fig:ucrvsresource}(b), the UCR performance of all the algorithms improves as the minimum resolution decreases, with our proposed algorithm showing a significant improvement. The reason for this is that the high data volumes associated with high resolution can lead to higher energy consumption and system delay.
% , and even an increase in user's perceptual utility cannot compensate for the loss of subjective user experience.
The UCR of the proposed algorithm gradually plateaus when the minimum resolution reaches below $1920\times1080$ (1080p).
 \item 
\textit{\textbf{UCR versus transmission power.}}
Here we configure the maximum downlink transmission power from 0.03W to 100W. From Fig.~\ref{fig:ucrvsresource}(c), the UCR of all algorithms increases as the transmission power grows, since raising the transmission power   widens the search space for the optimization. When the transmission power is very small (e.g., 0.03W or 0.3W), the UCR of the proposed algorithm is slightly higher than other algorithms, but as the transmission power increases, the performance of the proposed algorithm far exceeds others. The UCR of all methods plateaus when the transmission power reaches 50W.
 \item 
\textit{\textbf{UCR versus computation resource.}}
% \subsubsection{UCR versus maximum server or VU CPU frequency}
We vary the maximum server CPU frequency from 0.5GHz to 60GHz and all VU's maximum  CPU frequencies from 10MHz to 10GHz. In Fig.~\ref{fig:ucrvsresource}(d) and Fig. \ref{fig:ucrvsresource}(e), the system UCR increases as the maximum CPU frequency grows, since the optimization problem has a wider search space.
% , thus reducing the negative impact of system delay and energy consumption on the user experience.
The proposed algorithm outperforms the average allocation, reaching a difference of 660.9\% and 653.2\% for server CPU frequency at 60GHz and VU CPU frequencies at 10GHz, respectively.
\item \textit{\textbf{Impact of user number on UCR.}}
We now amplify the bandwidth, transmission power, and server's CPU frequency by 10 times and fix other parameters to see the impact of user number $N$ on UCR. 
% The compression rate is set as 100 and hyper-parameter $\kappa^{\textnormal{MS}}$ and $\kappa^{\textnormal{VU}}$ is fixed as $10^{-12}$. The utility functions for users are all assumed to be the same (e.g., the utility for user 1 seated - FormationPace in Fig. \ref{fig:curvefitting} (a)). 
In Fig.~\ref{fig:resolution2}, as the user number increases from 10 to 160, the average UCR decreases. This is because the server allocates fewer resources to each VU and induces decreasing utility. In general,  the comfortable frame rate for VR applications is at least 90 \cite{VRsurvey}, i.e., at most 11ms for one frame. Note that the system delay in Fig.~\ref{fig:resolution2} is to complete each user's all frames. In all user scenarios shown in Fig.~\ref{fig:resolution2}, the delay for one frame (i.e., $\frac{\text{delay}}{\text{frame number}}$) is less than 7 ms, which satisfies the comfortable frame rate requirement.
\end{itemize}

All the above simulations compare our algorithm with the baselines. Below we provide additional simulation results to show the impact of other settings to our algorithm. 

%\subsection{Impact of communication resources}
\renewcommand{\thesubfigure}{\footnotesize(\alph{subfigure}}

% \subsection{Impact of cost weights}
% % from $\left\{(0.9, 0.1), (0.8, 0.2), \ldots, (0.1, 0.9)\right\}$

\textbf{Impact of cost weights on UCR.}
We configure different cost weights of energy and delay $(c_e, c_t)$ to see the effect on the system UCR, where we enforce $c_e + c_t=1$. 
% The user volume $N$ is fixed as five, the compression rate is set as 100, and hyper-parameter $\kappa^{\textnormal{MS}}$ and $\kappa^{\textnormal{VU}}$ is fixed as $10^{-12}$. The utility functions of the five users are also fixed. 
In Fig.~\ref{fig:ucrvsresource}(f), as $c_t$ rises to 0.8, the system UCR also increases, reflecting the importance of delay optimization for the whole system. However, as $c_t$ increases to 0.9, the system UCR instead drops significantly, since emphasizing the latency overwhelmingly while undervaluing the energy may enlarge the system cost.

% \subsection{Impact of user volume}

\begin{figure*}[t]
\renewcommand{\thesubfigure}{\footnotesize(\alph{subfigure}}
\centering
\begin{minipage}{.33\textwidth}
\centering
\includegraphics[width=1\linewidth]{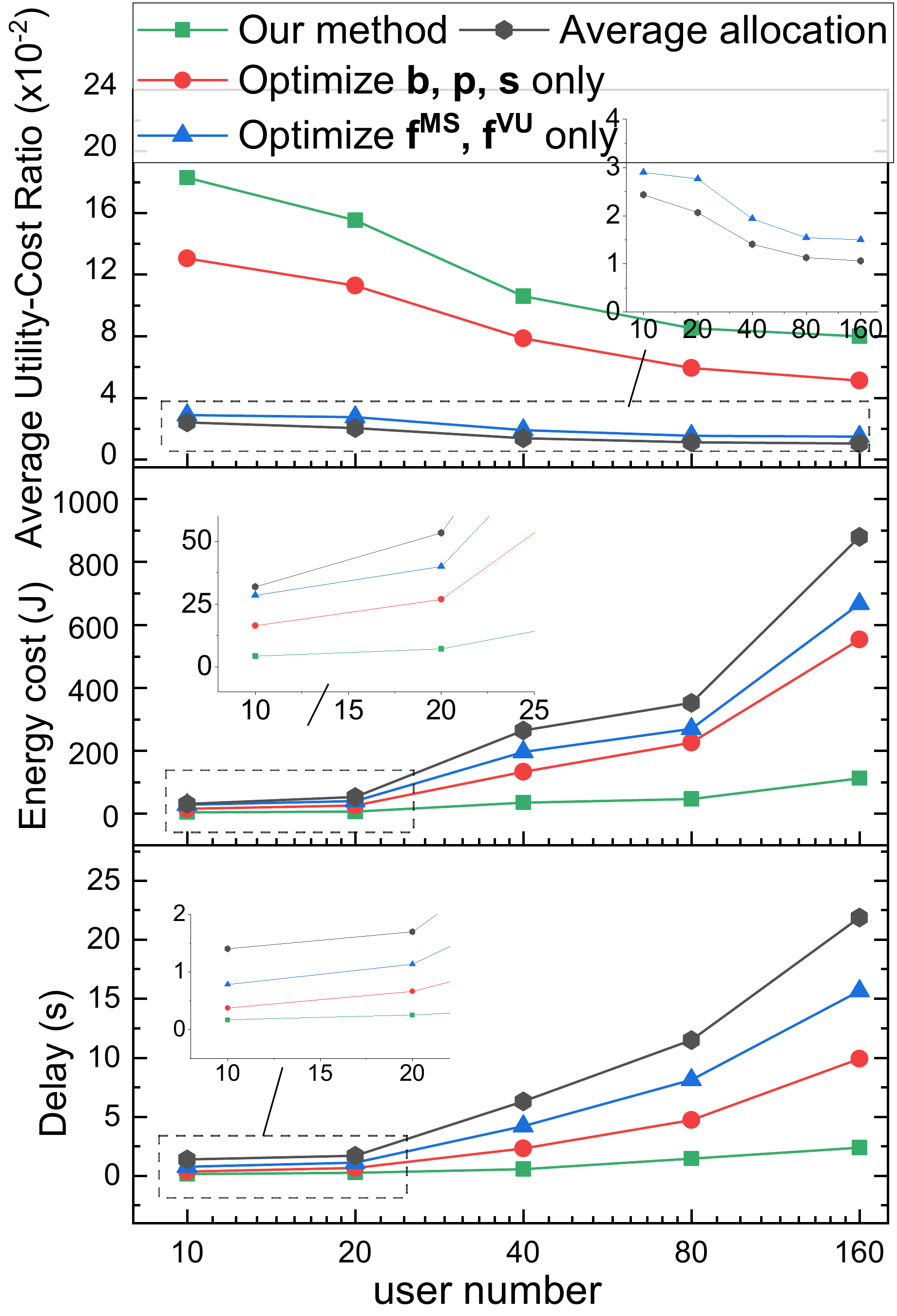}
\vspace{-5mm}
\caption{Metrics with respect to the number of users.}\label{fig:resolution2}\vspace{-2mm}
\end{minipage}
\begin{minipage}{.66\textwidth}
\begin{subfigure}{0.49\textwidth}
\centering
\includegraphics[width=1\linewidth]{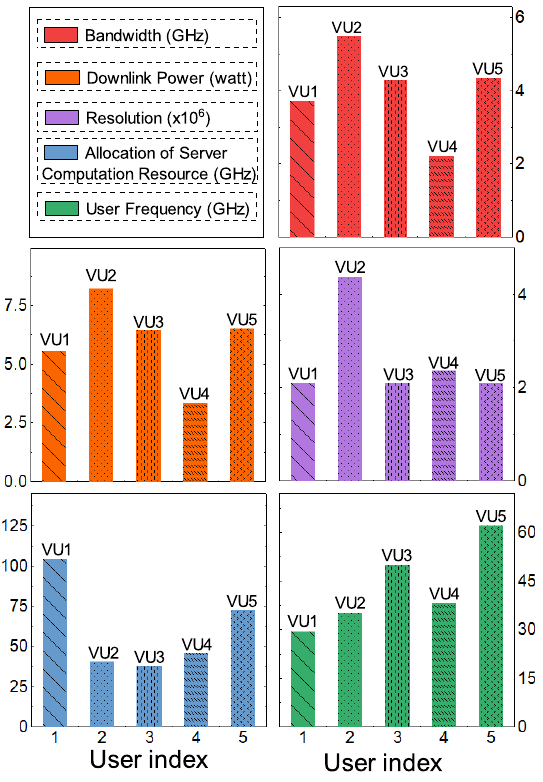}
\vspace{-7mm}
\subcaption{Allocated parameters for each user.}\vspace{-2mm}
\end{subfigure}%
\begin{subfigure}{0.49\textwidth}
\centering
\includegraphics[width=1\linewidth]{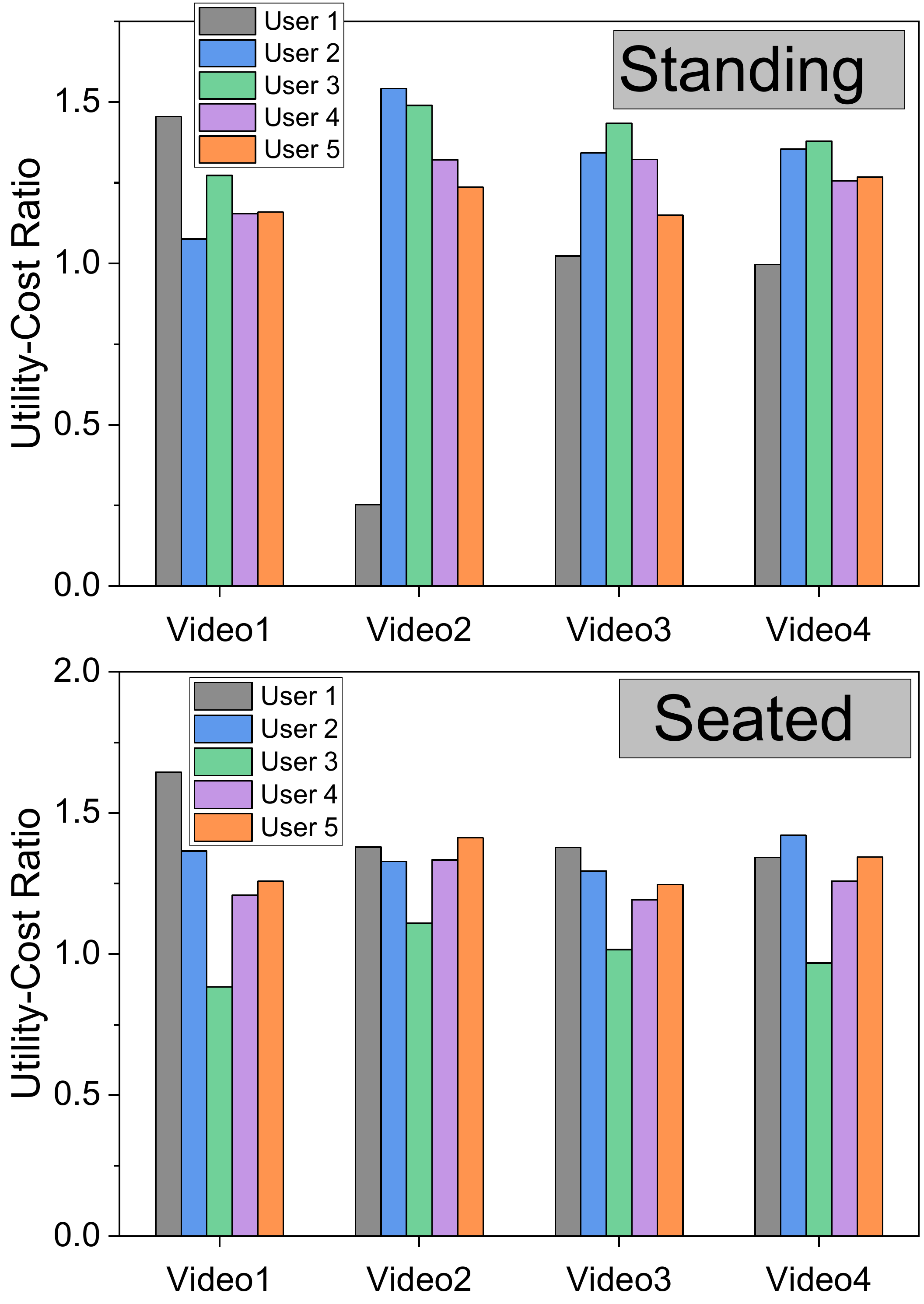}
\vspace{-7mm}
\subcaption{Comparing individual users' UCR.
%(for a single user) comparison under different positions and videos in SSV.
}\vspace{-2mm}
\end{subfigure}%

\caption{Reviewing the allocated results of different users and the impact of user scenarios on UCR.}\vspace{-6mm}
\label{fig:optimization}
\end{minipage}
\end{figure*}

\textbf{Reviewing different users' allocated results.}
% In this section, we fix user volume $N$ as five, five different utility functions and change wireless resources to see their impacts on UCR.?  Besides, in 
In Fig.~\ref{fig:optimization}(a), we present the optimized bandwidth $b_n$, transmit power $p_n$, resolution $s_n$, MS resource allocation $f^{\textnormal{MS}}_n$, and user CPU frequency $f^{\textnormal{VU}}_n$ of five users to visualize the resource allocation. Fig.~\ref{fig:optimization}(b) shows the impact of user scenarios on individual UCRs. Different user preferences and physical states affect subjective scores. For example, users who prefer high-quality videos may give stricter subjective scores than those who do not require high video quality. Generally, many users found sitting to provide more comfort than standing, as indicated by a higher UCR in most seated scenarios. However, it's important to note that individual differences were observed, and this trend did not hold true for all participants.

To summarize, extensive simulation results above confirm the effectiveness of our proposed\vspace{-5pt} algorithm.

\section{Conclusion\vspace{-5pt}} \label{sec: conclusion}
In this paper, we optimize the system utility-cost ratio (UCR) for the Metaverse over wireless networks. The optimization variables include the allocation of both communication and computation resources as well as the resolutions of virtual reality (VR) videos. Our human-centric utility measure represents users' subjective assessment of the VR video quality, and is supported by real datasets. 
% create a comprehensive mobile-edge QoE model for the user-centric Metaverse, taking subjective test results into consideration.
We tackle the non-convex system UCR optimization by proposing a novel technique for fractional programming. Our computationally efficient algorithm for the system UCR optimization is validated by extensive simulations. Three future directions are as follows. Firstly, since the current paper solves the optimization problem via alternating optimization (AO) of video frame resolution and other variables, a future task is to see whether we can optimize all variables simultaneously to obtain the globally optimal solution. Secondly, we may incorporate the priorities of different users into computing the system utility (e.g., using a weighted sum with weights representing users' priorities), and investigate the impact of such formulation on the optimization. Thirdly, while the current paper contains extensive simulation results to support the analysis, we can implement real-world systems to evaluate the performance of our proposed algorithm in practice.

% In this model, we jointly optimize the video resolutions, computing resources for both MS and VUs, and network resources including bandwidth and power. We propose a novel algorithm for tackling the sum of concave-convex ratios problem, effectively achieving a satisfactory UCR for the whole system. In the future, we plan to further consider a more complete system model and integrate the user sickness, and propose novel solutions.

\section*{Acknowledgement}

This research is partly supported by the Singapore Ministry of Education Academic Research Fund under Grant Tier 1 RG90/22, Grant Tier 1 RG97/20, Grant Tier 1 RG24/20 and Grant Tier 2 MOE2019-T2-1-176; and partly by the Nanyang Technological University (NTU)-Wallenberg AI, Autonomous Systems and Software Program (WASP) Joint Project.

\begin{spacing}{1.18}
\renewcommand{\refname}{~\\[-20pt]References\vspace{-6pt} }

% \bibliographystyle{IEEEtran}
% \bibliography{related}


\iffalse


\fi

\end{spacing}

\begin{IEEEbiography}
[{\includegraphics[width=1in,height=1.25in,clip,keepaspectratio]{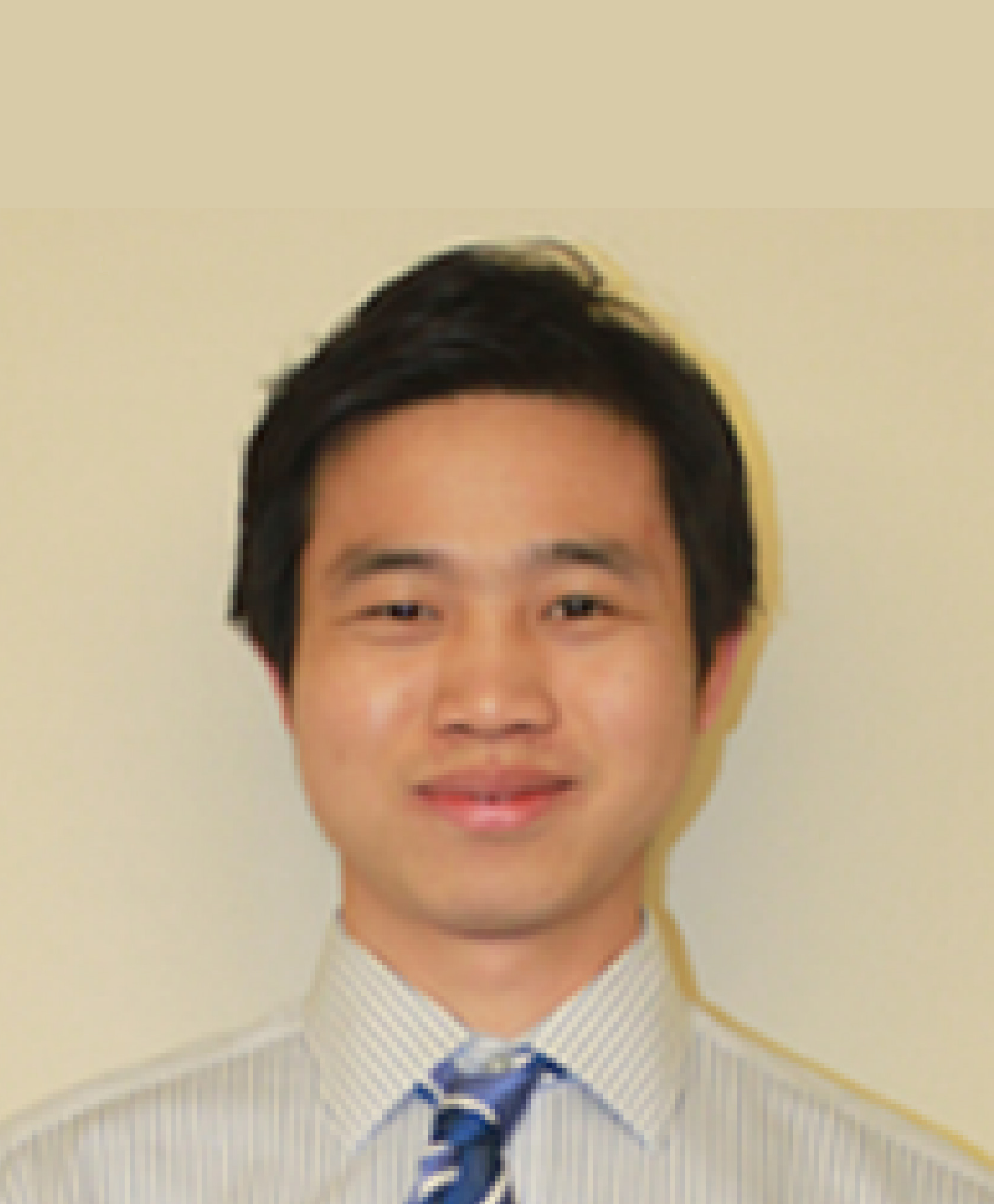}}]
{Jun Zhao} 
(Member, IEEE) received a bachelor’s degree in Information Engineering from Shanghai Jiao Tong University, China, in July 2010, and a Ph.D. degree in Electrical and Computer Engineering from Carnegie Mellon University (CMU), USA, in May 2015, (advisors: Virgil Gligor and Osman Yagan; collaborator: Adrian Perrig), affiliating with CMU’s renowned CyLab Security \& Privacy Institute. He is currently an Assistant Professor in the School of Computer Science and Engineering (SCSE), Nanyang Technological University (NTU), Singapore. Before joining NTU as a Faculty Member, he was a Postdoctoral Researcher under the supervision of Xiaokui Xiao at NTU. Before that, he was a Postdoctoral Researcher at Arizona State University as an Arizona Computing Post-Doctoral Researcher Best Practices Fellow (advisors: Junshan Zhang and Vincent Poor).
\end{IEEEbiography}
\begin{IEEEbiography}
[{\includegraphics[width=1in,height=1.25in,clip,keepaspectratio]{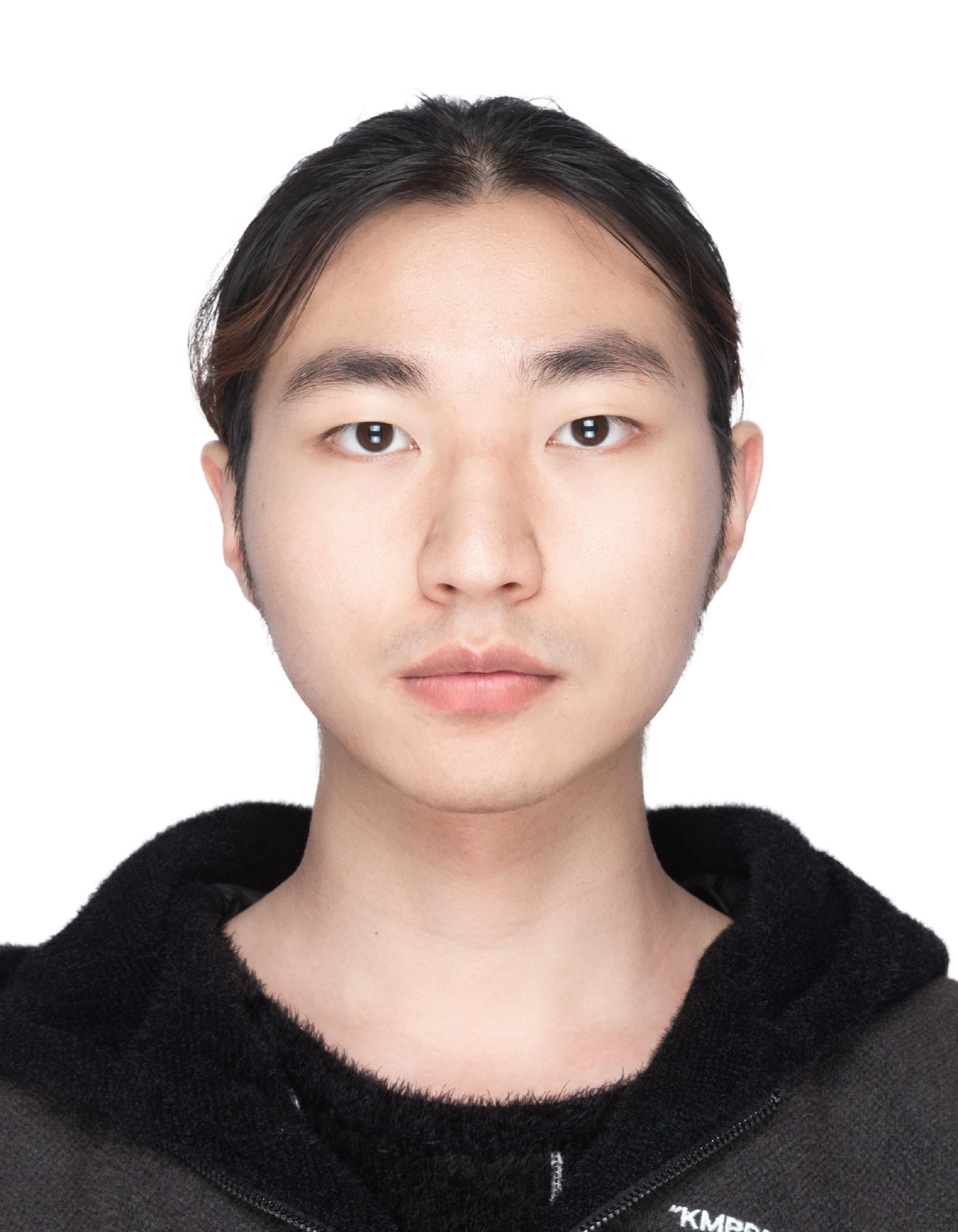}}]
{Liangxin Qian} received bachelor's and master's degrees in communication engineering from the University of Electronic Science and Technology of China, Chengdu, China, in 2019 and 2022, respectively. He is currently working toward his Ph.D. at the School of Computer Science and Engineering, Nanyang Technological University, Singapore. His research interests include Metaverse, mobile edge computing, and communication theory.
%\vspace{-1cm}
\end{IEEEbiography}
\begin{IEEEbiography}
[{\includegraphics[width=1in,height=1.25in,clip,keepaspectratio]{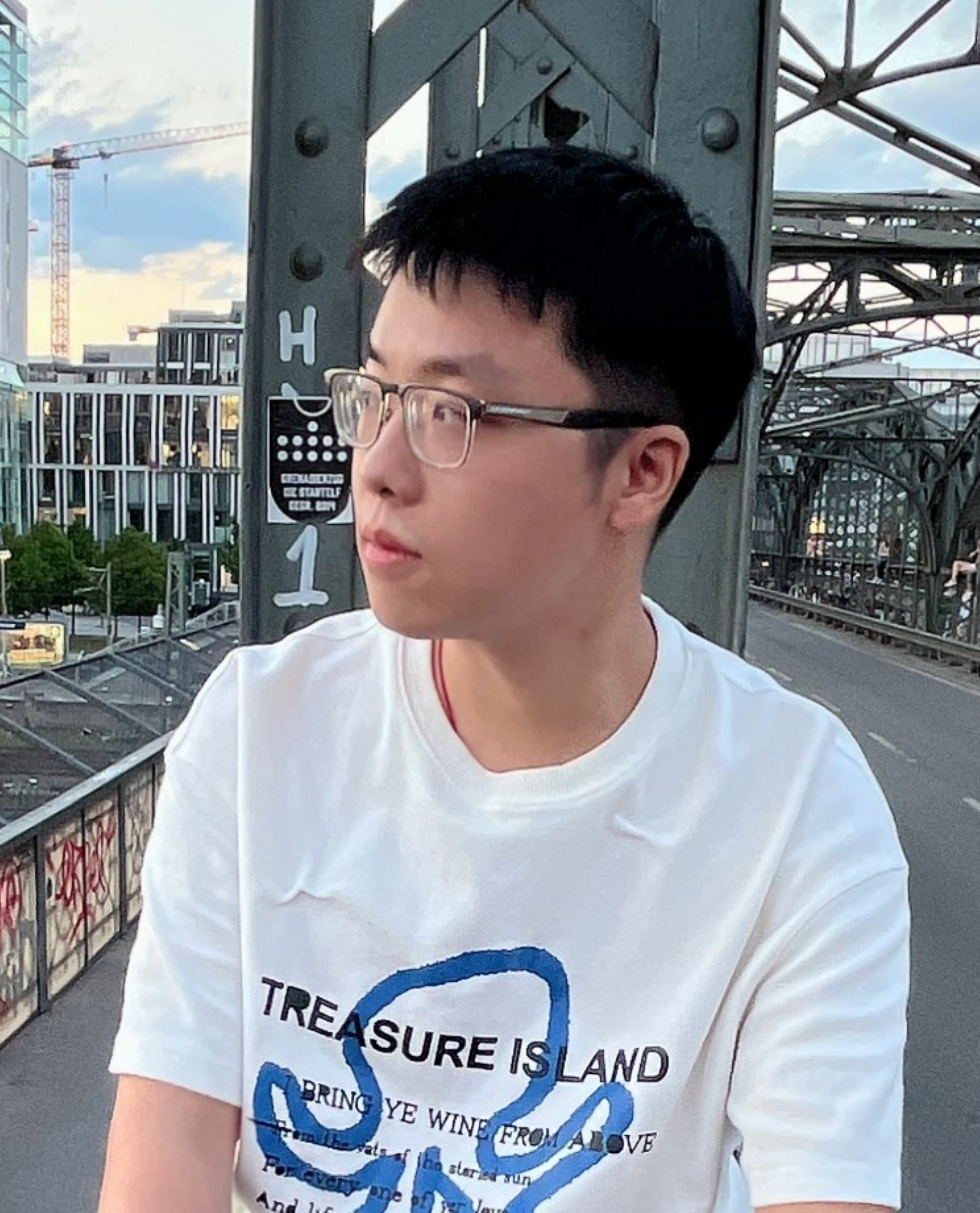}}]
{Wenhan Yu}
(Student Member, IEEE) received his B.S. degree in Computer Science and Technology from Sichuan University, Sichuan, China in 2021. He is currently pursuing a Ph.D. degree in the School of Computer Science and Engineering, Nanyang Technological University (NTU), Singapore. His research interests cover wireless communications, deep reinforcement learning, optimization, and Metaverse.
%\vspace{-1cm}
\end{IEEEbiography}

\appendix

\setlength{\abovedisplayskip}{3pt plus 3pt minus 0pt}
\setlength{\belowdisplayskip}{3pt plus 3pt minus 0pt}
\setlength\abovedisplayshortskip{3pt plus 0pt minus 0pt}
\setlength\belowdisplayshortskip{3pt plus 0pt minus 0pt}

\subsection[]{Proving the relationship between \mbox{\texttt{FP-minimization}} in~(\ref{FP-minimization}) and minimizing $W(\boldsymbol{x},\boldsymbol{y}):=G(\boldsymbol{x}) + \sum_{n=1}^N K_n(\boldsymbol{x},y_n)$ subject to $\boldsymbol{x} \in \mathcal{S}$ and $y_n \in \mathbb{R}^+$, stated in Section~\ref{opt-sum-of-ratios} on Pages~\pageref{opt-sum-of-ratios} and~\pageref{opt-sum-of-ratios2}}

Recall that \mbox{\texttt{FP-minimization}} in~(\ref{FP-minimization}) means minimizing $H(\boldsymbol{x}) : = G(\boldsymbol{x}) + \sum_{n=1}^N \frac{A_n(\boldsymbol{x})}{B_n(\boldsymbol{x})}$ subject to $\boldsymbol{x}$ in a convex set $\mathcal{S}$, for convex  $A_n(\boldsymbol{x})$ and concave $B_n(\boldsymbol{x})$.

We now consider alternating optimization (AO) of $\boldsymbol{x}$ and $\boldsymbol{y}$ to minimize $W(\boldsymbol{x},\boldsymbol{y}):=G(\boldsymbol{x}) + \sum_{n=1}^N K_n(\boldsymbol{x},y_n)$ subject to $\boldsymbol{x} \in \mathcal{S}$ and $\boldsymbol{y}:=[y_1,\ldots,y_N] \in (\mathbb{R}^+)^N$, where $K_n(\boldsymbol{x},y_n):=[A_n(\boldsymbol{x})]^2y_n + \frac{1}{4[B_n(\boldsymbol{x})]^2y_n} $, and $\mathbb{R}^+$ denotes the set of positive numbers. We will show that
\begin{align}
\begin{array}{l}
\text{if the above AO process of minimizing $W(\boldsymbol{x},\boldsymbol{y})$ converges}\\
\text{to $(\boldsymbol{x}^*,\boldsymbol{y}^*)$, $\boldsymbol{x}^*$ is a stationary point for \mbox{\texttt{FP-minimization}}} \\
\text{in~(\ref{FP-minimization}).}
\end{array}     \label{eqpartialKn0}
\end{align}

The AO process is as follows: We start with a randomly initialized $\boldsymbol{x}^{(0)} \in \mathcal{S}$. Then we optimize $\boldsymbol{y}$ with $\boldsymbol{x}$ being $\boldsymbol{x}^{(0)}$ to  minimize $W(\boldsymbol{x},\boldsymbol{y})$, and denote the obtained  $\boldsymbol{y} \in (\mathbb{R}^+)^N$ as $\boldsymbol{y}^{(1)}$. Given $\boldsymbol{y}$ as $\boldsymbol{y}^{(1)}$, we optimize $\boldsymbol{x}$ to  minimize $W(\boldsymbol{x},\boldsymbol{y})$, and denote the obtained  $\boldsymbol{x} \in \mathcal{S}$ as $\boldsymbol{x}^{(1)}$. Given $\boldsymbol{x}$ as $\boldsymbol{x}^{(1)}$, we optimize $\boldsymbol{y}$ to  minimize $W(\boldsymbol{x},\boldsymbol{y})$, and denote the obtained  $\boldsymbol{y} \in (\mathbb{R}^+)^N$ as $\boldsymbol{y}^{(2)}$. The above process continues iteratively. The $i$-th iteration includes the following two steps:
\begin{itemize}
\item  Given $\boldsymbol{x}$ as $\boldsymbol{x}^{(i-1)}$, we optimize $\boldsymbol{y}$ to  minimize $W(\boldsymbol{x},\boldsymbol{y})$, and denote the obtained  $\boldsymbol{y} \in (\mathbb{R}^+)^N$ as $\boldsymbol{y}^{(i)}$.
\item Given $\boldsymbol{y}$ as $\boldsymbol{y}^{(i)}$, we optimize $\boldsymbol{x}$ to  minimize $W(\boldsymbol{x},\boldsymbol{y})$, and denote the obtained  $\boldsymbol{x} \in \mathcal{S}$ as $\boldsymbol{x}^{(i)}$. 
\end{itemize}
The above AO process converges when the relative difference between $W(\boldsymbol{x}^{(i-1)},\boldsymbol{y}^{(i-1)})$ and $W(\boldsymbol{x}^{(i)},\boldsymbol{y}^{(i)})$ is smaller than a predefined small error tolerance.

To examine the AO process of minimizing $W(\boldsymbol{x},\boldsymbol{y})$, we will analyze 1) optimizing $\boldsymbol{y}$ given $\boldsymbol{x}$, and 2) optimizing $\boldsymbol{x}$ given $\boldsymbol{y}$. Optimizing $\boldsymbol{y}$ given $\boldsymbol{x}$ means minimizing $K_n(\boldsymbol{x},y_n):=[A_n(\boldsymbol{x})]^2y_n + \frac{1}{4[B_n(\boldsymbol{x})]^2y_n} $ with respect to  $y_n$ for each $n=1,2,\ldots,N$; i.e., letting $y_n $ be $y_n^{\#}(\boldsymbol{x}):=\frac{1}{2A_n(\boldsymbol{x})B_n(\boldsymbol{x})} $. If such $\boldsymbol{y}$ is substituted back to $K_n(\boldsymbol{x},y_n)$, then $K_n(\boldsymbol{x},y_n)$ will become the desired $\frac{A_n(\boldsymbol{x})}{B_n(\boldsymbol{x})}$. Moreover, we now show that the partial derivative of $W(\boldsymbol{x},\boldsymbol{y})$ with respect to $\boldsymbol{x}$ at $\boldsymbol{y}$ being $\boldsymbol{y}^{\#}(\boldsymbol{x})$ is the same as the  derivative of $H(\boldsymbol{x})$ with respect to $\boldsymbol{x}$. In fact, we have
\begin{align}
  \frac{\partial K_n(\boldsymbol{x},y_n)}{\partial \boldsymbol{x}}   & =   2A_n(\boldsymbol{x})y_n \frac{\partial A_n(\boldsymbol{x})}{\partial \boldsymbol{x}}  - \frac{1}{2[B_n(\boldsymbol{x})]^3y_n} \cdot \frac{\partial B_n(\boldsymbol{x})}{\partial \boldsymbol{x}} , \label{eqpartialKn}
\end{align}
and
\begin{align}
  \frac{\partial \big(\frac{A_n(\boldsymbol{x})}{B_n(\boldsymbol{x})}\big)}{\partial \boldsymbol{x}}   & =   \frac{\frac{\partial A_n(\boldsymbol{x})}{\partial \boldsymbol{x}}\cdot B_n(\boldsymbol{x}) - \frac{\partial B_n(\boldsymbol{x})}{\partial \boldsymbol{x}}\cdot A_n(\boldsymbol{x})}{(B_n(\boldsymbol{x}))^2} .\label{eqpartialAnBn}
\end{align}
From~(\ref{eqpartialKn}) and~(\ref{eqpartialAnBn}), it holds that
\begin{align}
\bigg(  \frac{\partial K_n(\boldsymbol{x},y_n)}{\partial \boldsymbol{x}} \bigg)|_{y_n=\frac{1}{2A_n(\boldsymbol{x})B_n(\boldsymbol{x})} }  & =    \frac{\partial \big(\frac{A_n(\boldsymbol{x})}{B_n(\boldsymbol{x})}\big)}{\partial \boldsymbol{x}},\label{eqpartialKn2}
\end{align}
which further implies
\begin{align}
\bigg(  \frac{\partial W(\boldsymbol{x},\boldsymbol{y})}{\partial \boldsymbol{x}} \bigg)|_{\boldsymbol{y} = \boldsymbol{y}^{\#}(\boldsymbol{x})}  & =    \frac{\partial H(\boldsymbol{x})}{\partial \boldsymbol{x}}.\label{eqpartialKn3}
\end{align}
Moreover, as explained, we have
\begin{align}
W(\boldsymbol{x},\boldsymbol{y})|_{\boldsymbol{y} = \boldsymbol{y}^{\#}(\boldsymbol{x})}  & =    H(\boldsymbol{x}).\label{eqpartialKn4}
\end{align}

Using~(\ref{eqpartialKn3}) and~(\ref{eqpartialKn4}), we now show~(\ref{eqpartialKn0}). The AO process of minimizing $W(\boldsymbol{x},\boldsymbol{y})$ is \mbox{non-decreasing}. Specifically, we have $W(\boldsymbol{x}^{(i)},\boldsymbol{y}^{(i)}) \leq W(\boldsymbol{x}^{(i-1)},\boldsymbol{y}^{(i)}) \leq W(\boldsymbol{x}^{(i-1)},\boldsymbol{y}^{(i-1)})$. For \mbox{lower-bounded} $W(\boldsymbol{x},\boldsymbol{y})$, we know $W(\boldsymbol{x}^{(i)},\boldsymbol{y}^{(i)})$ converges as the iteration number $i \to \infty$. Supposing the variable solution of the AO process converges to  $(\boldsymbol{x}^*,\boldsymbol{y}^*)$, we know that 
\begin{itemize}[leftmargin=15pt]
\item[\ding{172}] $\boldsymbol{y}^*$ is the optimal $\boldsymbol{y}$ for minimizing $W(\boldsymbol{x},\boldsymbol{y})$ given $\boldsymbol{x}$ as $\boldsymbol{x}^*$ (i.e., $y_n^*:=\frac{1}{2A_n(\boldsymbol{x}^*)B_n(\boldsymbol{x}^*)} $), and
\item[\ding{173}] $\boldsymbol{x}^*$ is the optimal $\boldsymbol{x}$ for minimizing $W(\boldsymbol{x},\boldsymbol{y})$ given $\boldsymbol{y}$ as $\boldsymbol{y}^*$.   
\end{itemize}

Result ``\ding{173}'' means that $\boldsymbol{x}^*$ satisfies the KKT conditions for optimizing $\boldsymbol{x}$ to minimize $W(\boldsymbol{x},\boldsymbol{y}^*)$ subject to $\boldsymbol{x} \in \mathcal{S}$. Suppose $\boldsymbol{x}$ is $M$-dimensional, and $\boldsymbol{x} \in \mathcal{S}$ means 
\begin{align}
\begin{cases}
\mathcal{Q}_q(\boldsymbol{x}) \leq 0, & q=1,2,\ldots,Q, \\
\mathcal{R}_r(\boldsymbol{x}) = 0, & r=1,2,\ldots,R,\\
\boldsymbol{x} \in \mathbb{R}^M.
\end{cases} \nonumber
\end{align}
Then with $\boldsymbol{\alpha} $ and $\boldsymbol{\beta}$ denoting  the multipliers,
% the Lagrange function given by 
% \begin{talign}
% L(\boldsymbol{x},
% \boldsymbol{\alpha}, \boldsymbol{\beta})   & = \sum_{q=1}^Q  \alpha_q \mathcal{Q}_q(\boldsymbol{x}) + \sum_{r=1}^R \beta_r \mathcal{R}_r(\boldsymbol{x}) , \nonumber
% \end{talign} 
% where $\boldsymbol{\alpha} $ and $\boldsymbol{\beta}$ denote  the multipliers,
the KKT conditions mean
\begin{subnumcases}{}
\text{Stationarity:}& 
\text{$\begin{array}{l}
\hspace{-8pt}~~~\frac{\partial }{\partial x_m} (W(\boldsymbol{x}^*,\boldsymbol{y}^*) \\\hspace{-8pt}+ \sum_{q=1}^Q  \alpha_q \mathcal{Q}_q(\boldsymbol{x}^*) 
\\\hspace{-8pt}+ \sum_{r=1}^R \beta_r \mathcal{R}_r(\boldsymbol{x}^*) ) = 0,  
 \\ \hspace{-8pt}\text{for $m=1,2,\ldots,M$,}
\end{array}$} \label{kkt1} \\ \text{Primal feasibility:}& \hspace{-8pt}\text{$\mathcal{Q}_q(\boldsymbol{x}^*) \leq 0,$} \nonumber
\\ &\hspace{-8pt}\text{for $ q=1,2,\ldots,Q,$} \label{kkt2}  \\ & \hspace{-8pt}\text{$\mathcal{R}_r(\boldsymbol{x}^*) = 0, $} \nonumber
\\ &\hspace{-8pt}\text{for $ r=1,2,\ldots,R,$} \label{kkt3}\\
\text{Dual feasibility:}& \hspace{-8pt}\text{$\alpha_q  \geq 0$,} \nonumber
\\ &\hspace{-8pt}\text{for $ q=1,2,\ldots,Q,$} \label{kkt4}\\ \text{Complementary slackness:}& \hspace{-8pt}\text{$\alpha_q \mathcal{Q}_q(\boldsymbol{x}^*) = 0$} \nonumber
\\ &\hspace{-8pt}\text{for $ q=1,2,\ldots,Q.$}\label{kkt5}
\end{subnumcases}
Using (\ref{eqpartialKn3}) in (\ref{kkt1}), we know that~(\ref{kkt1})--(\ref{kkt5}) are equivalent to
\begin{subnumcases}{}
\text{Stationarity:}&  \text{\hspace{-9pt}$\begin{array}{l}\hspace{-2pt}\frac{\partial }{\partial x_m} (H(\boldsymbol{x}^*) \\
+ \sum_{q=1}^Q  \alpha_q \mathcal{Q}_q(\boldsymbol{x}^*) \\
+ \sum_{r=1}^R \beta_r \mathcal{R}_r(\boldsymbol{x}^*) ) = 0, \\ 
\text{for $m=1,2,\ldots,  M$,}\end{array}$} \label{kkt11} \\ 
\text{Primal feasibility:}&\hspace{-8pt}\text{$\mathcal{Q}_q(\boldsymbol{x}^*) \leq 0,$} \nonumber \\
&\hspace{-8pt}\text{for $ q=1,2,\ldots,Q,$} \label{kkt21}  \\ 
&\hspace{-8pt}\text{$\mathcal{R}_r(\boldsymbol{x}^*) = 0, $} \nonumber \\
&\hspace{-8pt}\text{for $ r=1,2,\ldots,R,$} \label{kkt31}\\
\text{Dual feasibility:}&\hspace{-8pt}\text{$\alpha_q  \geq 0$} \nonumber \\
&\hspace{-8pt}\text{for $ q=1,2,\ldots,Q,$} \label{kkt41}\\ \text{Complementary slackness:}&\hspace{-8pt}\text{$\alpha_q \mathcal{Q}_q(\boldsymbol{x}^*) = 0$} \nonumber \\
&\hspace{-8pt}\text{for $ q=1,2,\ldots,Q.$}\label{kkt51}
\end{subnumcases}
The above (\ref{kkt11})--(\ref{kkt51}) mean that $\boldsymbol{x}^*$ is a stationary point for optimizing $\boldsymbol{x}$ to minimize $H(\boldsymbol{x})$ subject to $\boldsymbol{x} \in \mathcal{S}$; i.e., $\boldsymbol{x}^*$ is a stationary point of \mbox{\texttt{FP-minimization}} in~(\ref{FP-minimization}). Hence, we have proved~(\ref{eqpartialKn0}).\qed

% \begin{align}
% \frac{\partial W(\boldsymbol{x},\boldsymbol{y})}{\partial \boldsymbol{x}} & = \frac{\partial G(\boldsymbol{x})}{\partial \boldsymbol{x}} + \sum_{n=1}^N \frac{\partial K_n(\boldsymbol{x},y_n)}{\partial \boldsymbol{x}} \nonumber \\ & = \frac{\partial G(\boldsymbol{x})}{\partial \boldsymbol{x}} + \sum_{n=1}^N (2A_n(\boldsymbol{x})y_n \frac{\partial A_n(\boldsymbol{x})}{\partial \boldsymbol{x}}  - \frac{1}{2[B_n(\boldsymbol{x})]^3y_n}) \cdot \frac{\partial B_n(\boldsymbol{x})}{\partial \boldsymbol{x}} 
% \end{align}
 
\subsection{Proving that the left-hand side of~(\ref{fmseq}) is \mbox{non-increasing} with respect to $\gamma$}

With $v(x)$ defined by $\zeta_n \frac{\mathcal{A}_n(s_n,\Lambda_n)}{x^2} - y c_{\hspace{0.5pt}\textnormal{e}} \cdot2 \kappa^{\textnormal{MS}} \mathcal{F}_n(s_n,\Lambda_n)x$, clearly $v(x)$ is \mbox{non-increasing} with respect to $x$. Then $\text{PositiveRoot}(v(x)= \gamma)$, denoting the positive root satisfying $v(x)= \gamma$, is \mbox{non-increasing} with respect to $\gamma$. Thus, the left-hand side of~(\ref{fmseq}) is \mbox{non-increasing} with respect to $\gamma$. \qed

\subsection{Proving that the left-hand side of~(\ref{solvepngivenalphaandbeta}) is \mbox{non-increasing} with respect to $p_n$}

From~(\ref{definepsin}), we know (i) $\psi_n(p_n,\alpha,\beta \mid \textnormal{\small$\bigstar$})$ is positive and decreasing in $p_n$, where ``$\textnormal{\small$\bigstar$}$'' denotes ``$\boldsymbol{z},y, \boldsymbol{s}$''. In addition, (ii) the function $\log_2(1+x) - \frac{x}{(1+x) \ln 2}$ is increasing and positive   for $x>0$ since the derivative $\frac{x}{(1+x)^2 \ln 2}$ is positive for $x>0$, and $\log_2(1+x) - \frac{x}{(1+x) \ln 2}$ at $x=0$ equals $0$. From the above Results ``(i)'' and ``(ii)'', we obtain (iii) $\left[
    \log_2\big(1+\psi_n(p_n,\alpha,\beta \mid \textnormal{\small$\bigstar$})\big)
     -\frac{\psi_n(p_n,\alpha,\beta \mid \textnormal{\small$\bigstar$})}{(1+\psi_n(p_n,\alpha,\beta \mid \textnormal{\small$\bigstar$}))\ln2}\right]$ is positive and decreasing  in $p_n$.

We also have (iv) the function $\frac{\log_2(1+x)}{x}$ is decreasing for $x>0$ since the derivative $\frac{1}{x^2}[\frac{x}{(1+x) \ln 2} - \log_2(1+x)]  $ is negative due to Result ``(ii)'' above. From the above Results ``(i)'' and ``(iv)'', we obtain (v) $\overline{r}_n(p_n,\alpha,\beta \mid \textnormal{\small$\bigstar$})$ defined in~(\ref{eqdefinernoverline}) is increasing in $p_n$. Since the utility function $U_n(r_n, s_n)$ is concave and \mbox{non-decreasing} in $r_n$, $\frac{\partial U_n(r_n, s_n)}{\partial r_n}$ is \mbox{non-negative} and \mbox{non-increasing} in $r_n$. Then (vi) $ \frac{\partial U_n(r_n, s_n)}{\partial r_n}   + \frac{yc_{\hspace{0.5pt}\textnormal{e}}}{2z_n{r_n}^3{{\nu_n}^2}}  + \frac{\zeta_n s_n\textcolor{black}{\mu_n}  {\color{black}\Lambda_n}}{{r_n}^2{\nu_n}} $ is positive and \mbox{non-increasing} in $r_n$.\vspace{1pt} From the above Results ``(v)'' and ``(vi)'', we obtain (vii) $     \big(\frac{\partial U_n(r_n, s_n)}{\partial r_n}\big)|_{r_n=\overline{r}_n(p_n,\alpha,\beta \mid \textnormal{\small$\bigstar$})}   + \frac{y c_{\hspace{0.5pt}\textnormal{e}}}{2z_n \cdot [\overline{r}_n(p_n,\alpha,\beta \mid \textnormal{\small$\bigstar$})]^3\textcolor{black}{{\nu_n}^2}}   + \frac{\zeta_n s_n\textcolor{black}{\mu_n}  {\color{black}\Lambda_n}}{ [\overline{r}_n(p_n,\alpha,\beta \mid \textnormal{\small$\bigstar$})]^2\textcolor{black}{\nu_n}}       
$ is positive and \mbox{non-increasing} in $p_n$.

From the above Results ``(iii)'' and ``(vii)'', the left hand side of~(\ref{solvepngivenalphaandbeta}) is \mbox{decreasing} as $p_n$ increases.  \qed

\subsection[]{Further explanations of the step-by-step analysis in~(\ref{roadmap}) on Page~\pageref{roadmap} for Problem $\mathbb{P}_{5}(\boldsymbol{z},y,\boldsymbol{s})$, which will be useful for Appendices~\ref{Appendixbeta},~\ref{Appendixalpha}, and~\ref{AppendixT}}

In this part, we provide further explanations of the step-by-step analysis in~(\ref{roadmap}) on Page~\pageref{roadmap} for Problem $\mathbb{P}_{5}(\boldsymbol{z},y,\boldsymbol{s})$, which will be useful to prove the results in Appendices~\ref{Appendixbeta},~\ref{Appendixalpha}, and~\ref{AppendixT} later.

We recall Problem $\mathbb{P}_{5}(\boldsymbol{z},y,\boldsymbol{s})$ (i.e., $\mathbb{P}_{5}(\textnormal{\small$\bigstar$})$) from (\ref{eqP5}):
\begin{align}
\textnormal{$\mathbb{P}_{5}(\textnormal{\small$\bigstar$})$}:\hspace{-10pt}\max\limits_{\boldsymbol{b},\boldsymbol{p},\boldsymbol{f}^{\textnormal{MS}},\boldsymbol{f}^{\textnormal{VU}},T} \hspace{-5pt} & F(\boldsymbol{b},\hspace{-2pt}\boldsymbol{p},\hspace{-2pt}\boldsymbol{f}^{\textnormal{MS}},\hspace{-2pt}\boldsymbol{f}^{\textnormal{VU}},\hspace{-2pt}T  \mid \boldsymbol{s}, \hspace{-2pt}y) \nonumber \\
&- yc_{\hspace{0.5pt}\textnormal{e}} \hspace{-2pt}\cdot \hspace{-2pt}\sum_{n \in \mathcal{N}} \big\{[(p_n +p_n^{\textnormal{cir}})  s_n\textcolor{black}{\mu_n}{\color{black}\Lambda_n}]^2z_n \hspace{-2pt} \nonumber \\
&+\hspace{-2pt} \frac{1}{4(r_n(b_n,p_n)\textcolor{black}{\nu_n})^2z_n}\big\} \nonumber \\ 
\textrm{s.t.} \quad &  \textnormal{(\ref{constraintbn}): } \sum_{n \in \mathcal{N}} b_n \leq b_{\textnormal{max}},\nonumber \\  &\textnormal{(\ref{constraintpn}): }\sum_{n \in \mathcal{N}} p_n \leq p_{\textnormal{max}} , \nonumber \\ &\textnormal{(\ref{constraintfMS}): }\sum_{n \in \mathcal{N}} f_n^{\textnormal{MS}} \leq f_{\textnormal{max}}^{\textnormal{MS}},  \nonumber \\ &\textnormal{(\ref{constraintfn}): }f_n^{\textnormal{VU}} \leq f_{n,\textnormal{max}}^{\textnormal{VU}},~\forall n \in \mathcal{N},  \nonumber \\ &\textnormal{(\ref{constraintTtau}): } t_n(b_n,p_n,s_n,f_n^{\textnormal{MS}},f_n^{\textnormal{VU}}) \leq T,~\forall n \in \mathcal{N}.  \nonumber 
\end{align} 
As shown at the beginning of Section~\ref{kktproblem5} on Page~\pageref{kktproblem5}, $\mathbb{P}_{5}(\textnormal{\small$\bigstar$})$ belongs to convex optimization, and $\alpha, \beta, \gamma, \boldsymbol{\delta}, \boldsymbol{\zeta}$ denote  the Lagrange multipliers for \textnormal{(\ref{constraintbn})}, \textnormal{(\ref{constraintpn})}, \textnormal{(\ref{constraintfMS})}, \textnormal{(\ref{constraintfn})}, and  \textnormal{(\ref{constraintTtau})}, respectively.

Suppose we already know $\boldsymbol{\zeta}$. We move $\boldsymbol{\zeta}$ and \textnormal{(\ref{constraintTtau})} to the objective function, and construct the following problem (recall that for an optimization problem $\mathbb{P}_{i}$, we use  $H_{\mathbb{P}_{i}}$ to denote its objective function):
\begin{align}
\textnormal{$\mathbb{P}_{7}(\textnormal{\small$\bigstar$},\boldsymbol{\zeta})$}:\hspace{-10pt}\min\limits_{\boldsymbol{b},\boldsymbol{p},\boldsymbol{f}^{\textnormal{MS}},\boldsymbol{f}^{\textnormal{VU}},T} \hspace{-5pt} & - H_{\mathbb{P}_{5}}(\boldsymbol{b},\boldsymbol{p},\boldsymbol{f}^{\textnormal{MS}},\boldsymbol{f}^{\textnormal{VU}},T \mid \textnormal{\small$\bigstar$}) \nonumber \\
&+  \sum_{n \in \mathcal{N}} \big[\zeta_n \cdot (t_n(b_n,p_n,s_n,f_n^{\textnormal{MS}},f_n^{\textnormal{VU}}) \hspace{-2pt}-\hspace{-2pt} T)\big]    \nonumber \\ 
\textrm{s.t.} \quad &  \textnormal{(\ref{constraintbn}): } \sum_{n \in \mathcal{N}} b_n \leq b_{\textnormal{max}},\nonumber \\  &\textnormal{(\ref{constraintpn}): }\sum_{n \in \mathcal{N}} p_n \leq p_{\textnormal{max}} , \nonumber \\ &\textnormal{(\ref{constraintfMS}): }\sum_{n \in \mathcal{N}} f_n^{\textnormal{MS}} \leq f_{\textnormal{max}}^{\textnormal{MS}},  \nonumber \\ &\textnormal{(\ref{constraintfn}): }f_n^{\textnormal{VU}} \leq f_{n,\textnormal{max}}^{\textnormal{VU}},~\forall n \in \mathcal{N} .  \nonumber 
\end{align} 
We define
\begin{itemize}
\item Statement $\mathcal{V}_{\mathbb{P}_{5}}$:
\item[] $\left\{ \begin{array}{l}
[\boldsymbol{b}^{\#}(\textnormal{\small$\bigstar$}),\boldsymbol{p}^{\#}(\textnormal{\small$\bigstar$}),(\boldsymbol{f}^{\textnormal{MS}})^{\#}(\textnormal{\small$\bigstar$}),(\boldsymbol{f}^{\textnormal{VU}})^{\#}(\textnormal{\small$\bigstar$}), \\ 
T^{\#}(\textnormal{\small$\bigstar$}),\alpha^{\#}(\textnormal{\small$\bigstar$}),\beta^{\#}(\textnormal{\small$\bigstar$}),\gamma^{\#}(\textnormal{\small$\bigstar$}),\boldsymbol{\delta}^{\#}(\textnormal{\small$\bigstar$}),\boldsymbol{\zeta}^{\#}(\textnormal{\small$\bigstar$})] \\ 
\textnormal{is a solution to the KKT conditions $\mathcal{S}_{KKT}$ in (\ref{KKTP5})} \\ \textnormal{for Problem $\mathbb{P}_{5}(\textnormal{\small$\bigstar$})$.}
\end{array}  \right\}$, and
\item Statement $\mathcal{V}_{\mathbb{P}_{7}}$:
\item[] $\left\{ \begin{array}{l}
[\boldsymbol{b}^{\#}(\textnormal{\small$\bigstar$}),\boldsymbol{p}^{\#}(\textnormal{\small$\bigstar$}),(\boldsymbol{f}^{\textnormal{MS}})^{\#}(\textnormal{\small$\bigstar$}),(\boldsymbol{f}^{\textnormal{VU}})^{\#}(\textnormal{\small$\bigstar$}),\\ 
T^{\#}(\textnormal{\small$\bigstar$}), \alpha^{\#}(\textnormal{\small$\bigstar$}),\beta^{\#}(\textnormal{\small$\bigstar$}),\gamma^{\#}(\textnormal{\small$\bigstar$}),\boldsymbol{\delta}^{\#}(\textnormal{\small$\bigstar$})] \\ \textnormal{is a solution to the KKT conditions of} \\
\textnormal{Problem $\mathbb{P}_{7}(\textnormal{\small$\bigstar$},\boldsymbol{\zeta}^{\#}(\textnormal{\small$\bigstar$}))$.}   
\end{array}  \right\} $.
\end{itemize}

By checking the KKT conditions of $\mathbb{P}_{5}(\textnormal{\small$\bigstar$})$ and $\mathbb{P}_{7}(\textnormal{\small$\bigstar$},\boldsymbol{\zeta}^{\#}(\textnormal{\small$\bigstar$}))$, we build the following relationship between
$\mathbb{P}_{5}(\textnormal{\small$\bigstar$})$ and $\mathbb{P}_{7}(\textnormal{\small$\bigstar$},\boldsymbol{\zeta})$:
\begin{align}
\text{Statement $\mathcal{V}_{\mathbb{P}_{5}}$}   & \hspace{-2pt}\Leftrightarrow\hspace{-2pt} \left\{ \begin{array}{l}  \textnormal{Statement $\mathcal{V}_{\mathbb{P}_{7}}$ holds,}  \\
\text{and $[\boldsymbol{b}^{\#}(\textnormal{\small$\bigstar$}),\boldsymbol{p}^{\#}(\textnormal{\small$\bigstar$}),(\boldsymbol{f}^{\textnormal{MS}})^{\#}(\textnormal{\small$\bigstar$}),$}\\
\text{$(\boldsymbol{f}^{\textnormal{VU}})^{\#}(\textnormal{\small$\bigstar$}),T^{\#}(\textnormal{\small$\bigstar$})  ,\boldsymbol{\zeta}^{\#}(\textnormal{\small$\bigstar$})]$ satisfies}  \\
\text{$\mathcal{S}_{2.1} \cup \mathcal{S}_{2.2}:=\big\{\textnormal{(\ref{partialLpartialT}), (\ref{Complementarytime}),  (\ref{constraintTtau}), (\ref{Dualfeasibility}e)}\big\}$.}
\end{array}  \hspace{-5pt}\right\}. \label{statev5v7}
\end{align}

In $\mathbb{P}_{7}(\textnormal{\small$\bigstar$},\boldsymbol{\zeta})$, the optimizations of $[\boldsymbol{b},\boldsymbol{p}] $, $[\boldsymbol{f}^{\textnormal{MS}},\boldsymbol{f}^{\textnormal{VU}}]$ and $T$ are independent and thus separable. This independence holds because $\boldsymbol{\zeta}$ is already given for $\mathbb{P}_{7}(\textnormal{\small$\bigstar$},\boldsymbol{\zeta})$. We do not have such independence in optimizing $\mathbb{P}_{5}(\textnormal{\small$\bigstar$})$ where $\boldsymbol{\zeta}$ is not decided yet. Hence, $\mathbb{P}_{7}(\textnormal{\small$\bigstar$},\boldsymbol{\zeta})$ is equivalent to the combination of $\mathbb{P}_{8}(\textnormal{\small$\bigstar$},\boldsymbol{\zeta})$, $\mathbb{P}_{9}(\textnormal{\small$\bigstar$},\boldsymbol{\zeta})$, and $\mathbb{P}_{10}(\textnormal{\small$\bigstar$},\boldsymbol{\zeta})$ defined below:
\begin{align}
&\textnormal{Problem $\mathbb{P}_{8}(\textnormal{\small$\bigstar$},\boldsymbol{\zeta})$}: \hspace{-1pt}\min_{T}   \quad y c_{\hspace{0.8pt}\textnormal{t}} T - T \sum_{n \in \mathcal{N}} \zeta_n   \nonumber \\ 
&\textnormal{Problem $\mathbb{P}_{9}(\textnormal{\small$\bigstar$},\boldsymbol{\zeta})$}: \hspace{-1pt}\min_{\boldsymbol{b},\boldsymbol{p}}   \hspace{-1pt} \left\{\hspace{-5pt}  \begin{array}{l} -\mathcal{U}(\boldsymbol{b},\hspace{-2pt}\boldsymbol{p},\hspace{-2pt}\boldsymbol{s})   \\ +yc_{\hspace{0.5pt}\textnormal{e}} \hspace{-2pt}\cdot \hspace{-2pt}\sum\limits_{n \in \mathcal{N}} \big\{[(p_n +p_n^{\textnormal{cir}})  s_n\textcolor{black}{\mu_n}{\color{black}\Lambda_n}]^2z_n \hspace{-2pt}\\
+\hspace{-2pt} \frac{1}{4(r_n(b_n,p_n)\textcolor{black}{\nu_n})^2z_n}\big\} \\ +   \sum\limits_{n \in \mathcal{N}} (\zeta_n \cdot t_n^{\textnormal{Tx}}(b_n,p_n, s_n)) \end{array}  \hspace{-5pt} \right\} \label{eqP9} \\
&~~~~~~\textrm{s.t.} \quad   \textnormal{(\ref{constraintbn})}, \textnormal{(\ref{constraintpn})},  \nonumber
\\
&\textnormal{Problem $\mathbb{P}_{10}(\textnormal{\small$\bigstar$},\boldsymbol{\zeta})$}: \hspace{-1pt}\min_{\boldsymbol{f}^{\textnormal{MS}},\boldsymbol{f}^{\textnormal{VU}}}   \hspace{-1pt} \left\{\hspace{-5pt}  
\begin{array}{l} y \hspace{-2pt}\cdot\hspace{-2pt} [c_{\hspace{0.5pt}\textnormal{e}}\hspace{-2pt} \cdot \hspace{-2pt}(\sum_{n\in\mathcal{N}}E_n^{\textnormal{MS}:\textnormal{Pro}}(s_n,\hspace{-2pt}f_n^{\textnormal{MS}})\hspace{-2pt}\\
+\hspace{-2pt}\sum_{n \in \mathcal{N}}E_n^{\textnormal{VU}:\textnormal{Pro}}(s_n,\hspace{-2pt}f_n^{\textnormal{VU}}))\hspace{-1pt}]  \\  +   \sum_{n \in \mathcal{N}} (\zeta_n \cdot [t_n^{\textnormal{MS}:\textnormal{Pro}}(s_n,f_n^{\textnormal{MS}})\\
+t_n^{\textnormal{VU}:\textnormal{Pro}}(s_n,f_n^{\textnormal{VU}})] \end{array}  \hspace{-5pt} 
\right\} \label{eqP10} \\
&~~~~~~\textrm{s.t.} \quad   \textnormal{(\ref{constraintfMS})}, \textnormal{(\ref{constraintfn})},  \nonumber
\end{align} 
Then after defining
\begin{itemize}
\item Statement $\mathcal{V}_{\mathbb{P}_{8}}$:
\item[] $\left\{ \begin{array}{l}
\textnormal{$T^{\#}(\textnormal{\small$\bigstar$})$ is a solution to Problem $\mathbb{P}_{8}(\textnormal{\small$\bigstar$},\boldsymbol{\zeta}^{\#}(\textnormal{\small$\bigstar$}))$.}   
\end{array}  \right\} $,
\item Statement $\mathcal{V}_{\mathbb{P}_{9}}$:
\item[] $\left\{ \begin{array}{l}
\textnormal{$[\boldsymbol{b}^{\#}(\textnormal{\small$\bigstar$}),\boldsymbol{p}^{\#}(\textnormal{\small$\bigstar$}),\alpha^{\#}(\textnormal{\small$\bigstar$}),\beta^{\#}(\textnormal{\small$\bigstar$})]$} \\ \textnormal{is a solution to the KKT conditions of} \\
\textnormal{Problem $\mathbb{P}_{9}(\textnormal{\small$\bigstar$},\boldsymbol{\zeta}^{\#}(\textnormal{\small$\bigstar$}))$.}   
\end{array}  \right\} $, and
\item Statement $\mathcal{V}_{\mathbb{P}_{10}}$:
\item[] $\left\{ \begin{array}{l}
\textnormal{$(\boldsymbol{f}^{\textnormal{MS}})^{\#}(\textnormal{\small$\bigstar$}),(\boldsymbol{f}^{\textnormal{VU}})^{\#}(\textnormal{\small$\bigstar$}), \gamma^{\#}(\textnormal{\small$\bigstar$}),\boldsymbol{\delta}^{\#}(\textnormal{\small$\bigstar$})]$} \\ \textnormal{is a solution to the KKT conditions of}\\ 
\textnormal{Problem $\mathbb{P}_{10}(\textnormal{\small$\bigstar$},\boldsymbol{\zeta}^{\#}(\textnormal{\small$\bigstar$}))$.}   
\end{array}  \right\} $
\end{itemize}
we obtain
\begin{align}
&\text{Statement $\mathcal{V}_{\mathbb{P}_{5}}$}  \nonumber\\
& \Leftrightarrow \left\{ \begin{array}{l}  \textnormal{Statements $\mathcal{V}_{\mathbb{P}_{8}}$,  $\mathcal{V}_{\mathbb{P}_{9}}$, and  $\mathcal{V}_{\mathbb{P}_{10}}$ hold,}  \\
\text{and $[\boldsymbol{b}^{\#}(\textnormal{\small$\bigstar$}),\boldsymbol{p}^{\#}(\textnormal{\small$\bigstar$}),(\boldsymbol{f}^{\textnormal{MS}})^{\#}(\textnormal{\small$\bigstar$}),(\boldsymbol{f}^{\textnormal{VU}})^{\#}(\textnormal{\small$\bigstar$}),$}\\
\text{$T^{\#}(\textnormal{\small$\bigstar$})  ,\boldsymbol{\zeta}^{\#}(\textnormal{\small$\bigstar$})]$ satisfies}  \\
\text{$\mathcal{S}_{2.1} \cup \mathcal{S}_{2.2}:=\big\{\textnormal{(\ref{partialLpartialT}), (\ref{Complementarytime}),  (\ref{constraintTtau}), (\ref{Dualfeasibility}e)}\big\}$.}
\end{array}  \right\}, \nonumber \\  & \Leftrightarrow \left\{ \begin{array}{l}  \textnormal{Statements $\mathcal{V}_{\mathbb{P}_{9}}$ and  $\mathcal{V}_{\mathbb{P}_{10}}$ hold,}  \\
\text{and $[\boldsymbol{b}^{\#}(\textnormal{\small$\bigstar$}),\boldsymbol{p}^{\#}(\textnormal{\small$\bigstar$}),(\boldsymbol{f}^{\textnormal{MS}})^{\#}(\textnormal{\small$\bigstar$}),(\boldsymbol{f}^{\textnormal{VU}})^{\#}(\textnormal{\small$\bigstar$}),$}\\
\text{$T^{\#}(\textnormal{\small$\bigstar$})  ,\boldsymbol{\zeta}^{\#}(\textnormal{\small$\bigstar$})]$ satisfies}  \\
\text{$\mathcal{S}_{2.1} \cup \mathcal{S}_{2.2}:=\big\{\textnormal{(\ref{partialLpartialT}), (\ref{Complementarytime}),  (\ref{constraintTtau}), (\ref{Dualfeasibility}e)}\big\}$.}
\end{array}  \right\} \label{statev5v8910}
\end{align}
where the last step means $\mathbb{P}_{8}(\textnormal{\small$\bigstar$},\boldsymbol{\zeta})$ can be neglected since (\ref{partialLpartialT}) induces the objective function of $\mathbb{P}_{8}(\textnormal{\small$\bigstar$},\boldsymbol{\zeta})$ to be always $0$. 

 Now we analyze Problem $\mathbb{P}_{9}(\textnormal{\small$\bigstar$},\boldsymbol{\zeta})$. 
Note that $\beta$ is  the Lagrange multiplier for \textnormal{(\ref{constraintpn})}. Suppose we already know $\beta$.  We move $\beta$ and \textnormal{(\ref{constraintpn})} to the objective function, and construct the following problem:
\begin{align}
&\textnormal{Problem $\mathbb{P}_{11}(\textnormal{\small$\bigstar$},\boldsymbol{\zeta},\beta)$}: \nonumber\\
&\hspace{-1pt}\min_{\boldsymbol{b},\boldsymbol{p}}   \begin{array}{l} \hspace{-1pt} \left\{ H_{\mathbb{P}_{9}}(\boldsymbol{b},\boldsymbol{p} \mid \textnormal{\small$\bigstar$},\boldsymbol{\zeta}) 
+ \beta \cdot \big( \sum_{n \in \mathcal{N}} p_n - p_{\textnormal{max}}\big)  \right\} \end{array}\\
&~~~~~~~~\textrm{s.t.} \quad  \textnormal{(\ref{constraintbn})}, \label{eqP11}
\end{align} 
where $H_{\mathbb{P}_{9}}(\boldsymbol{b},\boldsymbol{p} \mid \textnormal{\small$\bigstar$},\boldsymbol{\zeta})$ denotes the objective function of Problem $\mathbb{P}_{9}$.
Then after defining 
\begin{itemize}
% \item Statement $\mathcal{V}_{\mathbb{P}_{5}}$:
% \item[] $\left\{ \begin{array}{l}
% \textnormal{$[\boldsymbol{b}^{\#}(\textnormal{\small$\bigstar$}),\boldsymbol{p}^{\#}(\textnormal{\small$\bigstar$}),(\boldsymbol{f}^{\textnormal{MS}})^{\#}(\textnormal{\small$\bigstar$}),(\boldsymbol{f}^{\textnormal{VU}})^{\#}(\textnormal{\small$\bigstar$}),T^{\#}(\textnormal{\small$\bigstar$}), \alpha^{\#}(\textnormal{\small$\bigstar$}),\beta^{\#}(\textnormal{\small$\bigstar$}),\gamma^{\#}(\textnormal{\small$\bigstar$}),\boldsymbol{\delta}^{\#}(\textnormal{\small$\bigstar$})  ,\boldsymbol{\zeta}^{\#}(\textnormal{\small$\bigstar$})]$} \\ \textnormal{is a solution to the KKT conditions $\mathcal{S}_{KKT}$ in (\ref{KKTP5}) for Problem $\mathbb{P}_{5}(\textnormal{\small$\bigstar$})$.}
% \end{array}  \right\}$
\item Statement $\mathcal{V}_{\mathbb{P}_{11}}$:
\item[] $\left\{ \begin{array}{l}
\textnormal{$[\boldsymbol{b}^{\#}(\textnormal{\small$\bigstar$}),\boldsymbol{p}^{\#}(\textnormal{\small$\bigstar$}),\alpha^{\#}(\textnormal{\small$\bigstar$})]$} \\ \textnormal{is a solution to the KKT conditions of}\\
\textnormal{Problem $\mathbb{P}_{11}(\textnormal{\small$\bigstar$},\boldsymbol{\zeta}^{\#}(\textnormal{\small$\bigstar$}),\beta^{\#}(\textnormal{\small$\bigstar$}))$.}   
\end{array}  \right\} $
\end{itemize}
and checking the KKT conditions of $\mathbb{P}_{9}(\textnormal{\small$\bigstar$},\boldsymbol{\zeta}^{\#}(\textnormal{\small$\bigstar$})$ and $\mathbb{P}_{11}(\textnormal{\small$\bigstar$},\boldsymbol{\zeta}^{\#}(\textnormal{\small$\bigstar$}),\beta^{\#}(\textnormal{\small$\bigstar$}))$, we have
\begin{align} 
\text{Statement $\mathcal{V}_{\mathbb{P}_{9}}$}   & \Leftrightarrow \left\{ \begin{array}{l}  \textnormal{Statement $\mathcal{V}_{\mathbb{P}_{11}}$ holds,}  \\
\text{and $[\beta^{\#}(\textnormal{\small$\bigstar$}),\boldsymbol{p}^{\#}(\textnormal{\small$\bigstar$})]$ satisfies}\\
\text{$\mathcal{S}_{1.2.2.2}:=\big\{\textnormal{(\ref{Complementarybeta})}, \textnormal{(\ref{constraintpn})}, \textnormal{(\ref{Dualfeasibility}b)}\big\}$.}
\end{array}  \right\} \label{statev8v10} 
\end{align} 
 Now we analyze Problem $\mathbb{P}_{11}(\textnormal{\small$\bigstar$},\boldsymbol{\zeta},\beta)$. 
Note that $\alpha$ is  the Lagrange multiplier for \textnormal{(\ref{constraintbn})}. Suppose we already know $\alpha$.  We move $\alpha$ and \textnormal{(\ref{constraintbn})} to the objective function, and construct the following problem:
\begin{align}
&\textnormal{Problem $\mathbb{P}_{12}(\textnormal{\small$\bigstar$},\boldsymbol{\zeta},\beta,\alpha)$}: \nonumber\\
&\hspace{-1pt}\min_{\boldsymbol{b},\boldsymbol{p}}   \hspace{-1pt} \left\{\hspace{-5pt}  \begin{array}{l} H_{\mathbb{P}_{11}}(\boldsymbol{b},\boldsymbol{p} \mid \textnormal{\small$\bigstar$},\boldsymbol{\zeta},\beta)  + \alpha \cdot \big( \sum_{n \in \mathcal{N}} b_n - b_{\textnormal{max}}\big)   \end{array}  \hspace{-5pt} \right\},  \label{eqP12}
\end{align} 
where $H_{\mathbb{P}_{11}}(\boldsymbol{b},\boldsymbol{p} \mid \textnormal{\small$\bigstar$},\boldsymbol{\zeta},\beta)$ denotes the objective function of Problem $\mathbb{P}_{11}$.
Then after defining
\begin{itemize}
\item  Statement $\mathcal{V}_{\mathbb{P}_{12}}$:
\item[] \textnormal{$[\boldsymbol{b}^{\#}(\textnormal{\small$\bigstar$}),\boldsymbol{p}^{\#}(\textnormal{\small$\bigstar$})]$ is a globally optimal solution to Problem $\mathbb{P}_{12}(\textnormal{\small$\bigstar$},\boldsymbol{\zeta}^{\#}(\textnormal{\small$\bigstar$}),\beta^{\#}(\textnormal{\small$\bigstar$}),\alpha^{\#}(\textnormal{\small$\bigstar$}))$,}
\end{itemize}
and checking the KKT conditions of $\mathbb{P}_{11}(\textnormal{\small$\bigstar$},\boldsymbol{\zeta}^{\#}(\textnormal{\small$\bigstar$}),\beta^{\#}(\textnormal{\small$\bigstar$}))$ and $\mathbb{P}_{12}(\textnormal{\small$\bigstar$},\boldsymbol{\zeta}^{\#}(\textnormal{\small$\bigstar$}),\beta^{\#}(\textnormal{\small$\bigstar$}),\alpha^{\#}(\textnormal{\small$\bigstar$}))$, we get
\begin{align} \text{Statement $\mathcal{V}_{\mathbb{P}_{11}}$}   & \Leftrightarrow \left\{ \begin{array}{l}  \textnormal{Statement $\mathcal{V}_{\mathbb{P}_{12}}$ holds,}  \\
\text{and $[\alpha^{\#}(\textnormal{\small$\bigstar$}),\boldsymbol{b}^{\#}(\textnormal{\small$\bigstar$})]$ satisfies }\\ 
\text{$\mathcal{S}_{1.2.2.1}:=\big\{\textnormal{(\ref{Stationaritypn3alpha}), (\ref{Complementaryalpha2})}\big\}$.}
\end{array}  \right\} \label{statev10v11}
\end{align}

\subsection{Proving that the left-hand side of~(\ref{betahat}) is \mbox{non-increasing} with respect to $\beta$} \label{Appendixbeta}

From the conditions of Proposition~1.2.1 and Proposition~1.2.2.1, setting $[\boldsymbol{b},\boldsymbol{p},\alpha]$ as 
$[\widetilde{\boldsymbol{b}}(\breve{\alpha}(\beta,\iffalse T,\fi\boldsymbol{\zeta}\mid \textnormal{\small$\bigstar$}),\beta,\boldsymbol{\zeta}\mid \textnormal{\small$\bigstar$}),\widetilde{\boldsymbol{p}}(\breve{\alpha}(\beta,\iffalse T,\fi\boldsymbol{\zeta}\mid \textnormal{\small$\bigstar$}),\beta,\boldsymbol{\zeta}\mid \textnormal{\small$\bigstar$}),\breve{\alpha}(\beta,\iffalse T,\fi\boldsymbol{\zeta}\mid \textnormal{\small$\bigstar$})]$ satisfies $\mathcal{S}_{1.2.1} \cup \mathcal{S}_{1.2.2.1}=\big\{\textnormal{(\ref{Stationaritybn}), 
(\ref{Stationaritypn}), (\ref{Stationaritypn3alpha}), (\ref{Complementaryalpha2})}\big\}$; i.e., the KKT conditions 
% (\ref{Stationaritybn})--(\ref{partialLpartialT}) and (\ref{Complementarygamma})--(\ref{Dualfeasibility}) 
of convex optimization $\mathbb{P}_{11}(\textnormal{\small$\bigstar$},\boldsymbol{\zeta},\beta)$. 
Hence,
\begin{align}
\begin{array}{l}
\text{$[\widetilde{\boldsymbol{b}}(\breve{\alpha}(\beta,\iffalse T,\fi\boldsymbol{\zeta}\mid \textnormal{\small$\bigstar$}),\beta,\boldsymbol{\zeta}\mid \textnormal{\small$\bigstar$}),\widetilde{\boldsymbol{p}}(\breve{\alpha}(\beta,\iffalse T,\fi\boldsymbol{\zeta}\mid \textnormal{\small$\bigstar$}),\beta,\boldsymbol{\zeta}\mid \textnormal{\small$\bigstar$})]$} \\  \text{is a globally optimal solution to $\mathbb{P}_{11}(\textnormal{\small$\bigstar$},\boldsymbol{\zeta},\beta)$.}    
\end{array} \label{P6result}   
\end{align}

To prove the desired result, we consider the case where $\beta$ equals $\beta_1$, and the case where $\beta$ equals $\beta_2$, respectively, for arbitrarily chosen $\beta_1$ and $\beta_2$. Due to Result~(\ref{P6result}) above, for $H_{\mathbb{P}_{11}}(\boldsymbol{b},\boldsymbol{p} \mid \textnormal{\small$\bigstar$},\boldsymbol{\zeta},\beta)$ denoting the objective function of Problem $\mathbb{P}_{11}$, we obtain
\begin{align}
&H_{\mathbb{P}_{11}}(\widetilde{\boldsymbol{b}}(\breve{\alpha}(\beta_1,\iffalse T,\fi\boldsymbol{\zeta}\hspace{-2pt}\mid\hspace{-2pt} \textnormal{\small$\bigstar$}),\beta_1,\boldsymbol{\zeta}\hspace{-2pt}\mid\hspace{-2pt} \textnormal{\small$\bigstar$}),\widetilde{\boldsymbol{p}}(\breve{\alpha}(\beta_1,\iffalse T,\fi\boldsymbol{\zeta}\hspace{-2pt}\mid\hspace{-2pt} \textnormal{\small$\bigstar$}),\beta_1,\boldsymbol{\zeta}\hspace{-2pt}\mid\hspace{-2pt} \textnormal{\small$\bigstar$} \hspace{-2pt}\mid\hspace{-2pt} \textnormal{\small$\bigstar$},\boldsymbol{\zeta},\beta_1)) \nonumber \\ & \leq \nonumber \\
&H_{\mathbb{P}_{11}}(\widetilde{\boldsymbol{b}}(\breve{\alpha}(\beta_2,\iffalse T,\fi\boldsymbol{\zeta}\hspace{-2pt}\mid\hspace{-2pt} \textnormal{\small$\bigstar$}),\beta_2,\boldsymbol{\zeta}\hspace{-2pt}\mid\hspace{-2pt} \textnormal{\small$\bigstar$}),\widetilde{\boldsymbol{p}}(\breve{\alpha}(\beta_2,\iffalse T,\fi\boldsymbol{\zeta}\hspace{-2pt}\mid\hspace{-2pt} \textnormal{\small$\bigstar$}),\beta_2,\boldsymbol{\zeta}\hspace{-2pt}\mid\hspace{-2pt} \textnormal{\small$\bigstar$} \hspace{-2pt}\mid\hspace{-2pt} \textnormal{\small$\bigstar$},\boldsymbol{\zeta},\beta_1)  , \label{eqbeta1beta21}
\end{align}
and
\begin{align}
&H_{\mathbb{P}_{11}}(\widetilde{\boldsymbol{b}}(\breve{\alpha}(\beta_2,\iffalse T,\fi\boldsymbol{\zeta}\mid \textnormal{\small$\bigstar$}),\beta_2,\boldsymbol{\zeta}\hspace{-2pt}\mid\hspace{-2pt} \textnormal{\small$\bigstar$}),\widetilde{\boldsymbol{p}}(\breve{\alpha}(\beta_2,\iffalse T,\fi\boldsymbol{\zeta}\hspace{-2pt}\mid\hspace{-2pt} \textnormal{\small$\bigstar$}),\beta_2,\boldsymbol{\zeta}\hspace{-2pt}\mid\hspace{-2pt} \textnormal{\small$\bigstar$} \hspace{-2pt}\mid\hspace{-2pt} \textnormal{\small$\bigstar$},\boldsymbol{\zeta},\beta_2) \nonumber \\ & \leq \nonumber\\
&H_{\mathbb{P}_{11}}(\widetilde{\boldsymbol{b}}(\breve{\alpha}(\beta_1,\iffalse T,\fi\boldsymbol{\zeta}\hspace{-2pt}\mid\hspace{-2pt} \textnormal{\small$\bigstar$}),\beta_1,\boldsymbol{\zeta}\hspace{-2pt}\mid\hspace{-2pt} \textnormal{\small$\bigstar$}),\widetilde{\boldsymbol{p}}(\breve{\alpha}(\beta_1,\iffalse T,\fi\boldsymbol{\zeta}\hspace{-2pt}\mid\hspace{-2pt} \textnormal{\small$\bigstar$}),\beta_1,\boldsymbol{\zeta}\hspace{-2pt}\mid\hspace{-2pt} \textnormal{\small$\bigstar$} \hspace{-2pt}\mid\hspace{-2pt} \textnormal{\small$\bigstar$},\boldsymbol{\zeta},\beta_2) .\label{eqbeta1beta22}  
\end{align}
From~(\ref{eqbeta1beta21}) and~(\ref{eqbeta1beta22}), it follows that
\begin{align}
& \left[ \begin{array}{l}
H_{\mathbb{P}_{11}}(\widetilde{\boldsymbol{b}}(\breve{\alpha}(\beta_1,\iffalse T,\fi\boldsymbol{\zeta}\mid \textnormal{\small$\bigstar$}),\beta_1,\boldsymbol{\zeta}\mid \textnormal{\small$\bigstar$}),\\
\widetilde{\boldsymbol{p}}(\breve{\alpha}(\beta_1,\iffalse T,\fi\boldsymbol{\zeta}\mid \textnormal{\small$\bigstar$}),\beta_1,\boldsymbol{\zeta}\mid \textnormal{\small$\bigstar$} \mid \textnormal{\small$\bigstar$},\boldsymbol{\zeta},\beta_1))   \\   - H_{\mathbb{P}_{11}}(\widetilde{\boldsymbol{b}}(\breve{\alpha}(\beta_1,\iffalse T,\fi\boldsymbol{\zeta}\mid \textnormal{\small$\bigstar$}),\beta_1,\boldsymbol{\zeta}\mid \textnormal{\small$\bigstar$}),\\
\widetilde{\boldsymbol{p}}(\breve{\alpha}(\beta_1,\iffalse T,\fi\boldsymbol{\zeta}\mid \textnormal{\small$\bigstar$}),\beta_1,\boldsymbol{\zeta}\mid \textnormal{\small$\bigstar$} \mid \textnormal{\small$\bigstar$},\boldsymbol{\zeta},\beta_2))  
\end{array} \right] \nonumber \\ & + \left[ \begin{array}{l}
H_{\mathbb{P}_{11}}(\widetilde{\boldsymbol{b}}(\breve{\alpha}(\beta_2,\iffalse T,\fi\boldsymbol{\zeta}\mid \textnormal{\small$\bigstar$}),\beta_2,\boldsymbol{\zeta}\mid \textnormal{\small$\bigstar$}),\\
\widetilde{\boldsymbol{p}}(\breve{\alpha}(\beta_2,\iffalse T,\fi\boldsymbol{\zeta}\mid \textnormal{\small$\bigstar$}),\beta_2,\boldsymbol{\zeta}\mid \textnormal{\small$\bigstar$} \mid \textnormal{\small$\bigstar$},\boldsymbol{\zeta},\beta_2))   \\   - H_{\mathbb{P}_{11}}(\widetilde{\boldsymbol{b}}(\breve{\alpha}(\beta_2,\iffalse T,\fi\boldsymbol{\zeta}\mid \textnormal{\small$\bigstar$}),\beta_2,\boldsymbol{\zeta}\mid \textnormal{\small$\bigstar$}),\\
\widetilde{\boldsymbol{p}}(\breve{\alpha}(\beta_2,\iffalse T,\fi\boldsymbol{\zeta}\mid \textnormal{\small$\bigstar$}),\beta_2,\boldsymbol{\zeta}\mid \textnormal{\small$\bigstar$} \mid \textnormal{\small$\bigstar$},\boldsymbol{\zeta},\beta_1)   
\end{array} \right] \leq 0.  \label{eqbeta1beta2} 
\end{align}
Since $H_{\mathbb{P}_{11}}(\boldsymbol{b},\boldsymbol{p},\boldsymbol{f}^{\textnormal{MS}},\boldsymbol{f}^{\textnormal{VU}},T \mid \beta,\iffalse T,\fi\boldsymbol{\zeta},\boldsymbol{z},y, \boldsymbol{s})$ equals $H_{\mathbb{P}_{9}}(\boldsymbol{b},\boldsymbol{p} \mid \textnormal{\small$\bigstar$},\boldsymbol{\zeta})  + \beta \cdot \big( \sum_{n \in \mathcal{N}} p_n - p_{\textnormal{max}}\big)  $ from~(\ref{eqP11}), the term inside the first ``$[\cdot]$'' of~(\ref{eqbeta1beta2}) equals $(\beta_1 - \beta_2) \cdot \big( \sum_{n \in \mathcal{N}} \widetilde{p}_n(\breve{\alpha}(\beta_1,\iffalse T,\fi\boldsymbol{\zeta}\mid \textnormal{\small$\bigstar$}),\beta_1,\boldsymbol{\zeta}\mid \textnormal{\small$\bigstar$}) - p_{\textnormal{max}}\big)$, and the term inside the second ``$[\cdot]$'' of~(\ref{eqbeta1beta2}) equals $(\beta_2 - \beta_1) \cdot \big( \sum_{n \in \mathcal{N}}\widetilde{p}_n(\breve{\alpha}(\beta_2,\iffalse T,\fi\boldsymbol{\zeta}\mid \textnormal{\small$\bigstar$}),\beta_2,\boldsymbol{\zeta}\mid \textnormal{\small$\bigstar$}) - p_{\textnormal{max}}\big)$. Then we obtain
\begin{align}
&\left\{ \hspace{-5pt}\begin{array}{l}
      (\beta_1  - \beta_2) \cdot \big( \sum_{n \in \mathcal{N}} \widetilde{p}_n(\breve{\alpha}(\beta_1,\iffalse T,\fi\boldsymbol{\zeta}\mid \textnormal{\small$\bigstar$}),\beta_1,\boldsymbol{\zeta}\mid \textnormal{\small$\bigstar$}) - p_{\textnormal{max}}\big) \\
      + (\beta_2 - \beta_1) \cdot \big( \sum_{n \in \mathcal{N}} \widetilde{p}_n(\breve{\alpha}(\beta_2,\iffalse T,\fi\boldsymbol{\zeta}\mid \textnormal{\small$\bigstar$}),\beta_2,\boldsymbol{\zeta}\mid \textnormal{\small$\bigstar$}) - p_{\textnormal{max}}\big)
\end{array}\hspace{-5pt} \right\} \nonumber \\
& \leq 0;   
\end{align}
i.e., $(\beta_1 - \beta_2) \cdot \big( \sum_{n \in \mathcal{N}} \widetilde{p}_n(\breve{\alpha}(\beta_1,\iffalse T,\fi\boldsymbol{\zeta}\mid \textnormal{\small$\bigstar$}),\beta_1,\boldsymbol{\zeta}\mid \textnormal{\small$\bigstar$}) - \sum_{n \in \mathcal{N}} \widetilde{p}_n(\breve{\alpha}(\beta_2,\iffalse T,\fi\boldsymbol{\zeta}\mid \textnormal{\small$\bigstar$}),\beta_2,\boldsymbol{\zeta}\mid \textnormal{\small$\bigstar$})  \big) \leq 0$. Hence, $\sum_{n \in \mathcal{N}} \widetilde{p}_n(\breve{\alpha}(\beta,\iffalse T,\fi\boldsymbol{\zeta}\mid \textnormal{\small$\bigstar$}),\beta,\boldsymbol{\zeta}\mid \textnormal{\small$\bigstar$}) $, i.e., the left-hand side of~(\ref{betahat}), is \mbox{non-increasing} as $\beta$ increases. \qed

\subsection{Proving that the left-hand side of~(\ref{brevealpha}) is \mbox{non-increasing} with respect to $\alpha$}\label{Appendixalpha}
 
From Proposition~1.2.1's condition, setting $[\boldsymbol{b},\boldsymbol{p}]$ as
$[\widetilde{\boldsymbol{b}}(\alpha, \beta,\iffalse T,\fi\boldsymbol{\zeta}\hspace{-2pt}\mid\hspace{-2pt} \textnormal{\small$\bigstar$}),\widetilde{\boldsymbol{p}}(\alpha, \beta,\iffalse T,\fi\boldsymbol{\zeta}\hspace{-2pt}\mid\hspace{-2pt} \textnormal{\small$\bigstar$})]$ satisfies $\mathcal{S}_{1.2.1}=\big\{\textnormal{(\ref{Stationaritybn}), 
(\ref{Stationaritypn})}\big\}$; i.e., the KKT conditions 
% (\ref{Stationaritybn})--(\ref{partialLpartialT}) and (\ref{Complementarygamma})--(\ref{Dualfeasibility}) 
of convex optimization $\mathbb{P}_{12}(\textnormal{\small$\bigstar$},\boldsymbol{\zeta},\beta,\alpha)$. 
Hence, 
\begin{align}
\begin{array}{l}
\text{$[\widetilde{\boldsymbol{b}}(\alpha, \beta,\iffalse T,\fi\boldsymbol{\zeta}\mid \textnormal{\small$\bigstar$}),\widetilde{\boldsymbol{p}}(\alpha, \beta,\iffalse T,\fi\boldsymbol{\zeta}\mid \textnormal{\small$\bigstar$})]$ is a globally optimal solution } \\
\text{to $\mathbb{P}_{12}(\textnormal{\small$\bigstar$},\boldsymbol{\zeta},\beta,\alpha)$.}    
\end{array} \label{P7result}   
\end{align} 
 
To prove the desired result, we consider the case where $\alpha$ equals $\alpha_1$, and the case where $\alpha$ equals $\alpha_2$, respectively, for arbitrarily chosen $\alpha_1$ and $\alpha_2$. Due to Result~(\ref{P7result}) above, for $H_{\mathbb{P}_{12}}(\boldsymbol{b},\boldsymbol{p} \mid \textnormal{\small$\bigstar$},\boldsymbol{\zeta},\beta,\alpha)$ denoting the objective function of Problem $\mathbb{P}_{12}$, we obtain
\begin{align} 
& H_{\mathbb{P}_{12}}(\widetilde{\boldsymbol{b}}(\alpha_1, \beta,\iffalse T,\fi\boldsymbol{\zeta}\mid \textnormal{\small$\bigstar$}),\widetilde{\boldsymbol{p}}(\alpha_1, \beta,\iffalse T,\fi\boldsymbol{\zeta}\mid \textnormal{\small$\bigstar$}) \mid \textnormal{\small$\bigstar$},\boldsymbol{\zeta},\beta,\alpha_1) \nonumber \leq\\
&H_{\mathbb{P}_{12}}(\widetilde{\boldsymbol{b}}(\alpha_2, \beta,\iffalse T,\fi\boldsymbol{\zeta}\mid \textnormal{\small$\bigstar$}),\widetilde{\boldsymbol{p}}(\alpha_2, \beta,\iffalse T,\fi\boldsymbol{\zeta}\mid \textnormal{\small$\bigstar$}) \mid \textnormal{\small$\bigstar$},\boldsymbol{\zeta},\beta,\alpha_1)
   , \label{alphaeqalpha1alpha21}
\end{align}
and
\begin{align}
& H_{\mathbb{P}_{12}}(\widetilde{\boldsymbol{b}}(\alpha_2, \beta,\iffalse T,\fi\boldsymbol{\zeta}\mid \textnormal{\small$\bigstar$}),\widetilde{\boldsymbol{p}}(\alpha_2, \beta,\iffalse T,\fi\boldsymbol{\zeta}\mid \textnormal{\small$\bigstar$}) \mid \textnormal{\small$\bigstar$},\boldsymbol{\zeta},\beta,\alpha_2) \leq \nonumber \\
&H_{\mathbb{P}_{12}}(\widetilde{\boldsymbol{b}}(\alpha_1, \beta,\iffalse T,\fi\boldsymbol{\zeta}\mid \textnormal{\small$\bigstar$}),\widetilde{\boldsymbol{p}}(\alpha_1, \beta,\iffalse T,\fi\boldsymbol{\zeta}\mid \textnormal{\small$\bigstar$}) \mid \textnormal{\small$\bigstar$},\boldsymbol{\zeta},\beta,\alpha_2)
 . \label{alphaeqalpha1alpha22}  
\end{align}
From~(\ref{alphaeqalpha1alpha21}) and~(\ref{alphaeqalpha1alpha22}), it follows that
\begin{align}
& [ H_{\mathbb{P}_{12}}(\widetilde{\boldsymbol{b}}(\alpha_1, \beta,\iffalse T,\fi\boldsymbol{\zeta}\mid \textnormal{\small$\bigstar$}),\widetilde{\boldsymbol{p}}(\alpha_1, \beta,\iffalse T,\fi\boldsymbol{\zeta}\mid \textnormal{\small$\bigstar$}) \mid \textnormal{\small$\bigstar$},\boldsymbol{\zeta},\beta,\alpha_1) \nonumber\\
&- H_{\mathbb{P}_{12}}(\widetilde{\boldsymbol{b}}(\alpha_1, \beta,\iffalse T,\fi\boldsymbol{\zeta}\mid \textnormal{\small$\bigstar$}),\widetilde{\boldsymbol{p}}(\alpha_1, \beta,\iffalse T,\fi\boldsymbol{\zeta}\mid \textnormal{\small$\bigstar$}) \mid \textnormal{\small$\bigstar$},\boldsymbol{\zeta},\beta,\alpha_2)] \nonumber \\ & +\hspace{-2pt}  [ H_{\mathbb{P}_{12}}(\widetilde{\boldsymbol{b}}(\alpha_2,\hspace{-.5pt}  \beta,\hspace{-.5pt} \iffalse T,\fi\boldsymbol{\zeta}\hspace{-.5pt} \mid \hspace{-.5pt} \textnormal{\small$\bigstar$}),\widetilde{\boldsymbol{p}}(\alpha_2, \hspace{-.5pt} \beta,\hspace{-.5pt} \iffalse T,\fi\boldsymbol{\zeta}\hspace{-.5pt} \mid\hspace{-.5pt}  \textnormal{\small$\bigstar$}) \mid \textnormal{\small$\bigstar$},\hspace{-.5pt} \boldsymbol{\zeta},\hspace{-.5pt} \beta,\hspace{-.5pt} \alpha_2)  \nonumber \\
&- H_{\mathbb{P}_{12}}(\widetilde{\boldsymbol{b}}(\alpha_2,\hspace{-.5pt} \beta,\hspace{-.5pt} \iffalse T,\fi\boldsymbol{\zeta}\mid \textnormal{\small$\bigstar$}),\widetilde{\boldsymbol{p}}(\alpha_2, \beta,\iffalse T,\fi\boldsymbol{\zeta}\hspace{-.5pt} \mid\hspace{-.5pt}  \textnormal{\small$\bigstar$}) \mid \textnormal{\small$\bigstar$},\hspace{-.5pt} \boldsymbol{\zeta},\hspace{-.5pt} \beta,\hspace{-.5pt} \alpha_1) \hspace{-.5pt} ] \nonumber \\ & \leq 0.  \label{alphaeqalpha1alpha2} 
\end{align}
Since $H_{\mathbb{P}_{12}}(\boldsymbol{b},\boldsymbol{p} \mid \textnormal{\small$\bigstar$},\boldsymbol{\zeta},\beta,\alpha)$ equals $H_{\mathbb{P}_{11}}(\boldsymbol{b},\boldsymbol{p} \mid \textnormal{\small$\bigstar$},\boldsymbol{\zeta},\beta)  + \alpha \cdot \big( \sum_{n \in \mathcal{N}} b_n - b_{\textnormal{max}}\big)   $ from~(\ref{eqP12}),  the term inside the first ``$[\cdot]$'' of~(\ref{alphaeqalpha1alpha2}) equals $(\alpha_1 - \alpha_2) \cdot \big( \sum_{n \in \mathcal{N}} \widetilde{b}_n(\alpha_1,\beta,\iffalse T,\fi\boldsymbol{\zeta}\mid \textnormal{\small$\bigstar$}) - b_{\textnormal{max}}\big)$, \vspace{4pt} and the term inside the second ``$[\cdot]$'' of~(\ref{alphaeqalpha1alpha2}) equals $(\alpha_2 - \alpha_1) \cdot \big( \sum_{n \in \mathcal{N}} \widetilde{b}_n(\alpha_2,\beta,\iffalse T,\fi\boldsymbol{\zeta}\mid \textnormal{\small$\bigstar$}) - b_{\textnormal{max}}\big)$. Then we obtain
\begin{align}
&(\alpha_1 - \alpha_2) \cdot \big( \sum_{n \in \mathcal{N}} \widetilde{b}_n(\alpha_1,\beta,\iffalse T,\fi\boldsymbol{\zeta}\mid \textnormal{\small$\bigstar$}) - b_{\textnormal{max}}\big) \nonumber \\
+& (\alpha_2 - \alpha_1) \cdot \big( \sum_{n \in \mathcal{N}} \widetilde{b}_n(\alpha_2,\beta,\iffalse T,\fi\boldsymbol{\zeta}\mid \textnormal{\small$\bigstar$}) - b_{\textnormal{max}}\big) \leq 0;   
\end{align}
i.e., $(\alpha_1 - \alpha_2) \cdot \big( \sum_{n \in \mathcal{N}} \widetilde{b}_n(\alpha_1,\beta,\iffalse T,\fi\boldsymbol{\zeta}\mid \textnormal{\small$\bigstar$}) - \sum_{n \in \mathcal{N}} \widetilde{b}_n(\alpha_2,\beta,\iffalse T,\fi\boldsymbol{\zeta}\mid \textnormal{\small$\bigstar$}) \big) \leq 0$. Hence, $\sum_{n \in \mathcal{N}} \widetilde{b}_n(\alpha, \beta,\iffalse T,\fi\boldsymbol{\zeta}\mid \textnormal{\small$\bigstar$}) $, i.e., the left-hand side of~(\ref{brevealpha}), is \mbox{non-increasing} as $\alpha$ increases.  \qed

\subsection{Proving Page~\pageref{lemsolvepngivenalphaandbetazeta}'s Lemma~\ref{lemsolvepngivenalphaandbetazeta}}  

From Lemma~\ref{lemsolveacute} to be presented in  Appendix~\ref{seclemsolveacute}, $t_n(\acute{b}_n(\boldsymbol{\zeta}\mid \textnormal{\small$\bigstar$}),\acute{p}_n(\boldsymbol{\zeta}\mid \textnormal{\small$\bigstar$}),s_n,\acute{f}_n^{\textnormal{MS}}(\boldsymbol{\zeta}\mid \textnormal{\small$\bigstar$}),\acute{f}_n^{\textnormal{VU}}(\boldsymbol{\zeta}\mid \textnormal{\small$\bigstar$}))$ is \mbox{non-increasing} as $\zeta_n$ increases. For $h_n(\boldsymbol{\zeta}\mid T)$ defined in (\ref{definehn}), given $[\zeta_1, \ldots, \zeta_{n-1}, \zeta_{n+1}\ldots, \zeta_N]$ and $T$, we either have (\ref{definehn}a) or (\ref{definehn}b). In either case, ``$t_n(\acute{b}_n(\iffalse T,\fi\boldsymbol{\zeta}\mid \textnormal{\small$\bigstar$}),\acute{p}_n(\iffalse T,\fi\boldsymbol{\zeta}\mid \textnormal{\small$\bigstar$}),s_n,\acute{f}_n^{\textnormal{MS}}(\iffalse T,\fi\boldsymbol{\zeta}\mid \textnormal{\small$\bigstar$}),\acute{f}_n^{\textnormal{VU}}(\iffalse T,\fi\boldsymbol{\zeta}\mid \textnormal{\small$\bigstar$})) - T$'' or ``$-\zeta_n$''  defined for $h_n(\boldsymbol{\zeta}\mid T)$ is \mbox{non-increasing} as $\zeta_n$ increases. Hence, $h_n(\boldsymbol{\zeta}\mid T)$ is \mbox{non-increasing} as $\zeta_n$ increases, given $[\zeta_1, \ldots, \zeta_{n-1}, \zeta_{n+1}\ldots, \zeta_N]$ and $T$. \qed

\subsection{Proving that the left-hand side of~(\ref{zetah2}) is \mbox{non-increasing} with respect to $T$}\label{AppendixT}

We recall from~(\ref{definehn2r}) and~(\ref{zetah}) that
\begin{talign}
&\text{setting $\boldsymbol{\zeta}$ as $\grave{\boldsymbol{\zeta}}(T\mid \textnormal{\small$\bigstar$})$   ensures $h_n(\boldsymbol{\zeta}\mid T)=0$ for any $n \in \mathcal{N}$,} \label{zetah3}
\end{talign}
where $h_n(\boldsymbol{\zeta}\mid T)=0$ is defined in (\ref{definehn}).

From
Lemma~\ref{lemsolveacute} to be presented in Appendix~\ref{seclemsolveacute} below, we can prove that $\sum_{n \in \mathcal{N}} \grave{\zeta}_n(T\mid \textnormal{\small$\bigstar$})$; i.e., the left-hand side of~(\ref{zetah2}) is \mbox{non-increasing} with respect to $T$. The proof is similar to those in Appendices~\ref{Appendixbeta} and~\ref{Appendixalpha}. \qed

\subsection{Lemma~\ref{lemsolveacute} and its proof} \label{seclemsolveacute}

\begin{lemma} \label{lemsolveacute} 
% Given $[\iffalse T,\fi\boldsymbol{\zeta},\boldsymbol{z},y, \boldsymbol{s}]$, let  $[\acute{\boldsymbol{b}}(\iffalse T,\fi\boldsymbol{\zeta}\mid \boldsymbol{z},y, \boldsymbol{s}),\acute{\boldsymbol{p}}(\iffalse T,\fi\boldsymbol{\zeta}\mid \boldsymbol{z},y, \boldsymbol{s}),\acute{\boldsymbol{f}}^{\textnormal{MS}}(\iffalse T,\fi\boldsymbol{\zeta}\mid \boldsymbol{z},y, \boldsymbol{s}),\acute{\boldsymbol{f}}^{\textnormal{VU}}(\iffalse T,\fi\boldsymbol{\zeta}\mid \boldsymbol{z},y, \boldsymbol{s}),\acute{\alpha}(\iffalse T,\fi\boldsymbol{\zeta}\mid \boldsymbol{z},y, \boldsymbol{s}),\acute{\beta}(\iffalse T,\fi\boldsymbol{\zeta}\mid \boldsymbol{z},y, \boldsymbol{s}),\acute{\gamma}(\iffalse T,\fi\boldsymbol{\zeta}\mid \boldsymbol{z},y, \boldsymbol{s}),\acute{\boldsymbol{\delta}}(\iffalse T,\fi\boldsymbol{\zeta}\mid \boldsymbol{z},y, \boldsymbol{s})]$ be a solution of $[\boldsymbol{b},\boldsymbol{p},\boldsymbol{f}^{\textnormal{MS}},\boldsymbol{f}^{\textnormal{VU}},T, \alpha,\beta,\gamma,\boldsymbol{\delta}  ]$ to (\ref{Stationaritybn})--(\ref{Complementaryalpha}) and (\ref{Complementarygamma})--(\ref{Dualfeasibility}) (\ref{Dualfeasibility}a) (\ref{Dualfeasibility}c) (\ref{Dualfeasibility}d) (\ref{Dualfeasibility}e) (i.e., the KKT conditions (\ref{Stationaritybn})--(\ref{Dualfeasibility}) except for (\ref{Complementarybeta}) and (\ref{Dualfeasibility}b)).  
Given ``$\textnormal{\small$\bigstar$}$'' (i.e., $[\boldsymbol{z},y, \boldsymbol{s}]$) and $[\zeta_1, \ldots, \zeta_{n-1}, \zeta_{n+1}\ldots, \zeta_N]$, we have: \\
given $[\zeta_1, \ldots, \zeta_{n-1}, \zeta_{n+1}\ldots, \zeta_N]$, and ``$\textnormal{\small$\bigstar$}$'' (i.e., ``$\boldsymbol{z},y, \boldsymbol{s}$''), then as $\zeta_n$ increases,
\begin{itemize}
\item[i)]  $t_n^{\textnormal{Tx}}(\acute{b}_n(\boldsymbol{\zeta}\mid \textnormal{\small$\bigstar$}),\acute{p}_n(\boldsymbol{\zeta}\mid \textnormal{\small$\bigstar$}), s_n)$ is \mbox{non-increasing};
\item[ii)]  $t_n^{\textnormal{MS}:\textnormal{Pro}}(s_n,\acute{f}_n^{\textnormal{MS}}(\boldsymbol{\zeta}\mid \textnormal{\small$\bigstar$}))+t_n^{\textnormal{VU}:\textnormal{Pro}}(s_n,\acute{f}_n^{\textnormal{VU}}(\boldsymbol{\zeta}\mid \textnormal{\small$\bigstar$}))$ is \mbox{non-increasing}; and
\item[iii)]  $t_n(\acute{b}_n(\boldsymbol{\zeta}\mid \textnormal{\small$\bigstar$}),\acute{p}_n(\boldsymbol{\zeta}\mid \textnormal{\small$\bigstar$}),s_n,\acute{f}_n^{\textnormal{MS}}(\boldsymbol{\zeta}\mid \textnormal{\small$\bigstar$}),\acute{f}_n^{\textnormal{VU}}(\boldsymbol{\zeta}\mid \textnormal{\small$\bigstar$}))$ is \mbox{non-increasing}.
\end{itemize} 
\end{lemma}

\noindent\textbf{Proof of Lemma~\ref{lemsolveacute}:}

Below we prove Results ``i)'',  ``ii)'', and  ``iii)'', respectively.

\noindent\textbf{Proving Lemma~\ref{lemsolveacute}'s Result ``i)'':}

From Proposition~1.2's condition, setting $[\boldsymbol{b},\boldsymbol{p},\alpha,\beta]$ as
$[\acute{\boldsymbol{b}}(\boldsymbol{\zeta}\mid \textnormal{\small$\bigstar$}),\acute{\boldsymbol{p}}(\boldsymbol{\zeta}\mid \textnormal{\small$\bigstar$}),\acute{\alpha}(\boldsymbol{\zeta}\mid \textnormal{\small$\bigstar$}),\acute{\beta}(\boldsymbol{\zeta}\mid \textnormal{\small$\bigstar$})]$ satisfies $\mathcal{S}_{1.2.1} \cup \mathcal{S}_{1.2.2.1} \cup \mathcal{S}_{1.2.2.2}=\big\{\textnormal{(\ref{Stationaritybn}), (\ref{Stationaritypn}),  (\ref{Complementarybeta}), \textnormal{(\ref{constraintpn})}, (\ref{Dualfeasibility}b), (\ref{Stationaritypn3alpha}), (\ref{Complementaryalpha2})}\big\}$; i.e., the KKT conditions 
% (\ref{Stationaritybn})--(\ref{partialLpartialT}) and (\ref{Complementarygamma})--(\ref{Dualfeasibility}) 
of convex optimization $\mathbb{P}_{9}(\textnormal{\small$\bigstar$},\boldsymbol{\zeta})$. 
% It is straightforward to see that (\ref{Stationaritybn})--(\ref{partialLpartialT}) and (\ref{Complementarygamma})--(\ref{Dualfeasibility}) are exactly the KKT conditions to the following Problem $\mathbb{P}_{12}(\beta,\iffalse T,\fi\boldsymbol{\zeta},\textnormal{\small$\bigstar$})$: 
% \begin{align}
% \textnormal{Problem $\mathbb{P}_{12}(\textnormal{\small$\bigstar$},\boldsymbol{\zeta},\beta,\alpha)$}: \min_{\boldsymbol{b},\boldsymbol{p},\boldsymbol{f}^{\textnormal{MS}},\boldsymbol{f}^{\textnormal{VU}},T} \quad & -H_{\mathbb{P}_{6}}(\boldsymbol{b},\boldsymbol{p},\boldsymbol{f}^{\textnormal{MS}},\boldsymbol{f}^{\textnormal{VU}},T \mid \beta,\textnormal{\small$\bigstar$}) + \alpha \cdot \bigg( \sum_{n \in \mathcal{N}} b_n - b_{\textnormal{max}}\bigg)     \\
% \textrm{s.t.} \quad &   \textnormal{(\ref{constraintfMS})}, \textnormal{(\ref{constraintfn})},  \textnormal{(\ref{constraintTtau})} , \label{alphaeqP6}
% \end{align} 
% where $H_{\mathbb{P}_{6}}(\boldsymbol{b},\boldsymbol{p},\boldsymbol{f}^{\textnormal{MS}},\boldsymbol{f}^{\textnormal{VU}},T \mid \beta,\textnormal{\small$\bigstar$})$ denotes the objective function of $\mathbb{P}_{6}(\beta,\textnormal{\small$\bigstar$})$ in~(\ref{eqP6}).
% 
Hence, 
\begin{align}
\begin{array}{l}
\text{$[\acute{\boldsymbol{b}}(\boldsymbol{\zeta}\hspace{-2pt}\mid\hspace{-2pt} \textnormal{\small$\bigstar$}),\acute{\boldsymbol{p}}(\boldsymbol{\zeta}\hspace{-2pt}\mid\hspace{-2pt} \textnormal{\small$\bigstar$})]$ is a globally optimal solution to $\mathbb{P}_{9}(\textnormal{\small$\bigstar$},\boldsymbol{\zeta})$.}    
\end{array} \label{P7resultlem3i}   
\end{align} 
 
To prove the desired result, we consider the case where $\zeta_n$ equals $\zeta_n^{(1)}$, and the case where $\zeta_n$ equals $\zeta_n^{(2)}$, respectively, for arbitrarily chosen $\zeta_n^{(1)}$ and $\zeta_n^{(2)}$. Due to Result~(\ref{P7resultlem3i}) above, after
defining
\begin{align}
\boldsymbol{\zeta}^{(n,1)}&:=  [\zeta_1,\ldots,\zeta_{n-1},\zeta_n^{(1)},\zeta_{n+1},\ldots,\zeta_{N}] , \nonumber \\ \boldsymbol{\zeta}^{(n,2)}&:=  [\zeta_1,\ldots,\zeta_{n-1},\zeta_n^{(2)},\zeta_{n+1},\ldots,\zeta_{N}] ,\nonumber
\end{align}
then with $H_{\mathbb{P}_{9}}(\boldsymbol{b},\boldsymbol{p} \mid \textnormal{\small$\bigstar$},\boldsymbol{\zeta})$ denoting the objective function of Problem $\mathbb{P}_{9}$, we obtain
\begin{align} 
& H_{\mathbb{P}_{9}}(\acute{\boldsymbol{b}}(\boldsymbol{\zeta}^{(n,1)}\mid \textnormal{\small$\bigstar$}),\acute{\boldsymbol{p}}(\boldsymbol{\zeta}^{(n,1)}\mid \textnormal{\small$\bigstar$}) \mid \textnormal{\small$\bigstar$},\boldsymbol{\zeta}^{(n,1)}) \nonumber \\
&\leq H_{\mathbb{P}_{9}}(\acute{\boldsymbol{b}}(\boldsymbol{\zeta}^{(n,1)}\mid \textnormal{\small$\bigstar$}),\acute{\boldsymbol{p}}(\boldsymbol{\zeta}^{(n,1)}\mid \textnormal{\small$\bigstar$}) \mid \textnormal{\small$\bigstar$},\boldsymbol{\zeta}^{(n,2)})
   , \label{alphaeqalpha1alpha21lem3i}
\end{align}
and
\begin{align}
& H_{\mathbb{P}_{9}}(\acute{\boldsymbol{b}}(\boldsymbol{\zeta}^{(n,2)}\mid \textnormal{\small$\bigstar$}),\acute{\boldsymbol{p}}(\boldsymbol{\zeta}^{(n,2)}\mid \textnormal{\small$\bigstar$}) \mid \textnormal{\small$\bigstar$},\boldsymbol{\zeta}^{(n,2)}) \nonumber \\
&\leq H_{\mathbb{P}_{9}}(\acute{\boldsymbol{b}}(\boldsymbol{\zeta}^{(n,2)}\mid \textnormal{\small$\bigstar$}),\acute{\boldsymbol{p}}(\boldsymbol{\zeta}^{(n,2)}\mid \textnormal{\small$\bigstar$}) \mid \textnormal{\small$\bigstar$},\boldsymbol{\zeta}^{(n,1)})
 . \label{alphaeqalpha1alpha22lem3i}  
\end{align}
From~(\ref{alphaeqalpha1alpha21lem3i}) and~(\ref{alphaeqalpha1alpha22lem3i}), it follows that
\begin{align}
& \big[H_{\mathbb{P}_{9}}(\acute{\boldsymbol{b}}(\boldsymbol{\zeta}^{(n,1)}\mid \textnormal{\small$\bigstar$}),\acute{\boldsymbol{p}}(\boldsymbol{\zeta}^{(n,1)}\mid \textnormal{\small$\bigstar$}) \mid \textnormal{\small$\bigstar$},\boldsymbol{\zeta}^{(n,1)})  \nonumber \\
&- H_{\mathbb{P}_{9}}(\acute{\boldsymbol{b}}(\boldsymbol{\zeta}^{(n,1)}\mid \textnormal{\small$\bigstar$}),\acute{\boldsymbol{p}}(\boldsymbol{\zeta}^{(n,1)}\mid \textnormal{\small$\bigstar$}) \mid \textnormal{\small$\bigstar$},\boldsymbol{\zeta}^{(n,2)})\big] \nonumber \\ & +\hspace{-2pt}  \big[ H_{\mathbb{P}_{9}}(\acute{\boldsymbol{b}}(\boldsymbol{\zeta}^{(n,2)}\mid \textnormal{\small$\bigstar$}),\acute{\boldsymbol{p}}(\boldsymbol{\zeta}^{(n,2)}\mid \textnormal{\small$\bigstar$}) \mid \textnormal{\small$\bigstar$},\boldsymbol{\zeta}^{(n,2)})  \nonumber\\
& - H_{\mathbb{P}_{9}}(\acute{\boldsymbol{b}}(\boldsymbol{\zeta}^{(n,2)}\mid \textnormal{\small$\bigstar$}),\acute{\boldsymbol{p}}(\boldsymbol{\zeta}^{(n,2)}\mid \textnormal{\small$\bigstar$}) \mid \textnormal{\small$\bigstar$},\boldsymbol{\zeta}^{(n,1)})\big] \nonumber \\ & \leq 0.  \label{alphaeqalpha1alpha2lem3i} 
\end{align}
Note that $H_{\mathbb{P}_{9}}(\boldsymbol{b},\boldsymbol{p} \mid \textnormal{\small$\bigstar$},\boldsymbol{\zeta})$ is given by~(\ref{eqP9}). Then the term inside the first ``$[\cdot]$'' of~(\ref{alphaeqalpha1alpha2lem3i}) equals $(\zeta_n^{(1)} - \zeta_n^{(2)}) \cdot t_n^{\textnormal{Tx}}(\acute{b}_n(\boldsymbol{\zeta}^{(n,1)}\mid \textnormal{\small$\bigstar$}),\acute{p}_n(\boldsymbol{\zeta}^{(n,1)}\mid \textnormal{\small$\bigstar$}), s_n) $, and the term inside the second ``$[\cdot]$'' of~(\ref{alphaeqalpha1alpha2lem3i}) equals $(\zeta_n^{(2)} - \zeta_n^{(1)}) \cdot t_n^{\textnormal{Tx}}(\acute{b}_n(\boldsymbol{\zeta}^{(n,2)}\mid \textnormal{\small$\bigstar$}),\acute{p}_n(\boldsymbol{\zeta}^{(n,2)}\mid \textnormal{\small$\bigstar$}), s_n) $. Then we obtain
\begin{align}
&(\zeta_n^{(1)} - \zeta_n^{(2)}) \cdot t_n^{\textnormal{Tx}}(\acute{b}_n(\boldsymbol{\zeta}^{(n,1)}\mid \textnormal{\small$\bigstar$}),\acute{p}_n(\boldsymbol{\zeta}^{(n,1)}\mid \textnormal{\small$\bigstar$}), s_n) \nonumber \\
+ &(\zeta_n^{(2)} - \zeta_n^{(1)}) \cdot t_n^{\textnormal{Tx}}(\acute{b}_n(\boldsymbol{\zeta}^{(n,2)}\mid \textnormal{\small$\bigstar$}),\acute{p}_n(\boldsymbol{\zeta}^{(n,2)}\mid \textnormal{\small$\bigstar$}), s_n) \leq 0;   
\end{align} 
i.e., $(\zeta_n^{(1)} - \zeta_n^{(2)}) \cdot \big( t_n^{\textnormal{Tx}}(\acute{b}_n(\boldsymbol{\zeta}^{(n,1)}\mid \textnormal{\small$\bigstar$}),\acute{p}_n(\boldsymbol{\zeta}^{(n,1)}\mid \textnormal{\small$\bigstar$}), s_n) - t_n^{\textnormal{Tx}}(\acute{b}_n(\boldsymbol{\zeta}^{(n,2)}\mid \textnormal{\small$\bigstar$}),\acute{p}_n(\boldsymbol{\zeta}^{(n,2)}\mid \textnormal{\small$\bigstar$}), s_n)\big) \leq 0$. Hence, $t_n^{\textnormal{Tx}}(\acute{b}_n(\boldsymbol{\zeta}\mid \textnormal{\small$\bigstar$}),\acute{p}_n(\boldsymbol{\zeta}\mid \textnormal{\small$\bigstar$}), s_n) $ is \mbox{non-increasing} as $\zeta_n$ increases.

\noindent\textbf{Proving Lemma~\ref{lemsolveacute}'s Result ``ii)'':}

From Proposition~1.1's condition, setting $[\boldsymbol{f}^{\textnormal{MS}},\boldsymbol{f}^{\textnormal{VU}},\gamma,\boldsymbol{\delta}]$ as
$[\acute{\boldsymbol{f}}^{\textnormal{MS}}(\boldsymbol{\zeta}\mid \textnormal{\small$\bigstar$}),\acute{\boldsymbol{f}}^{\textnormal{VU}}(\boldsymbol{\zeta}\mid \textnormal{\small$\bigstar$}),\acute{\gamma}(\boldsymbol{\zeta}\mid \textnormal{\small$\bigstar$}),\acute{\boldsymbol{\delta}}(\boldsymbol{\zeta}\mid \textnormal{\small$\bigstar$})]$ satisfies $\mathcal{S}_{1.1}:=\big\{\textnormal{(\ref{partialLpartialfMS}), (\ref{partialLpartialfVU}), (\ref{Complementarygamma}), (\ref{Complementarydelta}), \textnormal{(\ref{constraintfMS})}, \textnormal{(\ref{constraintfn})}, (\ref{Dualfeasibility}c), (\ref{Dualfeasibility}d)}\big\}$; i.e., the KKT conditions 
% (\ref{Stationaritybn})--(\ref{partialLpartialT}) and (\ref{Complementarygamma})--(\ref{Dualfeasibility}) 
of convex optimization $\mathbb{P}_{10}(\textnormal{\small$\bigstar$},\boldsymbol{\zeta})$. 
% It is straightforward to see that (\ref{Stationaritybn})--(\ref{partialLpartialT}) and (\ref{Complementarygamma})--(\ref{Dualfeasibility}) are exactly the KKT conditions to the following Problem $\mathbb{P}_{12}(\beta,\iffalse T,\fi\boldsymbol{\zeta},\textnormal{\small$\bigstar$})$: 
% \begin{align}
% \textnormal{Problem $\mathbb{P}_{12}(\textnormal{\small$\bigstar$},\boldsymbol{\zeta},\beta,\alpha)$}: \min_{\boldsymbol{b},\boldsymbol{p},\boldsymbol{f}^{\textnormal{MS}},\boldsymbol{f}^{\textnormal{VU}},T} \quad & -H_{\mathbb{P}_{6}}(\boldsymbol{b},\boldsymbol{p},\boldsymbol{f}^{\textnormal{MS}},\boldsymbol{f}^{\textnormal{VU}},T \mid \beta,\textnormal{\small$\bigstar$}) + \alpha \cdot \bigg( \sum_{n \in \mathcal{N}} b_n - b_{\textnormal{max}}\bigg)     \\
% \textrm{s.t.} \quad &   \textnormal{(\ref{constraintfMS})}, \textnormal{(\ref{constraintfn})},  \textnormal{(\ref{constraintTtau})} , \label{alphaeqP6}
% \end{align} 
% where $H_{\mathbb{P}_{6}}(\boldsymbol{b},\boldsymbol{p},\boldsymbol{f}^{\textnormal{MS}},\boldsymbol{f}^{\textnormal{VU}},T \mid \beta,\textnormal{\small$\bigstar$})$ denotes the objective function of $\mathbb{P}_{6}(\beta,\textnormal{\small$\bigstar$})$ in~(\ref{eqP6}).
% 
Hence, 
\begin{align}
\begin{array}{l}
\text{$[\acute{\boldsymbol{f}}^{\textnormal{MS}}(\boldsymbol{\zeta}\mid \textnormal{\small$\bigstar$}),\acute{\boldsymbol{f}}^{\textnormal{VU}}(\boldsymbol{\zeta}\mid \textnormal{\small$\bigstar$})]$ is a globally optimal solution} \\
\textnormal{to $\mathbb{P}_{10}(\textnormal{\small$\bigstar$},\boldsymbol{\zeta})$.}    
\end{array} \label{P7resultlem3ii}   
\end{align} 
 
To prove the desired result, we consider the case where $\zeta_n$ equals $\zeta_n^{(1)}$, and the case where $\zeta_n$ equals $\zeta_n^{(2)}$, respectively, for arbitrarily chosen $\zeta_n^{(1)}$ and $\zeta_n^{(2)}$. Due to Result~(\ref{P7resultlem3ii}) above, after
defining
\begin{align}
\boldsymbol{\zeta}^{(n,1)}&:=  [\zeta_1,\ldots,\zeta_{n-1},\zeta_n^{(1)},\zeta_{n+1},\ldots,\zeta_{N}] , \nonumber \\ \boldsymbol{\zeta}^{(n,2)}&:=  [\zeta_1,\ldots,\zeta_{n-1},\zeta_n^{(2)},\zeta_{n+1},\ldots,\zeta_{N}] ,\nonumber
\end{align}
then with $H_{\mathbb{P}_{10}}(\boldsymbol{f}^{\textnormal{MS}},\boldsymbol{f}^{\textnormal{VU}} \mid \textnormal{\small$\bigstar$},\boldsymbol{\zeta})$ denoting the objective function of Problem $\mathbb{P}_{10}$, we obtain
\begin{align} 
& H_{\mathbb{P}_{10}}(\acute{\boldsymbol{f}}^{\textnormal{MS}}(\boldsymbol{\zeta}^{(n,1)}\mid \textnormal{\small$\bigstar$}),\acute{\boldsymbol{f}}^{\textnormal{VU}}(\boldsymbol{\zeta}^{(n,1)}\mid \textnormal{\small$\bigstar$}) \mid \textnormal{\small$\bigstar$},\boldsymbol{\zeta}^{(n,1)})\leq \nonumber \\
&H_{\mathbb{P}_{10}}(\acute{\boldsymbol{f}}^{\textnormal{MS}}(\boldsymbol{\zeta}^{(n,1)}\mid \textnormal{\small$\bigstar$}),\acute{\boldsymbol{f}}^{\textnormal{VU}}(\boldsymbol{\zeta}^{(n,1)}\mid \textnormal{\small$\bigstar$}) \mid \textnormal{\small$\bigstar$},\boldsymbol{\zeta}^{(n,2)})
   , \label{alphaeqalpha1alpha21lem3ii}
\end{align}
and
\begin{align}
& H_{\mathbb{P}_{10}}(\acute{\boldsymbol{f}}^{\textnormal{MS}}(\boldsymbol{\zeta}^{(n,2)}\mid \textnormal{\small$\bigstar$}),\acute{\boldsymbol{f}}^{\textnormal{VU}}(\boldsymbol{\zeta}^{(n,2)}\mid \textnormal{\small$\bigstar$}) \mid \textnormal{\small$\bigstar$},\boldsymbol{\zeta}^{(n,2)})\leq \nonumber \\
&H_{\mathbb{P}_{10}}(\acute{\boldsymbol{f}}^{\textnormal{MS}}(\boldsymbol{\zeta}^{(n,2)}\mid \textnormal{\small$\bigstar$}),\acute{\boldsymbol{f}}^{\textnormal{VU}}(\boldsymbol{\zeta}^{(n,2)}\mid \textnormal{\small$\bigstar$}) \mid \textnormal{\small$\bigstar$},\boldsymbol{\zeta}^{(n,1)})
 . \label{alphaeqalpha1alpha22lem3ii}  
\end{align}
From~(\ref{alphaeqalpha1alpha21lem3ii}) and~(\ref{alphaeqalpha1alpha22lem3ii}), it follows that
\begin{align}
& \big[H_{\mathbb{P}_{10}}(\acute{\boldsymbol{f}}^{\textnormal{MS}}(\boldsymbol{\zeta}^{(n,1)}\mid \textnormal{\small$\bigstar$}),\acute{\boldsymbol{f}}^{\textnormal{VU}}(\boldsymbol{\zeta}^{(n,1)}\mid \textnormal{\small$\bigstar$}) \mid \textnormal{\small$\bigstar$},\boldsymbol{\zeta}^{(n,1)})  \nonumber \\
&- H_{\mathbb{P}_{10}}(\acute{\boldsymbol{f}}^{\textnormal{MS}}(\boldsymbol{\zeta}^{(n,1)}\mid \textnormal{\small$\bigstar$}),\acute{\boldsymbol{f}}^{\textnormal{VU}}(\boldsymbol{\zeta}^{(n,1)}\mid \textnormal{\small$\bigstar$}) \mid \textnormal{\small$\bigstar$},\boldsymbol{\zeta}^{(n,2)})\big] \nonumber \\ & +\hspace{-2pt}  \big[ H_{\mathbb{P}_{10}}(\acute{\boldsymbol{f}}^{\textnormal{MS}}(\boldsymbol{\zeta}^{(n,2)}\mid \textnormal{\small$\bigstar$}),\acute{\boldsymbol{f}}^{\textnormal{VU}}(\boldsymbol{\zeta}^{(n,2)}\mid \textnormal{\small$\bigstar$}) \mid \textnormal{\small$\bigstar$},\boldsymbol{\zeta}^{(n,2)})   \nonumber \\
&- H_{\mathbb{P}_{10}}(\acute{\boldsymbol{f}}^{\textnormal{MS}}(\boldsymbol{\zeta}^{(n,2)}\mid \textnormal{\small$\bigstar$}),\acute{\boldsymbol{f}}^{\textnormal{VU}}(\boldsymbol{\zeta}^{(n,2)}\mid \textnormal{\small$\bigstar$}) \mid \textnormal{\small$\bigstar$},\boldsymbol{\zeta}^{(n,1)})\big] \nonumber \\ & \leq 0.  \label{alphaeqalpha1alpha2lem3ii} 
\end{align}
Note that $H_{\mathbb{P}_{10}}(\boldsymbol{f}^{\textnormal{MS}},\boldsymbol{f}^{\textnormal{VU}} \mid \textnormal{\small$\bigstar$},\boldsymbol{\zeta})$ is given by~(\ref{eqP10}). Then the term inside the first ``$[\cdot]$'' of~(\ref{alphaeqalpha1alpha2lem3ii}) equals $(\zeta_n^{(1)} - \zeta_n^{(2)}) \cdot [t_n^{\textnormal{MS}:\textnormal{Pro}}(s_n,\acute{f}_n^{\textnormal{MS}}(\boldsymbol{\zeta}^{(n,1)}\mid \textnormal{\small$\bigstar$}))+t_n^{\textnormal{VU}:\textnormal{Pro}}(s_n,\acute{f}_n^{\textnormal{VU}}(\boldsymbol{\zeta}^{(n,1)}\mid \textnormal{\small$\bigstar$}))] $, and the term inside the second ``$[\cdot]$'' of~(\ref{alphaeqalpha1alpha2lem3ii}) equals $(\zeta_n^{(2)} - \zeta_n^{(1)}) \cdot [t_n^{\textnormal{MS}:\textnormal{Pro}}(s_n,\acute{f}_n^{\textnormal{MS}}(\boldsymbol{\zeta}^{(n,2)}\mid \textnormal{\small$\bigstar$}))+t_n^{\textnormal{VU}:\textnormal{Pro}}(s_n,\acute{f}_n^{\textnormal{VU}}(\boldsymbol{\zeta}^{(n,2)}\mid \textnormal{\small$\bigstar$}))]$. Then we obtain 
\begin{align}
\left\{\begin{array}{l}
 (\zeta_n^{(1)} - \zeta_n^{(2)}) \cdot [t_n^{\textnormal{MS}:\textnormal{Pro}}(s_n,\acute{f}_n^{\textnormal{MS}}(\boldsymbol{\zeta}^{(n,1)}\mid \textnormal{\small$\bigstar$}))\\
 +t_n^{\textnormal{VU}:\textnormal{Pro}}(s_n,\acute{f}_n^{\textnormal{VU}}(\boldsymbol{\zeta}^{(n,1)}\mid \textnormal{\small$\bigstar$}))]  \\  
 + (\zeta_n^{(2)} - \zeta_n^{(1)}) \cdot [t_n^{\textnormal{MS}:\textnormal{Pro}}(s_n,\acute{f}_n^{\textnormal{MS}}(\boldsymbol{\zeta}^{(n,2)}\mid \textnormal{\small$\bigstar$}))\\
 +t_n^{\textnormal{VU}:\textnormal{Pro}}(s_n,\acute{f}_n^{\textnormal{VU}}(\boldsymbol{\zeta}^{(n,2)}\mid \textnormal{\small$\bigstar$}))]
\end{array}\right\}\leq 0;  
\end{align} 
namely, $(\zeta_n^{(1)} \hspace{-2pt}-\hspace{-2pt} \zeta_n^{(2)}) \cdot \left\{\begin{array}{l}
  [t_n^{\textnormal{MS}:\textnormal{Pro}}(s_n,\acute{f}_n^{\textnormal{MS}}(\boldsymbol{\zeta}^{(n,1)}\mid \textnormal{\small$\bigstar$}))\\
  +t_n^{\textnormal{VU}:\textnormal{Pro}}(s_n,\acute{f}_n^{\textnormal{VU}}(\boldsymbol{\zeta}^{(n,1)}\mid \textnormal{\small$\bigstar$}))] \\ 
  -[t_n^{\textnormal{MS}:\textnormal{Pro}}(s_n,\acute{f}_n^{\textnormal{MS}}(\boldsymbol{\zeta}^{(n,2)}\mid \textnormal{\small$\bigstar$}))\\
  +t_n^{\textnormal{VU}:\textnormal{Pro}}(s_n,\acute{f}_n^{\textnormal{VU}}(\boldsymbol{\zeta}^{(n,2)}\mid \textnormal{\small$\bigstar$}))]
\end{array}\right\} \hspace{-2pt}\leq\hspace{-2pt} 0$. Therefore,  $t_n^{\textnormal{MS}:\textnormal{Pro}}(s_n,\acute{f}_n^{\textnormal{MS}}(\boldsymbol{\zeta}\mid \textnormal{\small$\bigstar$}))+t_n^{\textnormal{VU}:\textnormal{Pro}}(s_n,\acute{f}_n^{\textnormal{VU}}(\boldsymbol{\zeta}\mid \textnormal{\small$\bigstar$}))$ is \mbox{non-increasing} as $\zeta_n$ increases.

\noindent\textbf{Proving Lemma~\ref{lemsolveacute}'s Result ``iii)'':} From~(\ref{eq:userdelay}), $t_n(\acute{b}_n(\boldsymbol{\zeta}\mid \textnormal{\small$\bigstar$}),\acute{p}_n(\boldsymbol{\zeta}\mid \textnormal{\small$\bigstar$}),s_n,\acute{f}_n^{\textnormal{MS}}(\boldsymbol{\zeta}\mid \textnormal{\small$\bigstar$}),\acute{f}_n^{\textnormal{VU}}(\boldsymbol{\zeta}\mid \textnormal{\small$\bigstar$}))$ is the sum of $t_n^{\textnormal{Tx}}(\acute{b}_n(\boldsymbol{\zeta}\mid \textnormal{\small$\bigstar$}),\acute{p}_n(\boldsymbol{\zeta}\mid \textnormal{\small$\bigstar$}), s_n)$ and $t_n^{\textnormal{MS}:\textnormal{Pro}}(s_n,\acute{f}_n^{\textnormal{MS}}(\boldsymbol{\zeta}\mid \textnormal{\small$\bigstar$}))+t_n^{\textnormal{VU}:\textnormal{Pro}}(s_n,\acute{f}_n^{\textnormal{VU}}(\boldsymbol{\zeta}\mid \textnormal{\small$\bigstar$}))$. Then the desired result clearly follows from Lemma~\ref{lemsolveacute}'s Results ``i)'' and ``ii)''. \qed

\end{document}